\documentclass[floatfix, superscriptaddress, twocolumn, showpacs, aps, pra, 10pt,amsmath,amssymb]{revtex4-2}
\usepackage{graphicx}
\usepackage{dcolumn}
\usepackage{bm}
\usepackage{stackrel}
\usepackage{amsmath}
\usepackage{amsfonts}
\usepackage{amssymb}
\usepackage{amsthm}
\usepackage{epstopdf}
\usepackage{longtable}
\usepackage{array}
\usepackage{pifont}
\usepackage{latexsym}
\usepackage[latin1]{inputenc}
\usepackage{bm}
\usepackage{textcomp}
\usepackage{subfigure}
\usepackage{braket}
\usepackage{wasysym}
\usepackage{ulem}
\usepackage{color, soul}
\usepackage{pifont}
\usepackage{multirow}
\usepackage[pdftex,pdfstartpage=1]{hyperref}

\begin{document}

\title{Generalized hyper-Ramsey-Bordé matter-wave interferometry:
\\quantum engineering of robust atomic sensors with composite pulses}
\author{T. Zanon-Willette}
\affiliation{Sorbonne Université, Observatoire de Paris, Université PSL, CNRS, LERMA, F-75005, Paris, France}
\email{thomas.zanon@sorbonne-universite.fr}
\affiliation{MajuLab, International Joint Research Unit IRL 3654, CNRS, Université Côte d'Azur, Sorbonne Université, National University of Singapore, Nanyang
Technological University, Singapore}
\affiliation{Centre for Quantum Technologies, National University of Singapore, 117543 Singapore, Singapore}
\author{D. Wilkowski}
\address{MajuLab, International Joint Research Unit IRL 3654, CNRS, Université Côte d'Azur, Sorbonne Université, National University of Singapore, Nanyang
Technological University, Singapore}
\address{Centre for Quantum Technologies, National University of Singapore, 117543 Singapore, Singapore}
\address{School of Physical and Mathematical Sciences, Nanyang Technological University, 637371 Singapore, Singapore}
\author{R. Lefevre}
\address{Department of Physics, Royal Holloway, University of London, Royal Holloway Egham Hill, Egham TW20 0EX, United Kingdom}
\author{A.V. Taichenachev}
\address{Novosibirsk State University, ul. Pirogova 2, 630090 Novosibirsk, Russia}
\address{Institute of Laser Physics, Siberian Branch, Russian Academy of Sciences, prosp. Akad. Lavrent'eva 15B, 630090 Novosibirsk, Russia}
\author{V.I. Yudin}
\address{Novosibirsk State University, ul. Pirogova 2, 630090 Novosibirsk, Russia}
\address{Institute of Laser Physics, Siberian Branch, Russian Academy of Sciences, prosp. Akad. Lavrent'eva 15B, 630090 Novosibirsk, Russia}
\address{Novosibirsk State Technical University, prosp. Karla Marksa 20, 630073 Novosibirsk, Russia}

\begin{abstract}
A new class of atomic interferences using ultra-narrow optical transitions are pushing quantum engineering control to a very high level of precision for a next generation of sensors and quantum gate operations. In such context, we propose a new quantum engineering approach to Ramsey-Bordé interferometry introducing multiple composite laser pulses with tailored pulse duration, Rabi field amplitude, frequency detuning and laser phase-step. We explore quantum metrology with hyper-Ramsey and hyper-Hahn-Ramsey clocks below the $10^{-18}$ level of fractional accuracy by a fine tuning control of light excitation parameters leading to spinor interferences protected against light-shift coupled to laser-probe field variation.
We review cooperative composite pulse protocols to generate robust Ramsey-Bordé, Mach-Zehnder and double-loop atomic sensors shielded against measurement distortion related to Doppler-shifts and light-shifts coupled to pulse area errors.
Fault-tolerant auto-balanced hyper-interferometers are introduced eliminating several technical laser pulse defects that can occur during the entire probing interrogation protocol.
Quantum sensors with composite pulses and ultra-cold atomic sources should offer a new level of high accuracy in detection of acceleration and rotation inducing phase-shifts, a strong improvement in tests of fundamental physics with hyper-clocks while paving the way to a new conception of atomic interferometers tracking space-time gravitational waves with a very high sensitivity.
\end{abstract}

\date{\today}

\preprint{APS/123-QED}

\maketitle

\section{INTRODUCTION}

\indent More than seventy years ago, Ramsey established the first quantum mechanical
description of an interferometric resonance with the method of separated oscillating fields~\cite{Ramsey:1950}.
Ramsey spectroscopy with coherent radiation and phase manipulation became an effective tool to investigate internal properties of nuclei,
atoms and molecules~\cite{Ramsey:1951,Ramsey:1956,Ramsey:1958} while opening a revolution in quantum metrology with atomic fountains as primary frequency standards reaching today a fractional frequency accuracy of $2×10^{-16}$~\cite{Abgrall:2015}.

By labeling internal states with external momentum, Bordé has extended the method of separated oscillating field to atomic interferometry with optical transitions realizing laser beam splitters and mirrors for matter-waves~\cite{Borde:1989,Borde:2002,Borde:2019}. Pioneering works were also made by Chebotayev and Dubetsky based on separated optical fields with standing waves~\cite{Dubetsky:1984,Chebotayev:1985}.
Ramsey-Bordé interferometers using cold and ultra-cold atoms have reached high sensitivity to rotation~\cite{Riehle:1991,Gustavson:1991}, local acceleration~\cite{Kasevich:1992,Peters:1999,Peters:2001}, accurate determination of the fine structure constant~\cite{Weiss:1993,Weiss:1994,Bouchendira:2011,Parker:2018,Clade:2019,Morel:2020-1} or optical clock realization with supersonic beams reaching a fractional frequency instability around $2×10^{-16}$~\cite{Olson:2019}. Mach-Zehnder type quantum sensors have thus been developed for gravitational field measurements~\cite{Peters:1999,Rosi:2014,Asenbaum:2020}.

In parallel to laser spectroscopy and atom interferometry, composite pulses in Nuclear Magnetic Resonance (NMR) became a powerful tool to compensate for several imperfections due to radio-frequency (rf) pulses applied on large samples of nuclear spins~\cite{Levitt:1986,Wimperis:1994}. Various signal distortions from rf field inhomogeneities,
off-resonance effects and field amplitude error were reduced to a very low order of correction by means of complex sequences
of pulses adapted to single or even dual compensation of these systematics. Composite pulses have also demonstrated to be useful for robust
error compensation in high-fidelity qubit gates dedicated to quantum computation~\cite{Vandersypen:2004,Jones:2011,Merrill:2014,Kyoseva:2014,Torosov:2022}.

So far, understanding how to improve the robustness of precision measurements while reducing laser-probe-induced systematics
still remains a critical goal for quantum sensing.
But contrary to recent composite NMR-like pulse techniques applied in interferometers~\cite{Butts:2013,Saywell:2018,Saywell:2020,Olivares-Renteria:2020}, composite laser pulses
are required to compensate or eliminate residual phase-shifts leading to distortions of interferometric resonances.
A major step in that direction was realized in 2010 with the introduction of the hyper-Ramsey scheme to experimentally
reduce laser-probe induced frequency shifts by several orders of magnitude in optical clocks requiring large probe
intensities~\cite{Yudin:2010,Yudin:2010-2,Huntemann:2012}. A sequence of two Ramsey pulses was used where an additional third one acts like a spin echo compensation of field amplitude error. This extra pulse can be inserted either before or after the free evolution time~\cite{Zanon-Willette:2015}.
Moreover, new generalized hyper-Ramsey protocols have extended robustness of probing clock transitions against
residual light-shifts coupled to dephasing effect~\cite{Zanon-willette:2017,Zanon-Willette:2018}.

\begin{figure}[t!!]
\center
\resizebox{8.7cm}{!}{\includegraphics[angle=0]{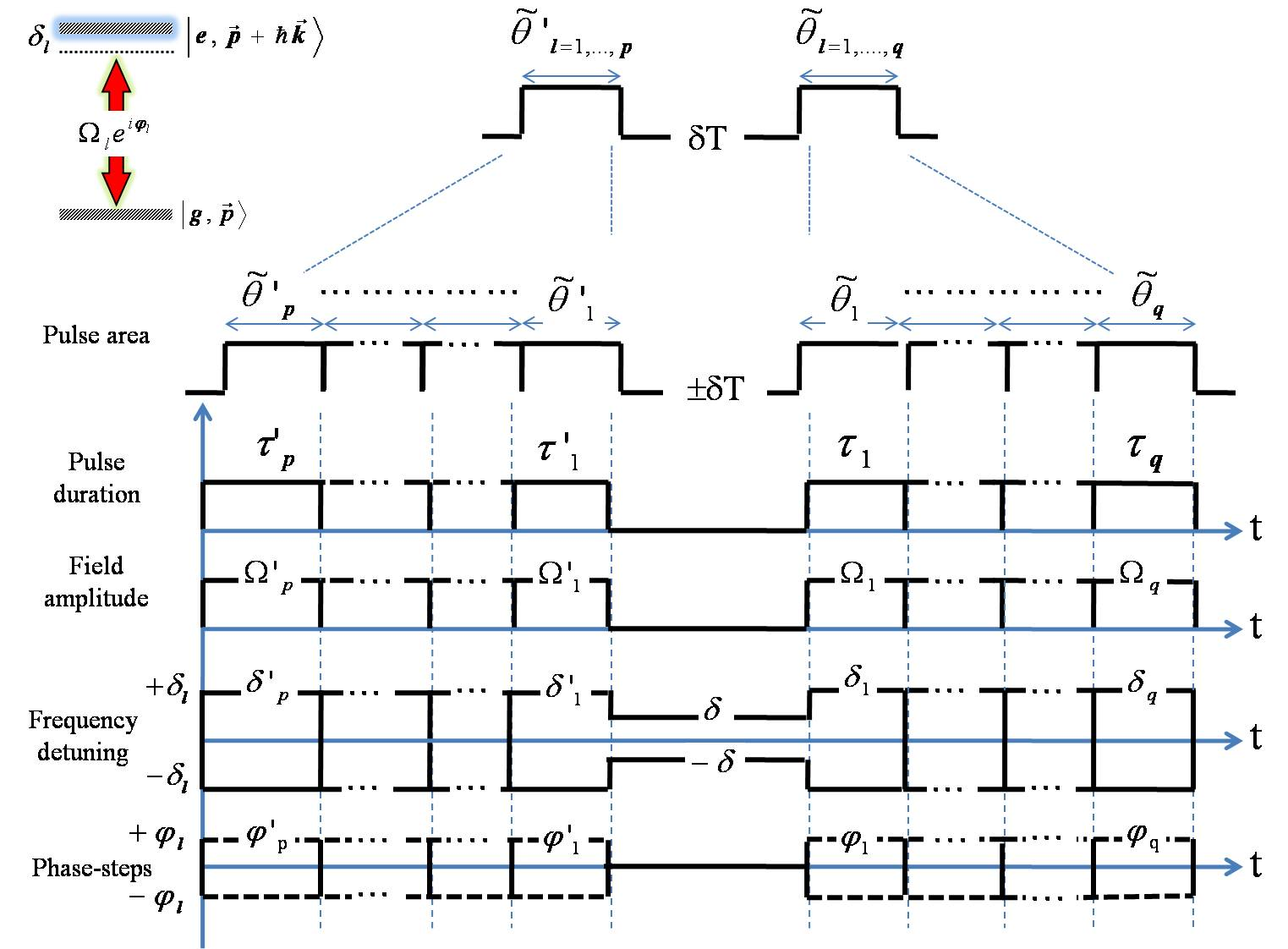}}
\caption{(color online). Generalized hyper-Ramsey-Bordé building-block denoted as $_{p}^{q}$M($\uparrow
$)($\downarrow$) for composite pulse matter-wave interferometry with optical traveling waves. Arbitrary composite pulses are introduced
 by pulse area $\widetilde{\vartheta}'_{l=1,....\textup{p}}$ and $\widetilde{\vartheta}_{l=1,....\textup{q}}$ around a single Ramsey free evolution time T. Pulse parameters are phases $\varphi'_{l},\varphi_{l}$, fields excitation $\Omega'_{l},\Omega_{l}$, pulses duration $\tau'_{l},\tau_{l}$ and frequency detunings $\delta'_{l},\delta_{l}$ including transverse Doppler-shifts $\textup{kv}_{z}$ related to arrow orientation $\uparrow\downarrow$, atomic recoil $\delta_{r}$ and a potential residual uncompensated part of light-shift $\Delta'_{l},\Delta_{l}$. During the free evolution time T, the detuning is $\delta'_{l}\equiv\delta_{l}=\pm\delta$ free from the light-shift induced by laser pulses.}
\label{fig1}
\end{figure}

\indent The main motivation of this original work is to bring optical composite pulses to matter-wave interferometry with efficient nonlinear compensation of pulse-defects induced phase-shifts while these methods are usually absent in modern treatment of atomic interferometry~\cite{Berman:1997,Kleinert:2015,Cadoret:2016}.
We will revisit the Ramsey-Bordé interferometry, light-shift, Doppler-shift and atomic recoil with arbitrary sequences of composite pulses around a single free evolution time.
A universal building-block with two-level operators, shown in Fig.\ref{fig1}, is developed through \textbf{section II}, offering an efficient computational algorithm to explore interferometric resonances and phase-shifts produced by composite pulses manipulating matter-waves.
We will review some robust composite pulse schemes including one or two free-evolution zones for robust hyper-clocks purpose in \textbf{section III}. Beside well-known results on clock interferometry with multi-pulses, the another goal of this section is to validate our computational method of elementary building-block decomposition when concatenated pulses are interleaved with several free evolution times. Robust generalized hyper-Ramsey and hyper-Hahn-Ramsey interrogation protocols are presented using appropriate laser phase-steps eventually associated to alternating clock detunings with opposite sign eliminating frequency drifts in the probe laser.
In \textbf{section IV}, we extend the method of composite pulses to hyper-interferometers with matter-waves which act against laser pulse errors induced by laser intensity variation during interrogation protocols. Robust Ramsey-Bordé, Mach-Zehnder and Butterfly (or double-loop) interferometer schemes are studied and the interferometric phase-shifts are derived. We demonstrate an upgrade of metrological performances for quantum sensing using particular phase-shifted interferences against detrimental errors in pulse parameters.
Finally in \textbf{section V}, we develop the concept of auto-balanced hyper-interferometers with composite pulses allowing an efficient compensation of systematics with a rapid convergence rate to fault-tolerance.
We conclude about the potential impact of applying composite pulses within Ramsey-Bordé matter-wave interferometry in \textbf{section VI}.

\section{HYPER RAMSEY-BORDÉ BUILDING-BLOCK}

\subsection{SU(2) QUANTUM ENGINEERING MODEL}

\indent We present here a universal framework relying on concepts of hyper-Ramsey probing schemes for clock interferometry using two-level operators that simplify the description of matter-waves propagating through several optical composite pulses. One can address first a large variety of multiple probe excitation pulses inducing technical shifts and defects that are leading to significant measurement errors in a conventional atomic interferometer. Then our computational algorithm allows us to explore clusters of multi-composite pulses interleaved with several free evolution times that may provide a better compensation of errors in laser parameters perturbing the matter-wave amplitude probability.

\indent The formal derivation of the generalized Ramsey-Bordé amplitude probability $\Psi(t)$ is based on Cayley-Klein
parametrization of rotation spinors as~\cite{Zanon-Willette:2019}:
\begin{equation}
	\begin{split}
		\textup{M}(\widetilde{\vartheta}_{l})=
		\left( \begin{array}{cc}
			\cos\widetilde{\vartheta}_{l}e^{i\phi_{l}} & -i e^{-i(\varphi_{l})}\sin\widetilde{\vartheta}_{l} \\
			-i e^{i(\varphi_{l})}\sin\widetilde{\vartheta}_{l} & \cos\widetilde{\vartheta}_{l}e^{-i\phi_{l}}\\
		\end{array}\right),
	\end{split}
	\label{eq:IIA-Cayley-Klein-Matrix}
\end{equation}
with the action of a phase $\varphi_{l}$ on the Rabi complex frequency $\Omega_{l}$.
Phase angles are introduced by:
\begin{subequations}
	\begin{align}
\widetilde{\vartheta}_{l}&=\arcsin\left[\frac{\Omega_{l}}{\omega_{l}}\sin\widetilde{\theta}_{l}\right],	\label{eq:IIA-Cayley-Klein-rotation-angles-1}\\
\phi_{l}&=\arctan\left[\frac{\delta_{l}}{\omega_{l}}\tan\widetilde{\theta}_{l}\right]	\label{eq:IIA-Cayley-Klein-rotation-angles-2}.
	\end{align}
\end{subequations}
Such a parametrization emphasizes the role of any residual light-shift correction as an additional phase factor acting on diagonal elements of the interaction matrix.
The effective pulse area is $\widetilde{\theta}_{l}=\theta_{l}/2=\omega_{l}\tau_{l}/2$ with a generalized Rabi frequency denoted as $\omega_{l}=\sqrt{\delta_{l}^{2}+\Omega_{l}^{2}}$.
The effective detuning $\delta_{l}=\delta+\Delta_{l}$ is a free clock detuning $\delta$ including a residual uncompensated part of the light-shift $\Delta_{l}$~\cite{Yudin:2010-2}.
It is replaced by $\delta_{l}\mapsto\delta_{l}\mp\textup{kv}_{z}-\delta_{r}$ if external Doppler-shift $\textup{kv}_{z}$ and quantized atomic recoil $\delta_{r}=\hbar\textup{k}^{2}/2\textup{m}$ are required for atomic interferometry~\cite{Borde:1984,Salomon:1984,Borde:1983}.

Our model is based on the exact description of a full composite wave-function with spinors~\cite{Borde:1984,Zanon-Willette:2019,Salomon:1984} incorporating independent control and fine tuning of coherent radiation parameters in the following form:
\begin{equation}
\begin{split}
\Psi(t)&=\left[\overleftarrow{\prod}_{l=\textup{1}}^{\textup{q}}\textup{M}(\widetilde{\vartheta}_{l})\right]\cdot\textup{M}(\delta\textup{T})\cdot
\left[\overrightarrow{\prod}_{l=\textup{1}}^{\textup{p}}\textup{M}(\widetilde{\vartheta}'_{l})\right]\Psi(0),
\end{split}
\label{composite-pulses}
\end{equation}
Each arrow indicates the direction to develop the product of matrices around a single free evolution time matrix $\textup{M}(\delta\textup{T})$ where laser fields, thus the light-shifts, are switch-off.

Complex amplitudes of $\Psi(t)$ for a two-level spin system being initially prepared in $\Psi(t=0)$, can be obtained by the application
of successive \textit{p} pulses before and \textit{q} pulses after the free evolution time (See Fig.~\ref{fig1}) leading to a complex matrix given by~\cite{Schwinger:1977}:
\begin{subequations}
\begin{align}
\Psi(t)&=_{p}^{q}\textup{M}\cdot\Psi(0)\label{eq:matrix-components-a}\\
\left(
  \begin{array}{c}
   C_{g}(t) \\
    C_{e}(t) \\
  \end{array}
\right)&=\left(
          \begin{array}{cc}
            _{p}^{q}C_{gg} &  _{p}^{q}C_{ge} \\
             _{p}^{q}C_{eg} &  _{p}^{q}C_{ee} \\
          \end{array}
        \right)\cdot\left(\begin{array}{c}
			C_{g}(0) \\
			C_{e}(0) \\
		\end{array}\right)\label{eq:matrix-components-b}.
\end{align}
\end{subequations}
The matrix $_{p}^{q}\textup{M}$ is a special unitary operator and relations between the matter-wave components are given by~\cite{Schwinger:1977}:
\begin{subequations}
\begin{align}
_{p}^{q}C_{gg}&=_{p}^{q}C_{ee}^{*}\\
_{p}^{q}C_{ge}&=-_{p}^{q}C_{eg}^{*}\\
|_{p}^{q}C_{gg}|^{2}+|_{p}^{q}C_{ge}|^{2}&=1
\end{align}
\end{subequations}

The complex probability amplitude associated to $_{p}^{q}C_{gg}$ and $_{p}^{q}C_{ge}$ can be recast
into a symmetric canonical form following refs~\cite{Zanon-Willette:2019,Zanon-Willette:2021}:
\begin{subequations}
\begin{align}
_{p}^{q}C_{gg}&=_{p}^{q}\alpha_{gg}e^{i\delta\textup{T}/2}\left[1-\left|_{p}^{q}\beta_{gg}\right|e^{-i(\delta\textup{T}+_{p}^{q}\Phi_{gg})}\right],\label{eq:Cgg}\\
_{p}^{q}C_{ge}&=_{p}^{q}\alpha_{ge}e^{i\delta\textup{T}/2}\left[1+\left|_{p}^{q}\beta_{ge}\right|e^{-i(\delta\textup{T}+_{p}^{q}\Phi_{ge})}\right],\label{eq:Cge}
\end{align}
\end{subequations}
Remarkably, we found that the complex parameters $\alpha$ and $\beta$ driving the overall envelop and composite phase-shifts can be separated in two independent contributions from \textit{p} pulses driven by pulse area $\widetilde{\vartheta}'_{l}$ and \textit{q} pulses driven by $\widetilde{\vartheta}_{l}$ as following:
\begin{subequations}
\begin{align}
_{p}^{q}\alpha_{gg}&=\alpha_{l}'^{p}(gg)\cdot\alpha_{l}^{q}(gg),\label{eq:env-gg}\\
_{p}^{q}\beta_{gg}&=\beta_{l}'^{p}(gg)\cdot\beta_{l}^{q}(gg),\label{eq:beta-gg}\\
_{p}^{q}\alpha_{ge}&=\alpha_{l}'^{p}(ge)\cdot\alpha_{l}^{q}(gg),\label{eq:env-ge}\\
_{p}^{q}\beta_{ge}&=\beta_{l}'^{p}(ge)\cdot\beta_{l}^{q}(gg).\label{eq:beta-ge}
\end{align}
\end{subequations}
Envelop terms $_{p}^{q}\alpha_{gg}$ and $_{p}^{q}\alpha_{ge}$ have been explicitly developed for arbitrary cases in the appendix section S0 following ref~\cite{Hardy:2017}.
From Eq.~\ref{eq:Cgg} and Eq.~\ref{eq:Cge}, it follows that interferometric phase-shifts affecting the central interference $_{p}^{q}\Phi_{gg}$ or $_{p}^{q}\Phi_{ge}$ are given by:
\begin{subequations}
\begin{align}
_{p}^{q}\Phi_{gg}&=\varphi_{L}+\phi_{L}-\textup{Arg}\left[_{p}^{q}\beta_{gg}\right],\label{eq:Phigg}\\
_{p}^{q}\Phi_{ge}&=\varphi_{L}+\phi_{L}-\textup{Arg}\left[_{p}^{q}\beta_{ge}\right],\label{eq:Phige}
\end{align}
\end{subequations}
with a remnant phase definition $\varphi_{L}\equiv\varphi_{1}-\varphi'_{1}$ corrected by a light-shifted contribution $\phi_{L}\equiv\phi'_{1}+\phi_{1}$ from pulses forming the original two-pulse Ramsey configuration. Note that phase-factors are now including a contribution from an arbitrary number of optical composite pulses extending previous results with three pulses~\cite{Zanon-Willette:2019}.
\begin{table}[t!!]
\renewcommand{\arraystretch}{1.7}
\begin{tabular}{|c|c|c|}
\hline
\hline
protocols & composite pulse building-block $_{p}^{q}$M \\
\hline
   \begin{tabular}{c}
       R1($\varphi$) \\
       ($\varphi=\pi/2$)
   \end{tabular}
 &
 \begin{tabular}{c}
$\boldsymbol{90}^{\circ'}_{\pm\varphi}\dashv\delta\textup{T}\vdash\boldsymbol{90}^{\circ}_{0}$ \\
$(\dagger)$ $\boldsymbol{90}^{\circ'}_{0}\dashv\delta\textup{T}\vdash\boldsymbol{90}^{\circ}_{\mp\varphi}$
                                \end{tabular}
\\
\hline
\hline
   \begin{tabular}{c}
       R2($\varphi$) \\
       ($\varphi=\pi/2$)
   \end{tabular}
 &
 \begin{tabular}{c}
$\boldsymbol{90}^{\circ'}_{\pm\varphi}\dashv\delta\textup{T}\vdash\boldsymbol{270}^{\circ}_{0}$ \\
$(\dagger)$ $\boldsymbol{270}^{\circ'}_{0}\dashv\delta\textup{T}\vdash\boldsymbol{90}^{\circ}_{\mp\varphi}$
                                \end{tabular}
\\
\hline
\hline

   \begin{tabular}{c}
       HR3$_{\pi}$($\varphi$) \\
       ($\varphi=\pi/2$)
   \end{tabular}  & \begin{tabular}{c}
$\boldsymbol{90}^{\circ'}_{\pm\varphi}\dashv\delta\textup{T}\vdash\boldsymbol{180}^{\circ}_{\pi}\boldsymbol{90}^{\circ}_{0}$ \\
$(\dagger)$ $\boldsymbol{90}^{\circ'}_{0}\boldsymbol{180}^{\circ'}_{\pi}\dashv\delta\textup{T}\vdash\boldsymbol{90}^{\circ}_{\mp\varphi}$
 \end{tabular}
\\
\hline
\hline
   \begin{tabular}{c}
       HR5$_{\pi}$($\varphi$) \\
       ($\varphi=\pi/2$)
   \end{tabular}  & \begin{tabular}{c}
$\boldsymbol{90}^{\circ'}_{\pm\varphi}\dashv\delta\textup{T}\vdash\boldsymbol{360}^{\circ}_{\pi}\boldsymbol{540}^{\circ}_{0}\boldsymbol{360}^{\circ}_{\pi}\boldsymbol{90}^{\circ}_{0}$ \\
$(\dagger)$ $\boldsymbol{90}^{\circ'}_{0}\boldsymbol{360}^{\circ'}_{\pi}\boldsymbol{540}^{\circ'}_{0}\boldsymbol{360}^{\circ'}_{\pi}\dashv\delta\textup{T}\vdash\boldsymbol{90}^{\circ}_{\mp\varphi}$
 \end{tabular}
\\
\hline
\hline
 \begin{tabular}{c}
       GHR($\varphi$) \\
       ($\varphi=\pi/4,3\pi/4$)
   \end{tabular} & \begin{tabular}{c}
$\boldsymbol{90}^{\circ'}_{0}\dashv\delta\textup{T}\vdash\boldsymbol{180}^{\circ}_{\pm\varphi}\boldsymbol{90}^{\circ}_{0}$ \\
$(\dagger)$ $\boldsymbol{90}^{\circ'}_{0}\boldsymbol{180}^{\circ'}_{\mp\varphi}\dashv\delta\textup{T}\vdash\boldsymbol{90}^{\circ}_{0}$
                                   \end{tabular}
\\
\hline
\hline
   \begin{tabular}{c}
      HHR$_{\pi}$($\varphi$) \\
      ($\varphi=\pi/2$)
   \end{tabular}
   & \begin{tabular}{c}
$\boldsymbol{90}^{\circ'}_{\pm\varphi}\boldsymbol{90}^{\circ'}_{0}\dashv\delta\textup{T}\vdash\boldsymbol{180}^{\circ}_{\pi}\boldsymbol{90}^{\circ}_{0}$ \\
$(\dagger)$ $\boldsymbol{90}^{\circ'}_{0}\boldsymbol{180}^{\circ'}_{\pi}\dashv\delta\textup{T}\vdash\boldsymbol{90}^{\circ}_{0}\vdash\boldsymbol{90}^{\circ}_{\mp\varphi}$
                                   \end{tabular}
\\
\hline
\hline
\end{tabular}
\centering%
\caption{Composite pulse protocols for hyper-clocks. Pulse area $\boldsymbol{\theta'_{l}}(\boldsymbol{\theta_{l}})$ is given in degrees $^{\circ}$ and phase-steps $\pm\varphi'_{l}(\varphi_{l})$ are indicated in subscript-brackets with radian unit. The standard Rabi frequency for all pulses is $\Omega=\pi/2\tau$ where $\tau$ is the pulse duration reference. The elementary block, used to perform calculation, is characterized by a $_{p}^{q}$M interaction matrix including $p,q$ pulses around a single free evolution zone $\delta$T. Reverse protocols in time are denoted by $(\dagger)$.}
\label{clock-protocol-table}
\end{table}
Let's now derive the formal expression of complex factors $\beta_{l}'^{p}(gg)$ and $\beta_{l}^{q}(gg)$ leading to a main distortion of matter-waves interferences. Composites phase-shifts $_{p}^{q}\Phi_{gg}$ and $_{p}^{q}\Phi_{ge}$ are driven by a truncated continued fraction expansion with ${p,q}$ pulses as following:
\begin{subequations}
\begin{align}
\beta_{l}'^{p}(gg)&=\frac{\tan\widetilde{\vartheta'}_{1}+e^{-i\Xi'_{12}}\frac{\tan\widetilde{\vartheta'}_{2}+e^{-i\Xi'_{23}}\frac{\tan\widetilde{\vartheta'}_{3}+...}{1-...}}
{1-e^{-i\Xi'_{23}}\tan\widetilde{\vartheta'}_{2}\frac{\tan\widetilde{\vartheta'}_{3}+...}{1-...}}}{1-e^{-i\Xi'_{12}}\tan\widetilde{\vartheta'}_{1}\frac{\tan\widetilde{\vartheta'}_{2}+e^{-i\Xi'_{23}}\frac{\tan\widetilde{\vartheta'}_{3}+...}{1-...}}
{1-e^{-i\Xi'_{23}}\tan\widetilde{\vartheta'}_{2}\frac{\tan\widetilde{\vartheta'}_{3}+...}{1-...}}},
\label{eq:phaseggp}\\
\beta_{l}^{q}(gg)&=\frac{\tan\widetilde{\vartheta}_{1}+e^{-i\Xi_{12}}\frac{\tan\widetilde{\vartheta}_{2}+e^{-i\Xi_{23}}\frac{\tan\widetilde{\vartheta}_{3}+...}{1-...}}
{1-e^{-i\Xi_{23}}\tan\widetilde{\vartheta}_{2}\frac{\tan\widetilde{\vartheta}_{3}+...}{1-...}}}{1-e^{-i\Xi_{12}}\tan\widetilde{\vartheta}_{1}\frac{\tan\widetilde{\vartheta}_{2}+e^{-i\Xi_{23}}\frac{\tan\widetilde{\vartheta}_{3}+...}{1-...}}
{1-e^{-i\Xi_{23}}\tan\widetilde{\vartheta}_{2}\frac{\tan\widetilde{\vartheta}_{3}+...}{1-...}}},
\label{eq:phaseggq}\\
\beta_{l}'^{p}(ge)&=\frac{1}{\left\{\beta_{l}'^{p}(gg)\right\}^{*}}.
\label{eq:phasegep}
\end{align}
\end{subequations}
where $\left\{\right\}^{*}$ means complex conjugate. Phase-factor expressions are respectively $\Xi'_{l,l+1}=\varphi'_{l}-\varphi'_{l+1}+\phi'_{l}+\phi'_{l+1}$ and $\Xi_{\textup{l,l+1}}=\varphi_{l+1}-\varphi_{l}+\phi_{l}+\phi_{l+1}$. Note that Eq.~\ref{eq:phaseggp} and Eq.~\ref{eq:phaseggq} have the same form than the Fresnel reflection complex coefficients of a plane wave propagating through multiple planar interfaces~\cite{Born:1999}.\\

\indent Turning to applications, we demonstrate the capacity of composite pulses to reduce or eliminate laser-probe induced systematics on matter-wave quantum interferences. We explore first clock interrogation protocols limited in accuracy by laser-probe-intensity drifts. Then we will extend our analysis on atomic interferometers limited in accuracy by residual Doppler-shift and light-shift induced by pulse area modification between sets of beam splitters from different Ramsey-Bordé building-blocks. Generalized transition probabilities related to laser pulse protocols are analytically derived in the appendix (section S1  for the hyper-clock interrogation scheme with one interaction matrix including $p=q=4$ pulses, section S2  for a generalized hyper-Hahn-Ramsey laser pulse protocol (GHHR) including a double free evolution zone, section S3 for a generalized hyper-Ramsey-Bordé (GHRB) interferometer with two interaction matrices including $p=1,q=2$ pulses and section S4 for hyper-Mach-Zehnder (HMZ) and hyper-Butterfly (HB) configurations).

\subsection{QUBIT TRAJECTORIES ON THE BLOCH SPHERE}

\indent Composite pulses can be analyzed by following the trajectory of the Bloch vector over a unitary sphere called the Bloch-sphere, starting from a given initial condition and verifying, visually or by geometric construction, which trajectories on a curved space are less sensitive to errors in the driving parameters. Further applications of a geometric approach, usually with the assistance of computer simulation, can produced a stream of composite pulses with different properties~\cite{Levitt:1986}, as those listed in Tab.~\ref{clock-protocol-table}. Still, the basic idea to test the efficiency of a given pulse sequence is to verify if, under non-ideal conditions like offset detuning error induced by a residual uncompensated light-shift or a small distortion applied during free evolution time, that the trajectories on the Bloch-sphere are refocusing closely to the ideal one, or deviate largely from it.
\begin{figure*}[t!!]
\center
\resizebox{7.5cm}{!}{\includegraphics[angle=0]{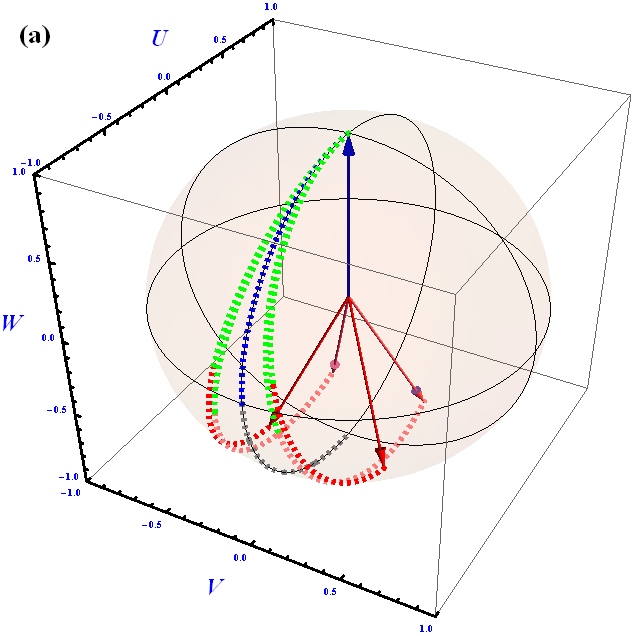}}\hspace{1cm}\resizebox{7.5cm}{!}{\includegraphics[angle=0]{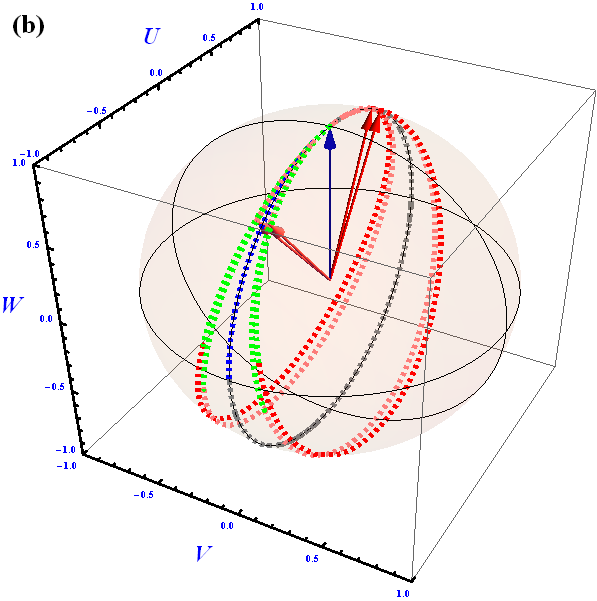}}
\resizebox{7.5cm}{!}{\includegraphics[angle=0]{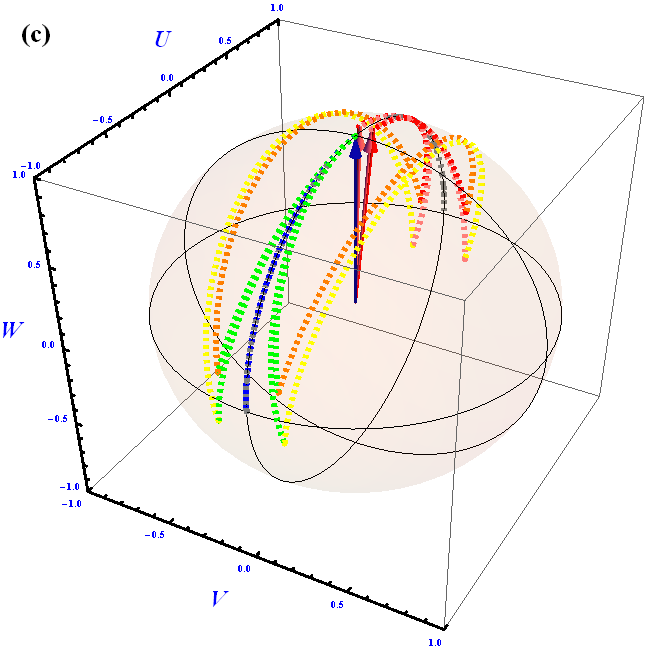}}\hspace{1cm}\resizebox{7.5cm}{!}{\includegraphics[angle=0]{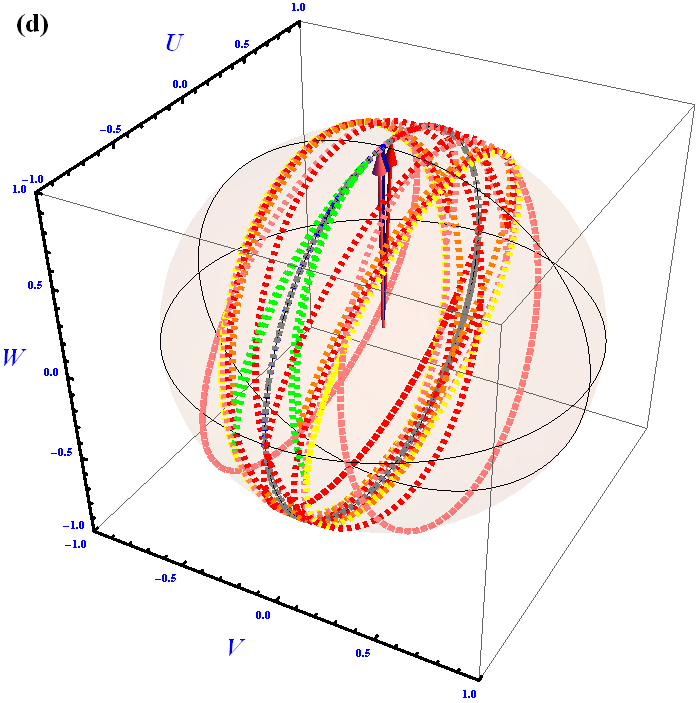}}
\caption{(color online). Trajectories of the $\textup{U}, \textup{V}, \textup{W}$ Bloch vector components for different laser pulse interrogation protocols reported in Tab.~\ref{clock-protocol-table}.
For visualization, the phase-step modulation and free evolution in the equatorial plane are ignored.
A small relative clock detuning offset about $\delta\Delta/\Delta=\pm15\%$ is splitting the Bloch vector components into two symmetrical trajectories after an initial $\boldsymbol{90}^{\circ'}_{0}$ pulse. A variation of the pulse area by $\Delta\theta/\theta=\pm10\%$ is visualized by different arrows that are pointing along different directions from the unperturbed trajectory followed by the central blue arrow. (a) R1, (b) R2, (c) HR3$_{\pi}$, (d) HR5$_{\pi}$. The efficient refocusing of final trajectories, merging all arrows, is clearly visible on panels (c) and (d).}
\label{fig:Bloch-sphere}
\end{figure*}
To describe any protocol from a geometrical point of view, we introduce a representation of the interaction matrix based on multiple laser pulses shown in Fig.~\ref{fig1}. The rotation of the state vector of a two-level system under the effect of composite pulses is described using the Feynman-Vernon-Hellwarth parametrization~\cite{Feynman:1957}.
We have applied this geometrical approach to test the sensitivity of different optical clock composite pulse protocols to small variations in the pulse parameters, and more precisely in the laser detuning and the pulse area. The $(\textup{U}, \textup{V}, \textup{W})$ components of the Bloch vector read~\cite{Feynman:1957}:
\begin{equation}
\left\{
\begin{split}
\textup{U}&= _{p}^{q}C_{\textup{gg}}^*~_{p}^{q}C_{\textup{ge}}+_{p}^{q}C_{\textup{gg}}~_{p}^{q}C_{\textup{ge}}^*, \\
\textup{V}&= i\left(_{p}^{q}C_{\textup{gg}}^*~_{p}^{q}C_{\textup{ge}}-_{p}^{q}C_{\textup{gg}}~_{p}^{q}C_{\textup{ge}}^*\right),  \\
\textup{W}&= \left|_{p}^{q}C_{\textup{gg}}\right|^2-\left|_{p}^{q}C_{\textup{ge}}\right|^2.
\end{split}
\right.
\end{equation}
A composite pulse sequence produces a temporal evolution on the unitary sphere starting from the upper pole with $\textup{W}=+1$ denoted by the blue arrow in
Fig.~\ref{fig:Bloch-sphere}.
For the ideal resonant case with  no detuning, fixing the laser phase-step as $\varphi=0$ for simplicity, the Bloch vector evolution takes place on the $(\textup{U},\textup{W})$ plane. A Ramsey double-pulse R1 protocol transfers the Bloch vector from the top to the bottom of the Bloch-sphere. The Bloch vector temporal evolution is
represented by blue dashed and gray dashed semicircle lines in Fig.~\ref{fig:Bloch-sphere}(a). When a small detuning error is added to the pulse protocol, trajectories move away from the unperturbed one and are represented by green lines like those in Fig.~\ref{fig:Bloch-sphere}(a) after application of the first $\boldsymbol{90}^{\circ'}_{0}$ Ramsey pulse. Finally a second $\boldsymbol{90}^{\circ}_{0}$ Ramsey pulse rotates the vector to the lower pole and due to variation in pulse area, different arrows are scattered away from the ideal inversion.

The R2 protocol is shown in Fig.~\ref{fig:Bloch-sphere}(b) where the second Ramsey pulse is three times longer replacing the second $\boldsymbol{90}^{\circ}_{0}$ by a $\boldsymbol{270}^{\circ}_{0}$ pulse. In such a case, the vector is forced to come back to the initial position indicated by the blue arrow orientation. The consequence of this modified Ramsey scheme is that the dispersion of misaligned arrows is reduced along the $(\textup{V},\textup{W})$ plane, moving closer to the ideal trajectory, but still suffering from pulse area variation indicated by shorter and longer paths along the $(\textup{U},\textup{V})$ plane.

The trajectories of the HR3$_{\pi}$ three-pulse and HR5$_{\pi}$ five-pulse composite protocols are presented in Fig.~\ref{fig:Bloch-sphere}(c) and (d). Those composite sequences, because of their total pulse area as a multiple integer of $\pi\equiv\boldsymbol{360}^{\circ}$, transfer the Bloch vector back to the initial orientation. Notice the different Bloch sphere trajectories for the HR3$_{\pi}$ and HR5$_{\pi}$ protocols, the later exploring both south and north hemispheres, while the first one is restricted to half part of the unitary sphere.\\
\begin{figure*}[t!!]
\center
\resizebox{9cm}{!}{\includegraphics[angle=0]{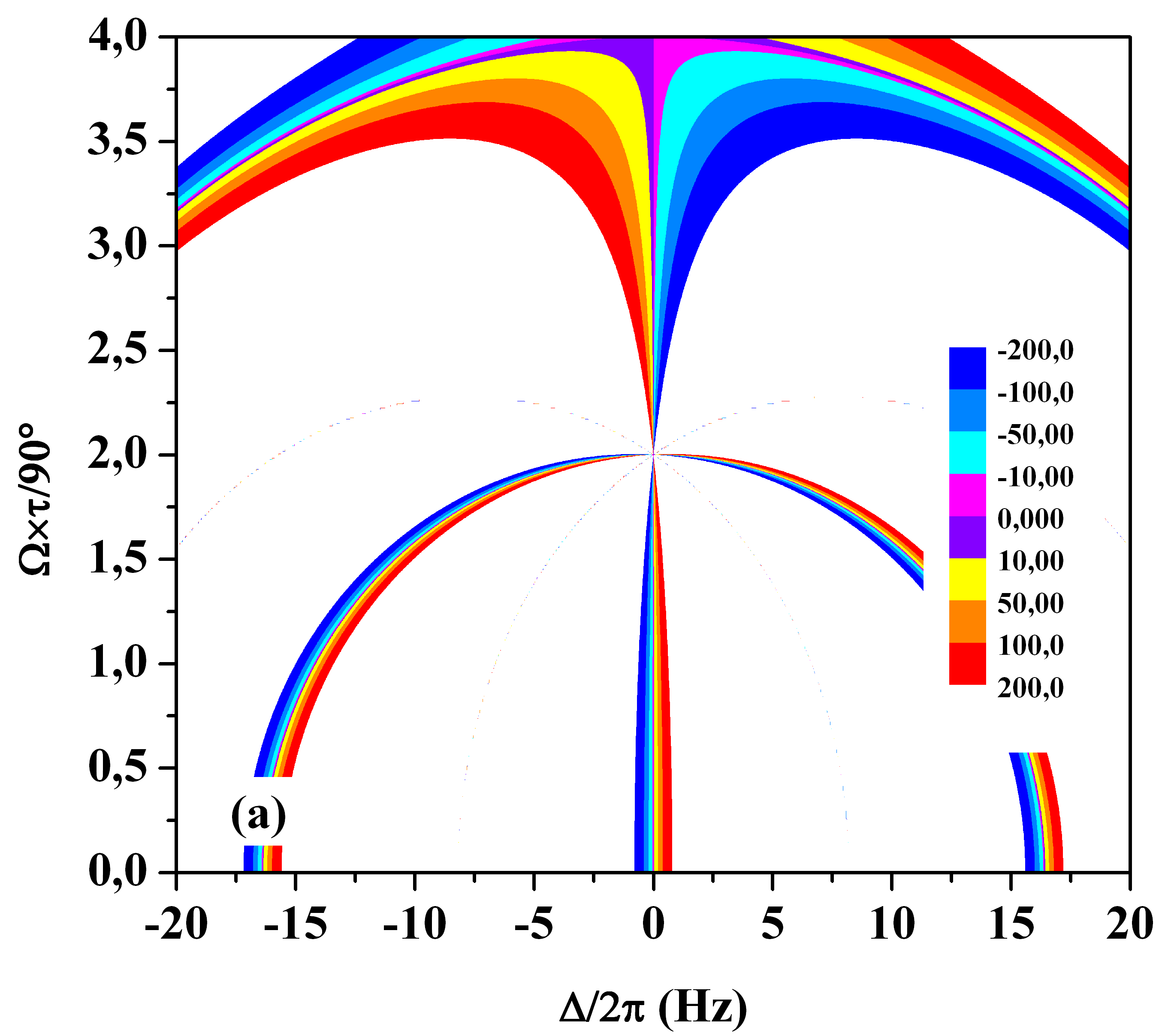}}\resizebox{9cm}{!}{\includegraphics[angle=0]{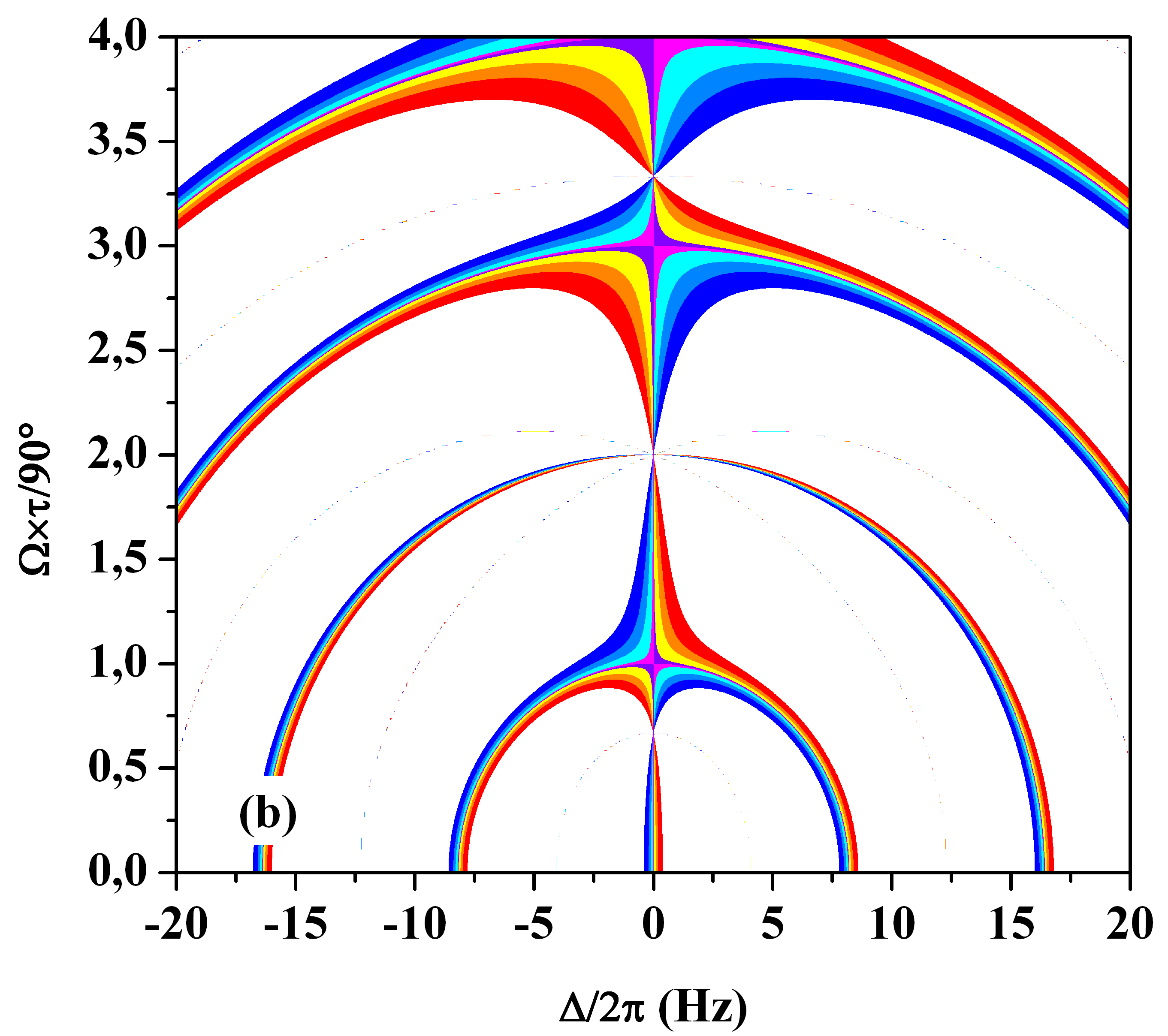}}
\resizebox{9cm}{!}{\includegraphics[angle=0]{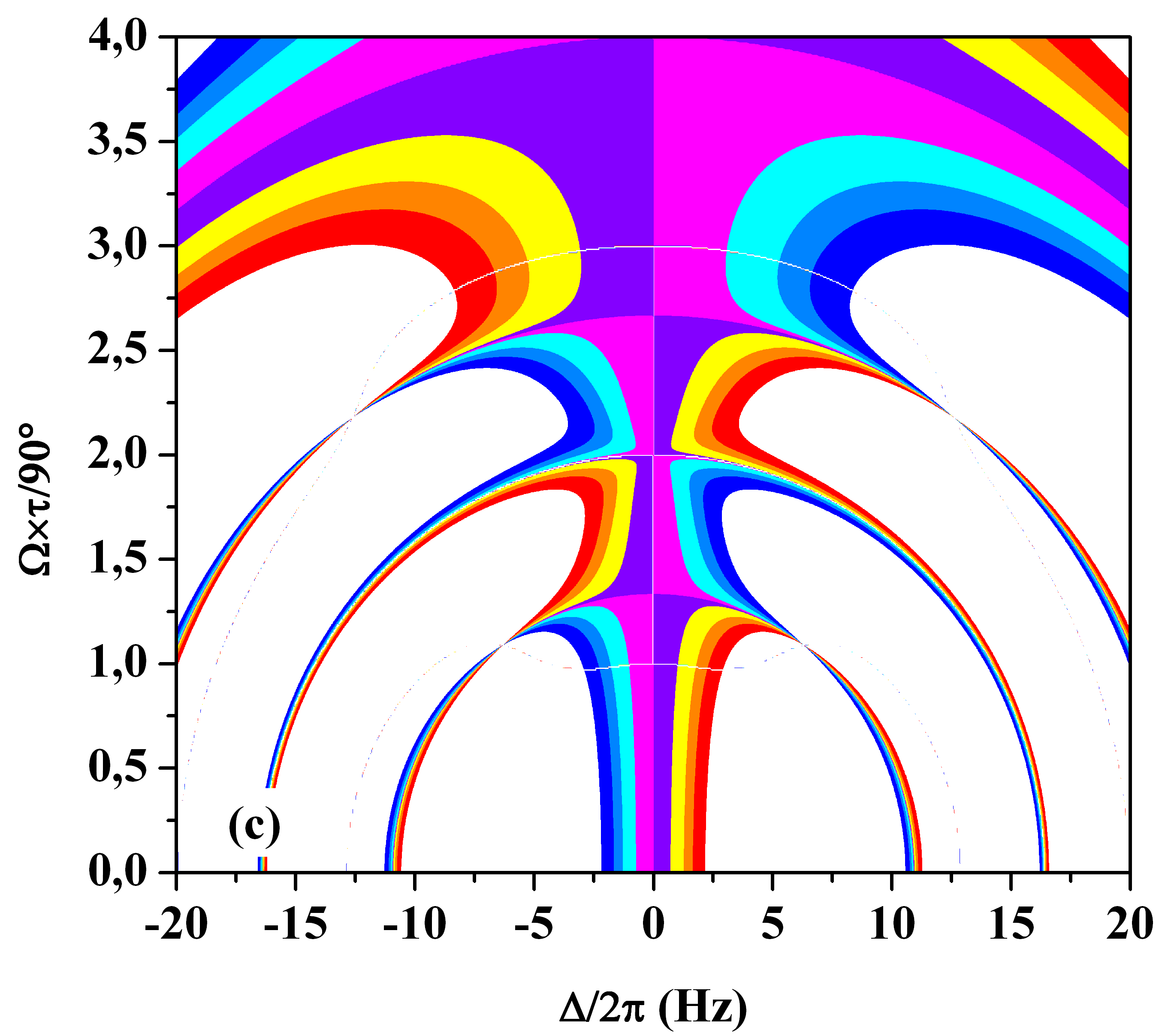}}\resizebox{9cm}{!}{\includegraphics[angle=0]{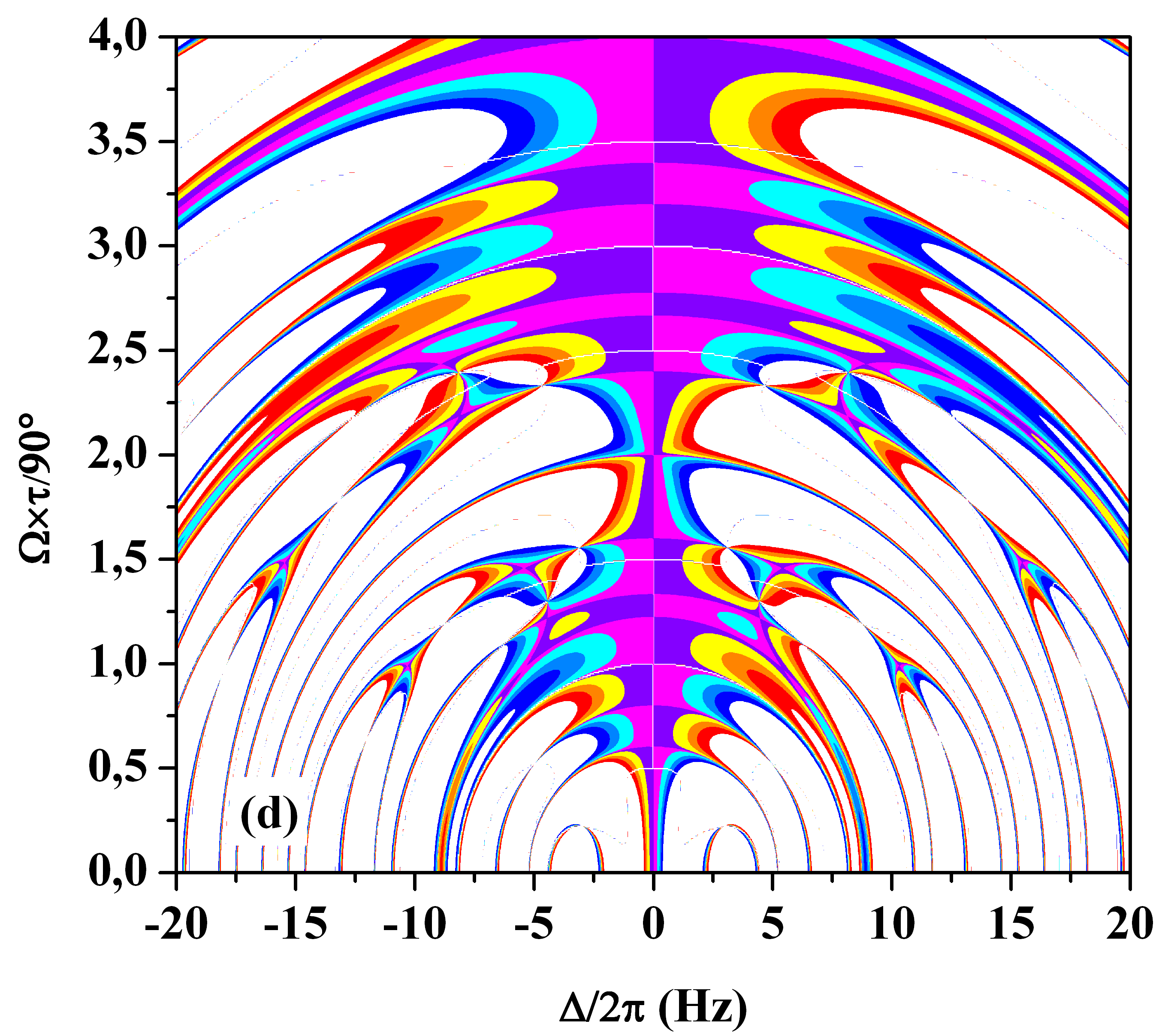}}
\caption{(color online). 2D clock frequency-shift diagrams of $\Delta\nu_{gg}$ based on Eq.~\ref{eq:clock-frequency-shift-a} (or equivalently $\Delta\nu_{ge}$ with Eq.~\ref{eq:clock-frequency-shift-b}) versus uncompensated part of a residual light-shift $\Delta/2\pi$ along the horizontal axis and pulse area $\Omega\tau$ variation along the vertical axis (see also~\cite{Zanon-Willette:2018}). (a) R1, (b) R2, (c) HR3$_{\pi}$ and (d) HR5$_{\pi}$.
Amplitude of the clock frequency-shift is indicated by a color graded scale between -200~mHz and 200~mHz on the right side of the first graph. The pulse reference is introduced as $\Omega\tau=n\times\pi/2$
($\pi/2\equiv\boldsymbol{90}^{\circ}$) where $\textup{n}$ is the parameter that is tuned along the vertical axis. Phase-shifts are evaluated modulo $\pm k\pi,~k\in\mathbb{N}$ (see also~\cite{Zanon-Willette:2019}). Typical parameters are $\tau=30.5$ms for the pulse duration with a free evolution time around T$=122$ms.}
\label{fig:2D-frequency-shift-diagrams}
\end{figure*}

Small variations in the detuning and pulse area laser parameters of each protocol transfer the Bloch vector to the final orientations denoted in the figure by multiple arrows along the trajectories indicated by the dashed lines. The spatial distributions of these arrows over the surface of the sphere are an indication of the instability of the vector temporal evolution against errors in the laser parameter settings. Notice the progressive reduction in the arrow spreading when progressing from the Ramsey R1 protocol, to the HR3$_{\pi}$ and HR5$_{\pi}$ ones. The difference between the last two ones is linked to the exploration by the Bloch vector of a limited Bloch sphere area around the starting point. Small variations in laser pulses parameters are compensated by generating sophisticated path trajectories on a Bloch's sphere reducing any distortion of vector components over the entire pulse interrogation protocol.
A systematic exploration of arrows orientations for different variations of the laser parameters allows to derive a quantitative comparison between protocols, for instance producing 2D-maps of the clock frequency-shift through contour plots with different laser parameters.

\subsection{2D CLOCK FREQUENCY-SHIFT DIAGRAMS}

\indent Composite pulses can be optimized by looking at 2D phase-shift or frequency-shift diagrams that are helpful to identify some key parameters (pulse area and laser phase-steps) in order to increase the robustness of clock interferometers to some detrimental effects from the probe laser itself. Such an optimization has been recently applied to hyper-clocks using three and five composite pulses~\cite{Zanon-Willette:2021}. Spinor rotation is described with a representation of the Pauli-spin matrices basis of the SU(2) group of rotations. It is extended to multiple excitation pulses by a recursive Euler-Rodrigues-Gibbs algorithm describing a composition of rotations with different rotation axis orientation. A general analytical formula for the phase-shift associated with the clock's interferometric signal is used to identify particular pulse areas that are either optimizing the clock signal amplitude, the clock frequency-shift or eventually both of them.

The same strategy can be applied within the SU(2) quantum engineering model presented in the previous sub-section. When the overall pulse duration is short relatively to the free evolution time, the clock frequency-shift associated to any small deviation of the Bloch vector trajectory from the ideal one can be reconstruct with the following relation:
\begin{subequations}
\begin{align}
\Delta\nu_{gg}&=_{p}^{q}\Phi_{gg}/(2\pi\textup{T})\label{eq:clock-frequency-shift-a},\\
\Delta\nu_{ge}&=_{p}^{q}\Phi_{ge}/(2\pi\textup{T})\label{eq:clock-frequency-shift-b}.
\end{align}
\end{subequations}
where phase-shifts expressions are given by Eq.~(\ref{eq:Phigg}) and Eq.~(\ref{eq:Phige}).

We have plotted, in Fig.~\ref{fig:2D-frequency-shift-diagrams}, the 2D clock frequency-shift diagrams versus an uncompensated part of the light-shift for some protocols reported in Tab.~\ref{clock-protocol-table}.
Amplitude of the clock frequency-shift is indicated by a color graded scale between -200~mHz to 200~mHz on the right side of the first graph. Better robustness against residual light-shift is achieved when graded regions, associated to a small frequency-shift, are growing up over a large horizontal axis scale inside the 2D diagram as shown in Fig.~\ref{fig:2D-frequency-shift-diagrams}(c) and (d).
By tuning the pulse area with a parameter $\textup{n}$, defining the pulse reference as $\Omega\tau=n\times\pi/2$ (where $\pi/2\equiv\boldsymbol{90}^{\circ}$ for a standard pulse area in degrees), we can look for particular pulse areas where the signal amplitude can be increased or the frequency-shift reduced.
We have thus presented in Fig.~\ref{fig:2D-frequency-shift-diagrams}(a)-(d) frequency-shifts associated to different laser protocols from Tab.~\ref{clock-protocol-table}.
Clearly, we can identify the important modification of the frequency-shift sensitivity  to residual light-shifts gradually reducing to a very low level of distortion when replacing the two-pulse Ramsey (R1) sequence by a composite pulse protocol with three pulses (HR3$_{\pi}$) or five pulses (HR5$_{\pi}$).
A comparison between Fig.~\ref{fig:2D-frequency-shift-diagrams}(b) and Fig.~\ref{fig:2D-frequency-shift-diagrams}(c) demonstrates that an efficient optimization of the clock frequency-shift against residual light-shift is reached by inserting an intermediate composite pulse as $\boldsymbol{270}^{\circ}_{0}\mapsto\boldsymbol{180}^{\circ}_{\pi}\boldsymbol{90}^{\circ}_{0}$. The hyper-Ramsey HR3$_{\pi}$ probing scheme was presented for the first time in~\cite{Yudin:2010}. The new HR5$_{\pi}$ protocol was recently introduced for a full optimization of metrological performances. The more sophisticated composite pulse is now given by the following modification through the replacement $\boldsymbol{180}^{\circ}_{\pi}\mapsto\boldsymbol{360}^{\circ}_{\pi}\boldsymbol{540}^{\circ}_{0}\boldsymbol{360}^{\circ}_{\pi}$~\cite{Zanon-Willette:2021}.
From Fig.~\ref{fig:2D-frequency-shift-diagrams}, any clock signal contrast is always maximized for odd values and vanishing for even values of the tuning $\textup{n}$ parameter.

By associating qubit trajectories on the Bloch-sphere from Fig.~\ref{fig:Bloch-sphere} with 2D diagrams from Fig.~\ref{fig:2D-frequency-shift-diagrams}, we observe that a very good refocusing of the Bloch vector orientation is always related to a strong reduction of the phase accumulation related to detuning errors and pulse area variations.
\begin{figure}[t!!]
\center
\resizebox{8.5cm}{!}{\includegraphics[angle=0]{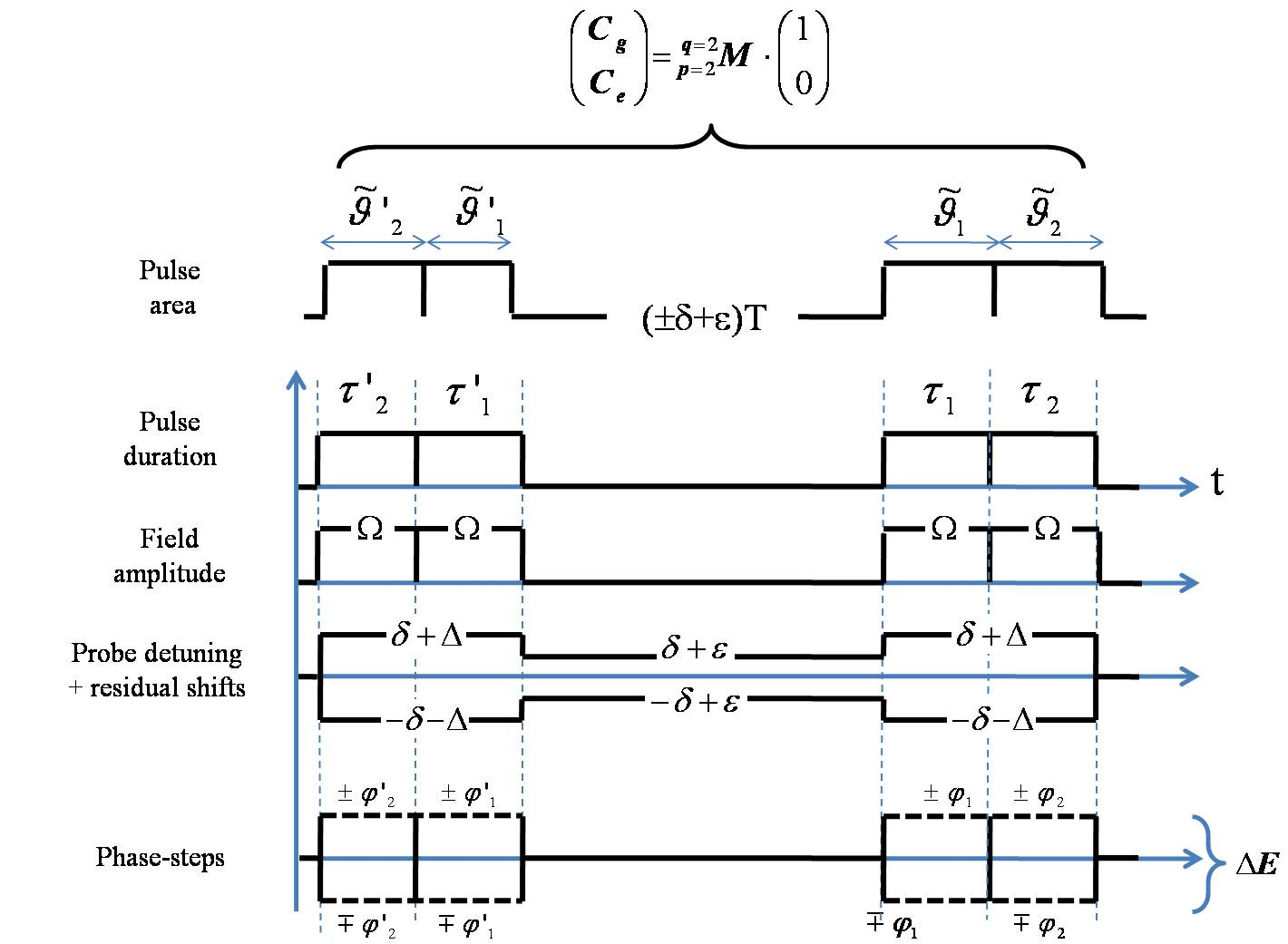}}
\caption{(color online). Four-pulse diagram with a single composite interaction matrix $_{2}^{2}$M to generate R, HR3$_{\pi}$, GHR, HHR$_{\pi}$, see Tab.~\ref{clock-protocol-table} for more details.
Each diagram introduces the action of pulse duration $\tau$, Rabi frequency $\Omega$, phase-step modulation $\pm\varphi$, uncompensated residual light-shift in the clock detuning $\pm(\delta+\Delta)$ and eventually a small distortion during the free evolution zone as $\pm\delta+\epsilon$.}
\label{fig:GHR-sequence}
\end{figure}
\begin{figure*}[t!!]
\center
\resizebox{9.cm}{!}{\includegraphics[angle=0]{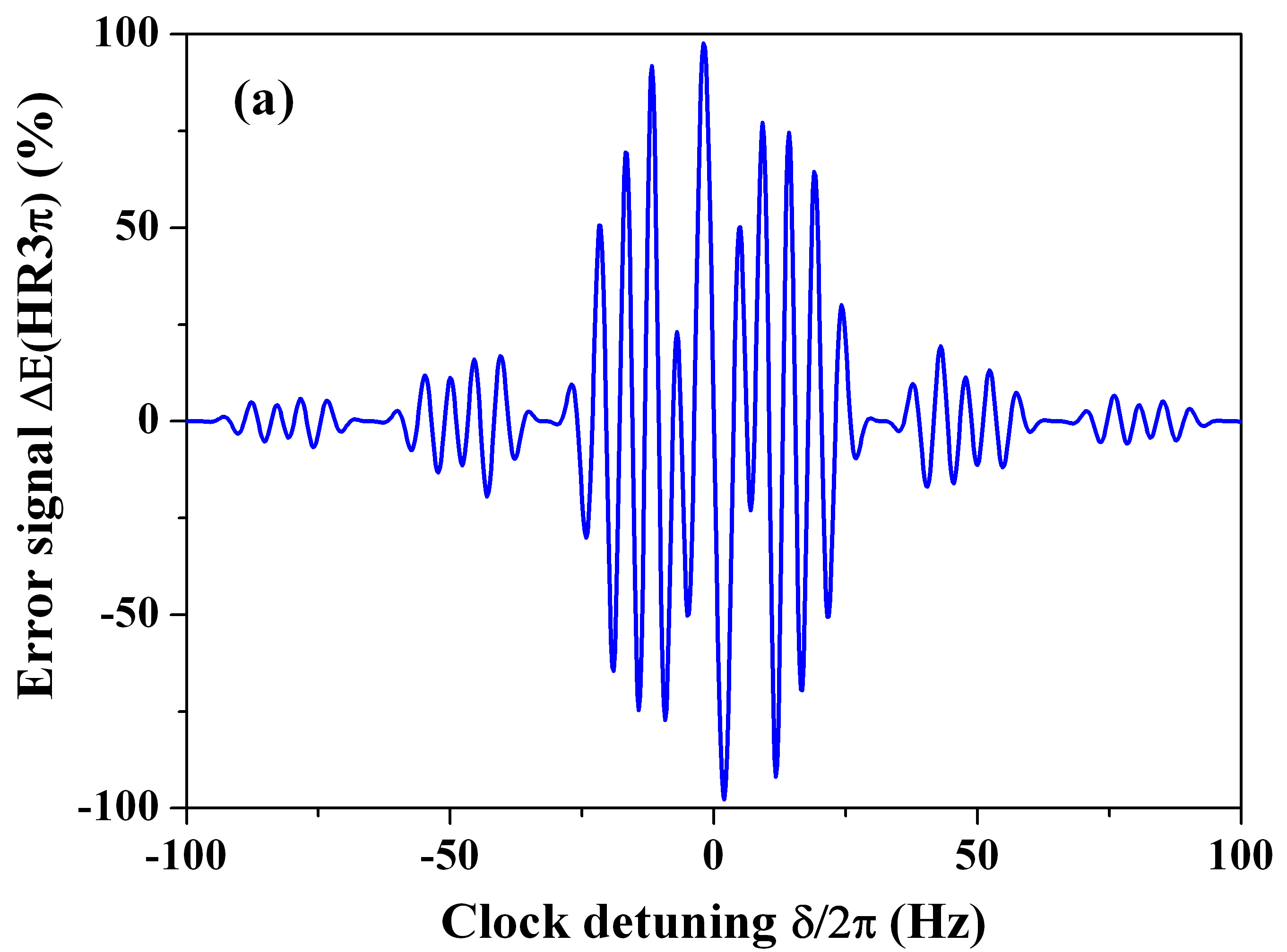}}\resizebox{9.cm}{!}{\includegraphics[angle=0]{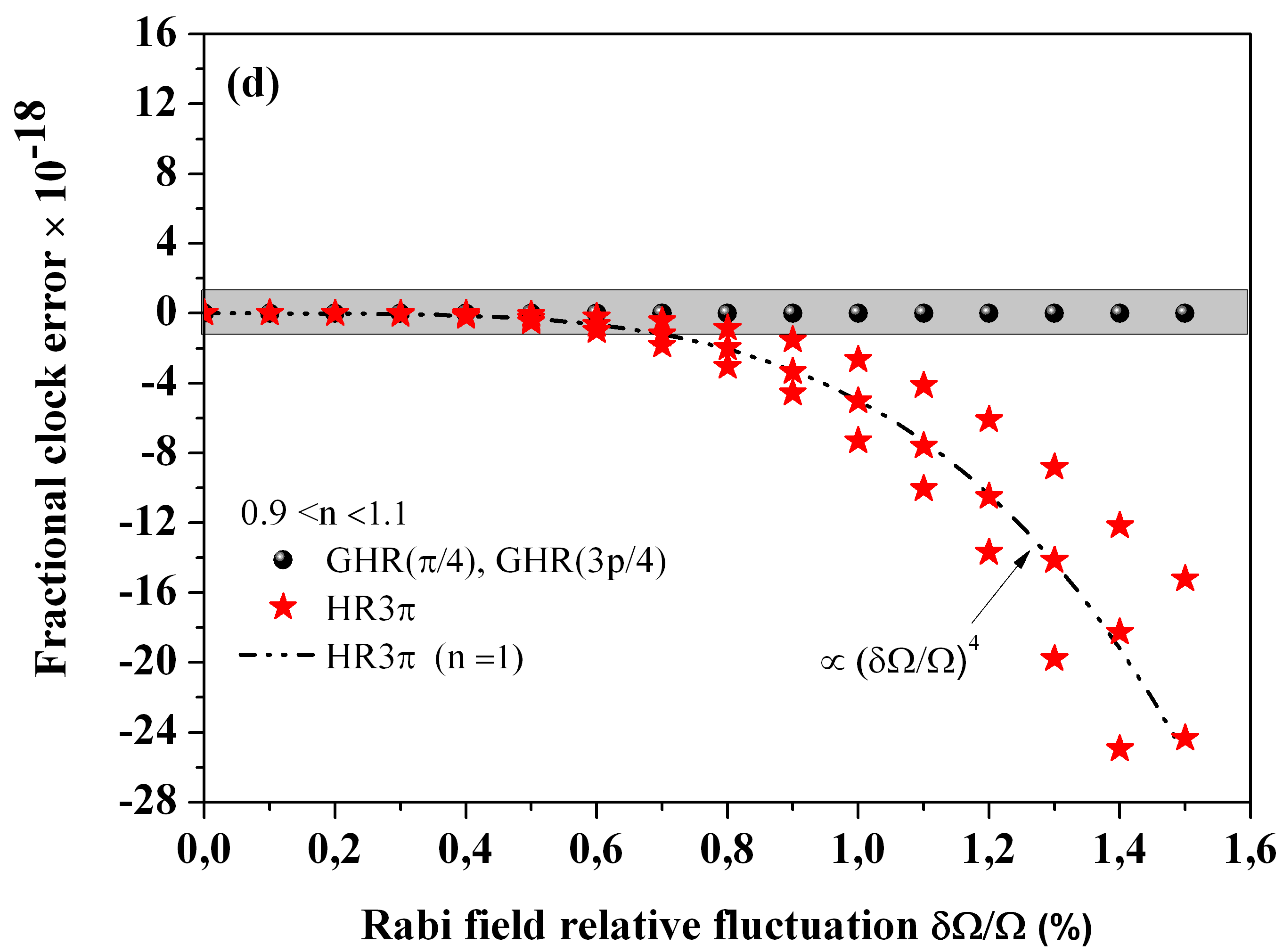}}
\resizebox{9.cm}{!}{\includegraphics[angle=0]{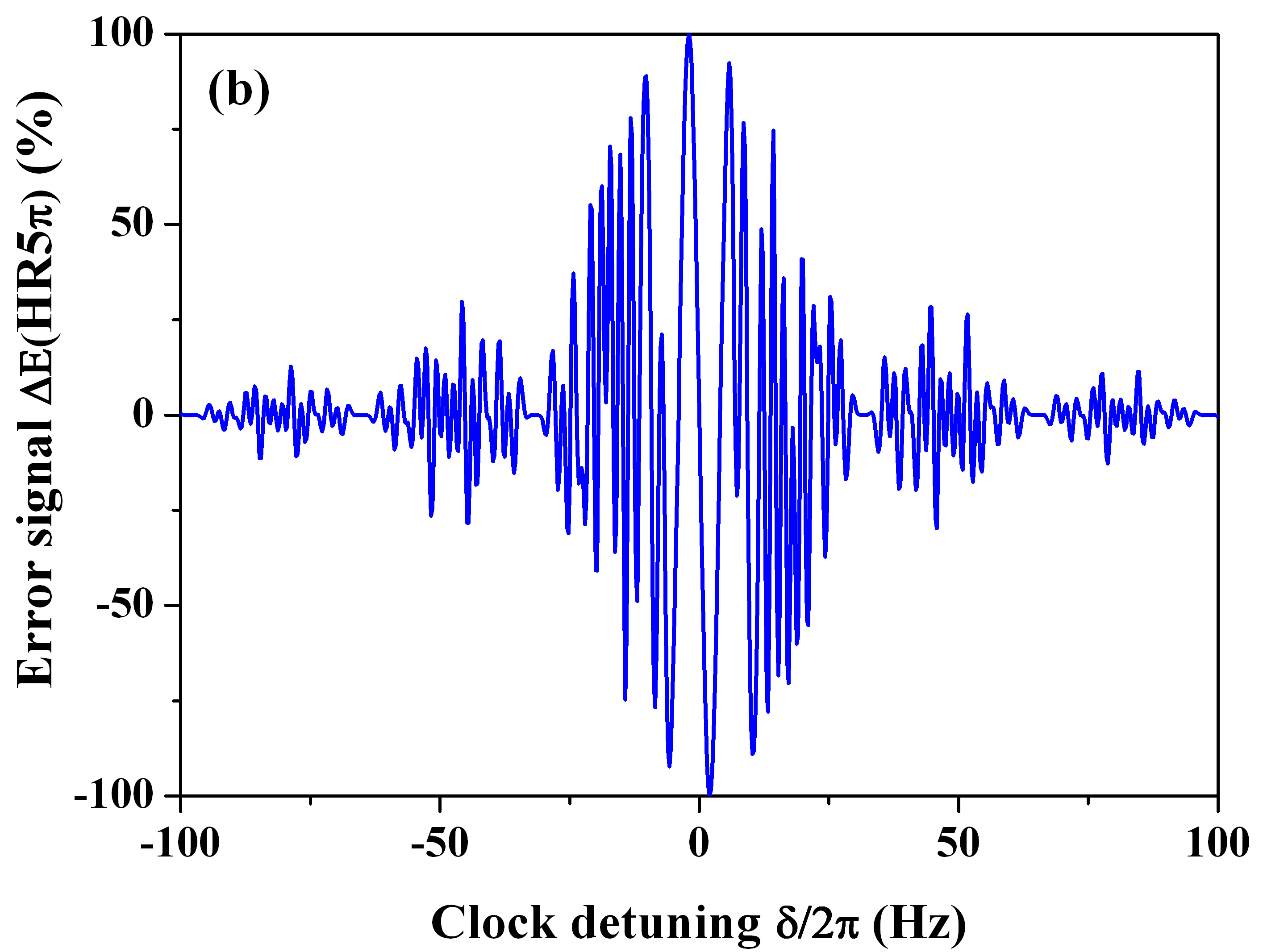}}\resizebox{9.cm}{!}{\includegraphics[angle=0]{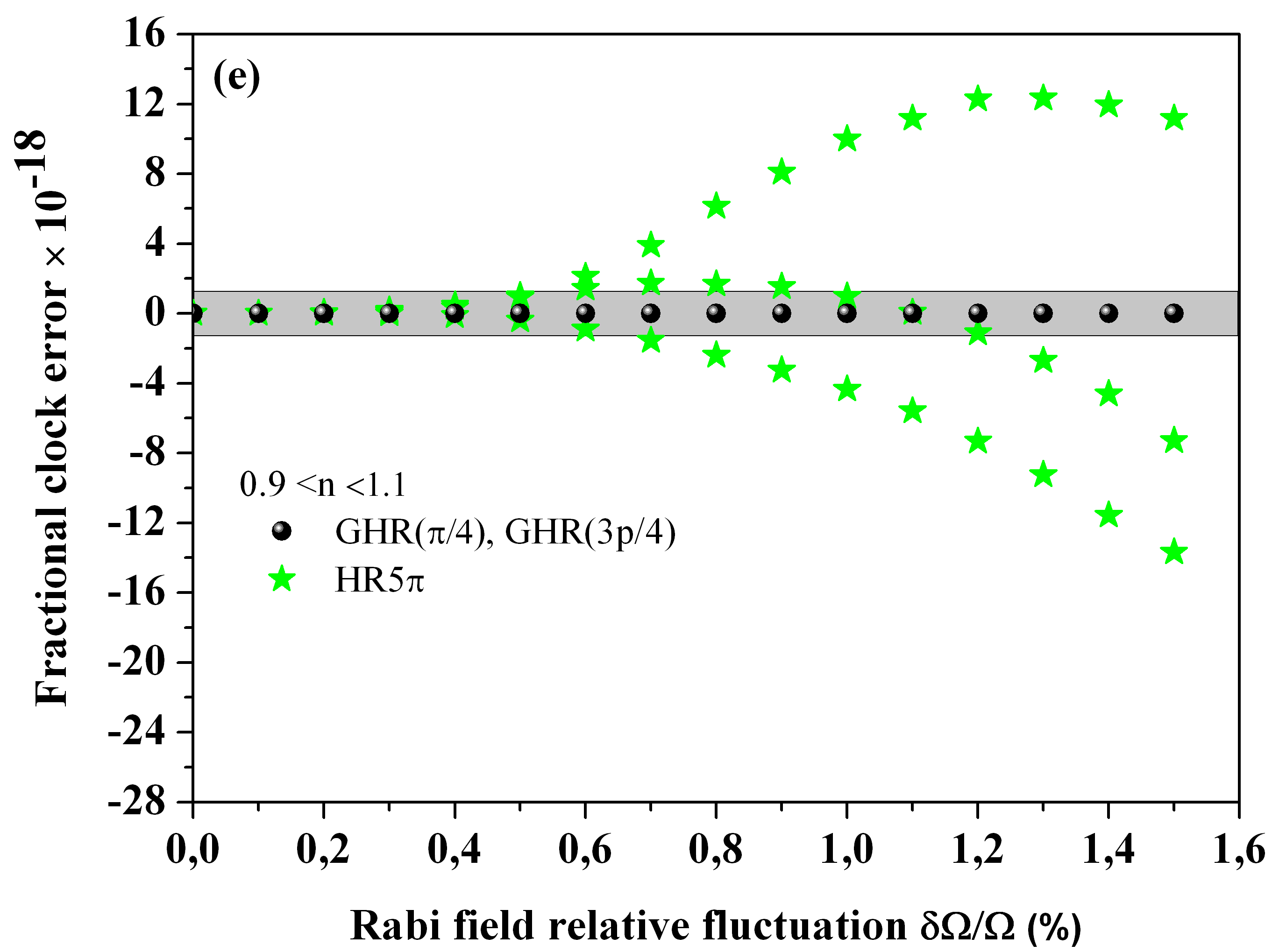}}
\resizebox{9.cm}{!}{\includegraphics[angle=0]{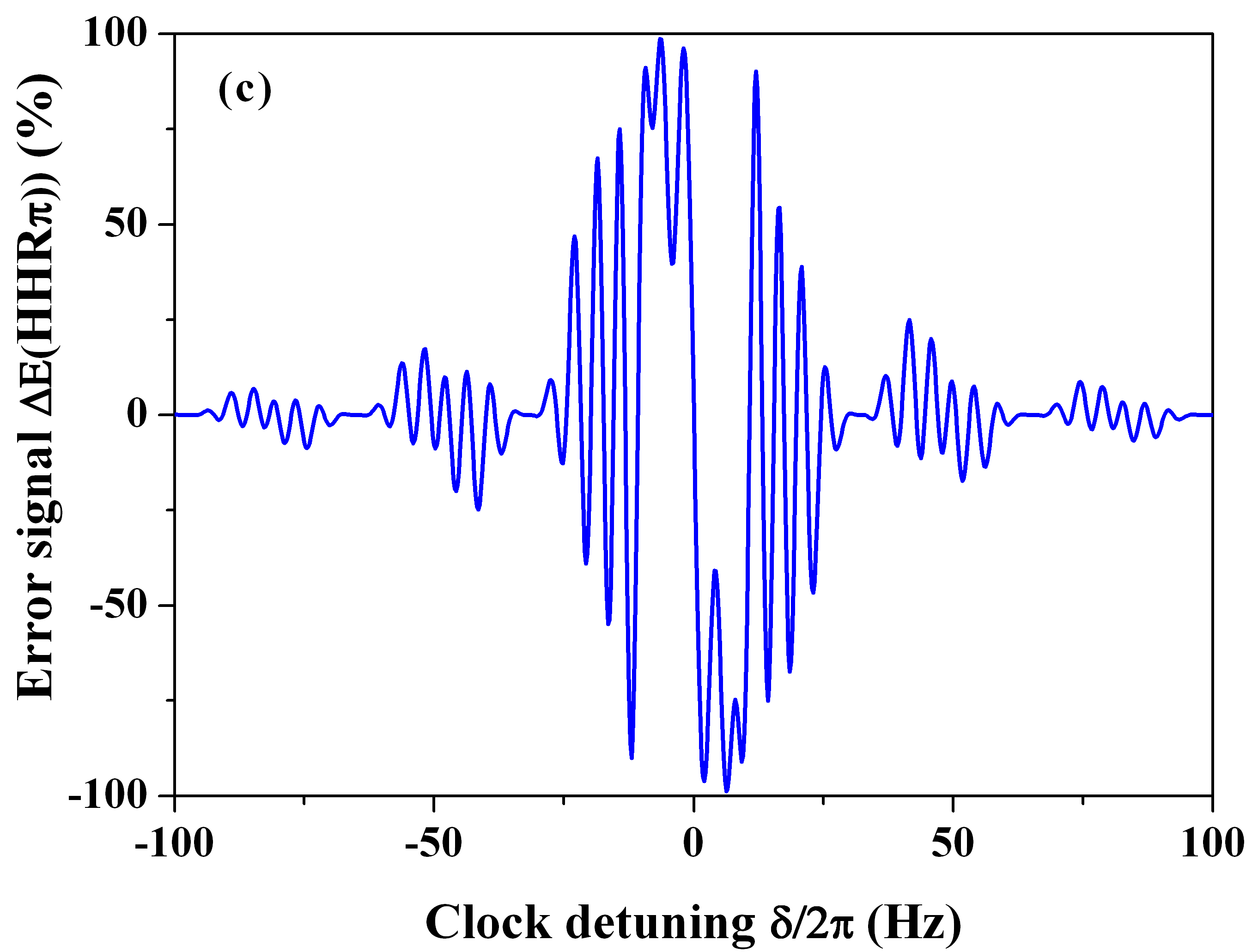}}\resizebox{9.cm}{!}{\includegraphics[angle=0]{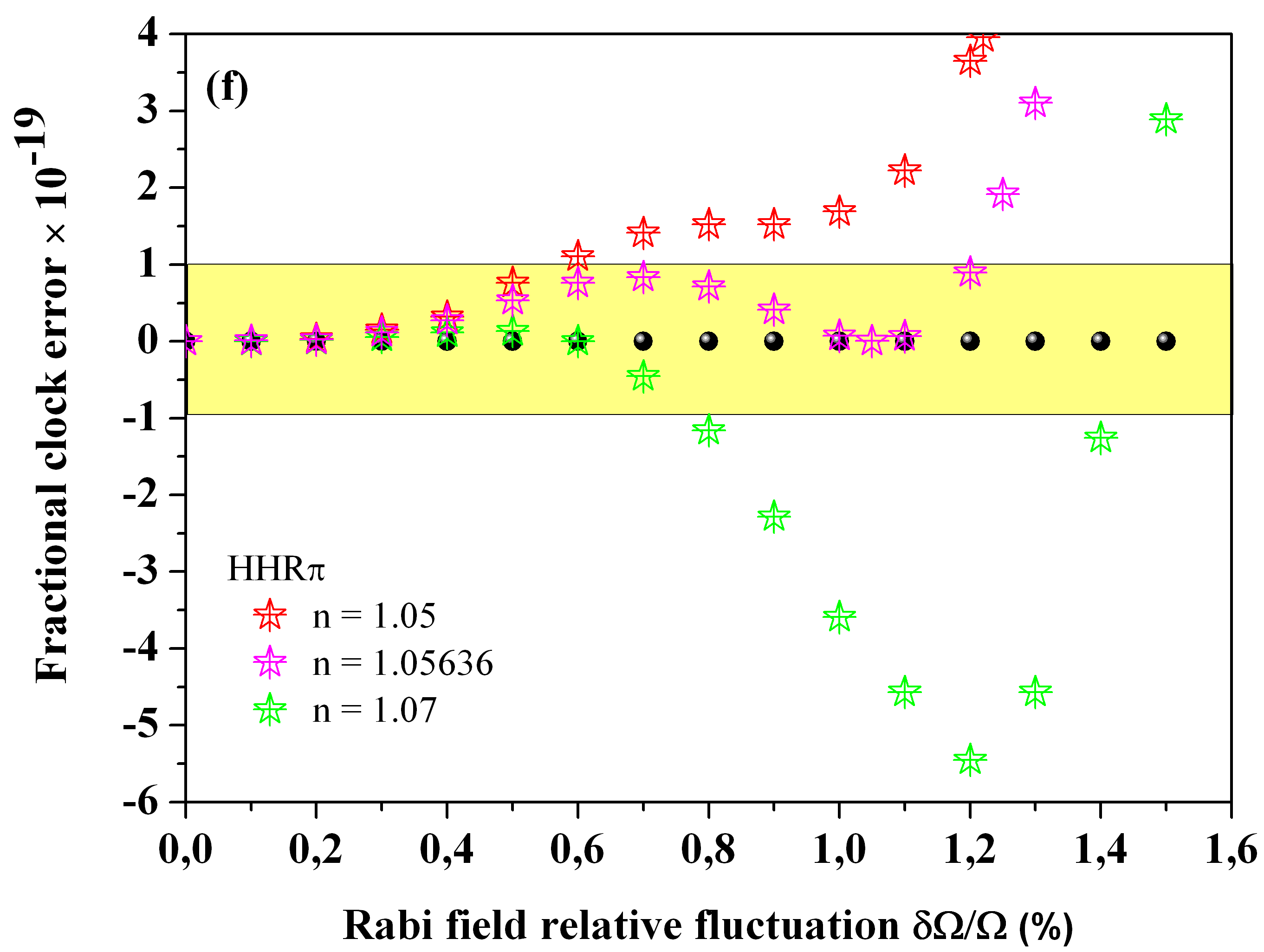}}
\caption{(color online). (Left panels) Dispersive error signals $\Delta\textup{E}$ generated with our computational algorithm based on a simulation of the $^{171}$Yb$^{+}$ ion clock interrogation with protocols from Tab.~\ref{clock-protocol-table}. (a) HR3$_{\pi}$ with 3 pulses. (b) HR5$_{\pi}$ with 5 pulses. (c) HHR$_{\pi}$ with 4 pulses. Laser pulse parameters are $\tau=30.5$ms, free evolution times around T$=122$ms under a mean compensated light-shift of $\Delta_{LS}/2\pi=95$Hz. A mean Rabi frequency $\Omega/2\pi=8.20$ Hz is fixed as in ref~\cite{Beloy:2018}. (Right panels) Corresponding fractional clock errors (d), (e) and (f) versus relative probe-laser fluctuation $\delta\Omega/\Omega$. We also report GHR$(\pi/4)$, GHR$(3\pi/4)$ protocols with $\bullet$ black dots for comparison. The pulse area $\Omega\tau=\textup{n}\times\pi/2$ ($\pi/2\equiv\boldsymbol{90}^{\circ}$) is driven by the parameter \textup{n} between 0.9 and 1.1 as described in~\cite{Beloy:2018}. Note the change in scale between the region in grey indicating a fractional clock-frequency shift correction at the $10^{-18}$ and the yellow region for a $10^{-19}$ level of relative accuracy.}
\label{fig2}
\end{figure*}

\section{HYPER-CLOCKS}

\subsection{HYPER-RAMSEY}

\indent In order to validate our composite pulse building-block simulator, we consider a four-pulse protocol for an hyper-clock with a fine tuning of the frequency shift of clock interferences. This general configuration allows us to retrieve any two-pulse or three-pulse schemes reported in Tab.~\ref{clock-protocol-table} by switching-off appropriate pulse parameters from diagrams shown in Fig.~\ref{fig:GHR-sequence}.
For a clock configuration, we consider a single trapped ion with an ultra-narrow optical transition confined into a Lamb-Dicke regime, i.e
where Doppler and recoil shifts are eliminated~\cite{Huntemann:2016,Sanner:2019,Brewer:2019}.

Dispersive error signals are produced to estimate precisely the center of the interferometric resonance eliminating any shape distortion~\cite{Ramsey:1951,Yudin:2010,Zanon-Willette:2015}.
They are generated by applying opposite phase-steps $\pm\varphi$ on particular laser pulses and are computed by taking the difference between two phase-shifted transition probabilities $_{p}^{q}\textup{P}_{e}(\pm\varphi)=1-_{p}^{q}\textup{P}_{g}$ as~\cite{Zanon-willette:2017,Zanon-Willette:2018}:
\begin{equation}
\begin{split}
\Delta\textup{E}(\varphi)=_{p}^{q}\textup{P}_{e}(\pm\varphi)-_{p}^{q}\textup{P}_{e}(\mp\varphi),
\label{eq:error-signal}
\end{split}
\end{equation}
\indent We establish explicitly two-level operator components from an interaction matrix $_{2}^{2}$M to evaluate all laser pulse hyper-clock protocols reported in Tab.~\ref{clock-protocol-table}.
The matrix coefficients $_{2}^{2}C_{gg},_{2}^{2}C_{ge}$ driving a four-pulse probe interrogation scheme are computed using the following elements:
\begin{subequations}
\begin{align}
\alpha_{1}'^{2}(gg)=&\left(\prod_{l=\textup{1}}^{p=\textup{2}}\cos\widetilde{\vartheta}'_{l}e^{i\phi'_{\textup{l}}}\right)\cdot\left(1-S'_{2,2}\right),\\
\alpha_{1}^{2}(gg)=&\left(\prod_{l=\textup{1}}^{q=\textup{2}}\cos\widetilde{\vartheta}_{l}e^{i\phi_{\textup{l}}}\right)\cdot\left(1-S_{2,2}\right),\\
\alpha_{1}'^{2}(ge)=&-ie^{-i(\phi'_{2}+\varphi'_{2}+\Xi'_{2})}\cdot\left(\prod_{l=\textup{1}}^{p=\textup{2}}\cos\widetilde{\vartheta}'_{l}e^{i\phi'_{\textup{l}}}\right)\cdot\left(S'_{2,1}\right),
\end{align}
\end{subequations}
where
\begin{subequations}
\begin{align}
S'_{2,2}(gg)=&e^{-i\Xi'_{12}}\tan\widetilde{\vartheta}'_{1}\tan\widetilde{\vartheta}'_{2},\\
S'_{2,1}(ge)=&\tan\widetilde{\vartheta}'_{1}+e^{i\Xi'_{2}}\tan\widetilde{\vartheta}'_{2},\\
S_{2,2}(gg)=&e^{-i\Xi_{12}}\tan\widetilde{\vartheta}_{1}\tan\widetilde{\vartheta}_{2}.
\end{align}
\end{subequations}
The corresponding complex phase factor read:
\begin{subequations}
\begin{align}
\beta_{1}'^{2}(gg)=&\frac{\tan\widetilde{\vartheta}'_{1}+e^{-i\Xi'_{12}}\tan\widetilde{\vartheta}'_{2}}
{1-e^{-i\Xi'_{12}}\tan\widetilde{\vartheta}'_{1}\tan\widetilde{\vartheta}'_{2}},\\
\beta_{1}^{2}(gg)=&\frac{\tan\widetilde{\vartheta}_{1}+e^{-i\Xi_{12}}\tan\widetilde{\vartheta}_{2}}
{1-e^{-i\Xi_{12}}\tan\widetilde{\vartheta}_{1}\tan\widetilde{\vartheta}_{2}},\\
\beta_{1}'^{2}(ge)=&\frac{1}{\{\beta_{1}'^{2}(gg)\}^{*}}.
\end{align}
\end{subequations}
We give the decomposition of phase factor expressions as following:
\begin{subequations}
\begin{align}
\Xi'_{1}=&0\\
\Xi'_{2}\equiv&\Xi'_{12}
\end{align}
\end{subequations}
\begin{subequations}
\begin{align}
\Xi'_{12}=&\varphi'_{1}-\varphi'_{2}+\phi'_{1}+\phi'_{2},\\
\Xi_{12}=&\varphi_{2}-\varphi_{1}+\phi_{1}+\phi_{2}.
\end{align}
\end{subequations}
See also~\cite{Footnote} as another computational way to obtain directly $_{2}^{2}C_{ge}$ from $_{2}^{2}C_{gg}$.\\

The transition probability can be deduced by measuring the atomic population fraction remaining in the ground state or pushed to the excited state with appropriate normalization.
Hyper-Ramsey HR3$_{\pi}$, HR5$_{\pi}$ and generalized hyper-Ramsey GHR($\pi/4$) or GHR($3\pi/4$) protocols are compared to a new hybrid hyper-Ramsey HHR$_{\pi}$ protocol using four pulses from Tab.~\ref{clock-protocol-table}. We employ the Beloy's model to analyze the laser-probe-intensity fluctuation between repetitive sequences of composite pulses for clock interrogation~\cite{Beloy:2018}. The recent hyper-Ramsey HR5$_{\pi}$ composite five pulse protocol has been proposed to reduce by two additional orders of magnitude the residual frequency-shift of the quantum interferences compared to the original Hyper-Ramsey HR3$_{\pi}$~\cite{Zanon-Willette:2021}. It has been demonstrated that the clock frequency-shift is a quintic function of the residual light-shift as $\propto\left(\Omega/\Delta\right)^{5}$ instead of the cubic dependence as $\propto\left(\Omega/\Delta\right)^{3}$ associated to the HR3$_{\pi}$ protocol~\cite{Zanon-Willette:2021}.

Typical dispersive shapes for $\Delta E$ are presented in Fig.~\ref{fig2} (a), (b), (c).
The fractional clock error, related to the specific electric octupole (E3) clock transition of a single trapped ion $^{171}$Yb$^{+}$, is reported in Fig.~\ref{fig2} (d), (e) and (f).
The robustness of the HR3$_{\pi}$ protocol exhibits a relative quartic sensitivity to
the Rabi field fluctuation $\delta\Omega/\Omega$ initially predicted by~\cite{Beloy:2018} while a GHR($\pi/4$) (or equivalently GHR($3\pi/4$)) protocol does not suffer from any residual light-shift correction even for a constrained $\pm10\%$ pulse area variation. As expected, the HR fractional clock correction requires a laser field control below $0.5\%$ (intensity below $1\%$) to reach a $5\times10^{-19}$ relative accuracy (grey region).
In Fig.~\ref{fig2}(b), a new hybrid HR5$_{\pi}$ protocol which replaces the $\boldsymbol{180}^{\circ}_{\pi}$ pulse by a composite $\boldsymbol{360}^{\circ}_{\pi}\boldsymbol{540}^{\circ}_{0}\boldsymbol{360}^{\circ}_{\pi}$ modifies the clock fractional error through a fine tuning control of the pulse area. This composite pulse sequence is not only able to remove the quartic dependance of the HR3$_{\pi}$ fractional clock error with the laser field fluctuation but is also capable of changing the sign of the clock frequency-shift. Similar results are obtained with the hybrid HHR$_{\pi}$ protocol which uses an additional $\boldsymbol{90}^{\circ'}_{0}$ laser pulse immediately after the first $\boldsymbol{90}^{\circ'}_{\pm\pi/2}$ phase-shifted Ramsey pulse.

Inspired by the precise tuning control of residual high-order light-shifts in optical lattices offered by the trap depth~\cite{Ushijima:2018}, we have also investigated a fine pulse area tuning control within the variation of the Rabi frequency through the \textit{n} parameter changed from 0.9 to 1.1 in Fig.\ref{fig2}(d) and Fig.\ref{fig2}(e) and from 1.05 to 1.07 in Fig.~\ref{fig2}(f) for different protocols.
The main result we obtain is, whatever any pulse area variation constraint at the $1\%$ has to be reached for a targeted fractional level of accuracy at the $10^{-19}$ level (yellow region), all GHR protocols with $\pi/4$ and $3\pi/4$ phase-steps remain very robust (as long as the phase-step is precisely controlled) since the residual part of the compensated light-shift is removed at all order in the clock detuning~\cite{Zanon-Willette:2016}.\\
\begin{figure*}[t!!]
\center
\resizebox{9.5cm}{!}{\includegraphics[angle=0]{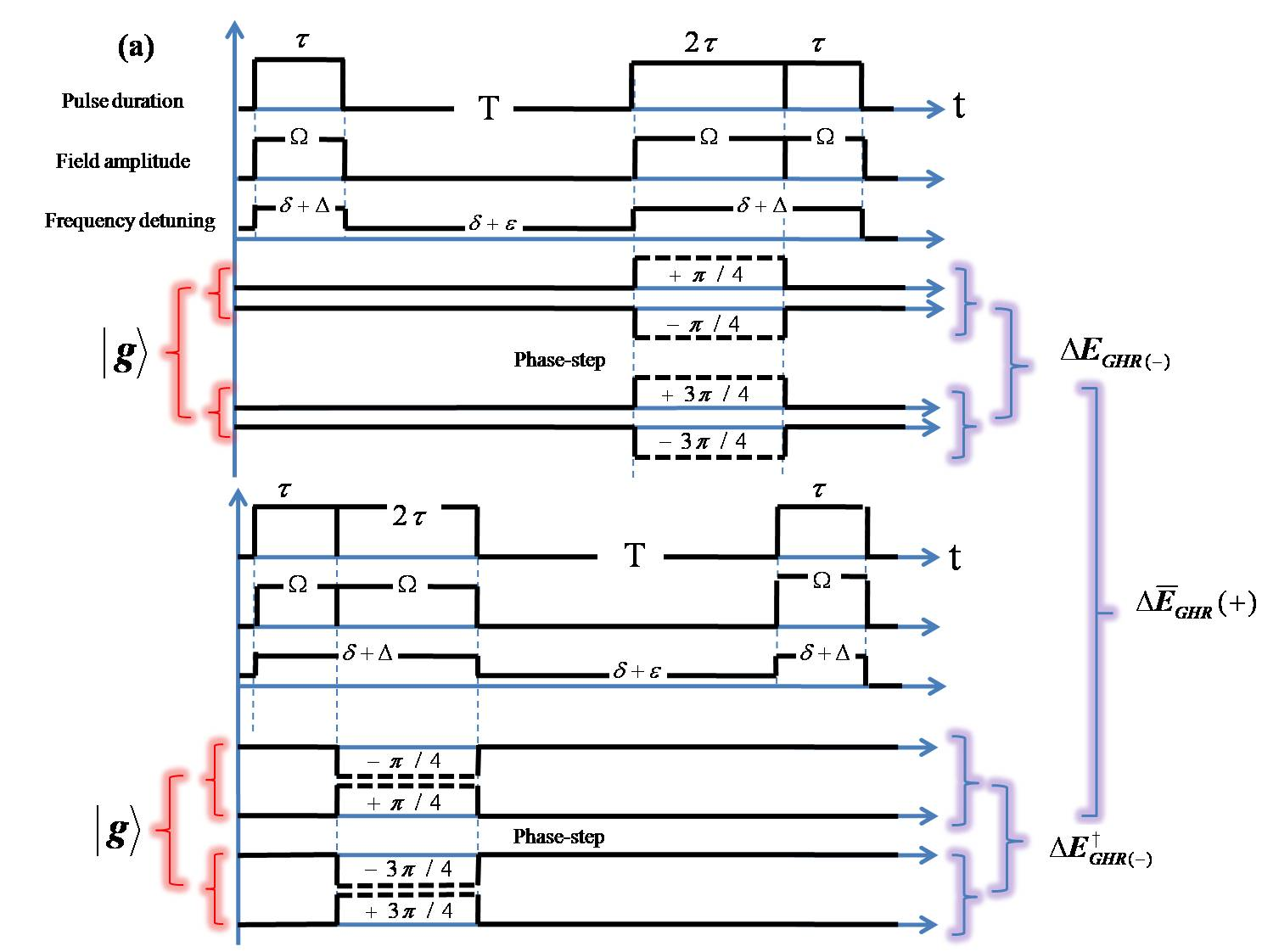}}\resizebox{8.5cm}{!}{\includegraphics[angle=0]{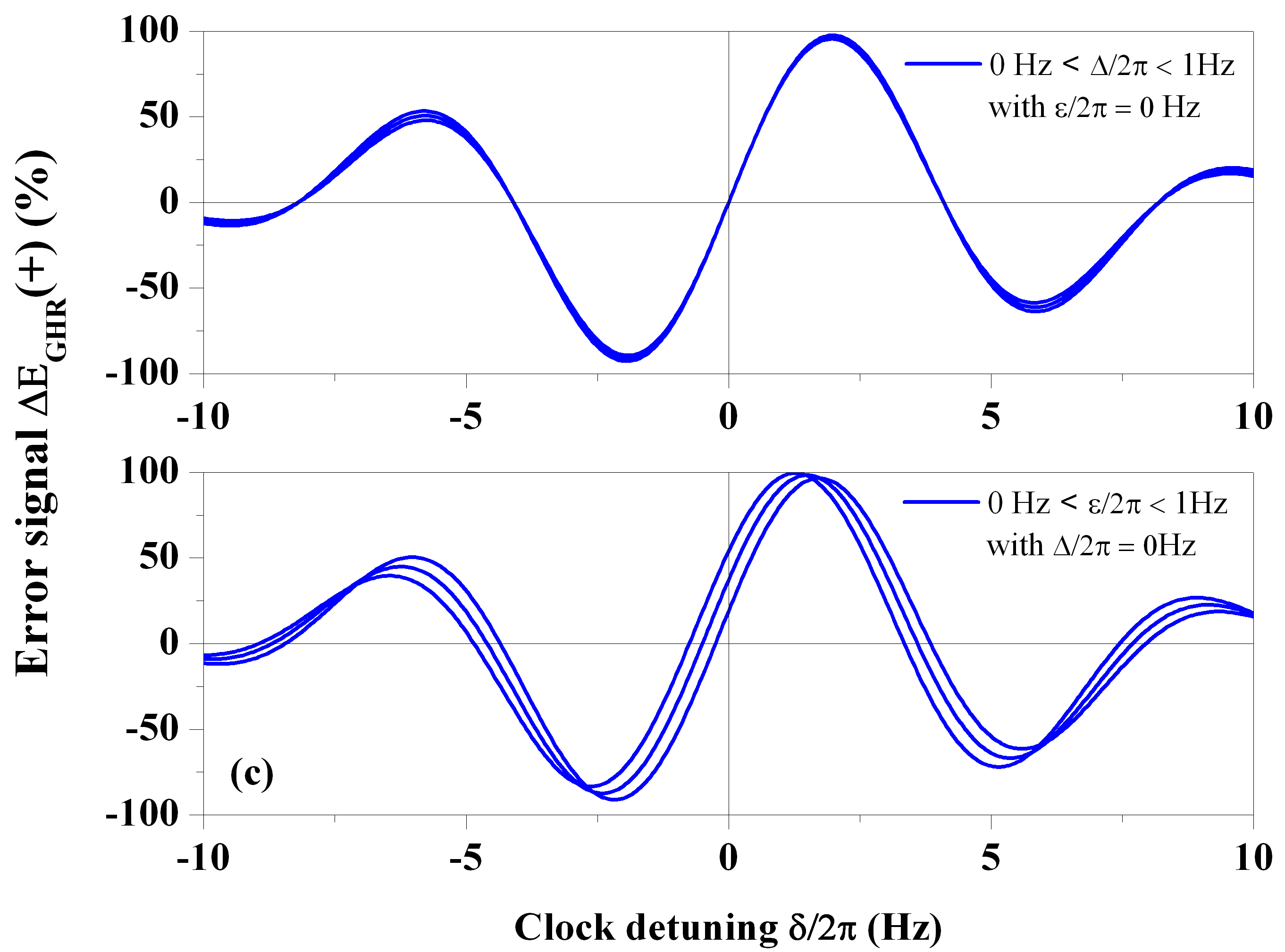}}
\resizebox{9.5cm}{!}{\includegraphics[angle=0]{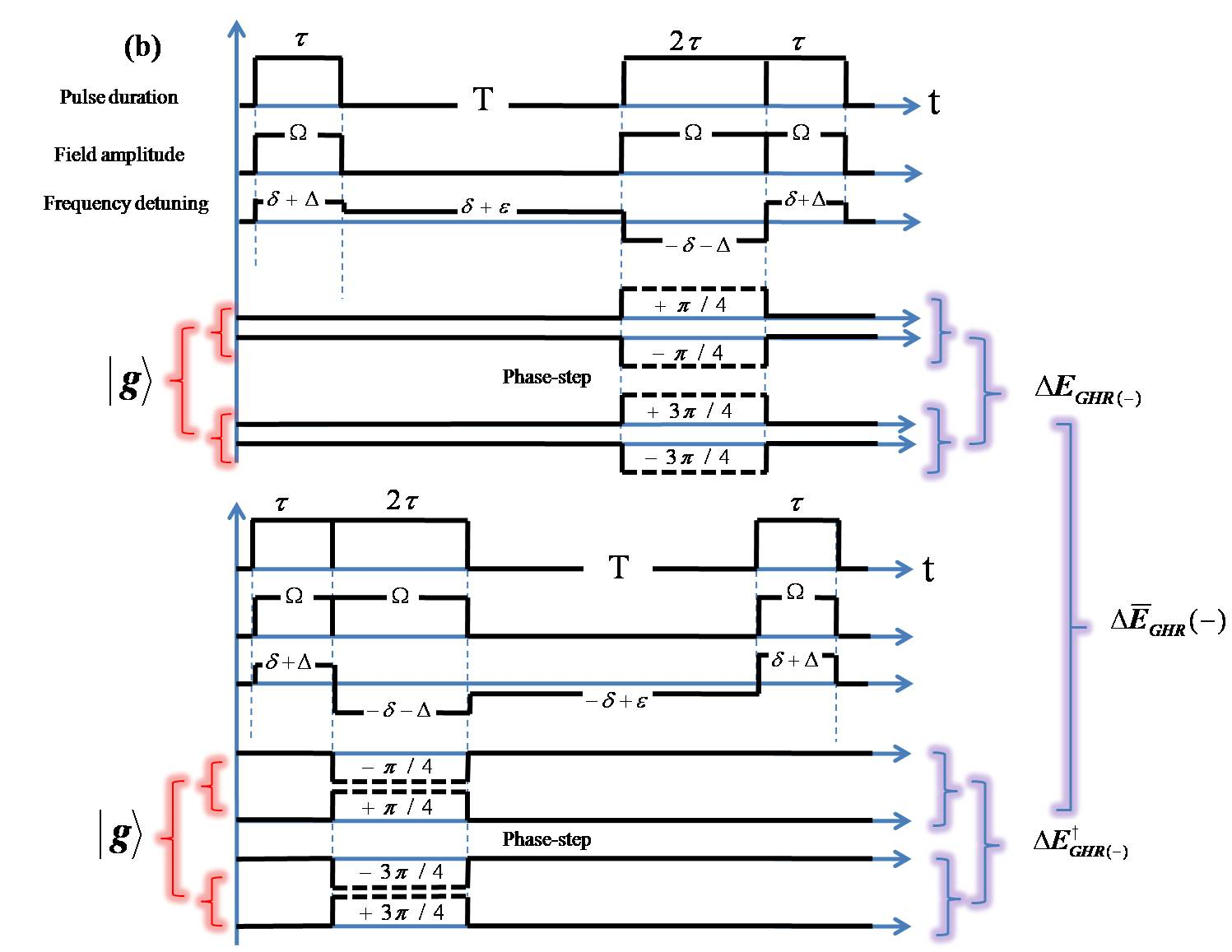}}\resizebox{8.5cm}{!}{\includegraphics[angle=0]{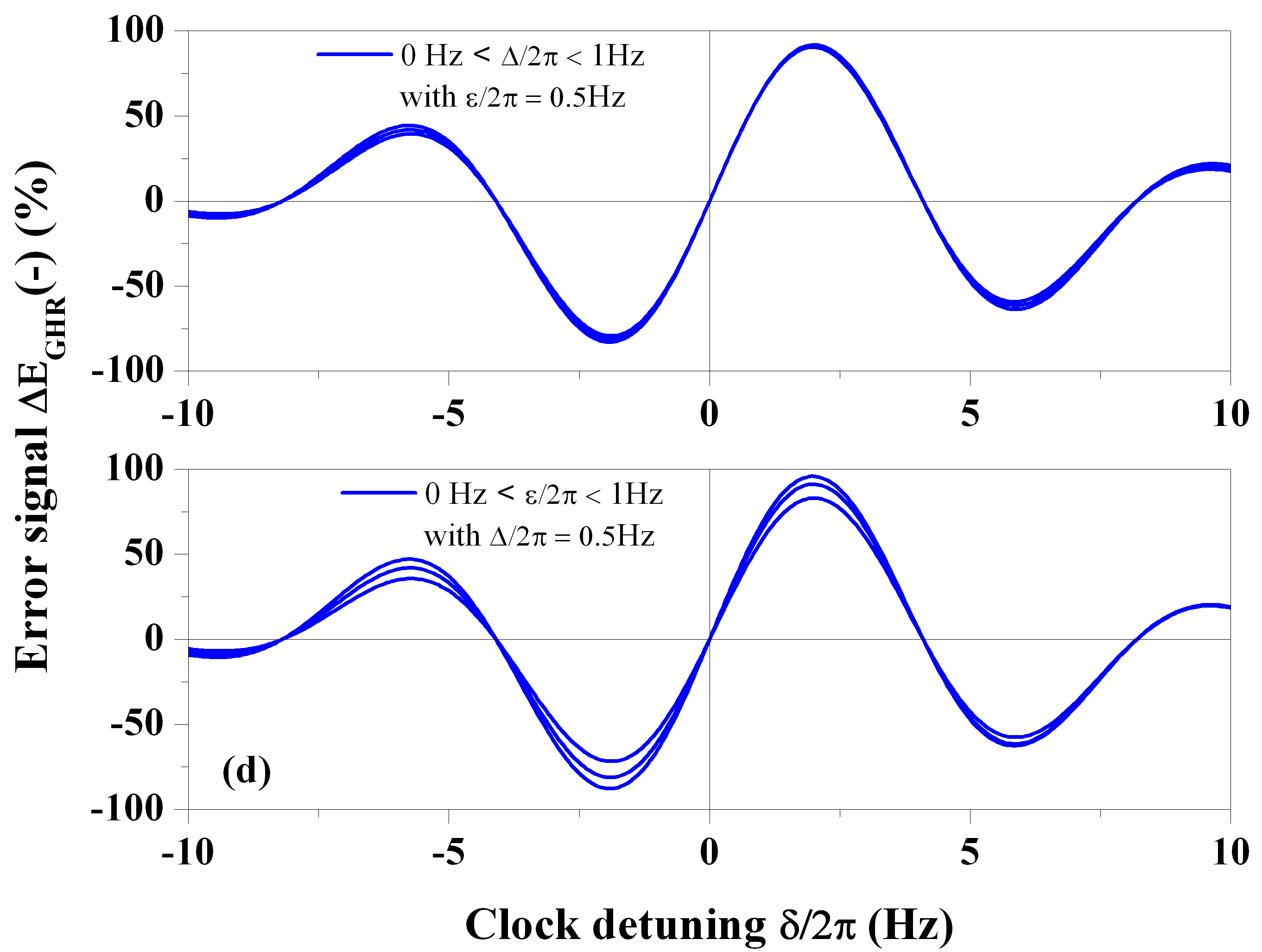}}
\caption{(color online). (a) and (b) GHR circuit-diagrams of laser parameters including a combination of pulse protocols to generate more robust error signals. Uncompensated part of a residual light-shift $\Delta/2\pi$ is present during pulses while a tiny pulse distortion $\epsilon/2\pi$ can be applied during the free evolution time.
(c) Dispersive error signal $\Delta\overline{\textup{E}}_{\textup{GHR}}(+)$ generated with Eq.~\ref{eq:GHR-error-signal-1} versus the clock detuning $\delta/2\pi$ for different residual uncompensated light-shifts ignoring the distortion during the free volution time (top panel) or with a small distortion applied during the free evolution time while ignoring uncompensated residual light-shifts (down panel). (d) Dispersive error signal $\Delta\overline{\textup{E}}_{\textup{GHR}}(-)$ generated with Eq.~\ref{eq:GHR-error-signal-2} versus the clock detuning $\delta/2\pi$ for different uncompensated parts of a residual light-shift including a fixed distortion during the free volution time (top panel) or with a tunable distortion applied during the free evolution time while fixing a residual uncompensated light-shifts (down panel).}
\label{fig-GHR}
\end{figure*}

\indent We switch to the additional concept of composite pulse cooperativity that is offered by GHR protocols to generate more robust error signals~\cite{Braun:2014,Zanon-willette:2017}. Composite pulses can provide significantly improved performance by compensating not only their pulse imperfections but also by reducing or eliminating additional distortion in a cooperative manner. For example, when dissipative processes can not be ignored, protocols that are relying on combination of $\pm\pi/4$ and $\pm3\pi/4$ laser phase-steps with their time reversal counterparts become very efficient~\cite{Zanon-willette:2017}.
To be more stringent, we shall now introduce another external weak perturbation $\epsilon$ during the single free evolution time. Such a perturbation is always adding constructively to the free clock detuning $\delta$ like a small drift of the laser probe frequency.

GHR circuit-diagrams of laser parameters and cooperative combination of error signals are reported in Fig.\ref{fig-GHR}(a) and (b).
They are produced with $\pm\pi/4$ or $\pm3\pi/4$ laser phase-steps, through Eq.~\ref{eq:error-signal} and are mixing error signals $\Delta\textup{E}_{\textup{GHR}(\pi/4)}$ and $\Delta\textup{E}_{\textup{GHR}(3\pi/4)}$ as following:
\begin{subequations}
\begin{align}
\Delta\textup{E}_{\textup{GHR}(-)}&=\frac{1}{2}\left(\Delta\textup{E}_{\textup{GHR}(\pi/4)}-\Delta\textup{E}_{\textup{GHR}(3\pi/4)}\right)\label{eq:GHR-error-signal-a},\\
\Delta\textup{E}^{\dagger}_{\textup{GHR}(-)}&=\frac{1}{2}\left(\Delta\textup{E}^{\dagger}_{\textup{GHR}(\pi/4)}-\Delta\textup{E}^{\dagger}_{\textup{GHR}(3\pi/4)}\right)\label{eq:GHR-error-signal-b},
\end{align}
\end{subequations}
Some additional cooperative protocols are generated by combining protocols with their time-reversal counterparts as following:
\begin{subequations}
\begin{align}
\Delta\overline{\textup{E}}_{\textup{GHR}(+)}&=\frac{1}{2}\left(\Delta\textup{E}_{\textup{GHR}(-)}+\Delta\textup{E}^{\dagger}_{\textup{GHR}(-)}\right)\label{eq:GHR-error-signal-1},\\
\Delta\overline{\textup{E}}_{\textup{GHR}(-)}&=\frac{1}{2}\left(\Delta\textup{E}_{\textup{GHR}(-)}-\Delta\textup{E}^{\dagger}_{\textup{GHR}(-)}\right)\label{eq:GHR-error-signal-2}.
\end{align}
\end{subequations}
The cooperativity of such pulse protocols provides more degrees of freedom in optimization of the pulse sequence against small distortions during free evolution time and laser pulses.
We have explored two configurations in Fig.\ref{fig-GHR}(a) $\Delta\overline{\textup{E}}_{\textup{GHR}}(+)$ and Fig.\ref{fig-GHR}(b) $\Delta\overline{\textup{E}}_{\textup{GHR}}(-)$ without and with alternating signs of clock detunings through the transformation $\delta+\Delta\rightarrow-\delta-\Delta$. The related error signal shifts are presented in Fig.\ref{fig-GHR}(c) and (d) respectively under a residual light-shift $\Delta/2\pi$ and a weak distortion $\epsilon/2\pi$ during the free evolution time. By combining error signals which also use clock detunings with opposite sign as in Fig.\ref{fig-GHR}(b), an additional suppression of the distortion-shift can be realize to compensate the central fringe shift as reported in Fig.\ref{fig-GHR}(d).

Combining multiple error signals are not only offering better robustness to small distortion during pulses and free evolution time but also to the signal distortion from ion motion heating in a single ion rf trap device. A potential fractional clock error below $5\times10^{-20}$ was estimated based on Eq.~\ref{eq:GHR-error-signal-a} in comparison to the original HR3$_{\pi}$ protocol~\cite{Kuznetsov:2019}. It has been recently demonstrated that GHR($\pi/4$) and GHR($3\pi/4$) schemes have a strong robustness against residual light-shift coupled to spontaneous emission in an optically dense medium of cold atoms~\cite{Barantsev:2020}.
Indeed, there is a large possibility of unexplored exotic composite pulse protocols for optical clocks due to the richness of the quantum Hilbert space engineering~\cite{Levitt:2008} including a manipulation of qubit detunings with opposite sign~\cite{Lin:2016}.
\begin{figure}[b!!]
\center
\resizebox{8.5cm}{!}{\includegraphics[angle=0]{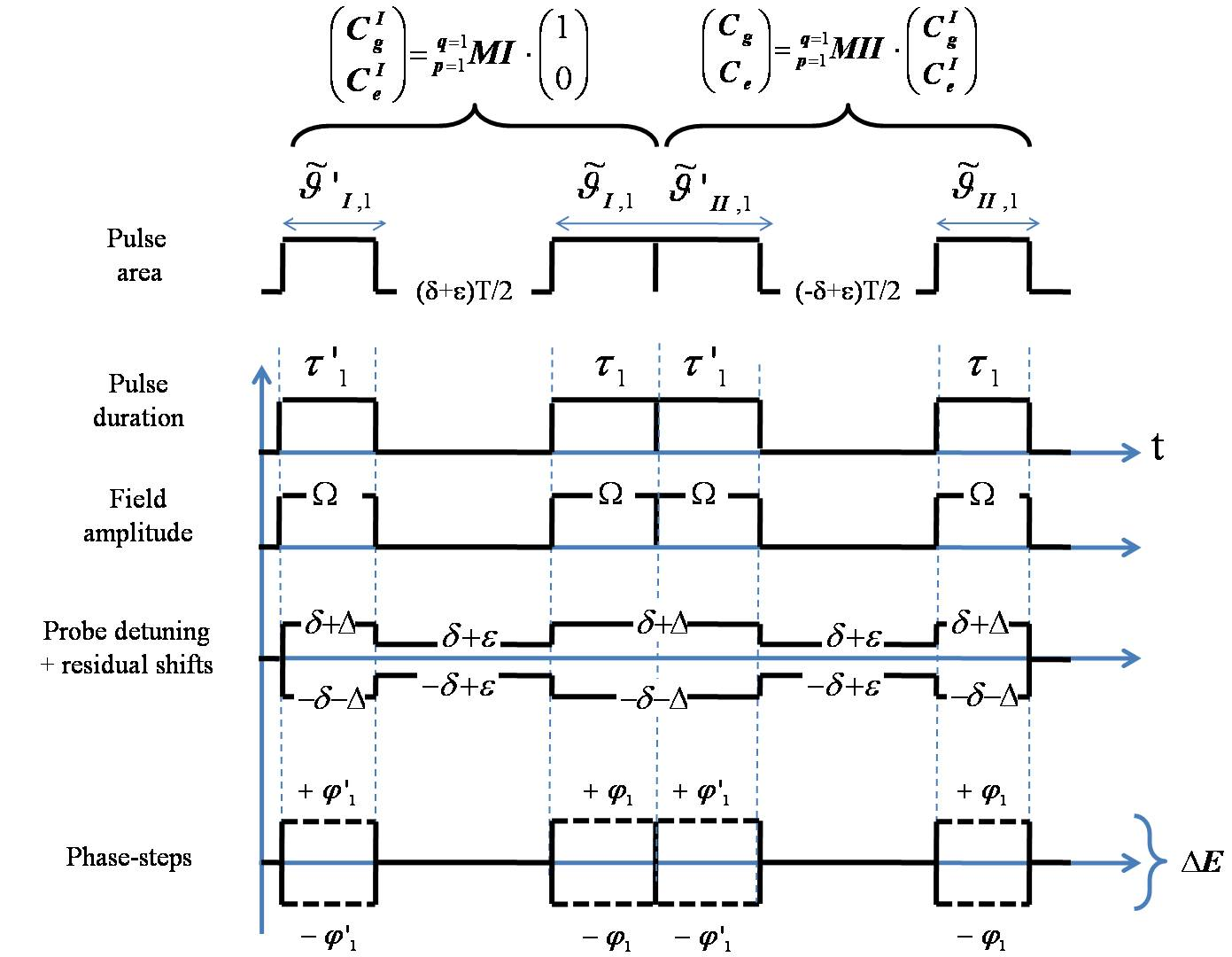}}
\caption{(color online). Four-pulse diagram generating the GHHR protocol. Same parameters as in Fig.~\ref{fig:GHR-sequence}. The protocol is divided into adjacent Ramsey two-zones with interaction matrices $_{1}^{1}$M\textbf{I} and $_{1}^{1}$M\textbf{II}. The free evolution time is $T\rightarrow T/2$ to obtain the same fringe periodicity as in Fig.~\ref{fig:GHR-sequence}.}
\label{fig:GHHR-sequence}
\end{figure}

\subsection{HYPER HAHN-RAMSEY}

\indent Another tool in clock interferometry is the fault-tolerant Hahn-Ramsey protocol also using frequency detunings with opposite sign~\cite{Vitanov:2015,Zlatanov:2020}. This scheme is designed to reduce any error in the carrier frequency of the driving oscillating field by applying opposite detunings during pulses and free evolution times in a spin-echo Hahn configuration as sketched in the panel of Fig.~\ref{fig:GHHR-sequence}.

We have extended the fault tolerant Hahn-Ramsey approach to a generalized hyper Hahn-Ramsey (GHHR) protocol which employs $\pi/4$ and $3\pi/4$ laser phase-steps in order to produce very robust error signals.
This is a straightforward modification of the generalized hyper-Ramsey spectroscopic scheme by adding a composite phase-modulated $\boldsymbol{90}^{\circ}_{\pm\varphi}\boldsymbol{90}^{\circ'}_{\pm\varphi}$ intermediate pulse exactly between $\boldsymbol{90}^{\circ'}$ and $\boldsymbol{90}^{\circ}$ pulses following the Hahn's spin-echo scheme~\cite{Hahn:1950}.
A reverse sign of the clock detuning, $\delta+\Delta\rightarrow-(\delta+\Delta)$ applied during the intermediate pulse, is preserving the shifted resonance condition by the residual light-shift.
Similarly, the free clock detuning, corrected by a small drift $\epsilon$ within the first evolution zone, becomes $\delta+\epsilon\rightarrow-\delta+\epsilon$ during the second free evolution zone as shown in Fig.~\ref{fig:GHHR-sequence}.
These manipulations of detunings offer an efficient decoupling of the static distortion $\epsilon$ from the residual light-shift $\Delta_{l}$ affecting the driving clock frequency of the probe pulses while
restoring Ramsey interferences associated to the free clock detuning $\delta$ at the output of the interferometer~\cite{Vitanov:2015}.
\begin{figure*}[t!!]
\center
\resizebox{9.5cm}{!}{\includegraphics[angle=0]{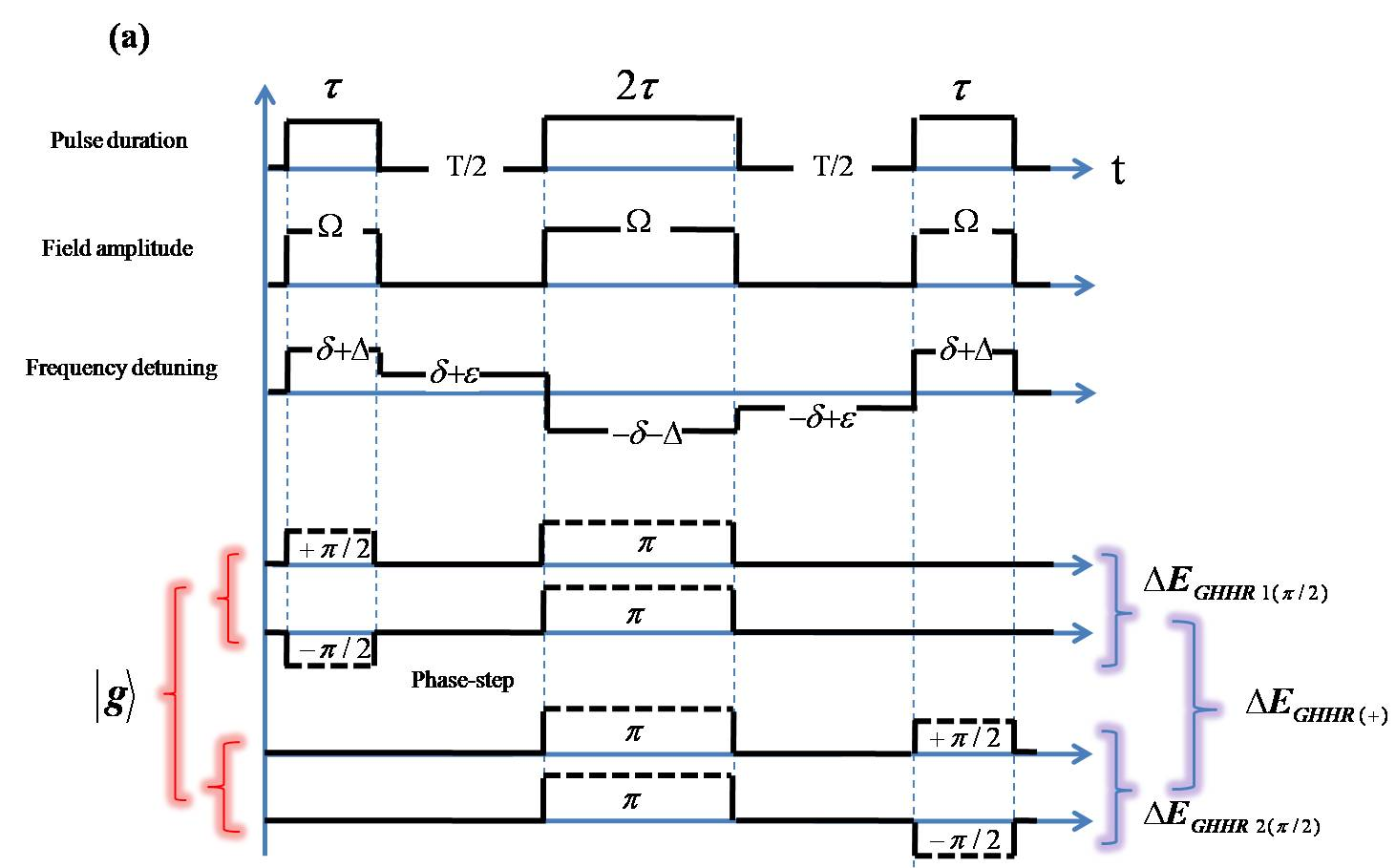}}\resizebox{8.5cm}{!}{\includegraphics[angle=0]{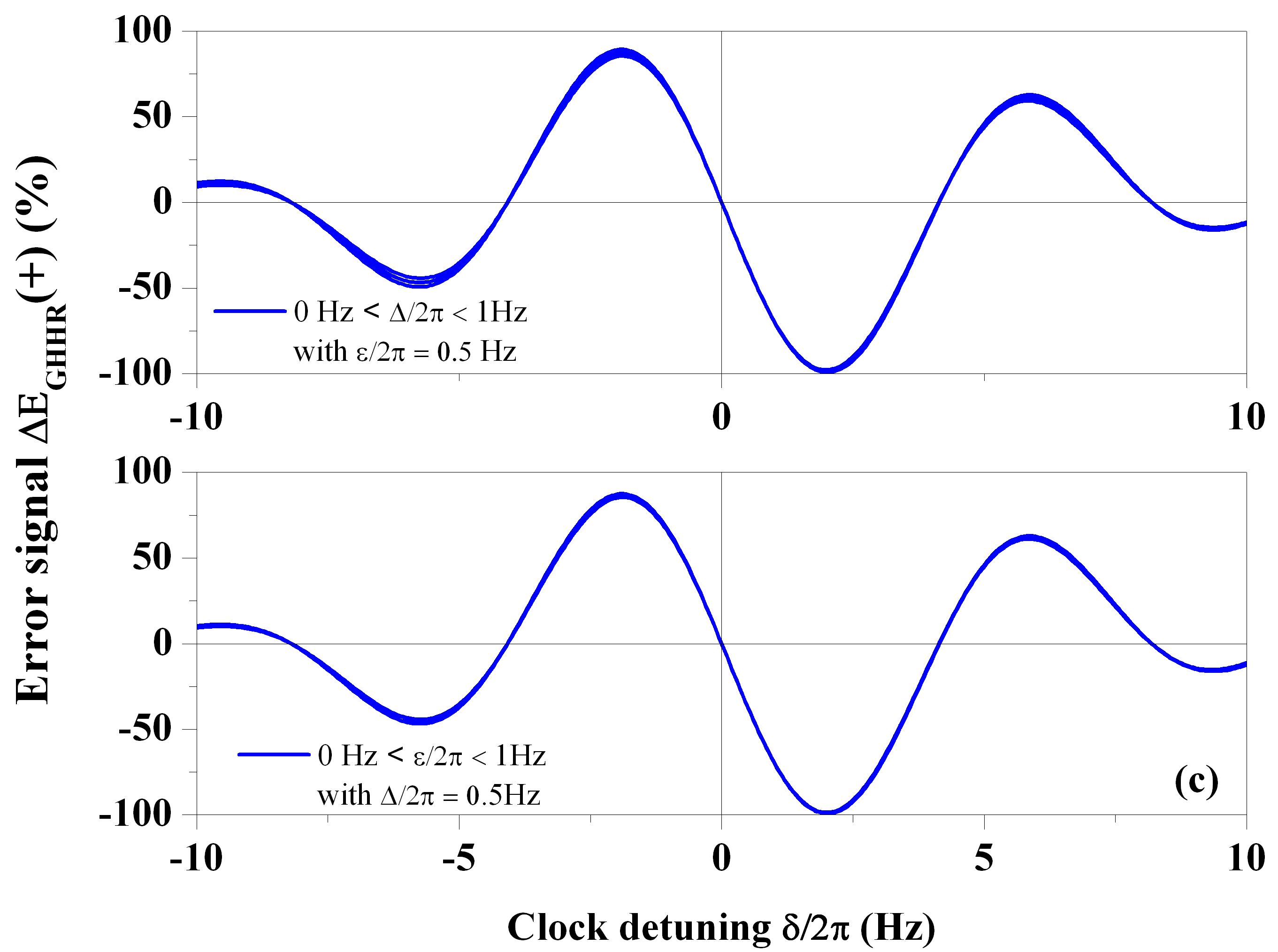}}
\resizebox{9.5cm}{!}{\includegraphics[angle=0]{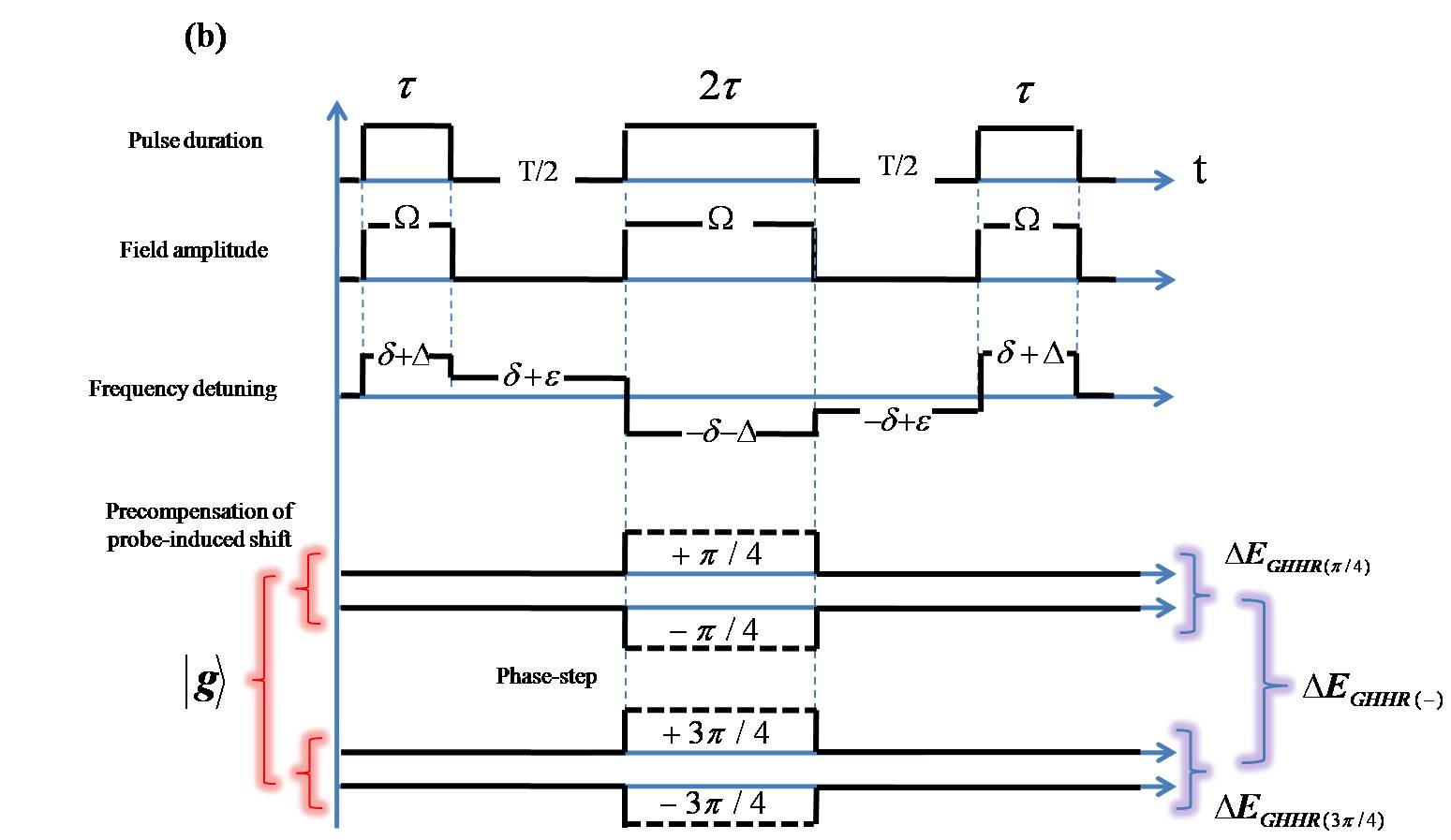}}\resizebox{8.5cm}{!}{\includegraphics[angle=0]{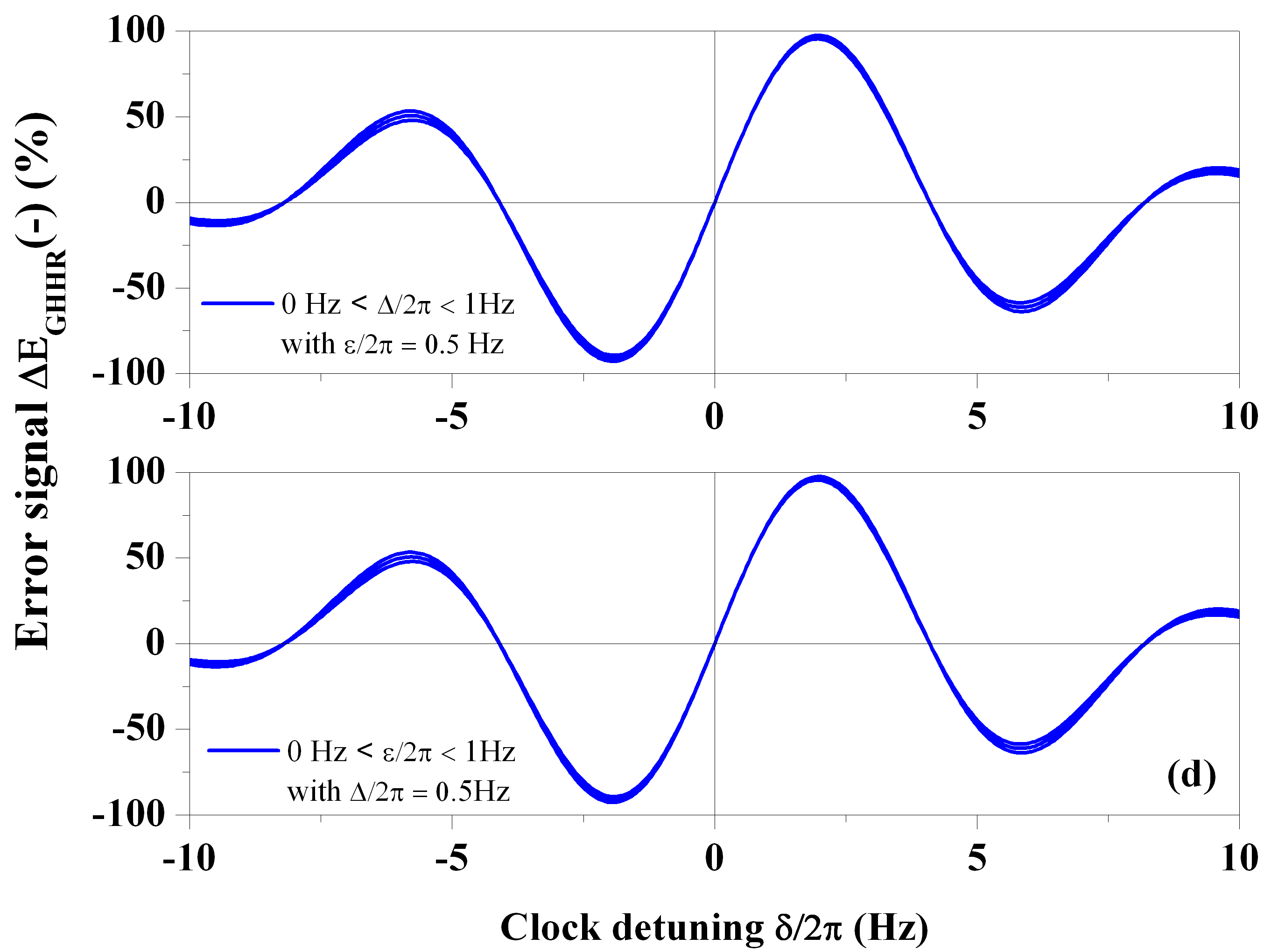}}
\caption{(color online). (a) and (b) GHHR circuit-diagrams of laser parameters generating robust error signals against uncompensated part of the residual light-shift $\Delta/2\pi$ coupled to a potential pulse distortion $\epsilon/2\pi$ during the free evolution time. (c) Dispersive error signal $\Delta\textup{E}_{\textup{GHHR}}(+)$ based on Eq.~\ref{eq:GHHR-error-signal-a} versus the clock detuning $\delta/2\pi$ for different residual uncompensated light-shifts including a fixed weak distortion during the free volution time (top panel) or with a tunable distortion applied during the free evolution time including a fixed residual uncompensated light-shift within pulses (down panel). (d) Dispersive error signal $\Delta\textup{E}_{\textup{GHHR}}(-)$ based on Eq.~\ref{eq:GHHR-error-signal-b} versus the clock detuning $\delta/2\pi$ for different residual uncompensated light-shifts including a fixed weak distortion during the free volution time (top panel) or with a tunable distortion applied during the free evolution time including a fixed residual uncompensated light-shift within pulses (down panel).}
\label{fig:GHHR-protocol}
\end{figure*}

To analytically derive the transition probability associated to the GHHR scheme, we choose to decompose the entire sequence of four pulses into two contiguous Ramsey interrogation zones $\boldsymbol{90}^{\circ'}_{0}\dashv\delta\textup{T}\vdash\boldsymbol{90}^{\circ}_{\pm\varphi}$ and $\boldsymbol{90}^{\circ'}_{\pm\varphi}\dashv\delta\textup{T}\vdash\boldsymbol{90}^{\circ}_{0}$ respectively modeled by two matrices $_{1}^{1}$M\textbf{I} and $_{1}^{1}$M\textbf{II} that are only differing in sign of their corresponding clock detunings during pulses (see Fig.~\ref{fig:GHHR-sequence}). The intermediate reversal pulse is changed into concatenated Ramsey pulses such that $\boldsymbol{180}^{\circ'}_{\pm\varphi}\equiv\boldsymbol{90}^{\circ}_{\pm\varphi}\boldsymbol{90}^{\circ'}_{\pm\varphi}$ as reported in Tab.~\ref{GHHR-protocol-table}. The full derivation can be achieved by applying twice Eq.~\ref{eq:matrix-components-a} (Eq.~\ref{eq:matrix-components-b}) first for evaluating the $_{1}^{1}$M\textbf{I} two-pulse Ramsey interrogation of the GHHR protocol then using wave-function solutions as initial conditions for the second $_{1}^{1}$M\textbf{II} two-pulse Ramsey interrogation.
\begin{table}[t!!]
\renewcommand{\arraystretch}{1.7}
\begin{tabular}{|c|c|c|}
\hline
   \begin{tabular}{c}
      GHHR1($\varphi$) \\
       ($\varphi=\pi/2$)\\
      GHHR2($\varphi$) \\
      ($\varphi=\pi/2$)
       \end{tabular}
    &  \begin{tabular}{c}
$\boldsymbol{90}^{\circ'}_{\pm\varphi}\dashv\frac{\delta\textup{T}}{2}\vdash\underbrace{\boldsymbol{90}^{\circ}_{\pi}\boldsymbol{90}^{\circ'}_{\pi}}\dashv-\frac{\delta\textup{T}}{2}\vdash\boldsymbol{90}^{\circ}_{0}$\\ $\boldsymbol{180}^{\circ'}_{\pi}$
\\
 $\boldsymbol{90}^{\circ'}_{0}\dashv\frac{\delta\textup{T}}{2}\vdash\underbrace{\boldsymbol{90}^{\circ}_{\pi}\boldsymbol{90}^{\circ'}_{\pi}}\dashv-\frac{\delta\textup{T}}{2}\vdash\boldsymbol{90}^{\circ}_{\pm\varphi}$\\
                                  $\boldsymbol{180}^{\circ'}_{\pi}$
\\
                                   \end{tabular}
\\
\hline
\hline
   \begin{tabular}{c}
      GHHR($\varphi$) \\
      ($\varphi=\pi/4,3\pi/4$)
       \end{tabular}
    &  \begin{tabular}{c}
$\boldsymbol{90}^{\circ'}_{0}\dashv\frac{\delta\textup{T}}{2}\vdash\underbrace{\boldsymbol{90}^{\circ}_{\pm\varphi}\boldsymbol{90}^{\circ'}_{\pm\varphi}}\dashv-\frac{\delta\textup{T}}{2}\vdash\boldsymbol{90}^{\circ}_{0}$\\
                                $\boldsymbol{180}^{\circ'}_{\pm\varphi}$
                                   \end{tabular}
\\
\hline
\end{tabular}
\centering%
\caption{Generalized hyper-Hahn-Ramsey (GHHR) protocols for hyper-clocks. The concatenated double pulse $\boldsymbol{90}^{\circ}_{\pm\varphi}\boldsymbol{90}^{\circ'}_{\pm\varphi}$ is equivalent to a $\boldsymbol{180}^{\circ'}_{\pm\varphi}$ pulse. See the text for explanation.}
\label{GHHR-protocol-table}
\end{table}
We give, in this subsection, some matrix elements that are necessary to evaluate the full transition probability associated to the generation of Hahn-Ramsey quantum interferences. We explicitly have for $_{1}^{1}$M$\textbf{I}$, $_{1}^{1}$M$\textbf{II}$ interaction zones labeled by $m=I,II$:
\begin{subequations}
\begin{align}
\alpha_{1}'^{1}(gg)^{m}=&\cos\widetilde{\vartheta}'_{\textup{m,1}}e^{i\phi'_{\textup{m,1}}},\\
\alpha_{1}^{1}(gg)^{m}=&\cos\widetilde{\vartheta}_{\textup{m,1}}e^{i\phi_{\textup{m,1}}},\\
\alpha_{1}'^{1}(ge)^{m}=&-ie^{-i(\phi'_{\textup{m,1}}+\varphi'_{\textup{m,1}})}\cos\widetilde{\vartheta}'_{\textup{m,1}}e^{i\phi'_{\textup{m,1}}}\tan\widetilde{\vartheta}'_{\textup{m,1}},
\end{align}
\label{eq:alpha-function-GHHR}
\end{subequations}
\begin{subequations}
\begin{align}
\beta_{1}'^{1}(gg)^{m}=&\tan\widetilde{\vartheta}'_{\textup{m,1}},\\
\beta_{1}^{1}(gg)^{m}=&\tan\widetilde{\vartheta}_{\textup{m,1}},\\
\beta_{1}'^{1}(ge)^{m}=&\frac{1}{\tan\widetilde{\vartheta}'_{\textup{m,1}}}.
\end{align}
\label{eq:beta-function-GHHR}
\end{subequations}
The coefficients $_{1}^{1}C^{m}_{gg},_{1}^{1}C^{m}_{ge}$ required to compute each $_{1}^{1}$M$\textbf{I}$ and $_{1}^{1}$M$\textbf{II}$ building-block components, are given by:
\begin{subequations}
\begin{align}
_{1}^{1}C^{m}_{gg}=&_{1}^{1}\alpha^{m}_{gg}e^{i\delta_{m}\textup{T}_{m}/2}\left[1-|_{1}^{1}\beta^{m}_{gg}|e^{-i(\delta_{m}\textup{T}_{m}+_{1}^{1}\Phi^{m}_{gg})}\right],\\
_{1}^{1}C^{m}_{ge}=&_{1}^{1}\alpha^{m}_{ge}e^{-i\delta_{m}\textup{T}_{m}/2}\left[1+|_{1}^{1}\beta^{m}_{ge}|e^{-i(\delta_{m}\textup{T}_{m}+_{1}^{1}\Phi^{m}_{ge})}\right],
\end{align}
\label{eq:coefficient-function-GHHR}
\end{subequations}
where phase-shifts are $\Phi^{m}_{gg}\equiv\Phi^{m}_{ge}=\varphi^{m}_{L}+\phi^{m}_{L}$ with $\varphi^{m}_{L}=\varphi_{m,1}-\varphi'_{m,1}$ and $\phi^{m}_{L}=\phi'_{m,1}+\phi_{m,1}$.
Final amplitudes of transition probability are easily computed and written as:
\begin{subequations}
\begin{align}
C_{g}(t)=&_{1}^{1}C^{I}_{eg}\cdot_{1}^{1}C^{II}_{ge}+_{1}^{1}C^{I}_{gg}\cdot_{1}^{1}C^{II}_{gg}\\
=&-_{1}^{1}C^{I*}_{ge}\cdot_{1}^{1}C^{II}_{ge}+_{1}^{1}C^{I}_{gg}\cdot_{1}^{1}C^{II}_{gg},\\
C_{e}(t)=&_{1}^{1}C^{I}_{eg}\cdot_{1}^{1}C^{II}_{ee}+_{1}^{1}C^{I}_{gg}\cdot_{1}^{1}C^{II}_{eg},\\
=&_{1}^{1}C^{I*}_{ge}\cdot_{1}^{1}C^{II*}_{gg}+_{1}^{1}C^{I}_{gg}\cdot_{1}^{1}C^{II*}_{ge},
\end{align}
\label{eq:amplitude-function-GHHR}
\end{subequations}
The set of previous equations from Eq.~\ref{eq:alpha-function-GHHR} to Eq.~\ref{eq:amplitude-function-GHHR} are general and can be also used to simulate GHR and GHHR error signals based on one or two free evolution zones. When two clock detunings are selected to be opposite in sign in each free evolution zone such that $\delta_{I}=-\delta_{II}$ while fixing $d_{+}\textup{T}_{+}=0$ and $d_{-}\textup{T}_{-}=2\delta\textup{T}$, an exact formula can be derived using some materials given in the S2 section from the appendix.
The exact GHHR amplitude is therefore reducing to an identical form adopted with Eq.~\ref{eq:Cgg} as:
\begin{equation}
\begin{split}
C_{g}(t)=\alpha_{gg}\left(1-\beta_{gg}(\textup{T})\left[1+\frac{\tan\widetilde{\vartheta}'_{\textup{I,1}}}{\tan\widetilde{\vartheta}_{\textup{II,1}}}e^{-i\left(2\delta\textup{T}+\Phi_{gg}\right)}\right]\right)
\end{split}
\label{eq:GHHR-amplitude-1}
\end{equation}
where envelops are now given by:
\begin{equation}
\begin{split}
\alpha_{gg}\equiv&\cos\widetilde{\vartheta}'_{\textup{I,1}}\cos\widetilde{\vartheta}'_{\textup{II,1}}\cos\widetilde{\vartheta}_{\textup{II,1}}e^{i(\phi'_{\textup{I,1}}+\phi'_{\textup{II,1}}+\phi_{\textup{II,1}})}\\
&\times\left(1-\tan\widetilde{\vartheta}'_{\textup{I,1}}\tan\widetilde{\vartheta}_{\textup{II,1}}e^{-i\Phi^{+}_{gg}}\right)\\
\beta_{gg}(\textup{T})\equiv&\frac{\tan\widetilde{\vartheta}'_{\textup{II,1}}\tan\widetilde{\vartheta}_{\textup{II,1}}}{\left(1-\tan\widetilde{\vartheta}'_{\textup{I,1}}\tan\widetilde{\vartheta}_{\textup{II,1}}e^{-i\Phi^{+}_{gg}}\right)}e^{i\left(\delta\textup{T}+\Phi^{II}_{gg}\right)}
\end{split}
\label{eq:GHHR-amplitude-2}
\end{equation}
The $\beta_{gg}(\textup{T})$ function is still modulated by the phase factor $d\textup{T}$. A good approximation of the previous expression eliminating the residual modulation is found to be:
\begin{equation}
\begin{split}
C_{g}(t)\approx\alpha_{gg}\left[1+\beta_{gg}e^{-i\left(2\delta\textup{T}+\Phi_{gg}\right)}\right]
\end{split}
\label{eq:GHHR-amplitude-simple}
\end{equation}
where:
\begin{equation}
\begin{split}
\alpha_{gg}=&\cos\widetilde{\vartheta}'_{\textup{I,1}}\sin\widetilde{\vartheta}'_{\textup{II,1}}\sin\widetilde{\vartheta}_{\textup{II,1}}e^{i(\phi'_{\textup{I,1}}+\phi'_{\textup{II,1}}+\phi_{\textup{II,1}})}\\
\beta_{gg}=&\frac{\tan\widetilde{\vartheta}'_{\textup{I,1}}}{\tan\widetilde{\vartheta}_{\textup{II,1}}}
\end{split}
\label{eq:GHHR-amplitude-2}
\end{equation}
and the phase-shift is:
\begin{equation}
\begin{split}
\Phi_{gg}=-\varphi'_{I,1}+2\varphi'_{II,1}-\varphi_{II,1}+\phi'_{\textup{I,1}}-\phi_{\textup{II,1}}
\end{split}
\label{eq:GHHR-phase-shift}
\end{equation}
Note that the Hahn-Ramsey phase-shift expression is formally equivalent to the Mach-Zehnder scheme applied in atomic interferometry.

\indent By producing error signals with either $\pm\pi/2$ phase-steps or $\pm\pi/4,\pm3\pi/4$ phase-steps, we generate hybrid combinations of GHHR error signals as following:
\begin{subequations}
\begin{align}
\Delta\textup{E}_{\textup{GHHR}(+)}&=\frac{1}{2}\left(\Delta\textup{E}_{\textup{GHHR1}(\pi/2)}+\Delta\textup{E}_{\textup{GHHR2}(\pi/2)}\right)\label{eq:GHHR-error-signal-a},\\
\Delta\textup{E}_{\textup{GHHR}(-)}&=\frac{1}{2}\left(\Delta\textup{E}_{\textup{GHHR}(\pi/4)}-\Delta\textup{E}_{\textup{GHHR}(3\pi/4)}\right)\label{eq:GHHR-error-signal-b}.
\end{align}
\end{subequations}
When comparing these GHHR error signals in term of robustness, they seem to be equivalent as shown in Fig.~\ref{fig:GHHR-protocol}. However the second cooperative protocol $\Delta\textup{E}_{\textup{GHHR}(-)}$ needs only two phase-steps either $\pm\pi/4$ or $\pm3\pi/4$ that are already simultaneously eliminating residual light-shift, weak distortion during free evolution time and eventually a small decoherence inducing a frequency-shift of interferences.
Compared to other error signals, a spin-echo like configuration, embedded into a Ramsey double pulse spectroscopic scheme with opposite detunings, eliminates very efficiently some systematics that can not be canceled by generalized hyper-Ramsey schemes with asymmetric position of the intermediate reversal pulse.

We have tested these configurations when a small distortion and a residual light-shift are present during pulses and free evolution zones as reported in Fig.~\ref{fig:GHHR-protocol}(a) and (b).
The strong robustness of the error signal $\Delta\textup{E}_{\textup{GHHR}}(\pm)$ frequency locked-point to residual light-shift is not compromised when a small distortion during free evolution is activated as shown in Fig.~\ref{fig:GHHR-protocol}(c) and (d). These GHHR error signal are still immune to residual light-shift while exhibiting an additional immunity against the small distortion which is active during the free evolution time. The Hahn-echo is canceling this low frequency perturbation by applying a refocusing pulse at the center of the composite pulse protocol reversing the spinor precession in the second half of the sequence~\cite{Vitanov:2015,Hahn:1950}. Note that a $\pm\pi/4,\pm3\pi/4$ phase-step modulation, applied during the intermediate pulse, is eliminating a residual tiny modulation of the fault-tolerant Hahn-Ramsey interferences leading to additional errors~\cite{Vitanov:2015}.

The inherent effect of the Hahn-Ramsey scheme with opposite detunings was already observed within a photon-echo pulse experiment reported in~\cite{Chaneliere:2015}. A detuned laser pulse, generating a distortion on the final photon-echo signal amplitude, is applied within a first free evolution zone $_{1}^{1}$M\textbf{I} while a second detuned optical pulse with the same detuning is applied in the second free evolution zone $_{1}^{1}$M\textbf{II}. The effect of the intermediate reversal pulse is to flip the sign of the second detuned optical pulse eliminating the overall effect of the light-shift distortion on the echo signal amplitude size~\cite{Chaneliere:2015}. In our case, such a distortion is simulated through the additonal $\epsilon$ parameter applied during free evolution zones.
Note the fault-tolerant scheme with opposite detunings was recently successfully applied to eliminate some low frequency noise and frequency drifts on a single trapped ion during free evolution time while recovering Ramsey fringes at the output of an effective Hahn-Ramsey interferometer~\cite{McCormick:2019}.

GHHR protocols might be considered for a class of optical lattice clocks using magnetically induced spectroscopy (MIS) with bosonic quantum systems,~\cite{Taichenachev:2006,Barber:2006,Baillard:2007,Kulosa:2015,Akatsuka:2008}. A Hahn spin-echo scheme embedded into a Ramsey interrogation protocol might be more resilient to inhomogeneous Doppler and spatially dependent ac Stark-shifts affecting clock transitions in a far detuned optical trap~\cite{Andersen:2003,Windpassinger:2008,Miyake:2019} or more robust to residual low fluctuations of the small mixing magnetic field not interrupted during a free evolution time~\cite{Hobson:2016}.
Generalized hyper-Ramsey and hyper-Hahn-Ramsey protocols combining $\pi/4$ and $3\pi/4$ phase-shifted interferences should remain very efficient to eliminate any kind of distortion inducing frequency-shifts that are always synchronized with laser pulses over a wide window of frequency detunings and large pulse area variations.

\section{HYPER-INTERFEROMETERS}

\indent We propose now to transfer composite pulse protocols, developed in section III for hyper-clocks, to spatial-domain interferometry.
We focus on the field of atomic matter-wave interferometers where the two-pulse Ramsey configuration is still an elementary building block of more elaborated successive interrogation schemes including external degrees of freedom with Doppler and recoil shifts and a residual uncompensated light-shift in the detuning.
\begin{figure}[b!!]
\center
\resizebox{8.5cm}{!}{\includegraphics[angle=0]{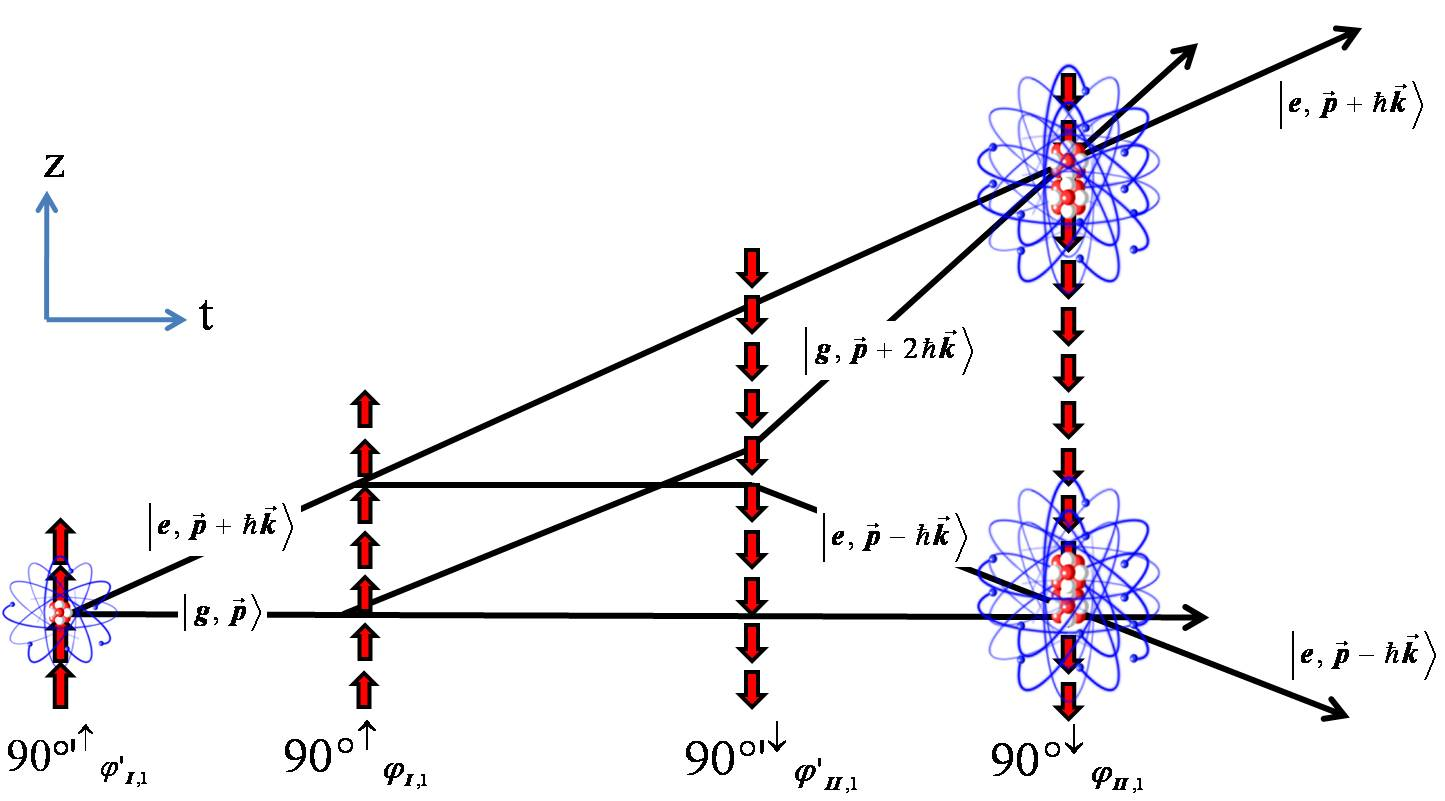}}
\caption{(color online). Original Ramsey-Bordé (RB) interferometer with four traveling waves~\cite{Borde:1984}. Two counter-propagating sets of two co-propagating laser pulses are introduced with $\uparrow\downarrow$ arrows that are corresponding to $\textup{k}v_{z}$ transverse Doppler wave-vector orientation. Laser pulse phases for each set are indicated respectively by $\varphi'_{I,1}(\varphi_{I,1})$ with $\uparrow$ and $\varphi'_{II,1}(\varphi_{II,1})$ with $\downarrow$.}
\label{fig-RB}
\end{figure}
The mechanical effects of light pulses were initially based on single photon transitions and exploited as beam-splitters or mirrors to spatially separate or recombine atomic or molecular wave-packets~\cite{Borde:1989,Borde:2019,Borde:1984}. Stimulated Raman transitions in a microwave excitation by optical two-photon processes were pioneered by Kasevich and Chu~\cite{Gustavson:1991,Kasevich:1992,Weiss:1993,Weiss:1994,Clade:2019} while Bragg diffraction was preferred to eliminate potential action of one photon light-shift in atomic states but requiring narrow momentum distribution to be very efficient~\cite{Abend:2019,Giese:2016,Szigeti:2012}.
Two-photon optical transitions in bosonic alkaline-earth quantum systems such as Yb, Hg, Sr and Mg have been already proposed as ultra-robust hyper clocks against detrimental light-shift and Zeeman effect~\cite{Zanon-Willette:2014}. All optical composite-pulse two-photon interferometry might thus be considered by inserting laser phase-steps, Doppler shift and atomic recoil state labeling in analogy with velocity-selective stimulated Raman transitions in alkali atoms~\cite{Moler:1992}. Indeed in such interferometers manipulating atom-light interaction, it is really primordial to avoid spontaneous emission that might destroyed the spatial coherence between wave-packets.

However, in all types of coherent Raman or Bragg manipulation of matter-waves, degraded performances of interferometers often rely on imperfect overlapping of wave-packets due to phase-shift accumulation during light pulses~\cite{Louchet-Chauvet:2011,Gillot:2016,Morel:2020,Cadoret:2009}.
Revisiting Ramsey-Bordé matter-wave interferometry is motivated by the application of composite phase-shifts from Eq.~\ref{eq:Phigg} and Eq.~\ref{eq:Phige} to compensate simultaneously for residual light-shifts associated to non vanishing Doppler-shifts when laser pulse area is drifting between pairs of atomic beam splitters.
\begin{figure*}[t!!]
\center
\resizebox{11cm}{!}{\includegraphics[angle=0]{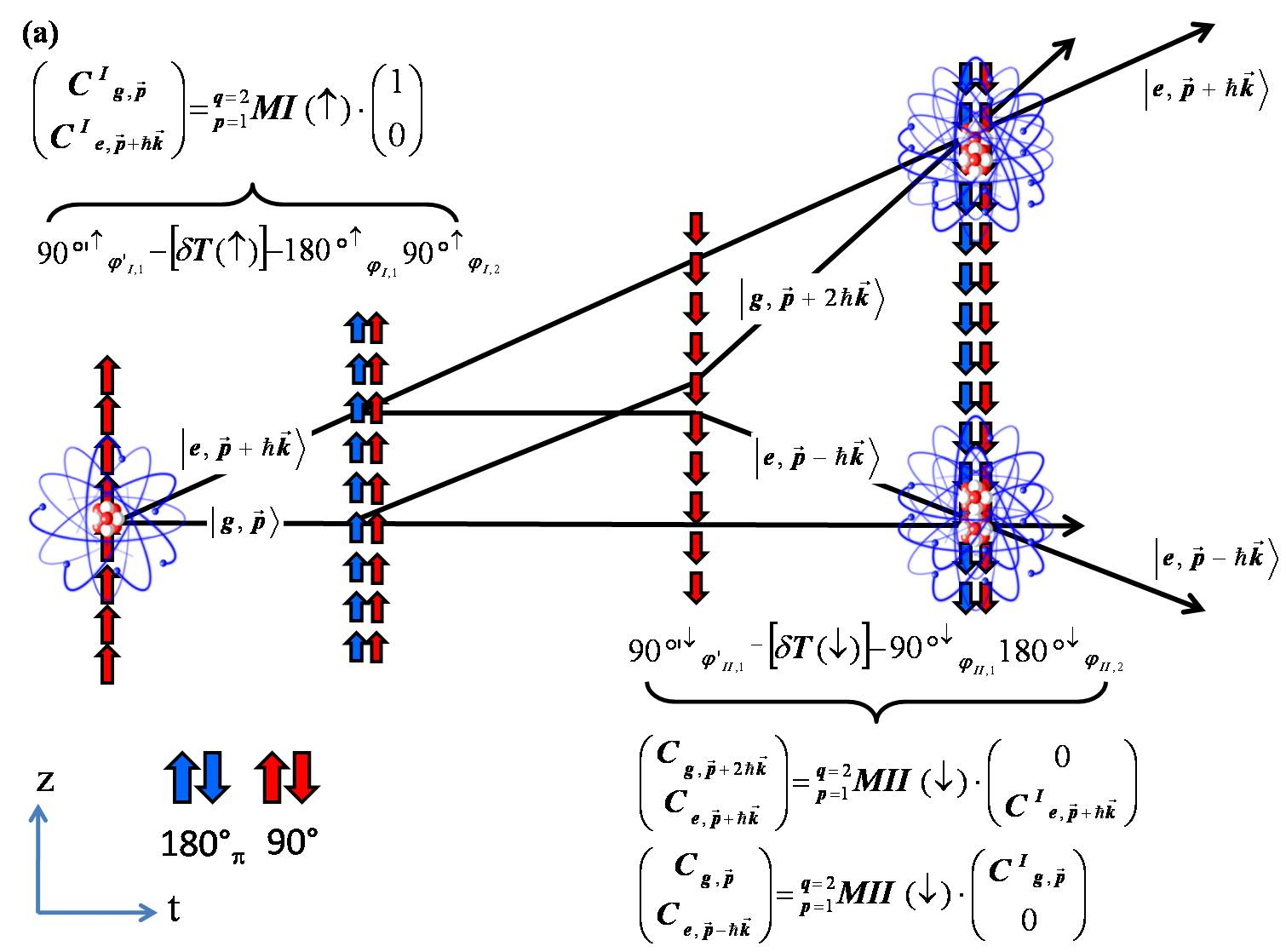}}
\resizebox{11cm}{!}{\includegraphics[angle=0]{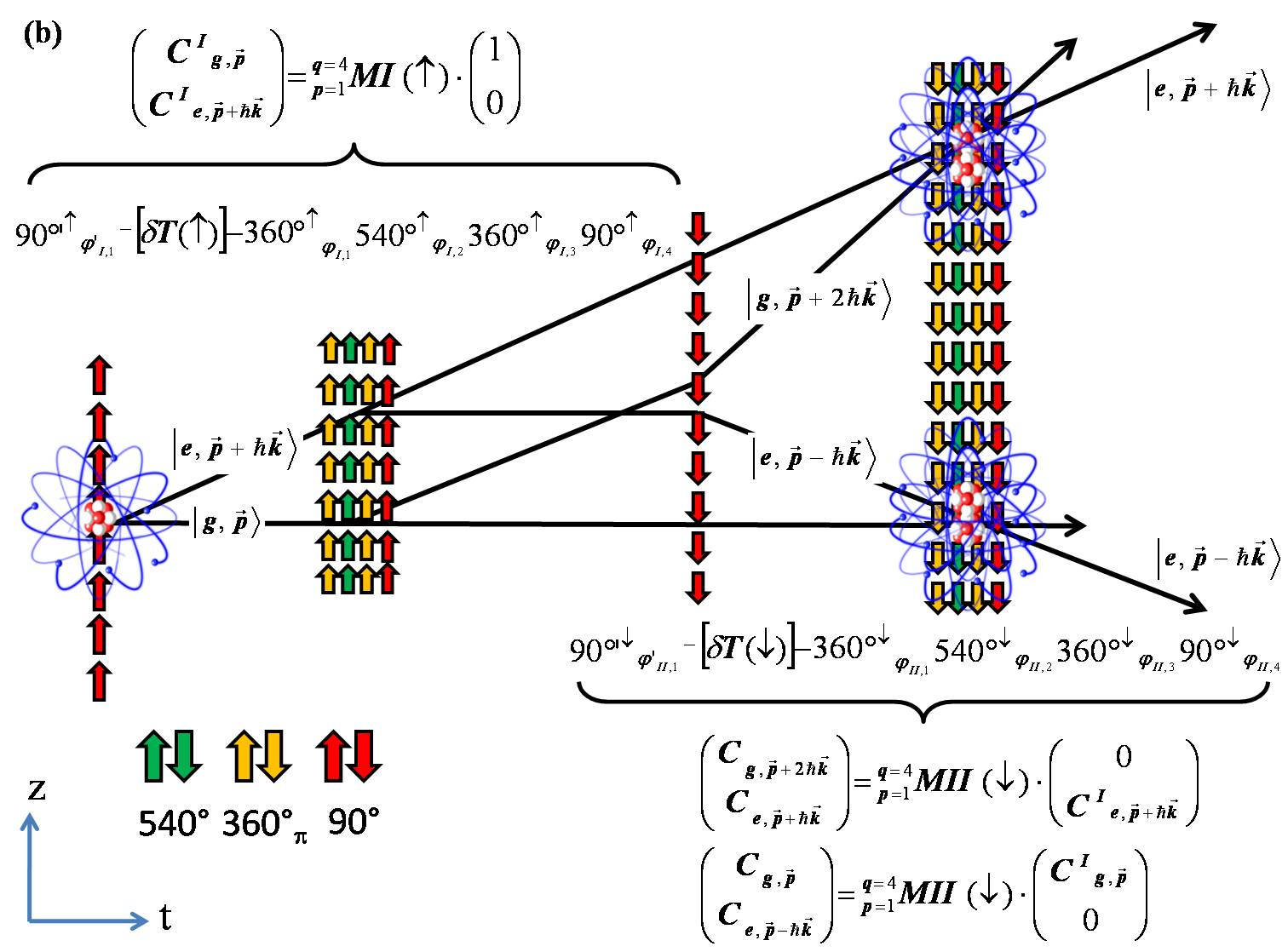}}
\caption{(color online). (a) and (b) Generalized hyper-Ramsey-Bordé (GHRB) interferometers with multiple traveling waves designed by \textit{p} and \textit{q} composite pulses. Two counter-propagating sets of $p=1$ and $q=2$ or $p=1$ and $q=4$ co-propagating composite laser pulses are introduced by interaction zones $_{p}^{q}$M\textbf{I$(\uparrow)$} and $_{p}^{q}$M\textbf{II$(\downarrow)$} where $\uparrow\downarrow$ arrows are corresponding to $\textup{k}v_{z}$ transverse Doppler wave-vector orientation. Laser pulse phases for each set are indicated respectively by $\varphi'_{I,l}(\varphi_{I,l})$ with $\uparrow$($\downarrow$) and $\varphi'_{II,l}(\varphi_{II,l})$ with $\downarrow$($\uparrow$). Colored red, blue, green and yellow $\Uparrow\Downarrow$ oriented arrows represent different phase-shifted pulse areas used to open and close the interferometers.}
\label{fig-GHRB-1}
\end{figure*}
\begin{figure*}[t!!]
\center
\resizebox{11cm}{!}{\includegraphics[angle=0]{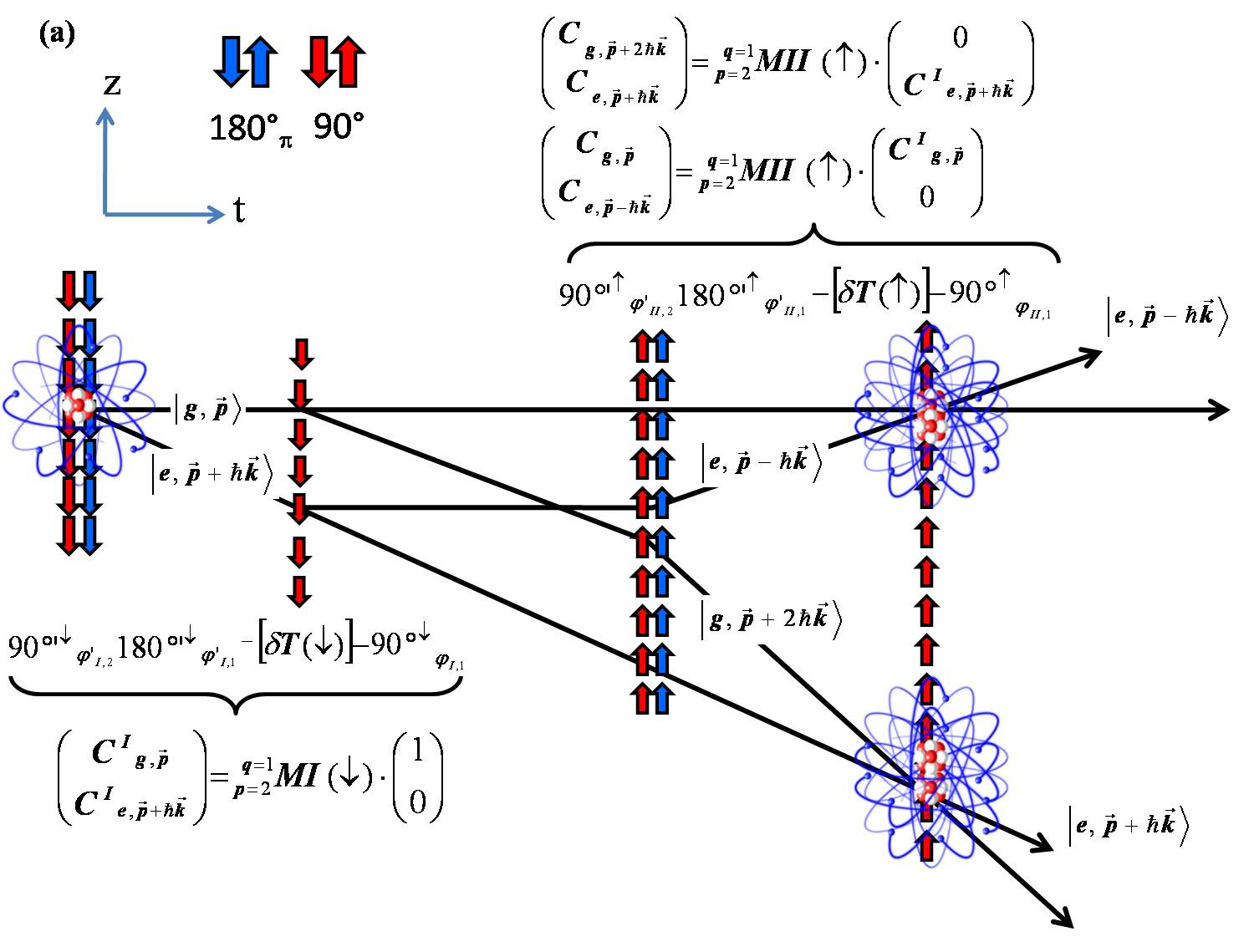}}
\resizebox{11cm}{!}{\includegraphics[angle=0]{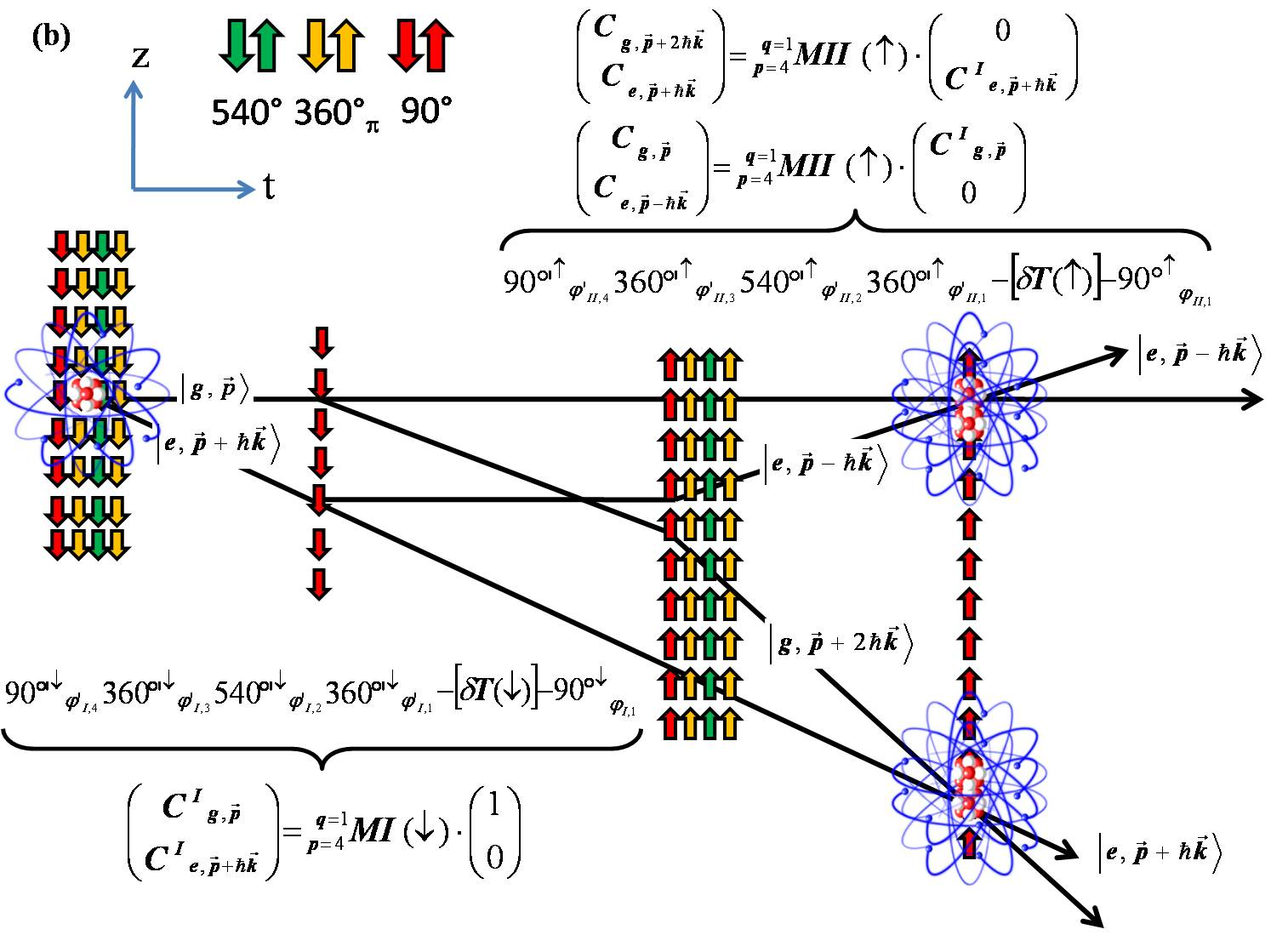}}
\caption{(color online). (a) and (b) Conjugate generalized hyper-Ramsey-Bordé (GHRB) interferometers with multiple traveling waves designed by \textit{p} and \textit{q} composite pulses. Two counter-propagating sets of $p=2$ and $q=1$ or $p=4$ and $q=1$ co-propagating composite laser pulses are introduced by interaction zones $_{p}^{q}$M\textbf{I$(\downarrow)$} and $_{p}^{q}$M\textbf{II$(\uparrow)$} where $\downarrow\uparrow$ arrows are corresponding to $\textup{k}v_{z}$ transverse Doppler wave-vector orientation. Laser pulse phases for each set are indicated respectively by $\varphi'_{I,l}(\varphi_{I,l})$ with $\downarrow$($\uparrow$) and $\varphi'_{II,l}(\varphi_{II,l})$ with $\uparrow$($\downarrow$). Colored red, blue, green and yellow $\Downarrow\Uparrow$ oriented arrows represent different phase-shifted pulse areas used to open and close the interferometers.}
\label{fig-GHRB-2}
\end{figure*}
Several types of interferometers have been developed from devices sensitive to recoil frequency or devices measuring rotation or acceleration.
Composite pulses with Bragg or stimulated Raman type transitions and butterfly geometry with four pulses have already been employed to improve current cold-atom gyroscopes in sensitivity and accuracy~\cite{Berg:2015,Savoie:2018}. Here, sophisticated sequences of pulses with phase-steps are rather proposed to shield matter-wave interferences against pulse defects inside atomic interferometers.

In this context, we study an asymmetric Ramsey-Bordé (RB) configuration used to determine the fine structure constant for fundamental test in QED~\cite{Borde:1984,Parker:2018} and Mach-Zehnder (MZ) and Butterfly (BU) interferometers for acceleration and rotation measurement~\cite{Peters:1999,Gillot:2016,Gustavson:2000,Cheinet:2006,Canuel:2006,Takase:2008,Gauguet:2009,Stockton:2011,Farah:2014,Dutta:2016,Menoret:2018,Geiger:2020,Li:2015,Tan:2020,Karcher:2020}.
We apply composite pulse protocols to realize a generalized hyper-Ramsey-Bordé (GHRB) interferometer reducing or eliminating residual corrections from light-shift
and sensitivity to residual transverse Doppler-shifts. We also present hyper-Mach-Zehnder (HMZ) and hyper-Butterfly (HBU) interferometers to strongly reduce the sensitivity against detrimental modification in pulse area variation between beam splitters investigated in~\cite{Gillot:2016} and more recently reported for a symmetric Ramsey-Bordé configuration~\cite{Morel:2020}.
\begin{table*}[t!!]
\renewcommand{\arraystretch}{1.7}
\begin{tabular}{|c|c|c|}
\hline
\hline
protocols & composite pulse building-blocks $_{p}^{q}$M\textbf{I}($\uparrow$)($\downarrow$), $_{p}^{q}$M\textbf{II}($\downarrow$)($\uparrow$) \\
\hline
\begin{tabular}{c}
     RB1($\varphi$) \\
     ($\varphi=\pi/4$)
\end{tabular}

 &

 \begin{tabular}{c}
           $\boldsymbol{90^{\circ'\uparrow}}_{\pm\varphi}\dashv\delta^{\uparrow}\textup{T}\vdash\boldsymbol{90^{\circ\uparrow}}_{0}\dashv\delta^{\circ\uparrow}\textup{T'}\vdash\boldsymbol{90^{\circ'\downarrow}}_{\pm\varphi}\dashv\delta^{\downarrow}\textup{T}\vdash\boldsymbol{90^{\circ\downarrow}}_{0}$ \\
           $(\dagger)$ $\boldsymbol{90^{\circ'\downarrow}}_{0}\dashv\delta^{\downarrow}\textup{T}\vdash\boldsymbol{90^{\circ\downarrow}}_{\mp\varphi}\dashv\delta^{\downarrow}\textup{T'}\vdash\boldsymbol{90^{\circ'\uparrow}}_{0}\dashv\delta^{\uparrow}\textup{T}\vdash\boldsymbol{90^{\circ\uparrow}}_{\mp\varphi}$
           \end{tabular}
\\

\hline
\hline
\begin{tabular}{c}
     RB2($\varphi$) \\
     ($\varphi=\pi/4$)
\end{tabular}

 &

 \begin{tabular}{c}
           $\boldsymbol{90^{\circ'\uparrow}}_{\pm\varphi}\dashv\delta^{\uparrow}\textup{T}\vdash\boldsymbol{270^{\circ\uparrow}}_{0}\dashv\delta^{\uparrow}\textup{T'}\vdash\boldsymbol{90^{\circ'\downarrow}}_{\pm\varphi}\dashv\delta^{\downarrow}\textup{T}\vdash\boldsymbol{270^{\circ\downarrow}}_{0}$ \\
           $(\dagger)$ $\boldsymbol{270^{\circ'\downarrow}}_{0}\dashv\delta^{\downarrow}\textup{T}\vdash\boldsymbol{90^{\circ\downarrow}}_{\mp\varphi}\dashv\delta^{\downarrow}\textup{T'}\vdash\boldsymbol{270^{\circ'\uparrow}}_{0}\dashv\delta^{\uparrow}\textup{T}\vdash\boldsymbol{90^{\circ\uparrow}}_{\mp\varphi}$
           \end{tabular}
\\

\hline
\hline
\begin{tabular}{c}
     HRB3$_{\pi}$($\varphi$)\\
     ($\varphi=\pi/4$)
     \end{tabular} & \begin{tabular}{c}
                           $\boldsymbol{90^{\circ'\uparrow}}_{\pm\varphi}\dashv\delta^{\uparrow}\textup{T}\vdash\boldsymbol{180^{\circ\uparrow}}_{\pi}\boldsymbol{90^{\circ\uparrow}}_{0}\dashv\delta^{\uparrow}\textup{T'}\vdash\boldsymbol{90^{\circ'\downarrow}}_{\pm\varphi}\dashv\delta^{\downarrow}\textup{T}\vdash\boldsymbol{180^{\circ\downarrow}}_{\pi}\boldsymbol{90^{\circ\downarrow}}_{0}$ \\
                          $(\dagger)$ $\boldsymbol{90^{\circ'\downarrow}}_{0}\boldsymbol{180^{\circ'\downarrow}}_{\pi}\dashv\delta^{\downarrow}\textup{T}\vdash\boldsymbol{90^{\circ\downarrow}}_{\mp\varphi}\dashv\delta^{\downarrow}\textup{T'}\vdash\boldsymbol{90^{\circ'\uparrow}}_{0}\boldsymbol{180^{\circ'\uparrow}}_{\pi}\dashv\delta^{\uparrow}\textup{T}\vdash\boldsymbol{90^{\circ\uparrow}}_{\mp\varphi}$

   \end{tabular}
\\
\hline
\hline

\begin{tabular}{c}
     HRB5$_{\pi}$($\varphi$)\\
     ($\varphi=\pi/4$)
     \end{tabular} & \begin{tabular}{c}
                           $\boldsymbol{90^{\circ'\uparrow}}_{\pm\varphi}\dashv\delta^{\uparrow}\textup{T}\vdash\boldsymbol{360^{\circ\uparrow}}_{\pi}\boldsymbol{540^{\circ\uparrow}}_{0}\boldsymbol{360^{\circ\uparrow}}_{\pi}\boldsymbol{90^{\circ\uparrow}}_{0}\dashv\delta^{\uparrow}\textup{T'}\vdash\boldsymbol{90^{\circ'\downarrow}}_{\pm\varphi}\dashv\delta^{\downarrow}\textup{T}\vdash\boldsymbol{360^{\circ\downarrow}}_{\pi}\boldsymbol{540^{\circ\downarrow}}_{0}\boldsymbol{360^{\circ\downarrow}}_{\pi}\boldsymbol{90^{\circ\downarrow}}_{0}$ \\
                          $(\dagger)$ $\boldsymbol{90^{\circ'\downarrow}}_{0}\boldsymbol{360^{\circ'\downarrow}}_{\pi}\boldsymbol{540^{\circ'\downarrow}}_{0}\boldsymbol{360^{\circ'\downarrow}}_{\pi}\dashv\delta^{\downarrow}\textup{T}\vdash\boldsymbol{90^{\circ\downarrow}}_{\mp\varphi}\dashv\delta^{\downarrow}\textup{T'}\vdash\boldsymbol{90^{\circ'\uparrow}}_{0}\boldsymbol{360^{\circ'\uparrow}}_{\pi}\boldsymbol{540^{\circ'\uparrow}}_{0}\boldsymbol{360^{\circ'\uparrow}}_{\pi}\dashv\delta^{\uparrow}\textup{T}\vdash\boldsymbol{90^{\circ\uparrow}}_{\mp\varphi}$

   \end{tabular}
\\
\hline
\hline

\begin{tabular}{c}
     GHRB($\varphi$) \\
      ($\varphi=\pi/8,3\pi/8$)
\end{tabular}  &    \begin{tabular}{c}
                                 $\boldsymbol{90^{\circ'\uparrow}}_{0}\dashv\delta^{\uparrow}\textup{T}\vdash\boldsymbol{180^{\circ\uparrow}}_{\pm\varphi}\boldsymbol{90^{\circ\uparrow}}_{0}\dashv\delta^{\uparrow}\textup{T'}\vdash\boldsymbol{90^{\circ'\downarrow}}_{0}\dashv\delta^{\downarrow}\textup{T}\vdash\boldsymbol{180^{\circ\downarrow}}_{\pm\varphi}\boldsymbol{90^{\circ\downarrow}}_{0}$ \\
                            $(\dagger)$ $\boldsymbol{90^{\circ'\downarrow}}_{0}\boldsymbol{180^{\circ'\downarrow}}_{\mp\varphi}\dashv\delta^{\downarrow}\textup{T}\vdash\boldsymbol{90^{\circ\downarrow}}_{0}\dashv\delta^{\downarrow}\textup{T'}\vdash\boldsymbol{90^{\circ'\uparrow}}_{0}\boldsymbol{180^{\circ'\uparrow}}_{\mp\varphi}\dashv\delta^{\uparrow}\textup{T}\vdash\boldsymbol{90^{\circ\uparrow}}_{0}$

                            \end{tabular}

 \\
\hline
\hline
 \begin{tabular}{c}
     GHRB$_{\pi}$($\varphi$) \\
     ($\varphi=\pi/8,3\pi/8$)
\end{tabular}  &  \begin{tabular}{c}

                           $\boldsymbol{90^{\circ'\uparrow}}_{\pi}\dashv\delta^{\uparrow}\textup{T}\vdash\boldsymbol{180^{\circ\uparrow}}_{\pm\varphi}\boldsymbol{90^{\circ\uparrow}}_{\pi}\dashv\delta^{\uparrow}\textup{T'}\vdash\boldsymbol{90^{\circ'\downarrow}}_{\pi}\dashv\delta^{\downarrow}\textup{T}\vdash\boldsymbol{180^{\circ\downarrow}}_{\pm\varphi}\boldsymbol{90^{\circ\downarrow}}_{\pi}$ \\
                                $(\dagger)$ $\boldsymbol{90^{\circ'\downarrow}}_{\pi}\boldsymbol{180^{\circ'\downarrow}}_{\mp\varphi}\dashv\delta^{\downarrow}\textup{T}\vdash\boldsymbol{90^{\circ\downarrow}}_{\pi}\dashv\delta^{\downarrow}\textup{T'}\vdash\boldsymbol{90^{\circ'\uparrow}}_{\pi}\boldsymbol{180^{\circ'\uparrow}}_{\mp\varphi}\dashv\delta^{\uparrow}\textup{T}\vdash\boldsymbol{90^{\circ\uparrow}}_{\pi}$

                                   \end{tabular}
 \\

 \hline
\hline

\end{tabular}
\centering%
\caption{Composite pulses interrogation protocols for hyper-Ramsey-Bordé atom interferometry. Pulse area $\boldsymbol{\theta'_{l}}(\boldsymbol{\theta_{l}})$ is given in degrees and phase-steps $\pm\varphi'_{l}(\varphi_{l})$ are indicated in subscript-brackets with radian unit. The standard Rabi frequency for all pulses is $\Omega=\pi/2\tau$ where $\tau$ is the pulse duration reference. Free evolution time regions are given by $\delta^{\uparrow}\textup{T}$ ($\delta^{\downarrow}\textup{T}$) where $\uparrow\downarrow$ denotes the transverse Doppler-shift orientation. Each elementary building-block $_{p}^{q}$M\textbf{I}($\uparrow$)($\downarrow$) and $_{p}^{q}$M\textbf{II}($\downarrow$)($\uparrow$) are separated by the intermediate $\delta^{\uparrow}\textup{T'}$ or $\delta^{\downarrow}\textup{T'}$ free evolution zone. Reverse protocols in time are denoted by $(\dagger)$.}
\label{protocol-table-2}
\end{table*}

\subsection{HYPER RAMSEY-BORDÉ}

\subsubsection{RB, HRB3$_{\pi}$, HRB5$_{\pi}$ and GHRB protocols}

\indent The original Ramsey-Bordé (RB) interferometer is based on a four-laser pulse configuration, as reported in first line of Tab.~\ref{protocol-table-2} and shown in Fig.~\ref{fig-RB}.
If the motion of atoms is taking into account, the first Ramsey two-zone setup denoted as $_{1}^{1}$M\textbf{I$(\uparrow)$}, recovers a strong sensitivity to the transverse first-order Doppler effect along the laser beam.
While, with microwaves or radio-frequencies, the atomic wave-packets are still interfering over a long time after a single Ramsey two-pulse interrogation, fringes are
rapidly destroyed by optical frequencies with large wave-vectors. Indeed, the splitting between wave-packets is velocity dependent and becomes spatially too large requiring a Doppler cancelation technique to close the interferometer avoiding the lost of quantum interferences.
Bordé proposed in the 1980's a configuration with four traveling waves consisting of two separated Ramsey two-zones $_{1}^{1}$M\textbf{I$(\uparrow)$} and $_{1}^{1}$M\textbf{II$(\downarrow)$} where arrows are describing two counter-propagating sets of co-propagating laser pulses~\cite{Borde:1989}. Within this interaction geometry, opposite sets of laser pulse wave-vectors cancel the Doppler-shift and Ramsey interferences are retrieved at the output of the interferometer~\cite{Borde:1984}.
\begin{figure*}[t!!]
\center
\resizebox{9cm}{!}{\includegraphics[angle=0]{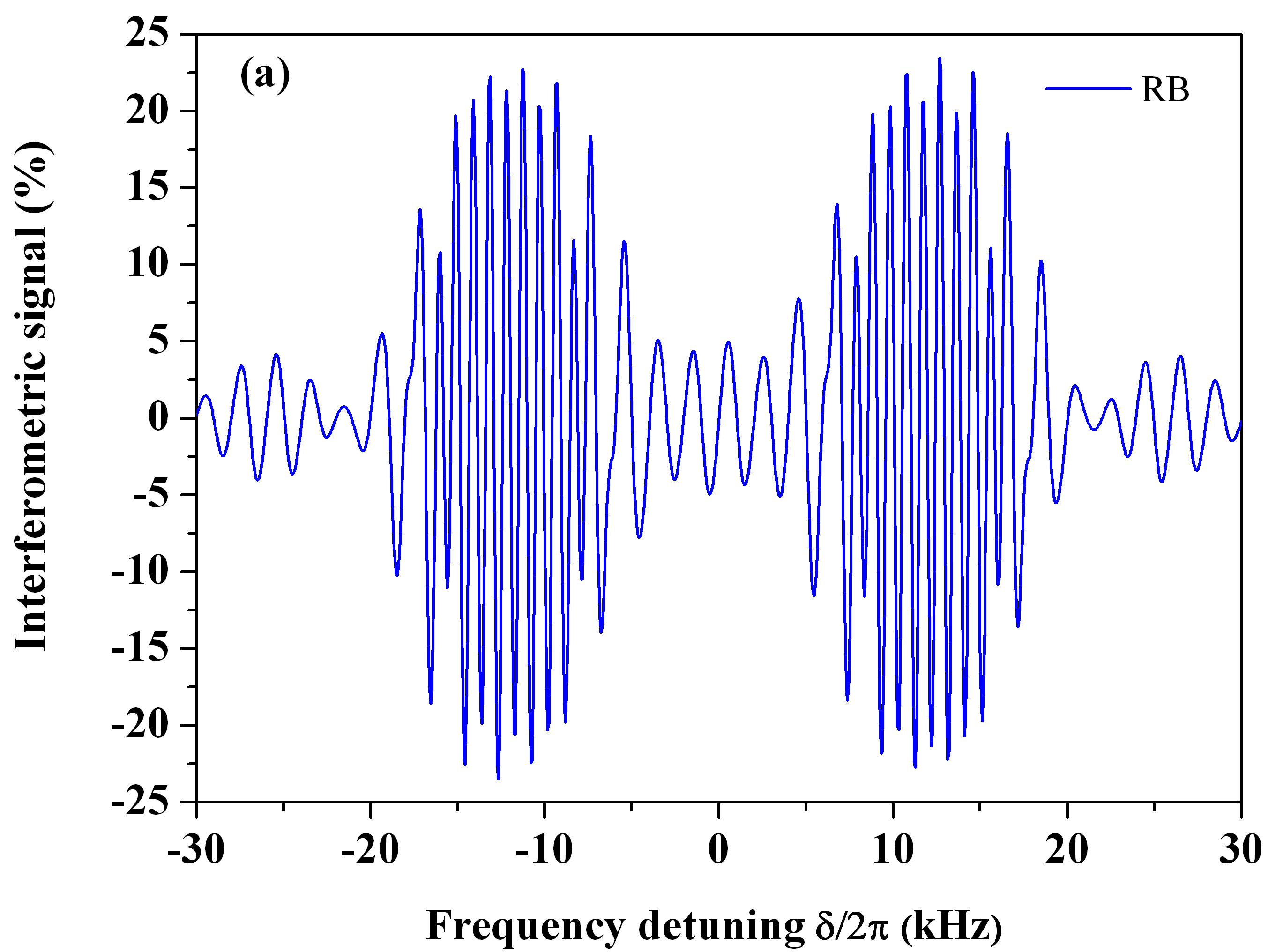}}\resizebox{9cm}{!}{\includegraphics[angle=0]{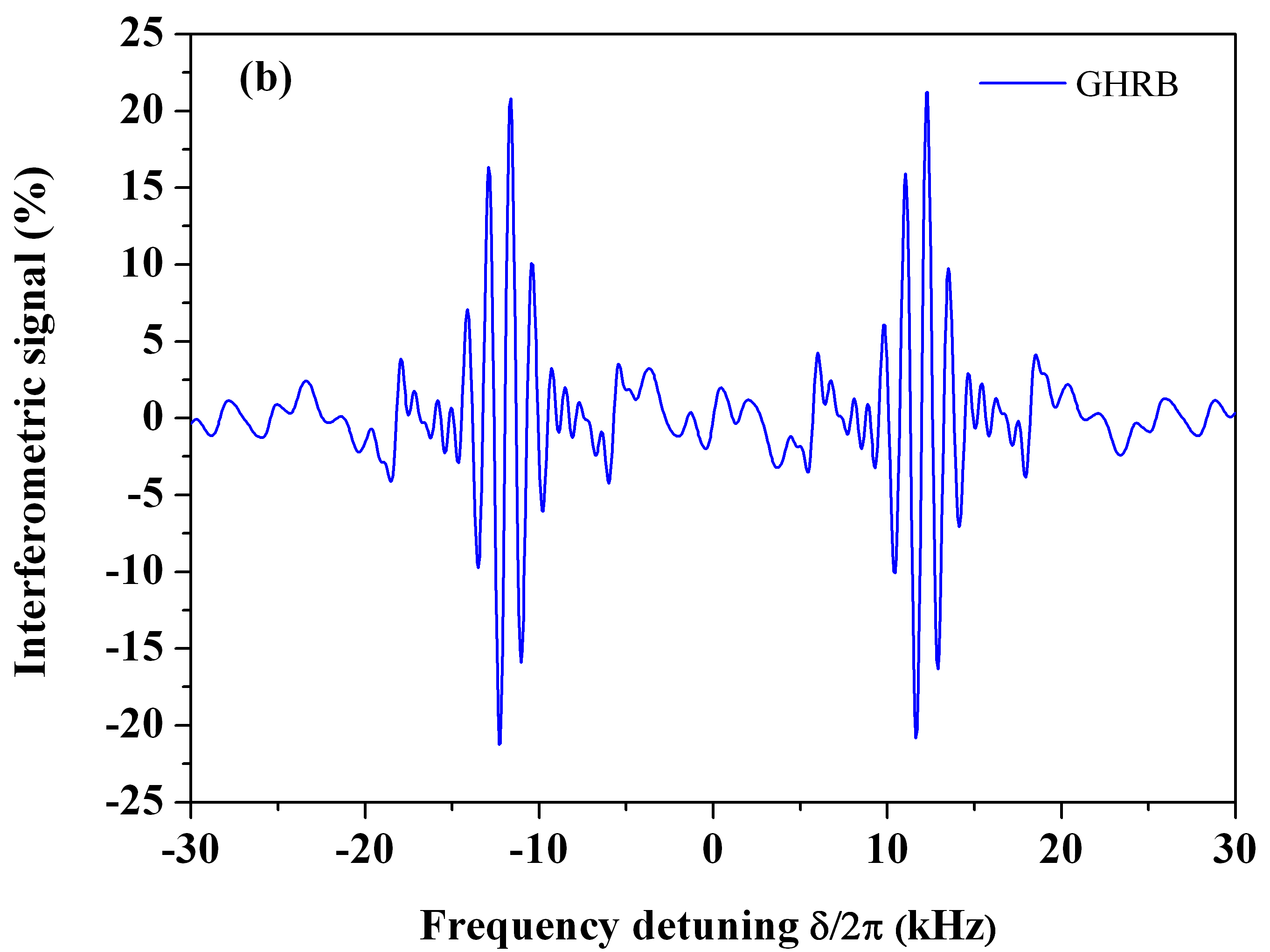}}
\resizebox{9cm}{!}{\includegraphics[angle=0]{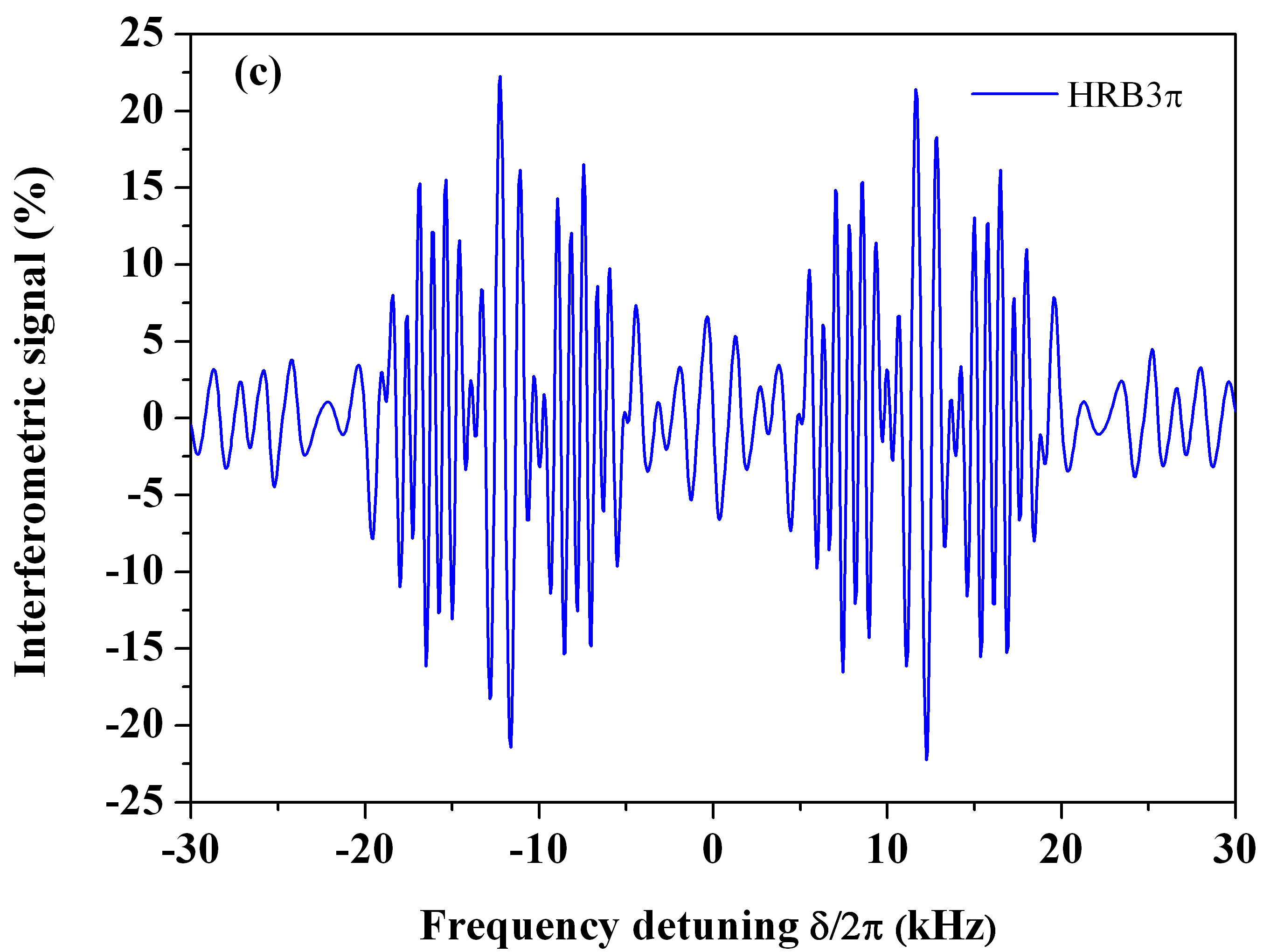}}\resizebox{9cm}{!}{\includegraphics[angle=0]{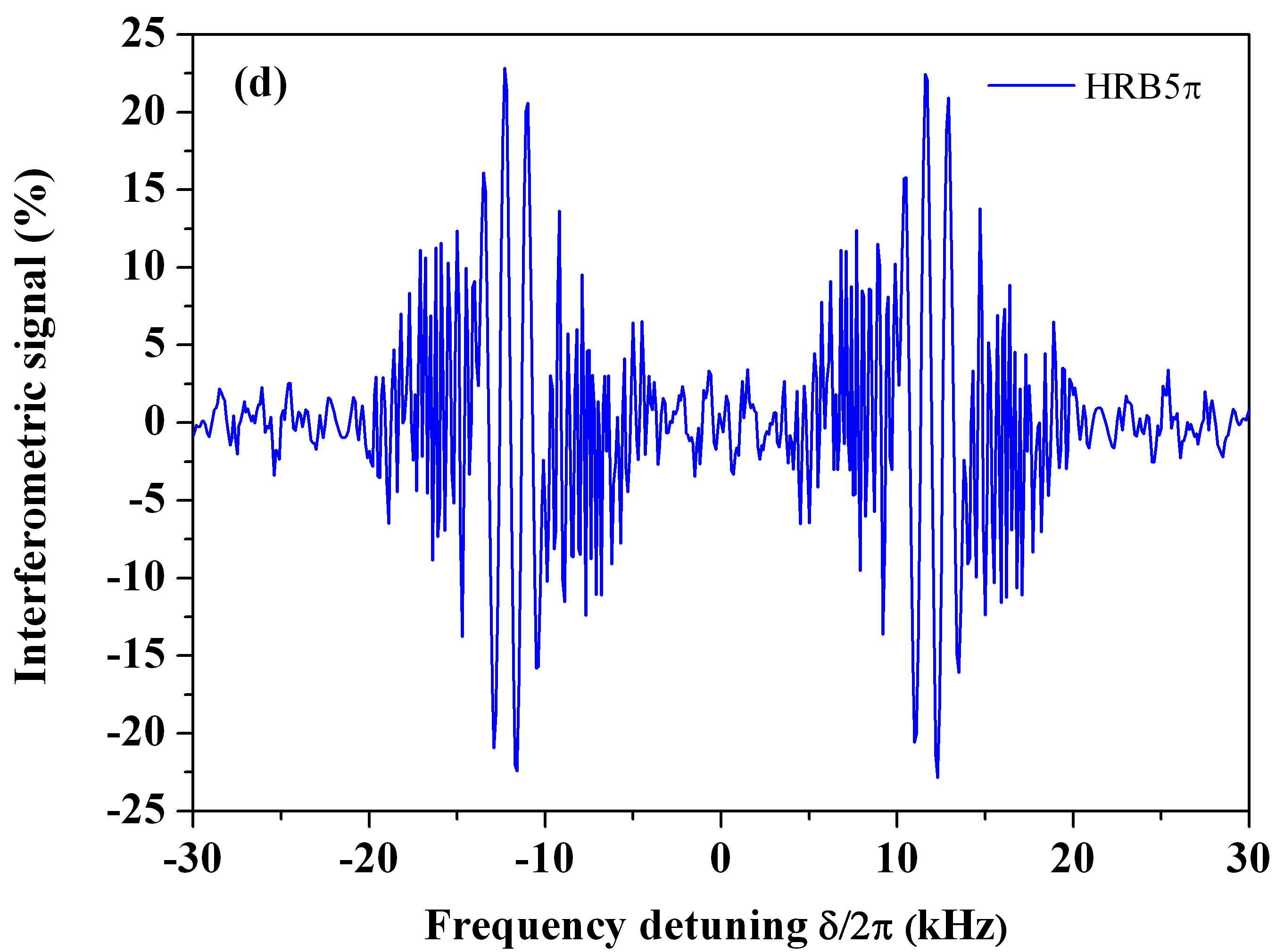}}
\caption{(color online). Resolving the $^{40}$Ca recoil doublet with matter-wave composite pulse interferometry integrated over a narrow gaussian transverse velocity distribution around T$=250$~pK versus frequency detuning $\delta/2\pi$. (a) RB interferences. (b) GHRB. (c) HRB3$_{\pi}$. (d) HRB5$_{\pi}$. Laser pulse duration is $\tau=0.1$ms and free evolution times around T$=30/\delta_{r}$ where we apply for the intermediate free evolution time T'$\mapsto0$ and $\delta_{r}=\hbar\textup{k}^{2}/2\textup{m}$.}
\label{fig:recoil-doublet}
\end{figure*}
Inserting composite pulses in atomic interferometry is motivated by eliminating potential uncompensated residual part of Doppler-shift sensitivity and light-shift on a recoil frequency determination when the pulse area changes over the whole pulse interrogation.
Indeed, for an original RB interferometer, a uniform distribution of laser amplitude over the full sequence will be rejected by a differential measurement between the two sets of shifted wave-packets. These terms drop out in any interferometric comparison between paths. However, if pulse area of pairs of beam splitters are modified between two Ramsey-Bordé interaction zones, a parasitic shift may be recovered.

We consider our two-level system interacting with an arbitrary number of traveling waves which may propagate in opposite direction. Following the decomposition rules with multiple interrogation zones from the previous section, the interaction geometry of a robust hyper-interferometer (HRB3$_{\pi}$, HRB5$_{\pi}$ and GHRB) is divided into two composite pulse building-blocks with interaction matrices $_{1}^{2}$M\textbf{I$(\uparrow)$}, $_{1}^{4}$M\textbf{I$(\uparrow)$} ($_{2}^{1}$M\textbf{I$(\downarrow)$}, $_{4}^{1}$M\textbf{I$(\downarrow)$}) and $_{1}^{2}$M\textbf{II$(\downarrow)$}, $_{1}^{4}$M\textbf{II$(\downarrow)$} ($_{2}^{1}$M\textbf{II$(\uparrow)$}, $_{4}^{1}$M\textbf{II$(\uparrow)$}) separated by an intermediate free evolution time T' as shown in Fig.~\ref{fig-GHRB-1} and in Fig.~\ref{fig-GHRB-2}, listed in Tab.~\ref{protocol-table-2}. The generalized components of interaction matrices required to compute each amplitude of probability associated to different path trajectories of wave-packets are given in section S1 and S3 from the appendix.

We evaluate first the complex coefficients $_{p}^{q}C^{I}_{g,\overrightarrow{p}}(t)$ and $_{p}^{q}C^{I}_{e,\overrightarrow{p}+\hbar\overrightarrow{k}}(t)$ within the first Ramsey zone $_{p}^{q}$M\textbf{I$(\uparrow)$} with $p$ and $q$ pulses. Then we evaluate the complex coefficients for the second interaction zone $_{p}^{q}$M\textbf{I$(\downarrow)$} starting from previous solutions of interfering trajectories closing the interferometer.
The first hyper-Ramsey-Bordé $_{p}^{q}$M\textbf{I$(\uparrow)$} building-block is thus computed taking $C_{g,\overrightarrow{p}}(0)=1,C_{e,\overrightarrow{p}+\hbar\overrightarrow{k}}(0)=0$ and gives:
\begin{subequations}
\begin{align}
_{p}^{q}C^{I}_{g,\overrightarrow{p}}(t)&=_{p}^{q}C^{I}_{gg},\\
_{p}^{q}C^{I}_{e,\overrightarrow{p}+\hbar\overrightarrow{k}}(t)&=_{p}^{q}C^{I}_{eg}=-_{p}^{q}C^{I*}_{ge},
\label{eq:complex-amplitudes-GHRB-1}
\end{align}
\end{subequations}
where $*$ means complex conjugate. The common laser detuning $\delta_{I}$ for all spinor matrix component in $_{p}^{q}$M\textbf{I$(\uparrow)$} zone is defined in section S2 from the appendix.
\begin{figure}[b!!]
\center
\resizebox{8.5cm}{!}{\includegraphics[angle=0]{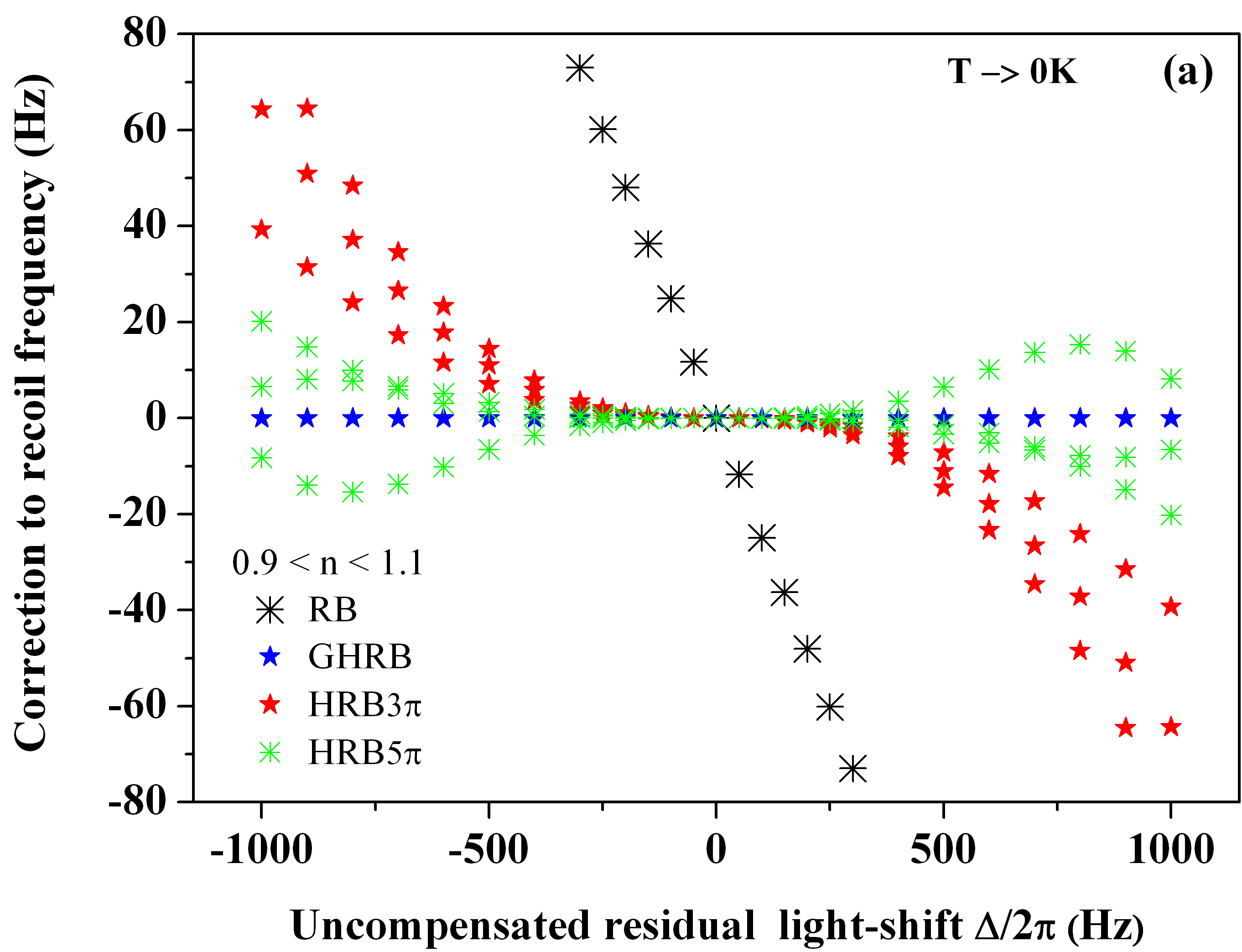}}
\resizebox{8.5cm}{!}{\includegraphics[angle=0]{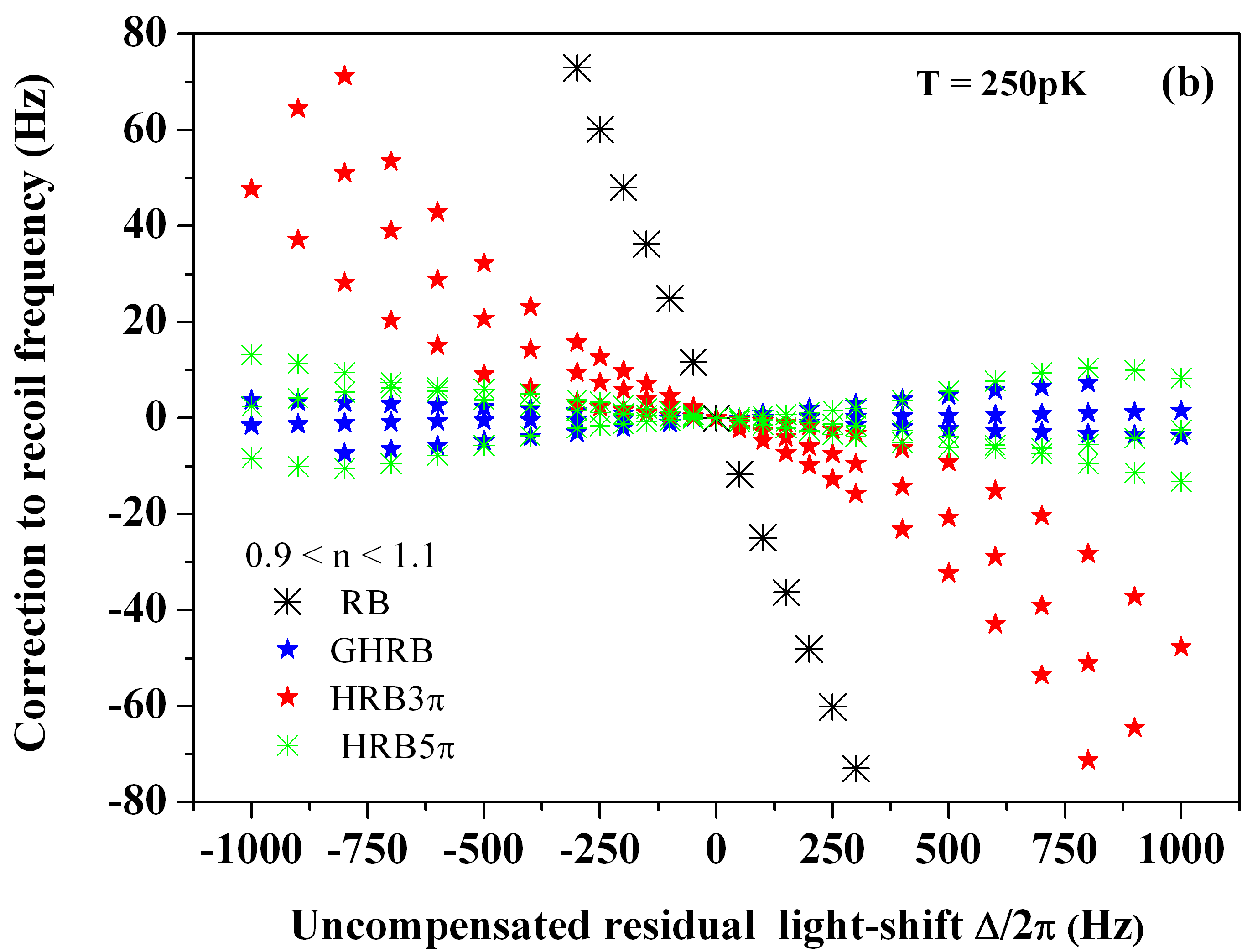}}
\caption{(color online).(a) Numerical tracking of the error signal frequency position for RB, HRB3$_{\pi}$, HRB5$_{\pi}$ and GHRB protocols around zero for various uncompensated part of the residual light-shift for T$\rightarrow0$~K. (b) Distortion effect of matter-waves due to a non zero temperature after integration over a narrow gaussian transverse velocity distribution around T$=250$~pK. All tracking points are generated with same parameters as in Fig.~\ref{fig:recoil-doublet} except pulse area variation $\Omega\tau=\textup{n}\times\pi/2$ ($\pi/2\equiv\boldsymbol{90}^{\circ}$) which is driven by the parameter \textup{n} from 0.9 to 1.1 between the two interaction zones $_{1}^{2}$M\textbf{I$(\uparrow)$}, $_{1}^{2}$M\textbf{II$(\downarrow)$}.}
\label{fig:RB-shift}
\end{figure}
At the end of this first GHRB $_{p}^{q}$M\textbf{I$(\uparrow)$} interaction zone, wave-packets are separated in several components in space during an intermediate T'
free evolution time. This is seen as a simple additional phase-factor of the form $e^{\pm i\delta\textup{T'}/2}$. In asymmetric or symmetric Ramsey-Bordé interferometers, this delay allows for Bloch-oscillations to transfer large number of photon momenta to the wave-packets~\cite{Parker:2018,Clade:2019}.

The final complex GHRB matter-wave amplitudes $_{p}^{q}C_{e,\overrightarrow{p}-\hbar\overrightarrow{k}}(t)$ and $_{p}^{q}C_{e,\overrightarrow{p}+\hbar\overrightarrow{k}}(t)$ after successive interaction with $_{p}^{q}$M\textbf{I$(\uparrow)$} and $_{p}^{q}$M\textbf{II$(\downarrow)$} regions are four overlapping wave-packets centered at the recoil frequency $\pm\delta_{r}$ ($\pm\hbar\overrightarrow{k}$)~\cite{Borde:1984}.
The final GHRB amplitude of the wave-function after successive interaction is now given by:
\begin{subequations}
\begin{align}
_{p}^{q}C_{e,\overrightarrow{p}-\hbar\overrightarrow{k}}(t)&=-_{p}^{q}C^{II*}_{ge}\cdot_{p}^{q}C^{I}_{gg},\\
_{p}^{q}C_{e,\overrightarrow{p}+\hbar\overrightarrow{k}}(t)&=-_{p}^{q}C^{II*}_{gg}\cdot_{p}^{q}C^{I*}_{ge},
\end{align}
\end{subequations}
After algebraic manipulation, one up-shifted wave-packet
expression $_{p}^{q}C_{e,\overrightarrow{p}-\hbar\overrightarrow{k}}(t)$ is expressed as:
\begin{subequations}
\begin{align}
_{p}^{q}C^{I}_{gg}=&_{p}^{q}\alpha^{I}_{gg}e^{i\delta_{I}\textup{T}/2}\left[1-|_{p}^{q}\beta^{I}_{gg}|e^{-i(\delta_{I}\textup{T}+_{p}^{q}\Phi^{I}_{gg})}\right],\\
_{p}^{q}C^{II*}_{ge}=&_{p}^{q}\alpha^{II*}_{ge}e^{-i\delta_{II}\textup{T}/2}\left[1+|_{p}^{q}\beta^{II*}_{ge}|e^{i(\delta_{II}\textup{T}+_{p}^{q}\Phi^{II*}_{ge})}\right],
\end{align}
\end{subequations}
where $\alpha^{II*}_{ge},\beta^{II*}_{ge}$ components from $_{p}^{q}C^{II*}_{ge}$ are analytically derived with the proper detuning definition $\delta_{II}$ from appendix section S2.
The other down-shifted wave-packet expression $_{p}^{q}C_{e,\overrightarrow{p}+\hbar\overrightarrow{k}}(t)$ is:
\begin{subequations}
\begin{align}
_{p}^{q}C^{I*}_{ge}=&_{p}^{q}\alpha^{I*}_{ge}e^{-i\delta_{I}\textup{T}/2}\left[1+|_{p}^{q}\beta^{I*}_{ge}|e^{i(\delta_{I}\textup{T}+_{p}^{q}\Phi^{I*}_{ge})}\right],\\
_{p}^{q}C^{II*}_{gg}=&_{p}^{q}\alpha^{II*}_{gg}e^{-i\delta_{II}\textup{T}/2}\left[1-|_{p}^{q}\beta^{II*}_{gg}|e^{i(\delta_{II}\textup{T}+_{p}^{q}\Phi^{II*}_{gg})}\right].
\end{align}
\end{subequations}
while $\alpha^{II*}_{gg},\beta^{II*}_{gg}$ components from $_{p}^{q}C^{II*}_{gg}$ are derived with the proper detuning definition $\delta_{II}$ also reported in appendix section S2.

Matter-wave interferences are centered around a high-frequency (HF) recoil term $_{p}^{q}C_{g,\overrightarrow{p}-\hbar\overrightarrow{k}}(t)$ and a low-frequency (LF) recoil term $_{p}^{q}C_{e,\overrightarrow{p}+\hbar\overrightarrow{k}}(t)$ where we have identified composite phase-shifts as:
\begin{subequations}
\begin{align}
_{p}^{q}\Phi_{e,\overrightarrow{p}-\hbar\overrightarrow{k}}&=\varphi_{L}+\phi_{L}-\textup{Arg}\left[_{p}^{q}\beta^{I}_{gg}\cdot_{p}^{q}\beta^{II}_{ge}\right]\label{eq:HF-recoil},\\
_{p}^{q}\Phi_{e,\overrightarrow{p}+\hbar\overrightarrow{k}}&=\varphi_{L}+\phi_{L}-\textup{Arg}\left[_{p}^{q}\beta_{ge}^{I}\cdot_{p}^{q}\beta_{gg}^{II}\right]\label{eq:LF-recoil},
\end{align}
\end{subequations}
with a phase composition $\varphi_{L}+\phi_{L}$ given by~\cite{Berman:1997}:
\begin{subequations}
\begin{align}
\varphi_{L}&\equiv\varphi_{I,L}+\varphi_{II,L}=\varphi_{I,1}-\varphi'_{I,1}+\varphi_{II,1}-\varphi'_{II,1}\label{eq:laser-phase},\\
\phi_{L}&\equiv\phi_{I,L}+\phi_{II,L}=\phi'_{I,1}+\phi_{I,1}+\phi'_{II,1}+\phi_{II,1}\label{eq:light-shift-pulses},
\end{align}
\end{subequations}
These composite phase-shift expressions are also consistent with a graphical representation of strong-field density matrix diagrams related to low and high frequency recoil peaks of Ramsey-Bordé fringes when additionnal interaction matrices are taken in to account~\cite{Borde:1984,Salomon:1984,Borde:1983}.
Our computational algorithm allows us to derive an analytical formulae of the composite phase-shift directly acting on matter-wave interferences based on asymmetrical generalized hyper-Ramsey-Bordé interferometer.
The overall effect of the residual uncompensated part of the light-shift remnant to the original Ramsey-Bordé scheme is finally encoded in Eq.~\ref{eq:light-shift-pulses}. Results based on Eq.~\ref{eq:HF-recoil} and Eq.~\ref{eq:LF-recoil} are generalizing the usual description of atom interferometers neglecting potential light-shift distortion.

Note that there is also the RB interferometer configuration where the last set of optical traveling waves used to close the interferometer are not reversed. This symmetric Ramsey-Bordé interferometer is also exploited for the fine structure determination~\cite{Clade:2019,Cadoret:2009}.
Such a geometry is not sensitive to the net frequency dependence of the interference signal such that the relative phase-shift accumulation between arms of this symmetrical RB configuration is given by ~\cite{Berman:1997}:
\begin{subequations}
\begin{align}
\varphi_{L}&\equiv\varphi_{I,L}-\varphi_{II,L}=\varphi_{I,1}-\varphi'_{I,1}+\varphi'_{II,1}-\varphi_{II,1}\label{eq:laser-phase-SRB},\\
\phi_{L}&\equiv\phi_{I,L}-\phi_{II,L}=\phi'_{I,1}+\phi_{I,1}-\phi'_{II,1}-\phi_{II,1}\label{eq:light-shift-pulses-SRB}.
\end{align}
\end{subequations}
It is interesting to note that the sensitivity to residual Doppler-shifts and light-shifts are equivalently coming from Eq.~\ref{eq:light-shift-pulses} for an asymmetric interferometer and from Eq.~\ref{eq:light-shift-pulses-SRB} for a symmetric configuration. These terms are responsible for a velocity-dependent phase-shift leading to an imperfect overlapping of wavepackets originally derived in~\cite{Morel:2020}.

Coming back to the asymmetric RB interferometer, the generalized hyper-Ramsey Bordé transition probability is thus given by:
\begin{equation}
_{p}^{q}P_{e,\overrightarrow{p}\pm\hbar\overrightarrow{k}}=\left|_{p}^{q}C_{e,\overrightarrow{p}\pm\hbar\overrightarrow{k}}(t)\right|^{2},
\end{equation}
Similar to the generation of error signals based on Eq.~\ref{eq:error-signal}, we also generate dispersive fringes as following:
\begin{equation}
\Delta E=_{p}^{q}P_{e,\overrightarrow{p}\pm\hbar\overrightarrow{k}}(\varphi)-_{p}^{q}P_{e,\overrightarrow{p}\pm\hbar\overrightarrow{k}}(-\varphi).
\label{eq:GHRB-error-signal}
\end{equation}
where we can apply phase-step protocols reported in Tab.~\ref{protocol-table-2}. Combination of phase-step protocols within two-successive building-blocks following Eq.~\ref{eq:laser-phase}, required to produce dispersive error signals, are using half values needed for an hyper-clock interrogation scheme based on a single building-block.
\begin{figure*}[t!!]
\center
\resizebox{9.5cm}{!}{\includegraphics[angle=0]{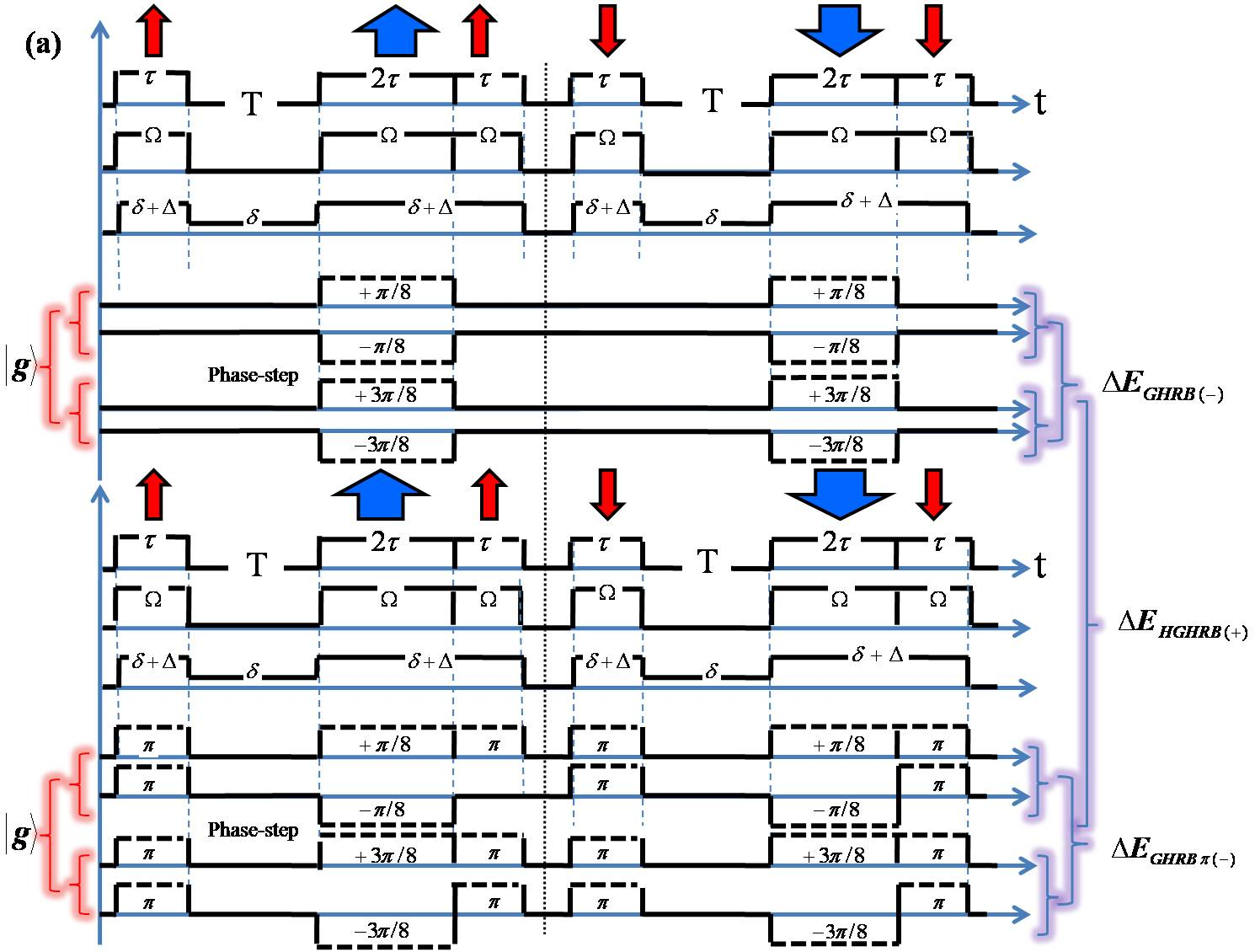}}\resizebox{8.5cm}{!}{\includegraphics[angle=0]{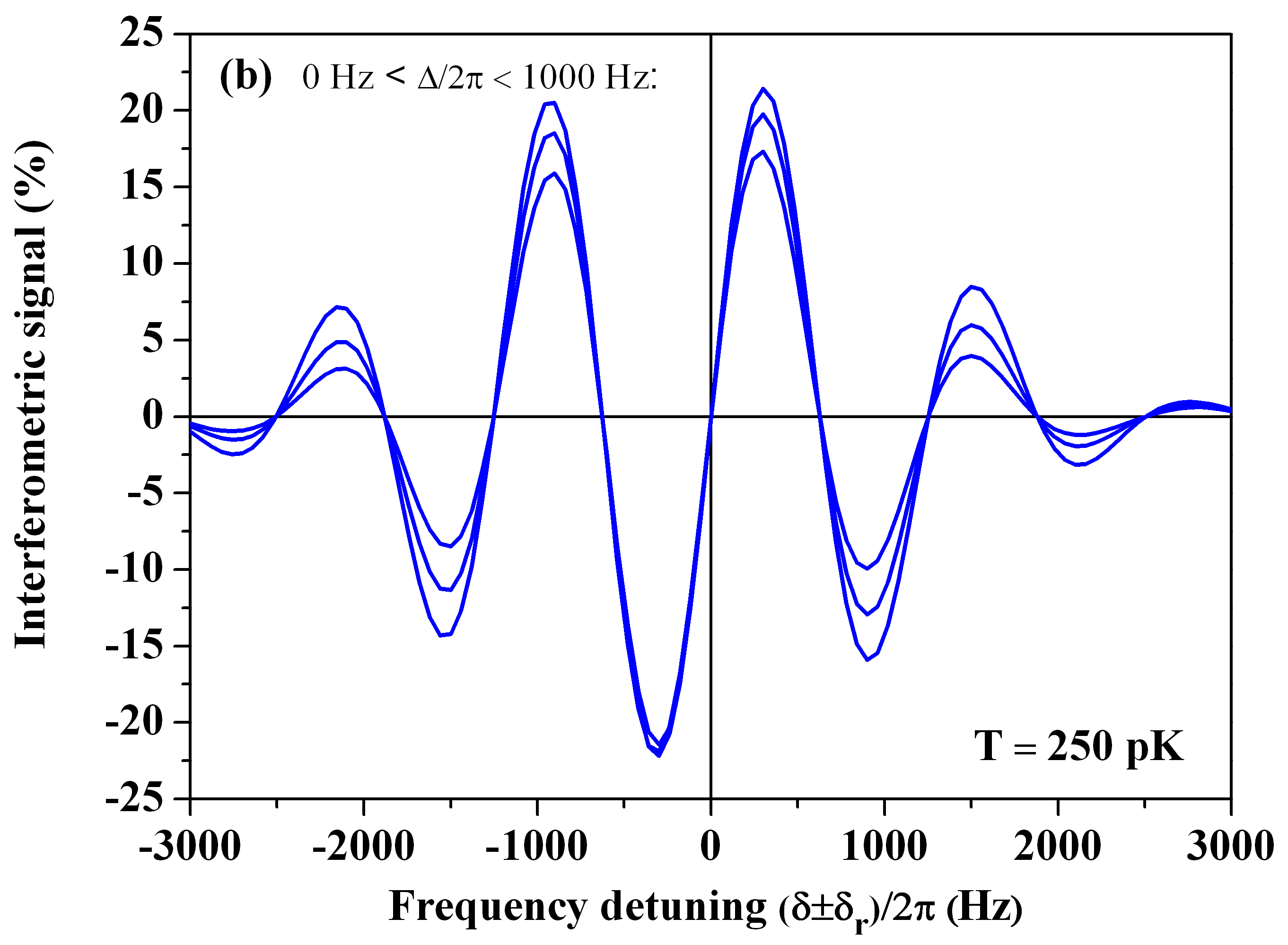}}
\caption{(color online). (a) Hybrid-GHRB (HGHRB) circuit-diagram of laser parameters including a cooperative protocol to generate a robust error signal against residual uncompensated part of the light-shift $\Delta/2\pi$ coupled to transverse velocity $\textup{kv}_{z}$. (b) Dispersive error signal $\Delta\textup{E}_{\textup{HGHRB}(+)}$ based on Eq.~\ref{eq:GHRB-error-signal-1} (or equivalently with Eq.~\ref{eq:GHRB-error-signal-2}) versus uncompensated part of the residual light-shift $\Delta/2\pi$ with $T=250$~pK. All plotted lines are generated with same parameters as in Fig.~\ref{fig:recoil-doublet}.}
\label{fig-HGHRB}
\end{figure*}
\begin{figure}[t!!]
\center
\resizebox{8.5cm}{!}{\includegraphics[angle=0]{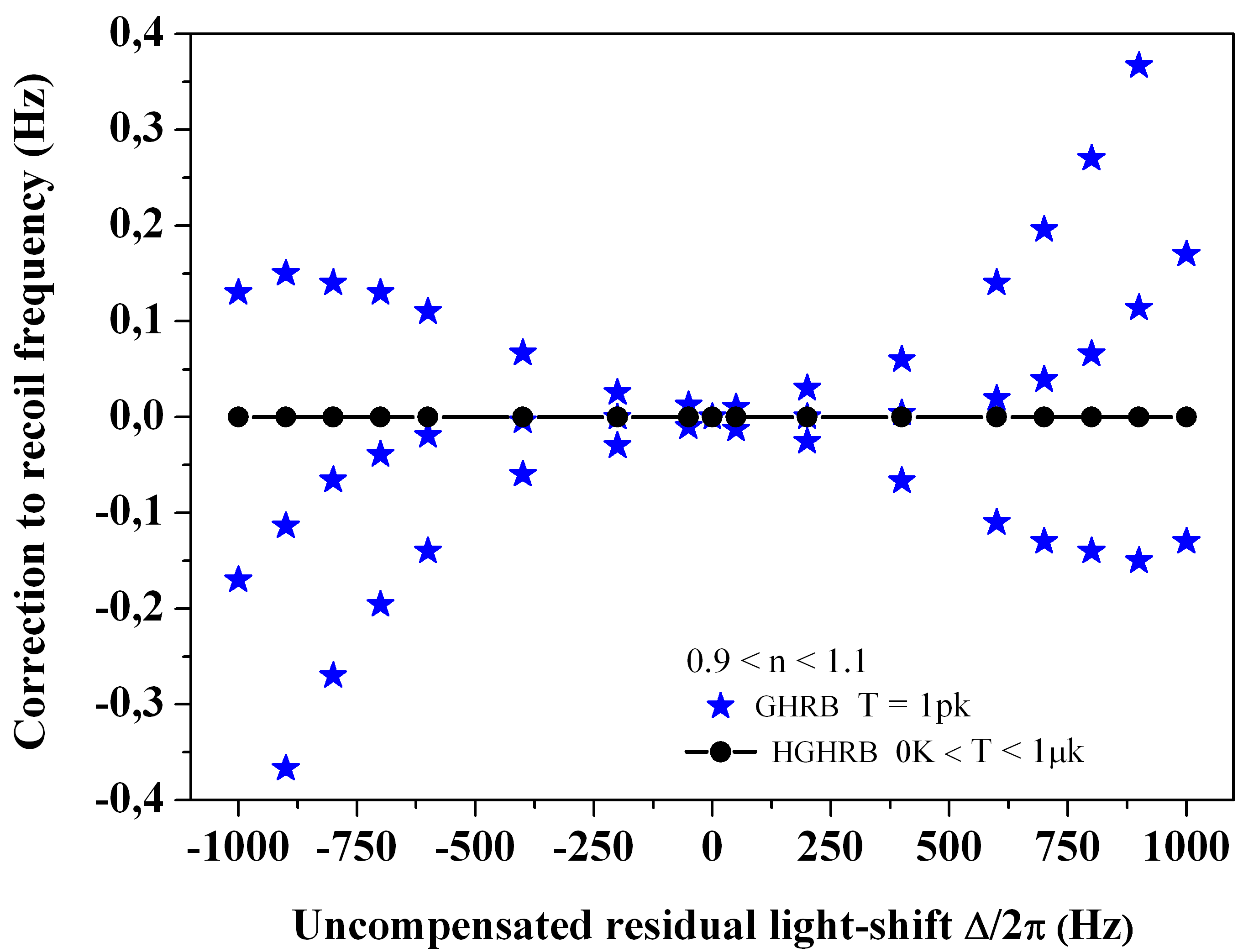}}
\caption{(color online). HGHRB error signal frequency-shift versus the residual light-shift $\Delta/2\pi$ for different temperatures. The pulse area variation $\Omega\tau=\textup{n}\times\pi/2$ ($\pi/2\equiv\boldsymbol{90}^{\circ}$) is driven by the parameter \textup{n} from 0.9 to 1.1 between the two interaction zones $_{1}^{2}$M\textbf{I$(\uparrow)$}, $_{1}^{2}$M\textbf{II$(\downarrow)$}. All plotted lines are generated with same other parameters as in Fig.~\ref{fig:recoil-doublet}.}
\label{fig-HGHRB-shift}
\end{figure}
The extraction of the composite phase-shift from analytical expressions of error signal shapes is not always an easy task and sometimes requires a numerical tracking of the central dispersive feature. In the following figures that we have produced, we have numerically plotted the error signal and associated frequency-shifts for an accurate evaluation of interference distortion. We have also checked that nonlinear effects leading to these distortions are effectively related in many part to the corrections resulting from Eq.~\ref{eq:HF-recoil} and Eq.~\ref{eq:LF-recoil}.

\indent Dispersive errors signals based on hyper-Ramsey-Bordé protocols from Tab.~\ref{protocol-table-2} are generated following Eq.~\ref{eq:GHRB-error-signal}.
We have reported typical dispersive error signals integrated over a narrow transverse gaussian velocity distribution around T$=250$~pK using Ca atomic parameters in Fig.~\ref{fig:recoil-doublet}.
 Dispersive error signal are shown related to the original RB interrogation scheme in Fig.~\ref{fig:recoil-doublet}(a), to the GHRB interferometer in (b), to the HRB3$_{\pi}$ and the HRB5$_{\pi}$ respectively in (c) and (d) figures. Matter-waves are all split in two wavepackets that are separated by the atomic doublet recoil doublet around $\sim23$~kHz for Ca.

We have also plotted the correction to the recoil due to residual light-shifts for two different transverse velocity distributions of the wave packet. In the ideal case T$\rightarrow0$~K presented in Fig.~\ref{fig:RB-shift}(a), the HRB3$_{\pi}$ and HRB5$_{\pi}$ interferometric schemes are exhibiting respectively a highly nonlinear cubic and quintic dependence of the recoil correction to the residual uncompensated part of the light-shift $\Delta/2\pi$. As expected, the GHRB scheme is still completely removing the dependence in the residual light-shift at all order in the detuning.
However, the assumption that a sample of trapped atoms are in the T$\rightarrow0$~K regime is unrealistic.

By integrating the interferometric error signal over a transverse gaussian distribution of velocities at T$=250$~K as shown in Fig.~\ref{fig:RB-shift}(b), the nonlinear compensation of the residual light-shift is lost and a small linear dependence of the recoil correction with $\Delta/2\pi$ is restored for both protocols. A small asymmetry in pulse area between the two sets of Ramsey-Bordé interaction zones also generates a small sensitivity to potential residual light-shifts.
However, let us remark that it is possible to cool atomic samples to ultra-cold temperatures relying on delta-kick techniques or sub-recoil cooling to
reach nK to pK temperatures with very narrow momentum dispersion~\cite{Kovachy:2015,Karcher:2018}. Reaching lower temperature is thus an additional benefit for robust matter-wave interferometry with composite pulses.

The hybrid GHRB circuit-diagram of laser parameters including a cooperative combination of error signals is reported in Fig.\ref{fig-HGHRB}(a).
They are produced with $\pm\pi/8$ or $\pm3\pi/8$ laser phase-steps, through Eq.~\ref{eq:GHRB-error-signal} and are mixing error signals $\Delta\textup{E}_{\textup{GHR}(\pi/8)}$ and $\Delta\textup{E}_{\textup{GHR}(3\pi/8)}$ as following:
\begin{subequations}
\begin{align}
\Delta\textup{E}_{\textup{GHRB}(-)}&=\frac{1}{2}\left(\Delta\textup{E}_{\textup{GHRB}(\pi/8)}-\Delta\textup{E}_{\textup{GHRB}(3\pi/8)}\right)\label{eq:GHRB-error-signal-a},\\
\Delta\textup{E}^{\dagger}_{\textup{GHRB}(-)}&=\frac{1}{2}\left(\Delta\textup{E}^{\dagger}_{\textup{GHRB}(\pi/8)}-\Delta\textup{E}^{\dagger}_{\textup{GHRB}(3\pi/8)}\right)\label{eq:GHRB-error-signal-b},
\end{align}
\end{subequations}
and equivalently
\begin{subequations}
\begin{align}
\Delta\textup{E}_{\textup{GHRB}_{\pi}(-)}&=\frac{1}{2}\left(\Delta\textup{E}_{\textup{GHRB}_{\pi}(\pi/8)}-\Delta\textup{E}_{\textup{GHRB}_{\pi}(3\pi/8)}\right)\label{eq:GHRB-PI-error-signal-a},\\
\Delta\textup{E}^{\dagger}_{\textup{GHRB}_{\pi}(-)}&=\frac{1}{2}\left(\Delta\textup{E}^{\dagger}_{\textup{GHRB}_{\pi}(\pi/8)}-\Delta\textup{E}^{\dagger}_{\textup{GHRB}_{\pi}(3\pi/8)}\right)\label{eq:GHRB-PI-error-signal-b},
\end{align}
\end{subequations}
The question arises if new dispersive error signals can be generated to be more robust to the Doppler transverse velocity of the cold atomic sample.
We have investigated the properties of a protocol against uncompensated light-shift coupled to pulse area variation introducing a phase-jump on laser pulses.
In order to make an error signal more robust to residual light-shifts and pulse area errors, even at relatively high temperatures, we present a robust interferometric error signal against transverse Doppler-shift.
The hybrid protocol, reported in Tab.~\ref{protocol-table-2} and shown in  Fig.~\ref{fig-HGHRB}(a), is based on a combination of GHRB error signals that are phase-shifted by $\pi$ on specific pulses as following:
\begin{subequations}
\begin{align}
\Delta\textup{E}_{\textup{HGHRB}(+)}=&\frac{1}{2}\left(\Delta\textup{E}_{\textup{GHRB}(-)}+\Delta\textup{E}_{\textup{GHRB}_{\pi}(-)}\right)\label{eq:GHRB-error-signal-1},\\
\Delta\textup{E}^{\dagger}_{\textup{HGHRB}(+)}=&\frac{1}{2}\left(\Delta\textup{E}^{\dagger}_{\textup{GHRB}(-)}+\Delta\textup{E}^{\dagger}_{\textup{GHRB}_{\pi}(-)}\right)\label{eq:GHRB-error-signal-2}.
\end{align}
\end{subequations}
Combination of phase-shifted signals are cooperatively working to completely cancel any residual light-shifts and transverse Doppler-shifts~\cite{Braun:2014,Zanon-willette:2017}. We have also checked that symmetric or asymmetric residual light-shifts equivalent to residual Doppler-shifts with opposite wave-vectors between $_{p}^{q}$M\textbf{I$(\uparrow)$} and $_{p}^{q}$M\textbf{II$(\downarrow)$} building-blocks are canceled when the pulse area is changing by $\pm10\%$ between the two free evolution zones.
We have plotted the dispersive error signal shape in Fig.~\ref{fig-HGHRB}(b) versus the clock detuning $\delta/2\pi$ centered around each recoil frequency component for different residual light-shifts and a fixed transverse temperature about  T$=250$~pK. The solid blue lines are the numerical tracking of the error signal dispersive shape near each recoil frequency component. Such a cooperative pulse protocol can mimic a spin-echo configuration without reversing the pulse order timing and changing the spatial orientation of optical traveling waves to the opposite direction.

\indent We have demonstrated in this section that coherent manipulation of quantum interferences with clock interferometers can be extended to robust atomic interferometers with the help of elaborated and cooperative sequences of composite pulses based on specific $\pi/8$ and $3\pi/8$ phase-steps.
The interferometric schemes we have analyzed in this section, based on Fig.~\ref{fig-GHRB-1} and Fig.~\ref{fig-GHRB-2}, are asymmetric because atoms are not spending the same amount of time in both arms along the interfering paths. An hyper-Ramsey-Bordé interferometer is thus sensitive to the global phase-shift accumulation related to the clock frequency detuning centered at the recoil components making possible to transfer the technique of composite pulse protocols from clocks to Ramsey-Bordé interferometers for recoil measurement.

\begin{table*}[t!!]
\renewcommand{\arraystretch}{1.7}
\begin{tabular}{|c|c|c|}
\hline
\hline
protocols & composite pulse building-block $_{p}^{q}$M\textbf{($\uparrow$)} \\
\hline
  MZ
    & \begin{tabular}{c}
  $\boldsymbol{90^{\circ'\uparrow}}_{0}\dashv\delta^{\uparrow}\textup{T}\vdash\boldsymbol{180^{\circ\uparrow}}_{0}\dashv\delta^{\uparrow}\textup{T}\vdash\boldsymbol{90^{\circ\uparrow}}_{0}$ \end{tabular}
 \\
 \hline
\hline
BU1
    & \begin{tabular}{c}
  $\boldsymbol{90^{\circ'\uparrow}}_{0}\dashv\delta^{\uparrow}\textup{T}\vdash\boldsymbol{180^{\circ\uparrow}}_{0}\dashv\delta^{\uparrow}\textup{T'}\vdash\boldsymbol{180^{\circ\uparrow}}_{0}\dashv\delta^{\uparrow}\textup{T}\vdash\boldsymbol{90^{\circ\uparrow}}_{0}$ \end{tabular}
   \\
\hline
\hline
BU2
    & \begin{tabular}{c}
  $\boldsymbol{90^{\circ'\uparrow}}_{0}\dashv\delta^{\uparrow}\textup{T}\vdash\boldsymbol{180^{\circ\uparrow}}_{0}\dashv\delta^{\uparrow}\textup{T'}\vdash\boldsymbol{180^{\circ\uparrow}}_{0}\dashv\delta^{\uparrow}\textup{T}\vdash\boldsymbol{270^{\circ\uparrow}}_{0}$\\
 $\boldsymbol{270^{\circ'\uparrow}}_{0}\dashv\delta^{\uparrow}\textup{T}\vdash\boldsymbol{180^{\circ\uparrow}}_{0}\dashv\delta^{\uparrow}\textup{T'}\vdash\boldsymbol{180^{\circ\uparrow}}_{0}\dashv\delta^{\uparrow}\textup{T}\vdash\boldsymbol{90^{\circ\uparrow}}_{0}$
\end{tabular}
   \\
\hline
\hline
BU3
    & \begin{tabular}{c}
  $\boldsymbol{90^{\circ'\uparrow}}_{0}\dashv\delta^{\uparrow}\textup{T}\vdash\boldsymbol{180^{\circ\uparrow}}_{0}\dashv\delta^{\uparrow}\textup{T'}\vdash\boldsymbol{180^{\circ\uparrow}}_{0}\dashv\delta^{\uparrow}\textup{T}\vdash\boldsymbol{180^{\circ\uparrow}}_{\pi}\boldsymbol{90^{\circ\uparrow}}_{0}$\\ $\boldsymbol{90^{\circ'\uparrow}}_{0}\boldsymbol{180^{\circ'\uparrow}}_{\pi}\dashv\delta^{\uparrow}\textup{T}\vdash\boldsymbol{180^{\circ\uparrow}}_{0}\dashv\delta^{\uparrow}\textup{T'}\vdash\boldsymbol{180^{\circ\uparrow}}_{0}\dashv\delta^{\uparrow}\textup{T}\vdash\boldsymbol{90^{\circ\uparrow}}_{0}$
\end{tabular}
   \\
 \hline
\hline
\begin{tabular}{c}
\\
  HMZ1($\varphi$)\\
  HMZ2($\varphi$)\\
   or  \\
    HMZ1($\varphi$)\\
  HMZ2($\varphi$)\\
      ($\varphi=\pi/4,3\pi/4$)\\

    \end{tabular}
       & \begin{tabular}{c}
                                                               $\boldsymbol{90^{\circ'\uparrow}}_{0}\dashv\delta^{\uparrow}\textup{T}\vdash\boldsymbol{180^{\circ\uparrow}}_{0}\dashv\delta^{\uparrow}\textup{T}\vdash\boldsymbol{180^{\circ\uparrow}}_{\pm\varphi}\boldsymbol{90^{\circ\uparrow}}_{0}$\\        $\boldsymbol{90^{\circ'\uparrow}}_{0}\boldsymbol{180^{\circ'\uparrow}}_{\pm\varphi}\dashv\delta^{\uparrow}\textup{T}\vdash\boldsymbol{180^{\circ\uparrow}}_{0}\dashv\delta^{\uparrow}\textup{T}\vdash\boldsymbol{90^{\circ\uparrow}}_{0}$\\
                                                               or  \\
                                                               $\boldsymbol{270^{\circ'\uparrow}}_{0}\dashv\delta^{\uparrow}\textup{T}\vdash\boldsymbol{180^{\circ\uparrow}}_{0}\dashv\delta^{\uparrow}\textup{T}\vdash\boldsymbol{180^{\circ\uparrow}}_{\pm\varphi}\boldsymbol{270^{\circ\uparrow}}_{0}$\\        $\boldsymbol{270^{\circ'\uparrow}}_{0}\boldsymbol{180^{\circ'\uparrow}}_{\pm\varphi}\dashv\delta^{\uparrow}\textup{T}\vdash\boldsymbol{180^{\circ\uparrow}}_{0}\dashv\delta^{\uparrow}\textup{T}\vdash\boldsymbol{270^{\circ\uparrow}}_{0}$
                                                               \end{tabular}
 \\
 \hline
\hline
\begin{tabular}{c}
\\
  HBU1($\varphi$)\\
    HBU2($\varphi$)\\
      or  \\
   HBU1($\varphi$)\\
    HBU2($\varphi$)\\
  ($\varphi=\pi/4,3\pi/4$)
    \end{tabular}
       & \begin{tabular}{c}
                                                              $\boldsymbol{90^{\circ'\uparrow}}_{0}\dashv\delta^{\uparrow}\textup{T}\vdash\boldsymbol{180^{\circ\uparrow}}_{0}\dashv\delta^{\uparrow}\textup{T'}\vdash\boldsymbol{180^{\circ\uparrow}}_{0}\dashv\delta^{\uparrow}\textup{T}\vdash\boldsymbol{180^{\circ\uparrow}}_{\pm\varphi}\boldsymbol{90^{\circ\uparrow}}_{0}$\\ $\boldsymbol{90^{\circ'\uparrow}}_{0}\boldsymbol{180^{\circ'\uparrow}}_{\mp\varphi}\dashv\delta^{\uparrow}\textup{T}\vdash\boldsymbol{180^{\circ\uparrow}}_{0}\dashv\delta^{\uparrow}\textup{T'}\vdash\boldsymbol{180^{\circ\uparrow}}_{0}\dashv\delta^{\uparrow}\textup{T}\vdash\boldsymbol{90^{\circ\uparrow}}_{0}$
                                                            \\   or  \\
                                                               $\boldsymbol{270^{\circ'\uparrow}}_{0}\dashv\delta^{\uparrow}\textup{T}\vdash\boldsymbol{180^{\circ\uparrow}}_{0}\dashv\delta^{\uparrow}\textup{T'}\vdash\boldsymbol{180^{\circ\uparrow}}_{0}\dashv\delta^{\uparrow}\textup{T}\vdash\boldsymbol{180^{\circ\uparrow}}_{\pm\varphi}\boldsymbol{270^{\circ\uparrow}}_{0}$\\ $\boldsymbol{270^{\circ'\uparrow}}_{0}\boldsymbol{180^{\circ'\uparrow}}_{\mp\varphi}\dashv\delta^{\uparrow}\textup{T}\vdash\boldsymbol{180^{\circ\uparrow}}_{0}\dashv\delta^{\uparrow}\textup{T'}\vdash\boldsymbol{180^{\circ\uparrow}}_{0}\dashv\delta^{\uparrow}\textup{T}\vdash\boldsymbol{270^{\circ\uparrow}}_{0}$

                                                              \end{tabular}

 \\
 \hline
\hline
\end{tabular}
\centering%
\caption{A few selected composite pulses interrogation protocols for (MZ), (BU) interferometers, hyper-Mach-Zehnder (HMZ) and hyper-Butterfly (HBU) interferometers. The laser phase is indicated as a subscript. Atomic trajectories accumulating phases that are sensitive to acceleration or rotation can be found in the appendix subsection S4-3. Arrow $\uparrow$ denotes the detuning corrected by Doppler-shift and recoil absorption within composite pulses and during free evolution time.}
\label{protocol-table-3}
\end{table*}
\begin{figure*}[t!!]
\center
\resizebox{9.5cm}{!}{\includegraphics[angle=0]{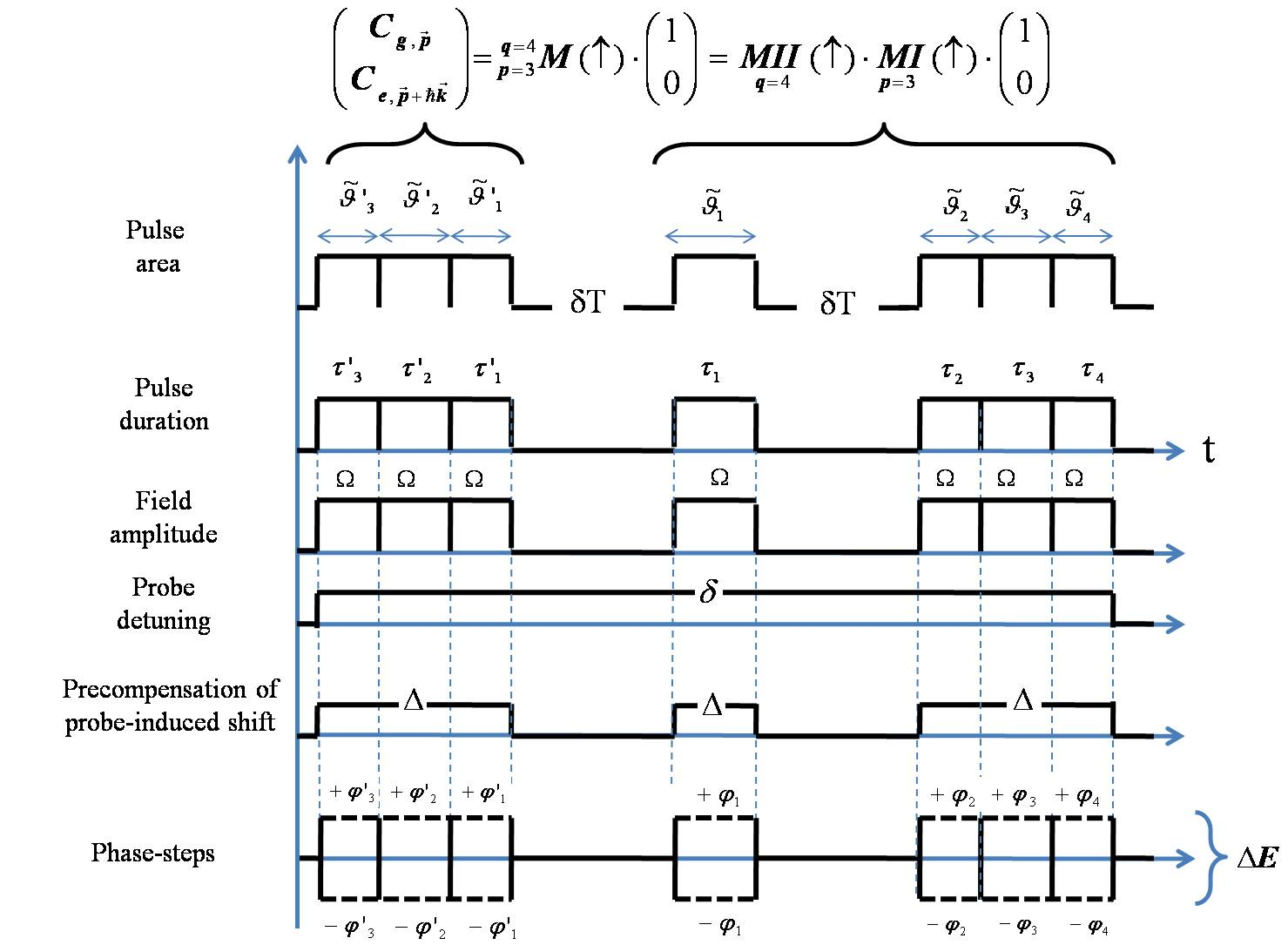}}\resizebox{8.5cm}{!}{\includegraphics[angle=0]{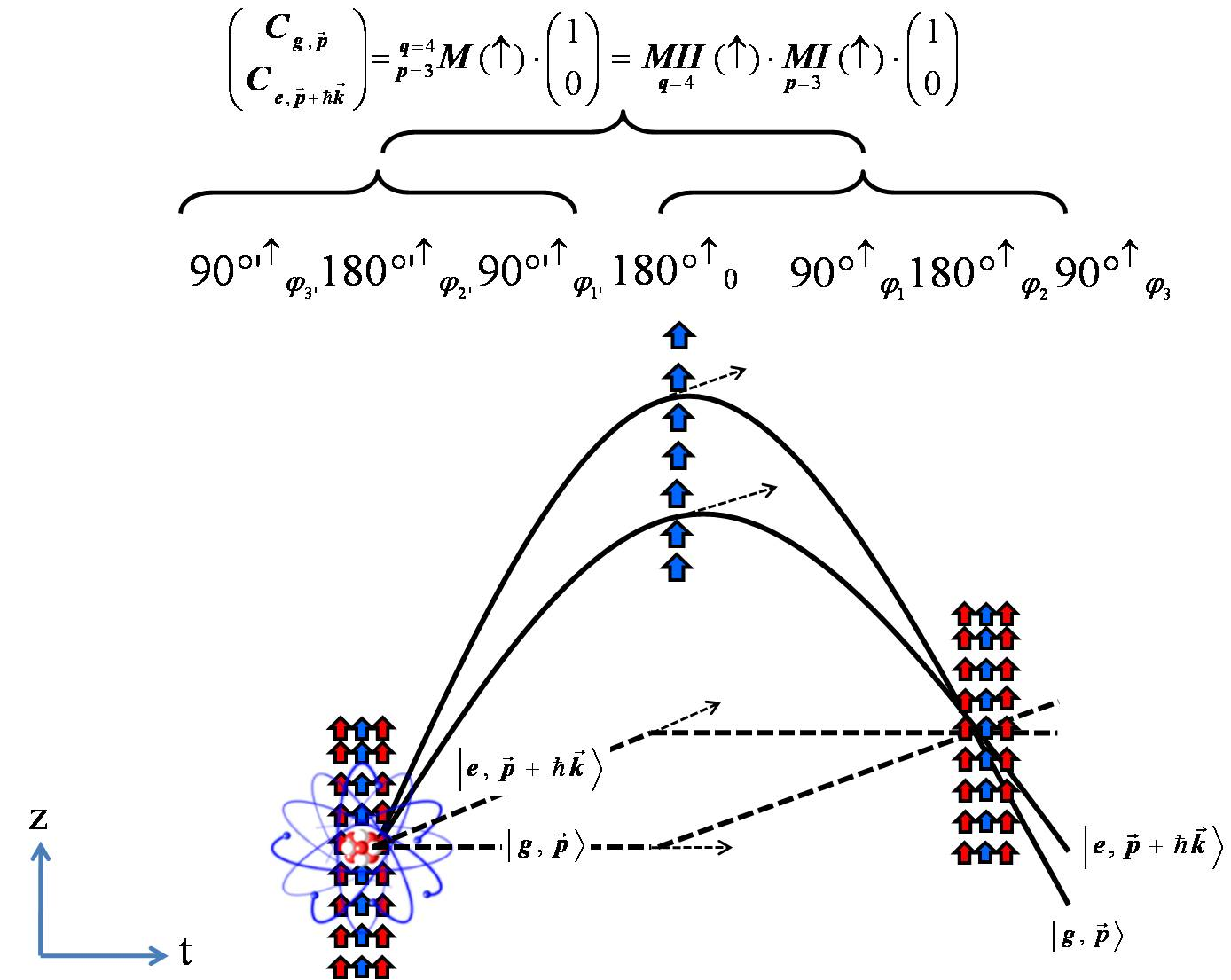}}
\resizebox{9.5cm}{!}{\includegraphics[angle=0]{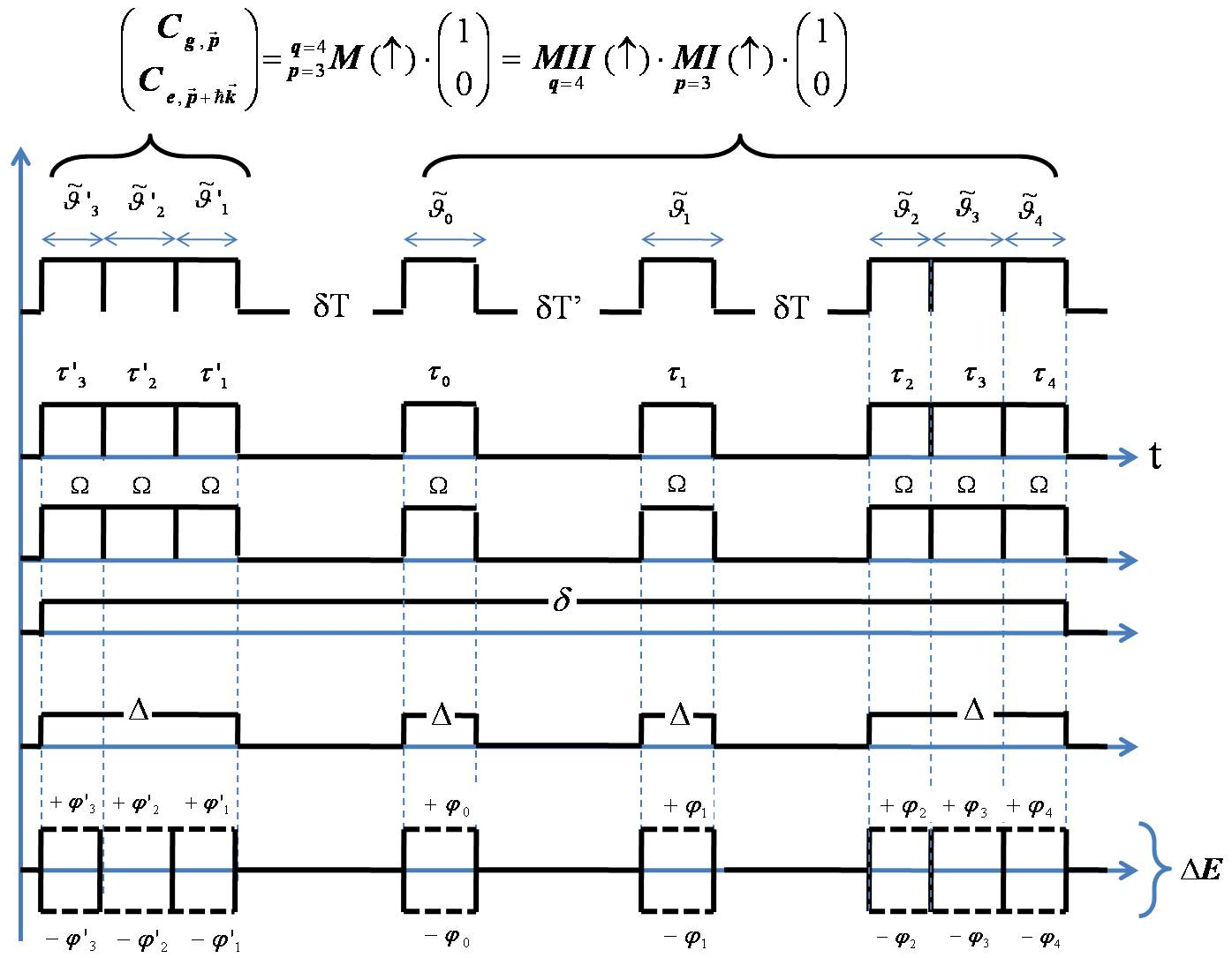}}\resizebox{8.5cm}{!}{\includegraphics[angle=0]{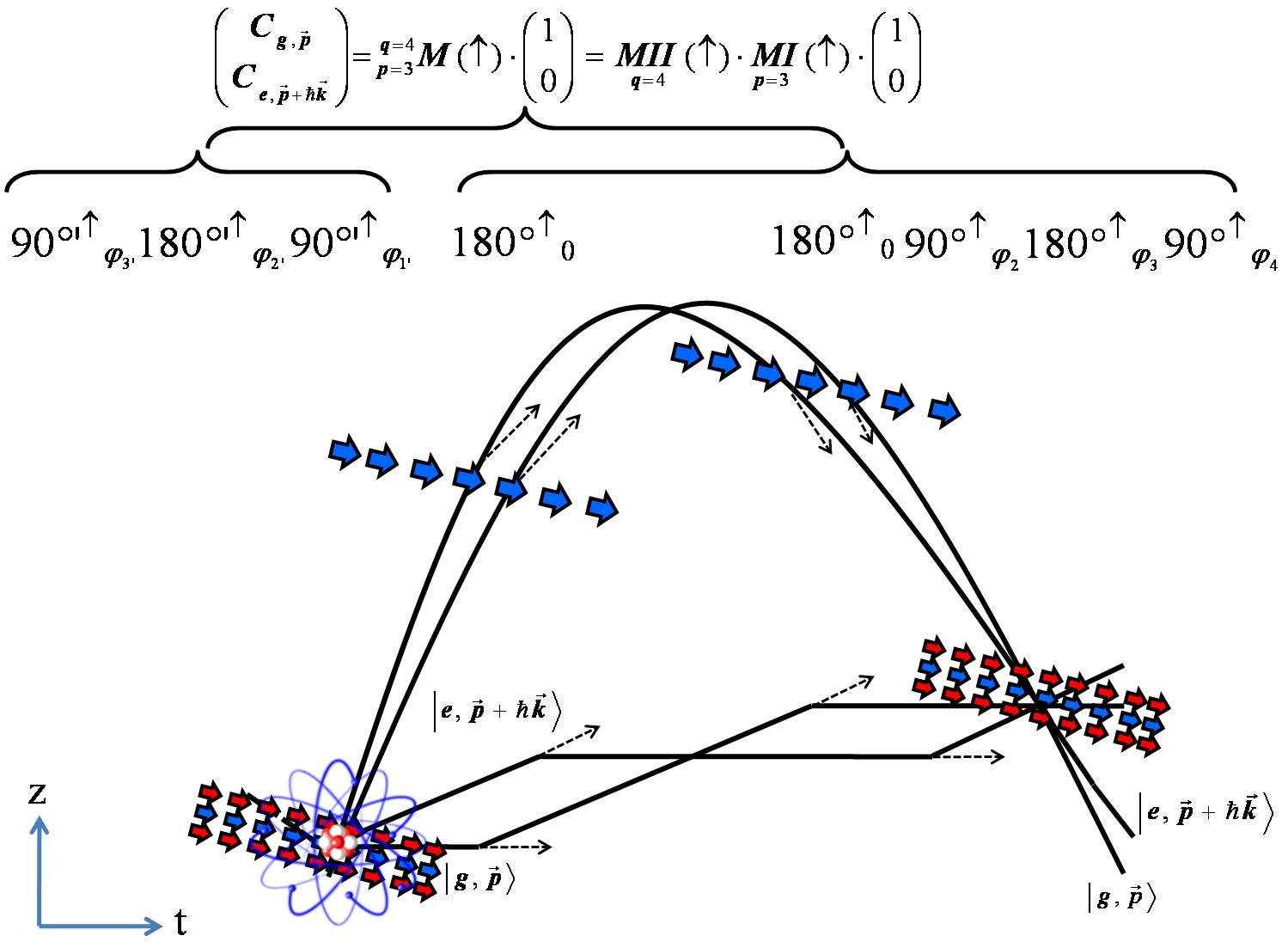}}
\caption{(color online). (Left top panel) Hyper-Mach-Zehnder (HMZ) and (left down panel) hyper-butterfly (GHB) interferometers with composite pulses. For these configurations, two sets of co-propagating composite laser pulses are separated by one single or two intermediate $\boldsymbol{180^{\uparrow}}_{0}$ pulses where $\uparrow$ arrows are corresponding to laser detuning modified by $\textup{k}v_{z}$ Doppler wave-vector orientation and atomic recoil (see Tab.~\ref{protocol-table-3}).
The full interaction geometry can be directly evaluated either from a single interaction matrix $_{3}^{4}$M\textbf{$(\uparrow)$} or from the product of two independent building-blocks M\textbf{I$(\uparrow)$} and M\textbf{II$(\uparrow)$} respectively with $p=3$ pulses and $q=4$ pulses (see appendix section S4).
Laser pulse phases are chosen to be for the first set of pulses denoted as $\varphi'_{3,2,1}$, for the reverse pulses $\varphi_{0}=\varphi_{1}=0$ and $\varphi_{2,3,4}$ for the last set of pulses.}
\label{fig-HMZ-HBU-diagram}
\end{figure*}

\subsection{HYPER-MACH-ZEHNDER AND HYPER-BUTTERFLY INTERFEROMETERS}

\indent We turn now to symmetric atomic Mach-Zehnder (MZ) and Butterfly or double-loop (BU) interferometers. Such geometrical configurations are insensitive to clock detunings and Doppler-shifts making them accurate and sensitive to acceleration and rotation~\cite{Berman:1997,Abend:2019}. However, variations of the laser field amplitude between sets of pulses acting as beam splitters may restore a parasitic distortion related to a residual Doppler-shift~\cite{Gillot:2016}. From an experimental view, the wavefront distortion from imperfect optics affects the phases and the amplitudes of the laser beams used to diffract the atomic wave-packet presenting some inhomogeneities during interrogation protocols and magnified by expansion of the atomic cloud over time~\cite{Schkolnik:2015,Parker:2016,Trimeche:2017,Bade:2018}.

So far, we will consider wave-packets trajectories modified by acceleration or rotation during pulses and the free flight.
We consider atoms interacting with traveling waves. The interaction with laser pulses introduce a change in the internal state accompanied by a change in transverse momentum. We should replace the laser phase expression from off-diagonal matrix coefficients $\textup{M}(\widetilde{\vartheta}_{l})$ by space-time dependent phase-shifts including atomic trajectories. Thus, the local laser phase that the atoms experience at the lth pulse is now given by~\cite{Cheinet:2006,Takase:2008,Li:2015,Storey:1994,Guery-Odelin:2011}:
\begin{subequations}
\begin{align}
\varphi_{l}\Rightarrow\varphi_{l}+\textbf{k}_{l}\cdot\textbf{z}_{l}(t_{l})-\omega_{l}t_{l}
\end{align}
\end{subequations}
where $k_{l}$ is the wave vector, $\omega_{l}$ the laser frequency at time $t_{l}$ and $z_{l}(t)$ is the classical path describing the atomic wave-packet motion into a frame containing local acceleration and rotation~\cite{Storey:1994,Guery-Odelin:2011}.

We propose to remove limits to the symmetry of the original MZ or BU type under pulse area variation between pulses.
We will consider, for simplicity, a HMZ symmetric configuration to only measure a gravitational acceleration phase-shift $\textbf{k}_{l}\cdot\textbf{g}\textup{T}^{2}$ and a HBU symmetric configuration to measure a rotation phase-shift $2\left(\textbf{k}_{l}\times\textbf{g}\right)\cdot\boldsymbol{\Omega}\textup{T}^{3}$ depending on wave vector orientation ($\textbf{g}$ is the local gravitational acceleration and $\boldsymbol{\Omega}$ is a rotation rate, see for example~\cite{Cheinet:2006,Takase:2008}). We have also ignored any asymmetric free evolution time interval between composite pulses to avoid a simultaneous combination of acceleration and rotation for the BU configuration~\cite{Kleinert:2015}.

Several options exist to reconstruct matter-wave interferences given access to acceleration or rotation phase-shifts.
From a quantum engineering perspective, we decide to focus on composite pulse protocols that are offering the best tailoring approach of atomic interferences
to produce, by pulse engineering methods, an optimization of some targeted performances, i.e frequency-shift and signal amplitude of interferences, making atomic sensors more robust to important
variations of relevant experimental parameters.

The present section starts by deriving analytically transition probabilities of beam splitters made of three composite pulses (see appendix, section S4, S4-1 and S4-2 with analytics used to derive $\textup{P}_{\textup{HMZ}}$ and $\textup{P}_{\textup{HBU}}$). Then, we will analyze the matter-wave interferences and associated phase-shifts for original MZ and BU atomic sensors.
Matter-wave interferences are recorded by chirping the laser frequency detuning to correct for the induced Doppler-shift during the free fall all over the pulsed sequence duration. It allows us to observe small distortions of fringes induced by residual uncompensated Doppler-shift and light-shift coupled to pulse area variation between beam splitters or set of composite beam splitters. We will then focus on new protocols based on $\pm\pi/4,3\pi/4$ laser phase-steps mixed with cooperative combinations of error signals that are able to eliminate distortions inducing residual phase-shifts. Finally, auto-balanced hyper-interferometers with composite pulses will be presented as a potential alternative to realize a fault-tolerant interferometer against phase and pulse area distortion that are arising along the wavepackets trajectories.

\subsubsection{HYPER-MACH-ZEHNDER (HMZ)}

\indent The composite pulse hyper-MZ (HMZ) interferometer is shown in the left top panel of Fig.~\ref{fig-HMZ-HBU-diagram}. In such a configuration also identical to the Hahn-echo scheme, the clock frequency dependence is removed~\cite{Gillot:2016}.
The exact hyper transition probability $\textup{P}_{\textup{HMZ}}$ is established fixing $C_{g}(0)=1,C_{e}(0)=0$. The analytic computation can be realized either reducing the sequence of pulses to a single interaction matrix $_{3}^{4}$M\textbf{$(\uparrow)$} or decomposing the interaction geometry in a product of two independent building-blocks M\textbf{I$(\uparrow)$} and M\textbf{II$(\uparrow)$} respectively with $p=3$ pulses and $q=4$ pulses. In all cases, we take $\delta^{\uparrow}\textup{T}\mapsto0$ simplifying the calculation while the intermediate $180^{\uparrow}_{0}$ reverse pulse is now played by index $l=1$ in the second set of \textit{q} pulses within the matrix M\textbf{II$(\uparrow)$} (see atomic trajectories associated to the HMZ configuration in the top right panel of Fig.~\ref{fig-HMZ-HBU-diagram}).

Here, we employ analytical results from the appendix section S4 evaluating the single interaction matrix $_{3}^{4}$M\textbf{$(\uparrow)$} for the HMZ configuration.
The matter-wave interferometric signal is computed leading to the following expression:
\begin{equation}
\textup{P}_{\textup{HMZ}}=\left|_{3}^{4}\alpha_{gg}\left[1-\left|_{3}^{4}\beta_{gg}\right|e^{-i_{3}^{4}\Phi_{\textup{HMZ}}}\right]\right|^{2}
\label{eq:HMZ}
\end{equation}
where $\alpha_{1}'^{3}(gg)$, $\alpha_{1}^{4}(gg)\mapsto\alpha_{2}^{4}(gg)$ and $\beta_{1}'^{3}(gg)$, $\beta_{1}^{4}(gg)\mapsto\beta_{2}^{4}(gg)$ (see appendix section S4-1 for laser index modification due to non overlapping wavepackets from the intermediate reversal pulse) are:
\begin{equation}
\begin{split}
\alpha_{1}'^{3}(gg)=&\left(\prod_{\textup{1}}^{3}\cos\widetilde{\vartheta}'_{l}e^{i\phi'_{\textup{l}}}\right)\cdot\left(1-S'_{3,2}\right)\\
S'_{3,2}=&e^{-i\Xi'_{12}}\tan\widetilde{\vartheta}'_{1}\tan\widetilde{\vartheta}'_{2}+e^{-i\Xi_{13}}\tan\widetilde{\vartheta}'_{1}\tan\widetilde{\vartheta}'_{3}\\
&+e^{-i\Xi_{23}}\tan\widetilde{\vartheta}'_{2}\tan\widetilde{\vartheta}_{3}\label{eq:modified-HMZalphagg-13}
\end{split}
\end{equation}
and
\begin{equation}
\begin{split}
\alpha_{2}^{4}(gg)=&\sin\widetilde{\vartheta}_{1}\left(\prod_{\textup{2}}^{q=\textup{4}}\cos\widetilde{\vartheta}_{l}e^{i\phi_{\textup{l}}}\right)\cdot\left(-S_{4,2}+S_{4,4}\right)\\
S_{4,2}=&e^{-i\Xi_{12}}\tan\widetilde{\vartheta}_{2}+e^{-i\Xi_{13}}\tan\widetilde{\vartheta}_{3}+e^{-i\Xi_{14}}\tan\widetilde{\vartheta}_{4}\\
S_{4,4}=&e^{-i\Xi_{1234}}\tan\widetilde{\vartheta}_{2}\tan\widetilde{\vartheta}_{3}\tan\widetilde{\vartheta}_{4}
\label{eq:modified-HMZalphagg-24}
\end{split}
\end{equation}
with
\begin{subequations}
\begin{align}
\beta_{1}'^{3}(gg)=&\frac{\tan\widetilde{\vartheta}'_{1}+e^{-i\Xi'_{12}}\frac{\tan\widetilde{\vartheta}'_{2}+e^{-i\Xi'_{23}}\tan\widetilde{\vartheta}'_{3}}
{1-e^{-i\Xi'_{23}}\tan\widetilde{\vartheta}'_{2}\tan\widetilde{\vartheta}'_{3}}}
{1-e^{-i\Xi'_{12}}\tan\widetilde{\vartheta}'_{1}\frac{\tan\widetilde{\vartheta}'_{2}+e^{-i\Xi'_{23}}\tan\widetilde{\vartheta}'_{3}}
{1-e^{-i\Xi'_{23}}\tan\widetilde{\vartheta}'_{2}\tan\widetilde{\vartheta}'_{3}}}\label{eq:modified-HMZbetagg-13}\\
\beta_{2}^{4}(gg)=&-\frac{1}{\frac{\tan\widetilde{\vartheta}_{2}+e^{-i\Xi_{23}}\frac{\tan\widetilde{\vartheta}_{3}+e^{-i\Xi_{34}}\tan\widetilde{\vartheta}_{4}}
{1-e^{-i\Xi_{34}}\tan\widetilde{\vartheta}_{3}\tan\widetilde{\vartheta}_{4}}}{1-e^{-i\Xi_{23}}\tan\widetilde{\vartheta}_{2}\frac{\tan\widetilde{\vartheta}_{3}+e^{-i\Xi_{34}}\tan\widetilde{\vartheta}_{4}}
{1-e^{-i\Xi_{34}}\tan\widetilde{\vartheta}_{3}\tan\widetilde{\vartheta}_{4}}}}\label{eq:modified-HMZbetagg-24}
\end{align}
\end{subequations}
The HMZ composite phase-shift is:
\begin{equation}
\begin{split}
_{3}^{4}\Phi_{\textup{HMZ}}=&-\varphi'_{1}+2\varphi_{1}-\varphi_{2}+\phi'_{1}-\phi_{2}\\
&-\textup{Arg}\left[\beta_{1}'^{3}(gg)\cdot\beta_{2}^{4}(gg)\right]
\label{eq:modified-HMZ-phase-shift}
\end{split}
\end{equation}
A direct use of the HMZ phase-shift expression $_{3}^{4}\Phi_{\textup{HMZ}}$ may not always correspond to the correct evaluation of the true central fringe phase-shift. The transition probability given by Eq.~\ref{eq:HMZ} is a non-trivial spectral function, where the terms $\alpha_{1}'^{3}(gg),\alpha_{2}^{4}(gg)$ and $\beta_{1}'^{3}(gg),\beta_{2}^{4}(gg)$ depend on laser phases. For this reason, the composite phase-shift of the central fringe can not be systematically associated to Eq.~\ref{eq:modified-HMZ-phase-shift}. The phase-shift related to the central interference is thus tracking numerically when fringes are recorded by scanning the laser phase $\varphi$ of the HMZ composite pulse beam splitters.

As a validation of previous analytical calculations, we re-derive the original three-pulse MZ configuration with the intermediate $180^{\uparrow}_{0}$ pulse still played with index $l=1$ in the second set of \textit{q} pulses (MZ from Tab.~\ref{protocol-table-3}).
We obtain, after straightforward simplification on envelops $\alpha_{1}'^{3}(gg),\alpha_{2}^{4}(gg)$ and complex terms $\beta_{1}'^{3}(gg),\beta_{2}^{4}(gg)$, the MZ transition probability expressed as:
\begin{equation}
\begin{split}
\textup{P}_{\textup{MZ}}=\left|-\cos\widetilde{\vartheta}'_{1}\sin\widetilde{\vartheta}_{1}\sin\widetilde{\vartheta}_{2}\cdot
\left[1+\frac{\tan\widetilde{\vartheta}'_{1}}{\tan\widetilde{\vartheta}_{2}}e^{-i\Phi_{MZ}}\right]\right|^{2}
\end{split}
\label{eq:}
\end{equation}
where, this time, the MZ phase-shift is easily identified to be:
\begin{equation}
\begin{split}
\Phi_{MZ}=-\varphi'_{1}+2\varphi_{1}-\varphi_{2}+\phi'_{1}-\phi_{2}
\end{split}
\label{eq:MZ-phase-shift}
\end{equation}
consistent with~\cite{Berman:1997,Kleinert:2015,Storey:1994,Guery-Odelin:2011} when $\phi'_{1}=\phi_{2}$.
An additional MZ phase sensitivity to residual Doppler shifts is retrieved when $\phi'_{1}\neq\phi_{2}$ due to the imbalance of Rabi fields between the first and the last beam splitter pulse as expected~\cite{Gillot:2016}.

We point out that a MZ interferometer is a particular case of the symmetric RB configuration. By fixing $\varphi'_{I,1}\equiv\varphi'_{1}$, $\varphi_{I,1}=\varphi'_{II,1}\equiv\varphi_{1}$, $\varphi_{II,1}\equiv\varphi_{2}$ in Eq.~\ref{eq:laser-phase-SRB} and $\phi'_{I,1}\equiv\phi'_{1}$, $\phi_{I,1}=\phi'_{II,1}$ and $\phi_{II,1}\equiv\phi_{2}$ in Eq.~\ref{eq:light-shift-pulses-SRB}, we retrieve the MZ phase-shift given by Eq.~\ref{eq:MZ-phase-shift}.

\subsubsection{HYPER-BUTTERFLY (HBU)}

Here, we employ some analytic results from appendix and section S4 used to evaluate the single interaction matrix $_{3}^{4}$M\textbf{$(\uparrow)$} of the HBU interferometer shown in the left down panel of Fig.~\ref{fig-HMZ-HBU-diagram}.
The matter-wave interferometric signal is computed leading to the following expression:
\begin{equation}
\textup{P}_{\textup{HBU}}=\left|_{3}^{4}\alpha_{gg}\left[1-\left|_{3}^{4}\beta_{gg}\right|e^{-i_{3}^{4}\Phi_{\textup{HBU}}}\right]\right|^{2}
\label{eq:HB}
\end{equation}
where $\alpha_{1}'^{3}(gg)$, $\alpha_{1}^{4}(gg)\mapsto\alpha_{2}^{4}(gg)$ and $\beta_{1}'^{3}(gg)$, $\beta_{1}^{4}(gg)\mapsto\beta_{2}^{4}(gg)$ (see appendix section S4-2 for laser index modification due to non overlapping wave-packets from the intermediate reversal pulse) are:
\begin{equation}
\begin{split}
\alpha_{1}'^{3}(gg)=&\left(\prod_{\textup{1}}^{3}\cos\widetilde{\vartheta}'_{l}e^{i\phi'_{\textup{l}}}\right)\cdot\left(1-S'_{3,2}\right)\\
S'_{3,2}=&e^{-i\Xi'_{12}}\tan\widetilde{\vartheta}'_{1}\tan\widetilde{\vartheta}'_{2}+e^{-i\Xi_{13}}\tan\widetilde{\vartheta}'_{1}\tan\widetilde{\vartheta}'_{3}\\
&+e^{-i\Xi_{23}}\tan\widetilde{\vartheta}'_{2}\tan\widetilde{\vartheta}_{3}\label{eq:modified-HBalphagg-13}
\end{split}
\end{equation}
and
\begin{equation}
\begin{split}
\alpha_{2}^{4}(gg)=&e^{i(\varphi_{1}-\varphi_{0})}\sin\widetilde{\vartheta}_{0}\sin\widetilde{\vartheta}_{1}\left(\prod_{\textup{2}}^{q=\textup{4}}\cos\widetilde{\vartheta}_{l}e^{i\phi_{\textup{l}}}\right)\cdot\left(1-S_{4,2}\right)\\
S_{4,2}=&e^{-i\Xi_{23}}\tan\widetilde{\vartheta}_{2}\tan\widetilde{\vartheta}_{3}+e^{-i\Xi_{34}}\tan\widetilde{\vartheta}_{3}\tan\widetilde{\vartheta}_{4}\\
&+e^{-i\Xi_{24}}\tan\widetilde{\vartheta}_{2}\tan\widetilde{\vartheta}_{4}
\label{eq:modified-HBalphagg-24}
\end{split}
\end{equation}
with
\begin{subequations}
\begin{align}
\beta_{1}'^{3}(gg)=&\frac{\tan\widetilde{\vartheta}'_{1}+e^{-i\Xi'_{12}}\frac{\tan\widetilde{\vartheta}'_{2}+e^{-i\Xi'_{23}}\tan\widetilde{\vartheta}'_{3}}
{1-e^{-i\Xi'_{23}}\tan\widetilde{\vartheta}'_{2}\tan\widetilde{\vartheta}'_{3}}}
{1-e^{-i\Xi'_{12}}\tan\widetilde{\vartheta}'_{1}\frac{\tan\widetilde{\vartheta}'_{2}+e^{-i\Xi'_{23}}\tan\widetilde{\vartheta}'_{3}}
{1-e^{-i\Xi'_{23}}\tan\widetilde{\vartheta}'_{2}\tan\widetilde{\vartheta}'_{3}}}\label{eq:modified-HBbetagg-13}\\
\beta_{2}^{4}(gg)=&\frac{\tan\widetilde{\vartheta}_{2}+e^{-i\Xi_{23}}\frac{\tan\widetilde{\vartheta}_{3}+e^{-i\Xi_{34}}\tan\widetilde{\vartheta}_{4}}
{1-e^{-i\Xi_{34}}\tan\widetilde{\vartheta}_{3}\tan\widetilde{\vartheta}_{4}}}{1-e^{-i\Xi_{23}}\tan\widetilde{\vartheta}_{2}\frac{\tan\widetilde{\vartheta}_{3}+e^{-i\Xi_{34}}\tan\widetilde{\vartheta}_{4}}
{1-e^{-i\Xi_{34}}\tan\widetilde{\vartheta}_{3}\tan\widetilde{\vartheta}_{4}}}\label{eq:modified-HBbetagg-24}
\end{align}
\end{subequations}
The HBU composite phase-shift is:
\begin{equation}
\begin{split}
_{3}^{4}\Phi_{\textup{HBU}}=&-\varphi'_{1}-2\varphi_{1}+2\varphi_{0}+\varphi_{2}+\phi'_{1}+\phi_{2}\\
&-\textup{Arg}\left[\beta_{1}'^{3}(gg)\cdot\beta_{2}^{4}(gg)\right]
\label{eq:HBU-phase-shift}
\end{split}
\end{equation}

The original four-pulse BU configuration (BU1 and BU2 from Tab.~\ref{protocol-table-3}), is obtained, after straightforward simplification, leading  to the simplified transition probability:
\begin{equation}
\begin{split}
\textup{P}_{\textup{BU}}=&\left|\cos\widetilde{\vartheta}'_{1}\sin\widetilde{\vartheta}_{0}\sin\widetilde{\vartheta}_{1}\cos\widetilde{\vartheta}_{2}\right.\\
&\left.\cdot\left[1-\tan\widetilde{\vartheta}'_{1}\tan\widetilde{\vartheta}_{2}e^{-i\Phi_{BU}}\right]\right|^{2}
\end{split}
\label{eq:BU}
\end{equation}
where the BU phase-shift is identified to be:
\begin{equation}
\begin{split}
\Phi_{BU}=-\varphi'_{1}+2\varphi_{0}-2\varphi_{1}+\varphi_{2}+\phi'_{1}+\phi_{2}
\end{split}
\label{eq:BU-phase-shift}
\end{equation}
that is a result consistent with~\cite{Kleinert:2015,Gustavson:2000,Takase:2008}. An additional phase-shift due to asymmetrical pulse area variation is found. Changing the geometry of the matter-wave interferometer from a Mach-Zehnder to a Butterfly device modifies the internal phase-shift contribution $\phi'_{1}-\phi_{2}\rightarrow\phi'_{1}+\phi_{2}$. Note that these two last contributions are remnant to Ramsey and Hahn-Ramsey protocols established in section III which can be exploited to modify the sensitivity of interferometers to residual Doppler-shifts and light-shifts as we will now demonstrate in the following subsection.

\subsection{HMZ AND HBU MATTER-WAVE INTERFERENCES}

\subsubsection{MZ AND BU INTERFERENCES}

\begin{figure}[t!!]
\center
\resizebox{9cm}{!}{\includegraphics[angle=0]{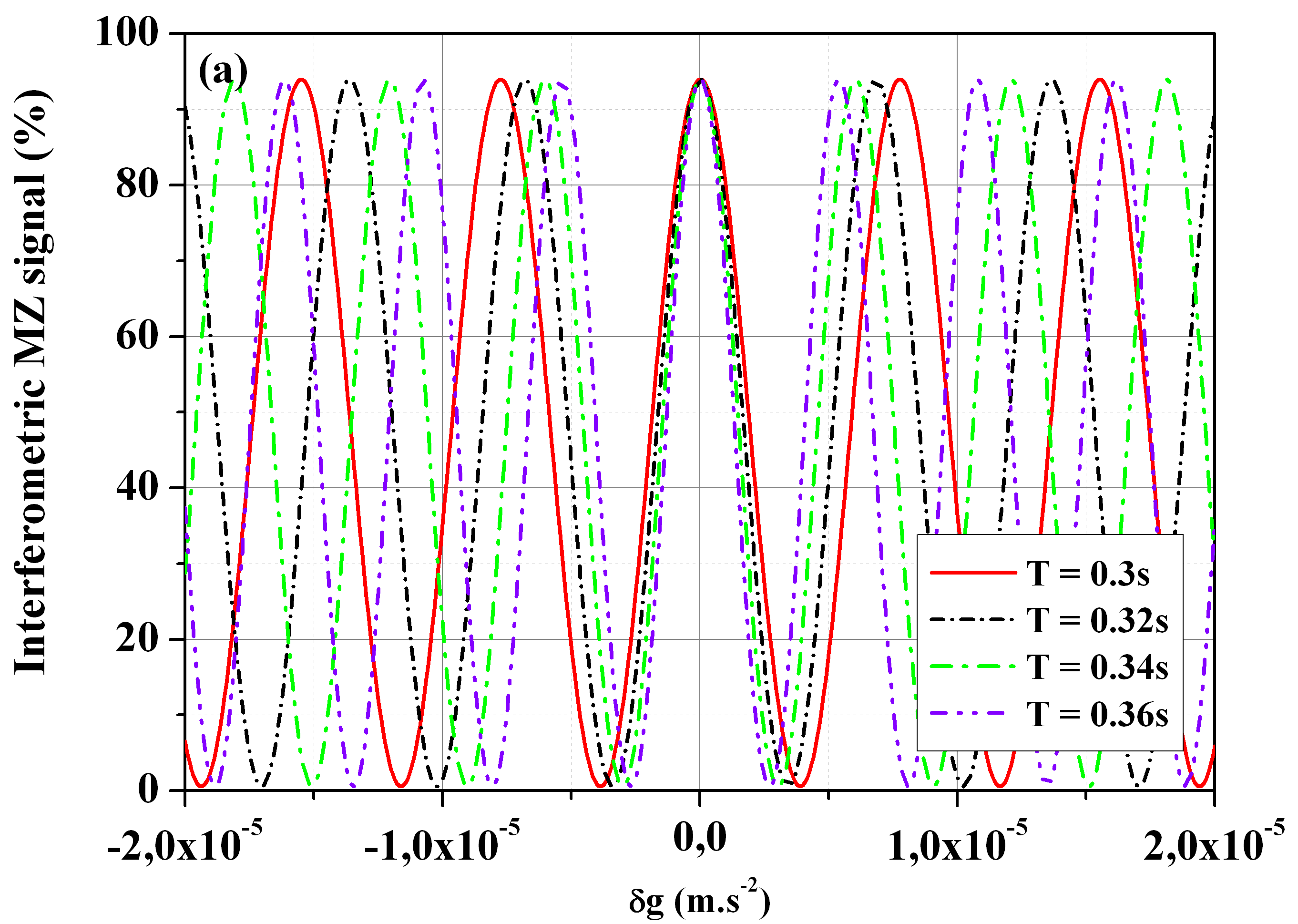}}
\resizebox{9cm}{!}{\includegraphics[angle=0]{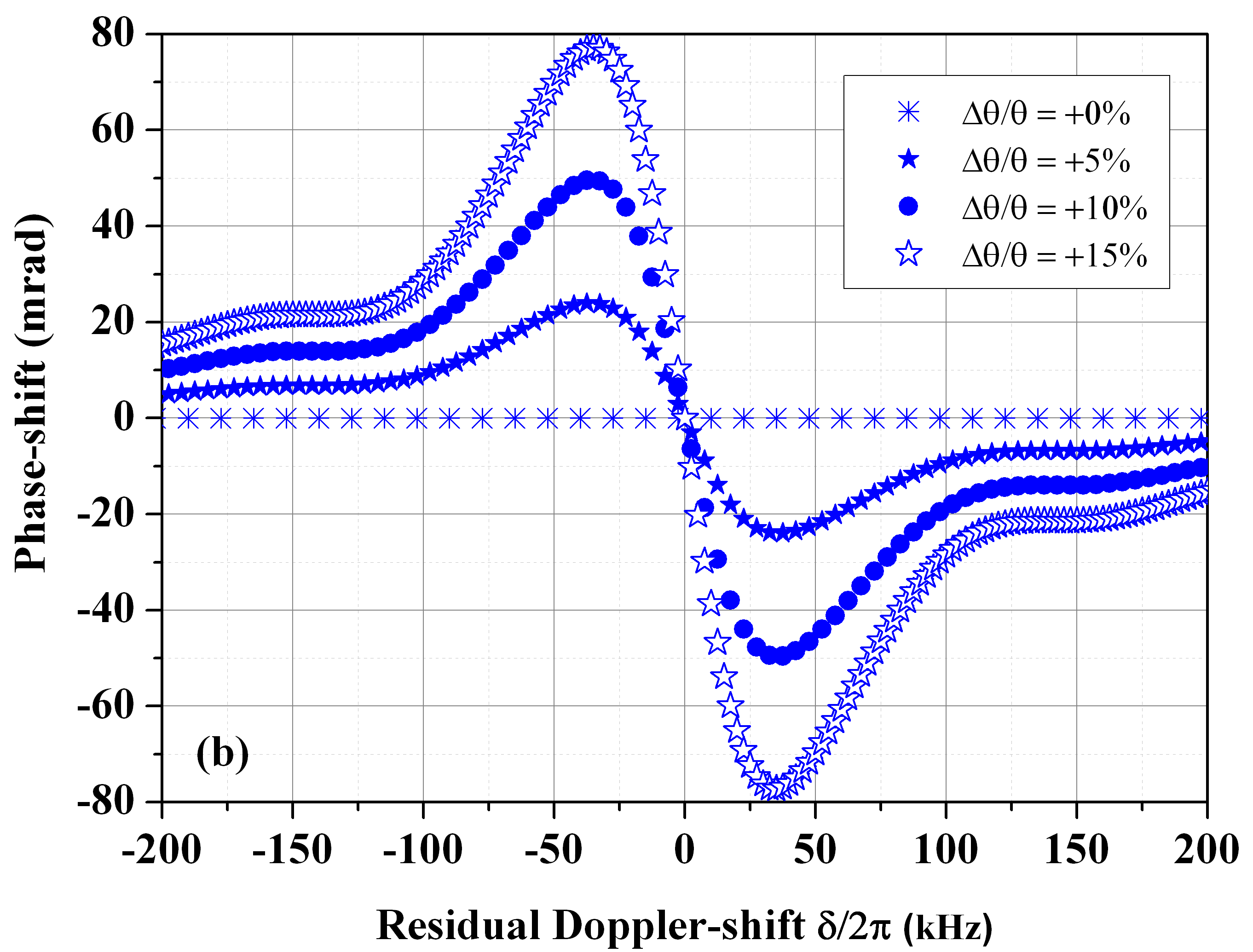}}
\caption{(color online). MZ pulse protocol versus the laser frequency chirp $\delta g=g-g_{chirp}$. (a) MZ matter-wave interferences with different free evolution times 0.3~s$\leq$T$\leq$0.36~s. Solid, dashed and dots-dashed lines are recorded with a residual Doppler-shift $\delta_{D}/2\pi=6$~kHz ($\delta_{D}\equiv\textup{kv}_{z}$). A Rabi field error of $10\%$ is fixed between the first beam splitter and the last one as in~\cite{Gillot:2016}. Note that the residual phase-shift induced by the pulse area error is too small to be seen on fringe maximum. (b) MZ phase-shift versus the residual Doppler-shift $\delta_{D}/2\pi$ for different pulse area variation $\Delta\theta/\theta\leq15\%$ between first and last pulses with a fixed free evolution time. Pulse duration is $\tau=10\mu$s and the Rabi frequency is $\Omega=\pi/2\tau$.}
\label{fig-MZ}
\end{figure}
\begin{figure*}[t!!]
\center
\resizebox{9cm}{!}{\includegraphics[angle=0]{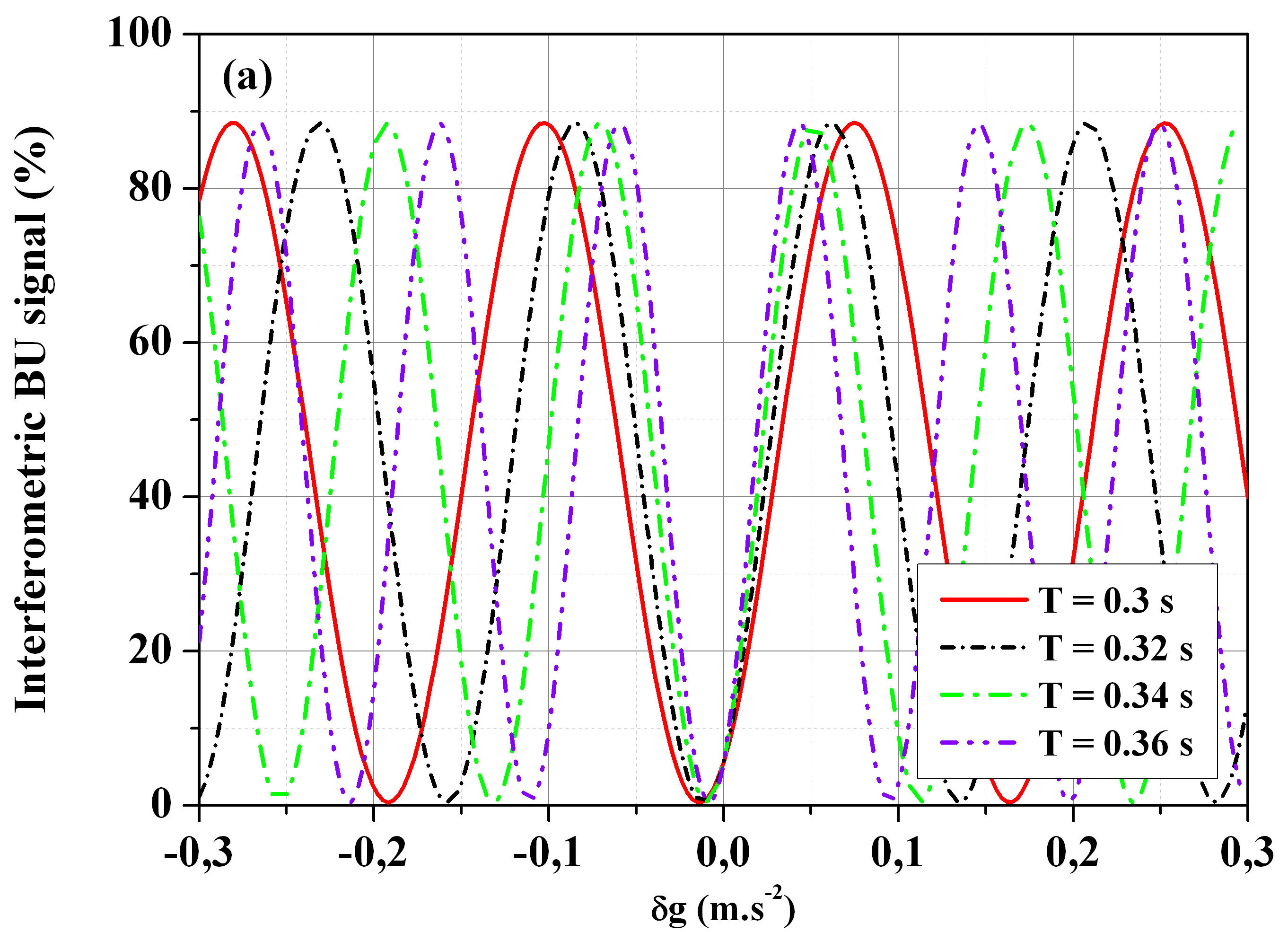}}\resizebox{9cm}{!}{\includegraphics[angle=0]{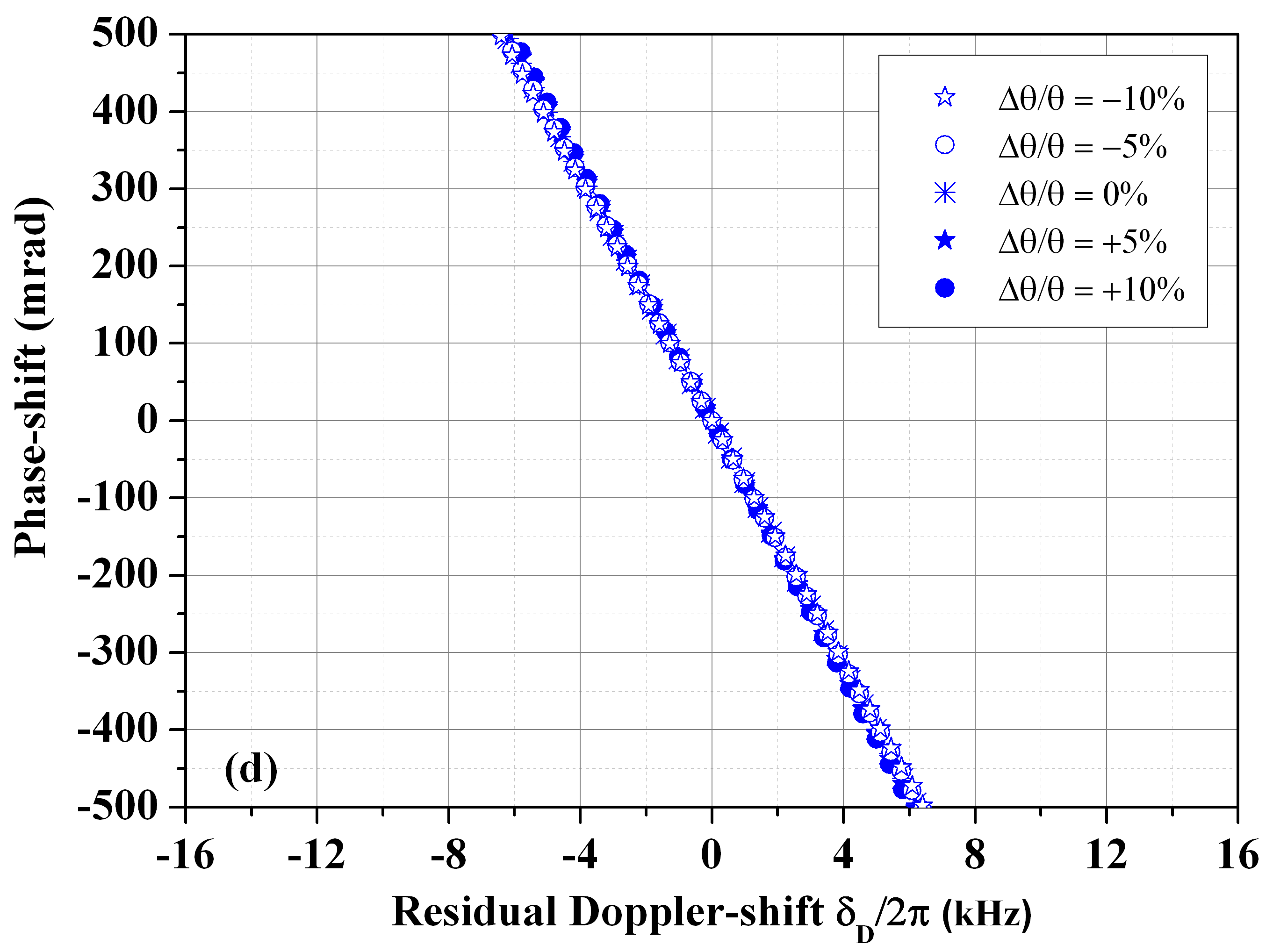}}
\resizebox{9cm}{!}{\includegraphics[angle=0]{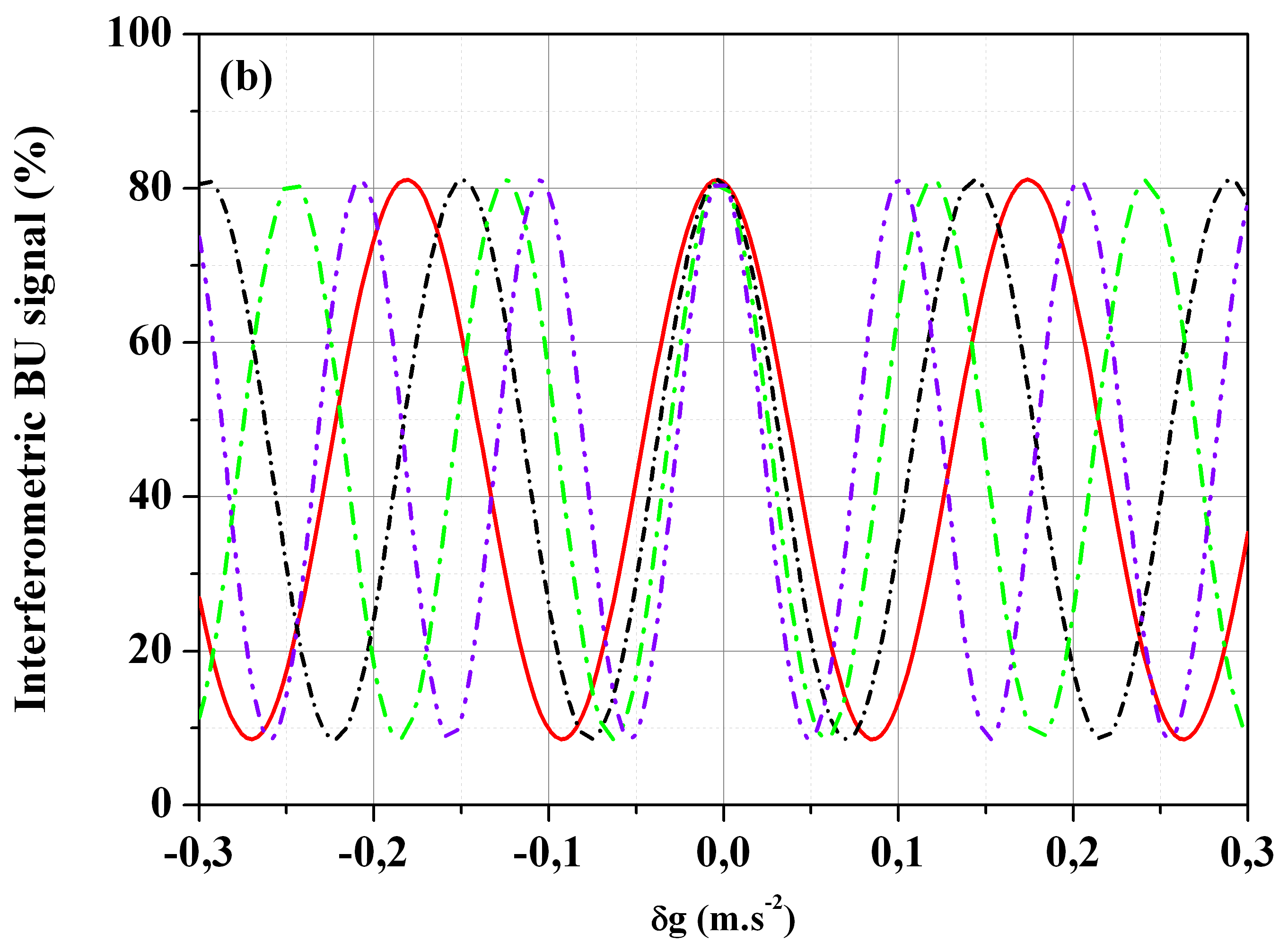}}\resizebox{9cm}{!}{\includegraphics[angle=0]{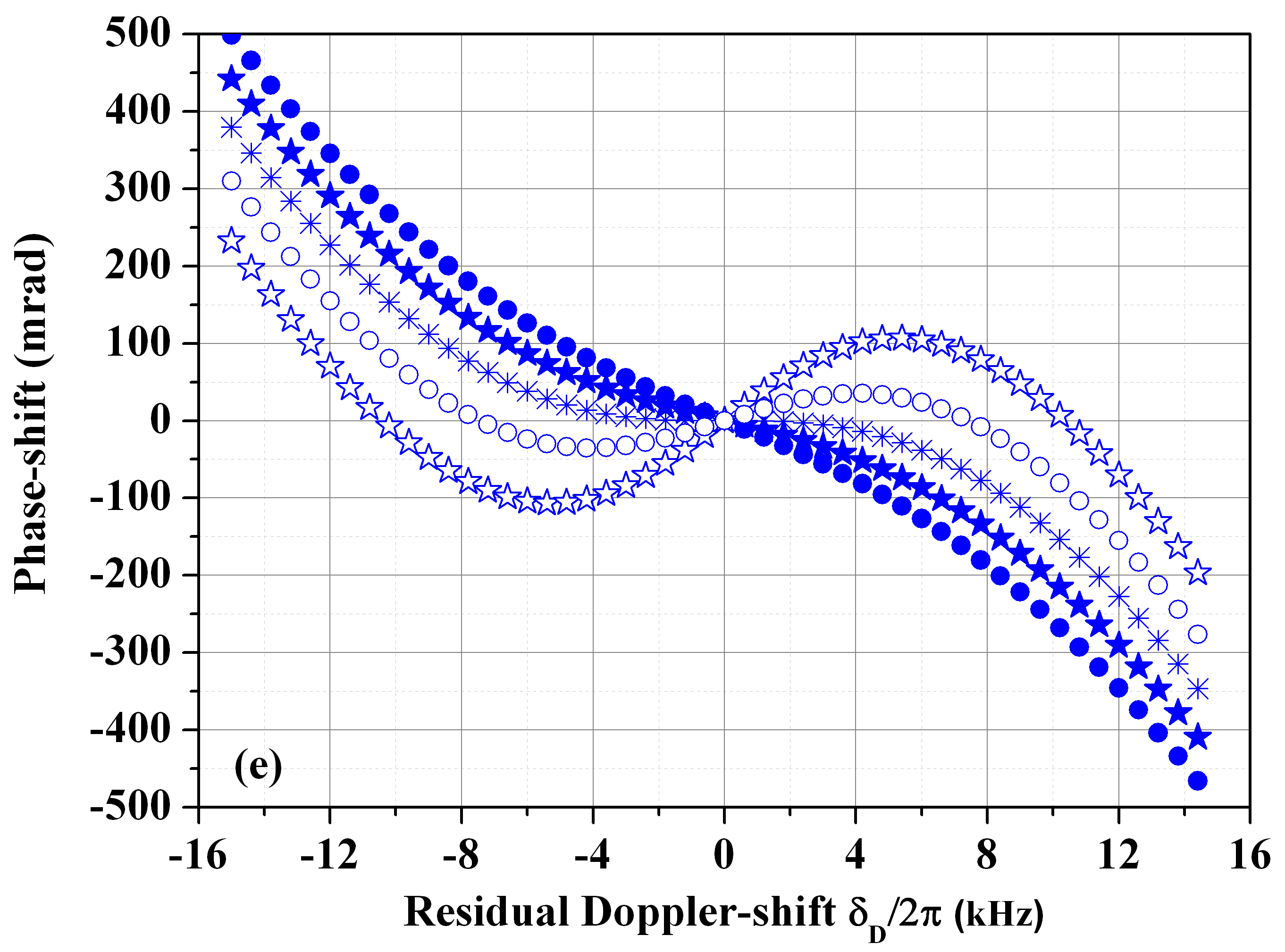}}
\resizebox{9cm}{!}{\includegraphics[angle=0]{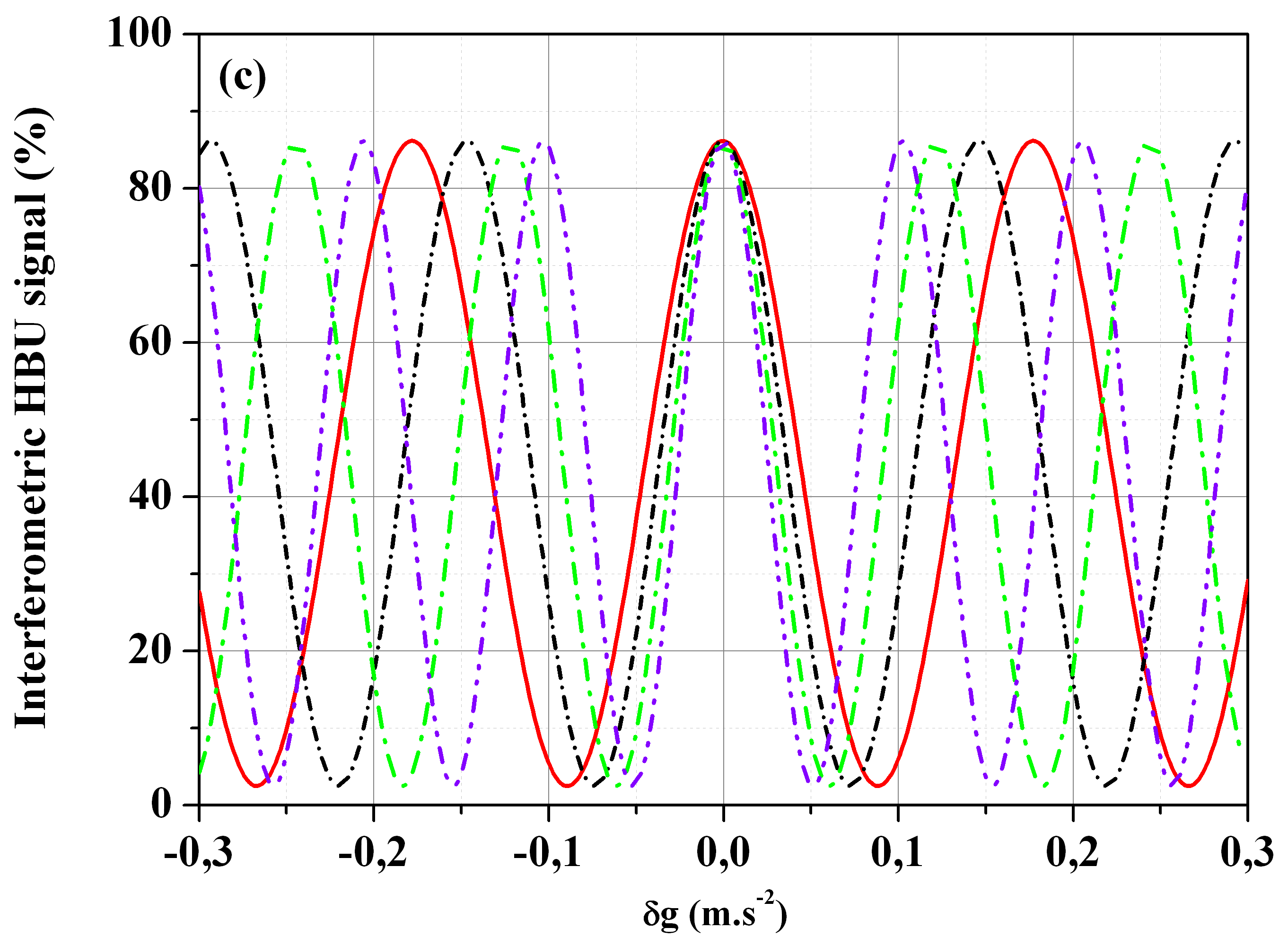}}\resizebox{9cm}{!}{\includegraphics[angle=0]{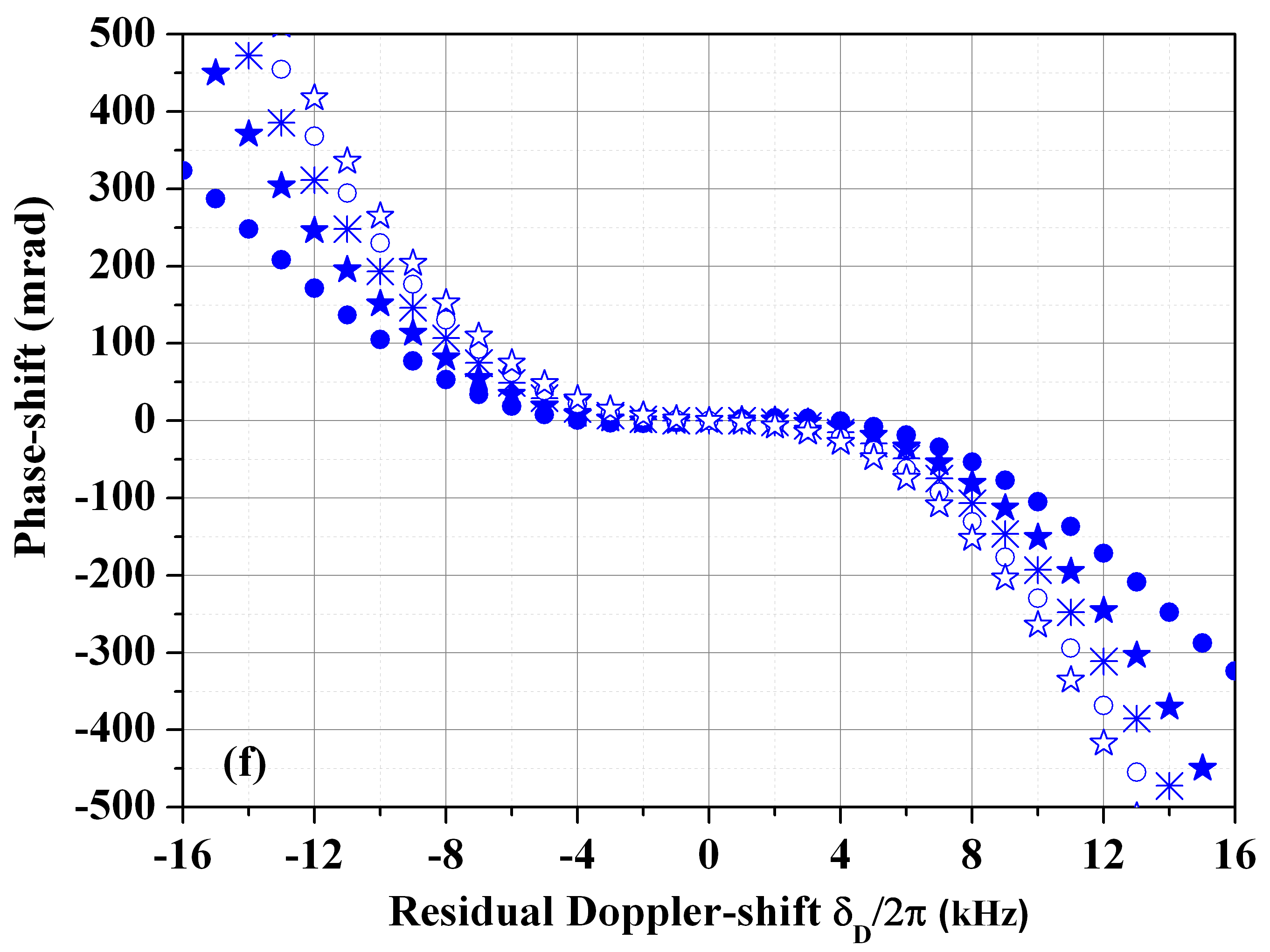}}
\caption{(color online). (Left panels) BU matter-wave interferences versus the laser frequency chirp $\delta g=g-g_{chirp}$ and (right panels) related phase-shifts versus a residual Doppler-shift $\delta_{D}/2\pi$ for three different protocols reported in Tab.~\ref{protocol-table-3}. (a) BU1 protocol with a $\boldsymbol{90^{\uparrow}}_{0}$ pulse, (b) BU2 protocol with a $\boldsymbol{270^{\uparrow}}_{0}$ pulse and (c) BU3 protocol using a $\boldsymbol{180^{\uparrow}}_{\pi}\boldsymbol{90^{\uparrow}}_{0}$ composite pulse. Solid, dashed and dots-dashed lines are recorded with a residual Doppler-shift $\delta_{D}/2\pi=6$~kHz ($\delta_{D}\equiv\textup{kv}_{z}$). (d), (e) and (f) are numerical plots tracking of the central interference position for different pulse area variation $-10\%\leq\Delta\theta/\theta\leq10\%$  with a fixed free evolution time from protocols (a), (b) and (c). Pulse duration is $\tau=10\mu$s and the standard Rabi frequency is $\Omega=\pi/2\tau$. The phase-shift can exhibit a nonlinear sensitivity to Doppler frequency-shifts from panel (d) to panel (f).}
\label{fig-BU}
\end{figure*}
\begin{figure*}[t!!]
\center
\resizebox{9cm}{!}{\includegraphics[angle=0]{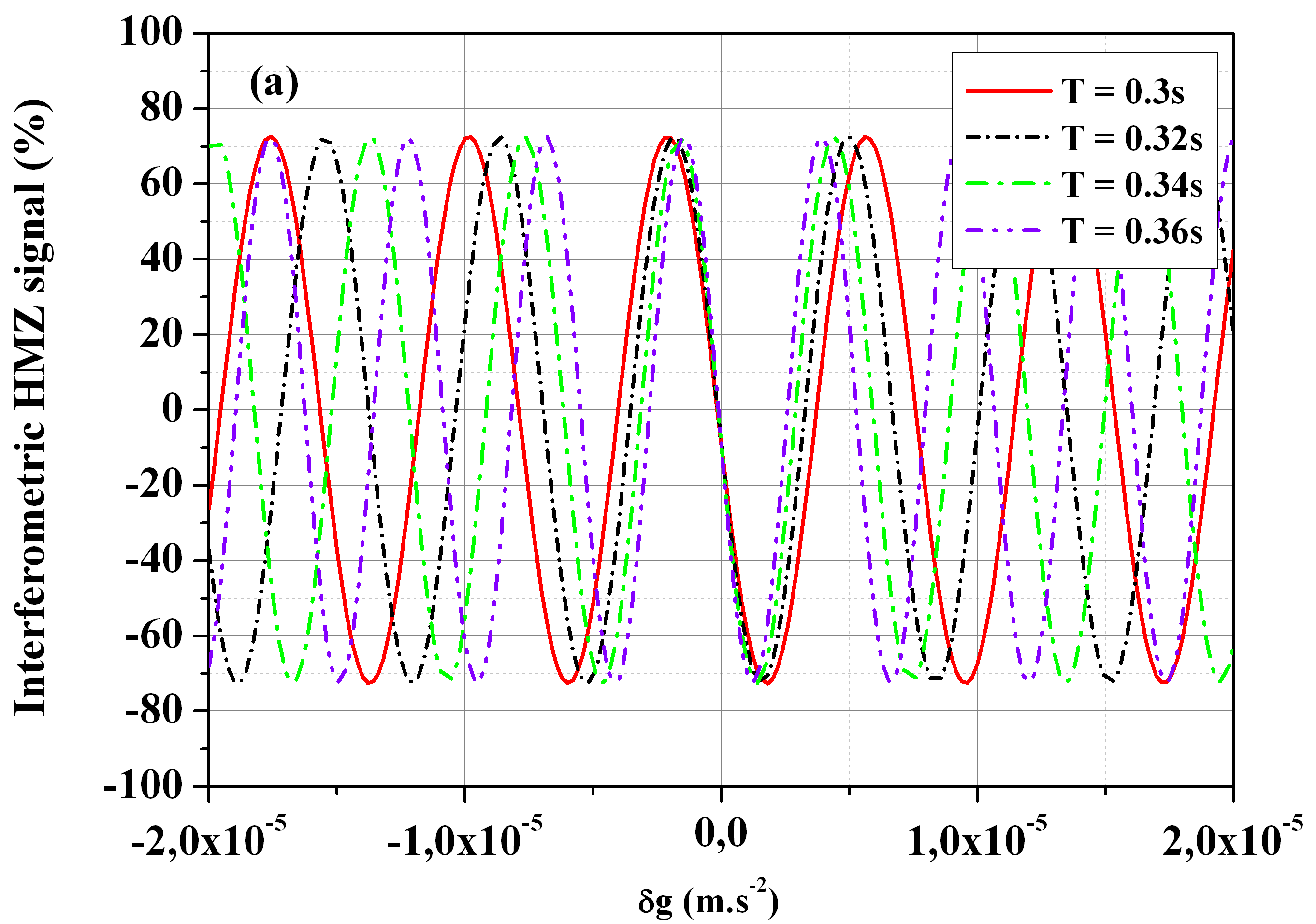}}\resizebox{9cm}{!}{\includegraphics[angle=0]{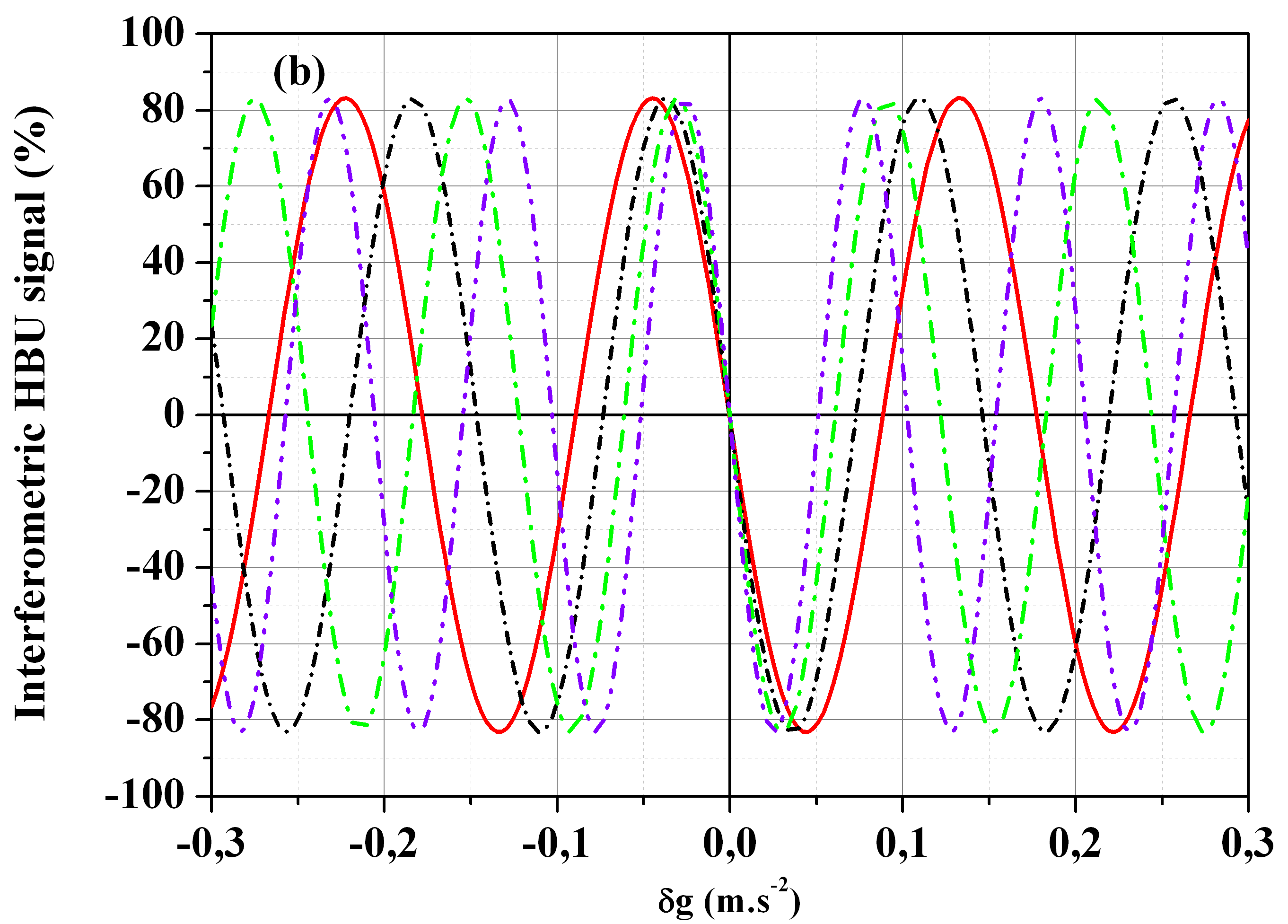}}
\resizebox{9cm}{!}{\includegraphics[angle=0]{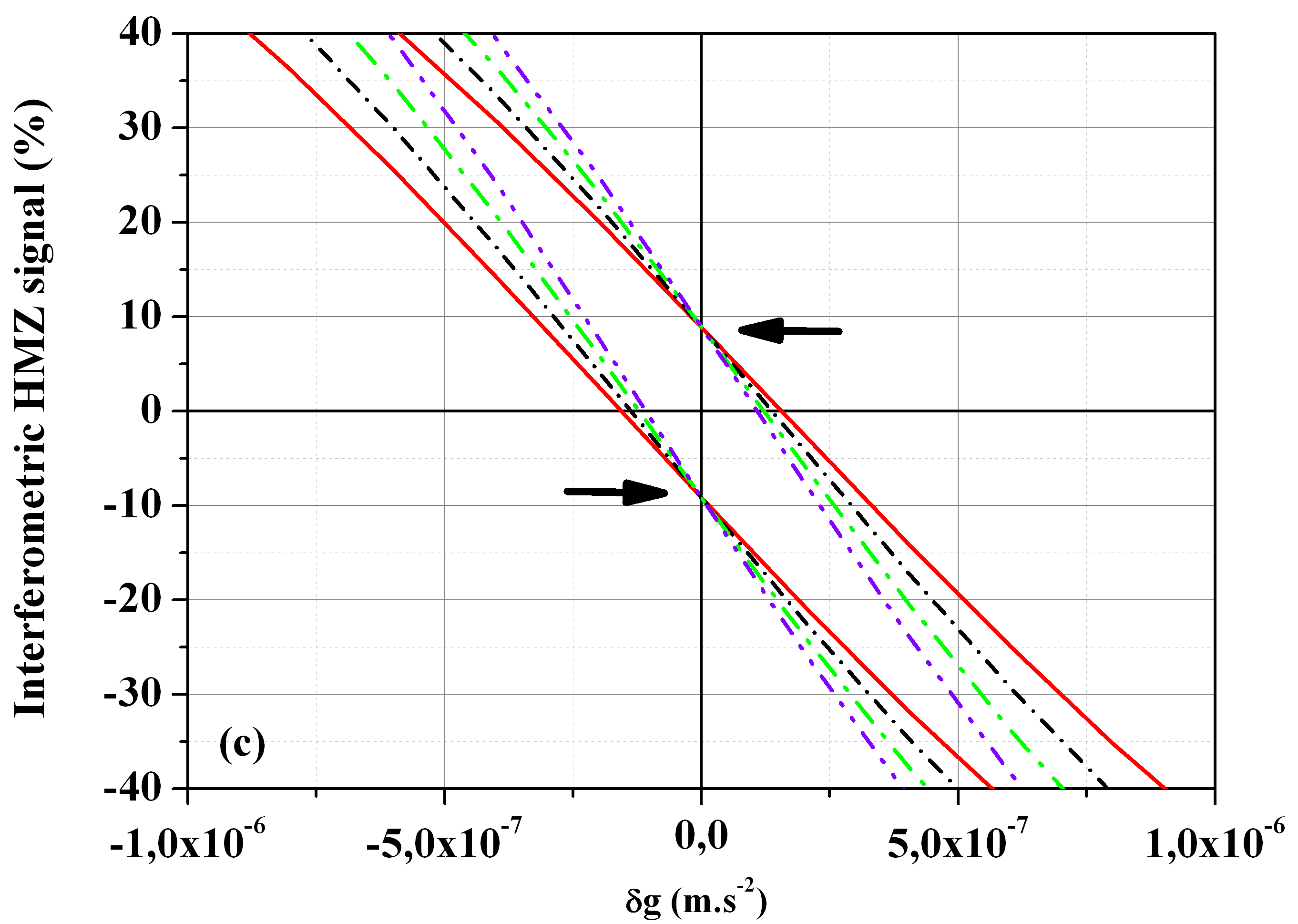}}\resizebox{9cm}{!}{\includegraphics[angle=0]{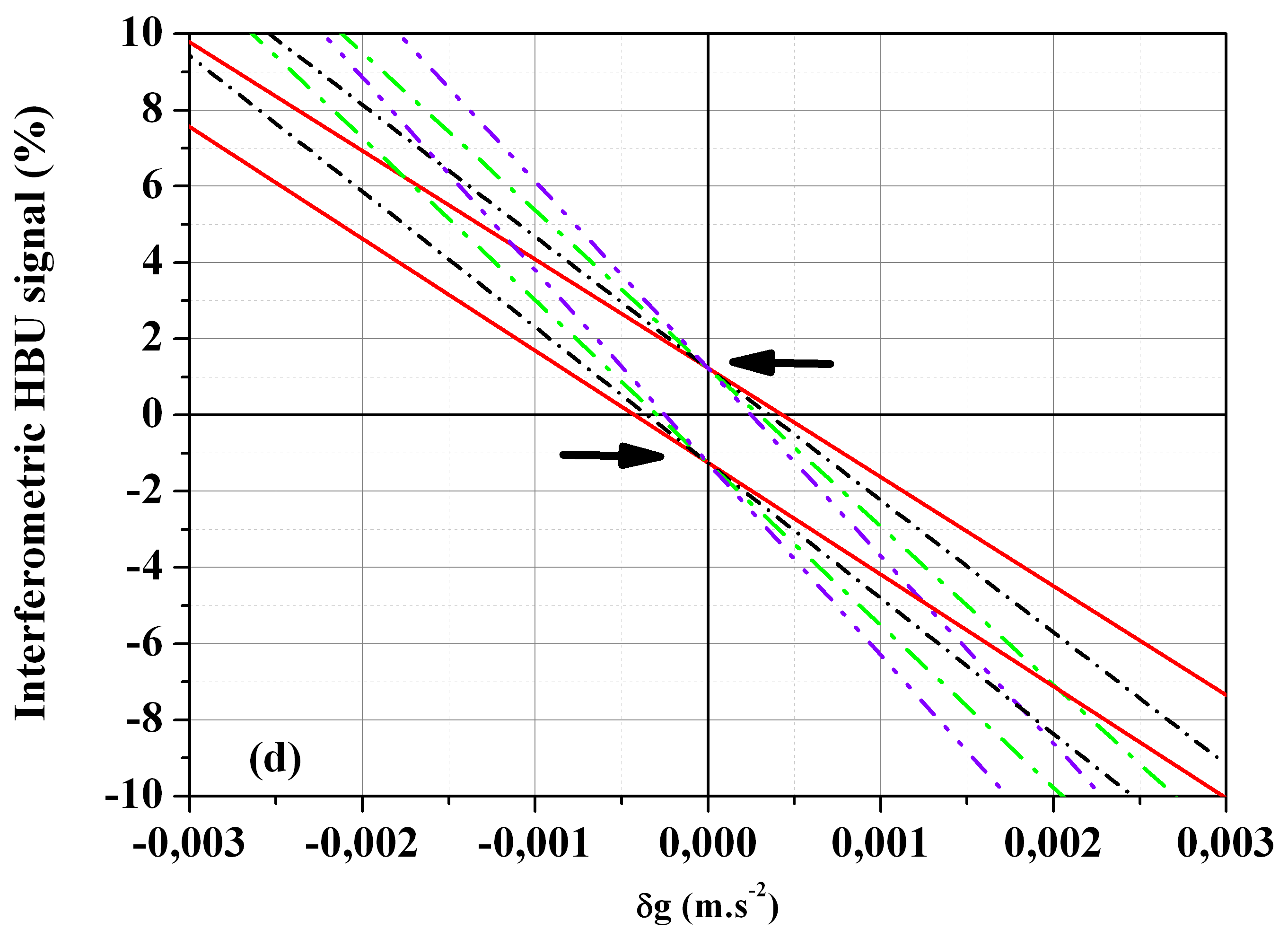}}
\caption{(color online). (a) $\Delta\textup{E}_{\textup{HMZ1}}(-)$ ($\Delta\textup{E}_{\textup{HMZ2}}(-)$) and (b) $\Delta\textup{E}_{\textup{HBU1}}(-)$ ($\Delta\textup{E}_{\textup{HBU2}}(-)$) phase-shifted matter-wave interferences based on Eq.~\ref{eq:coop-HMZ-a} (Eq.~\ref{eq:coop-HMZ-b}) and Eq.~\ref{eq:coop-HBU-a} (Eq.~\ref{eq:coop-HBU-b}) versus chirp of the laser frequency $\delta g=g-g_{chirp}$. Solid, dashed and dots-dashed lines are recorded with a residual a residual Doppler-shift $\delta_{D}/2\pi=6$kHz ($\delta_{D}\equiv\textup{kv}_{z}$). Pulse duration is $\tau=10\mu$s and the Rabi frequency is $\Omega=\pi/2\tau$. A Rabi pulse area variation of $0\%\leq\Delta\theta/\theta\leq10\%$ is tolerated between the first set of composite pulses and the last set of composite pulses as in~\cite{Gillot:2016}. (c) and (d) Zoom of dispersive shapes of error signals around the correct laser frequency chirp $\delta g=0$. Black arrows $\leftrightarrows$ are indicating the crossing point independent of the free evolution time allowing another determination of the correct laser frequency chirp canceling the Doppler-shift induced by the free fall.}
\label{fig-HMZ-HBU}
\end{figure*}
\begin{figure}[h!!]
\center
\resizebox{8.5cm}{!}{\includegraphics[angle=0]{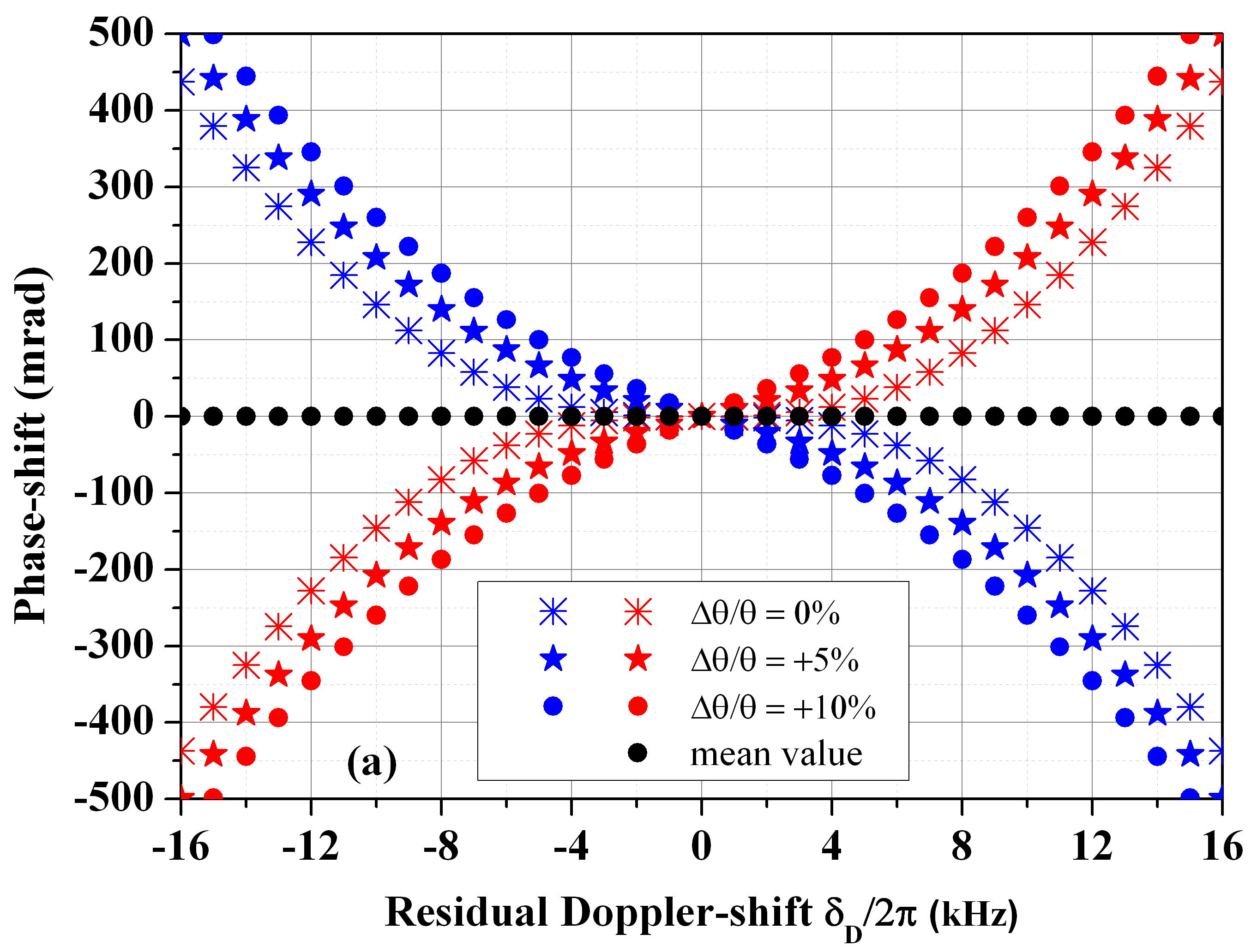}}
\resizebox{8.5cm}{!}{\includegraphics[angle=0]{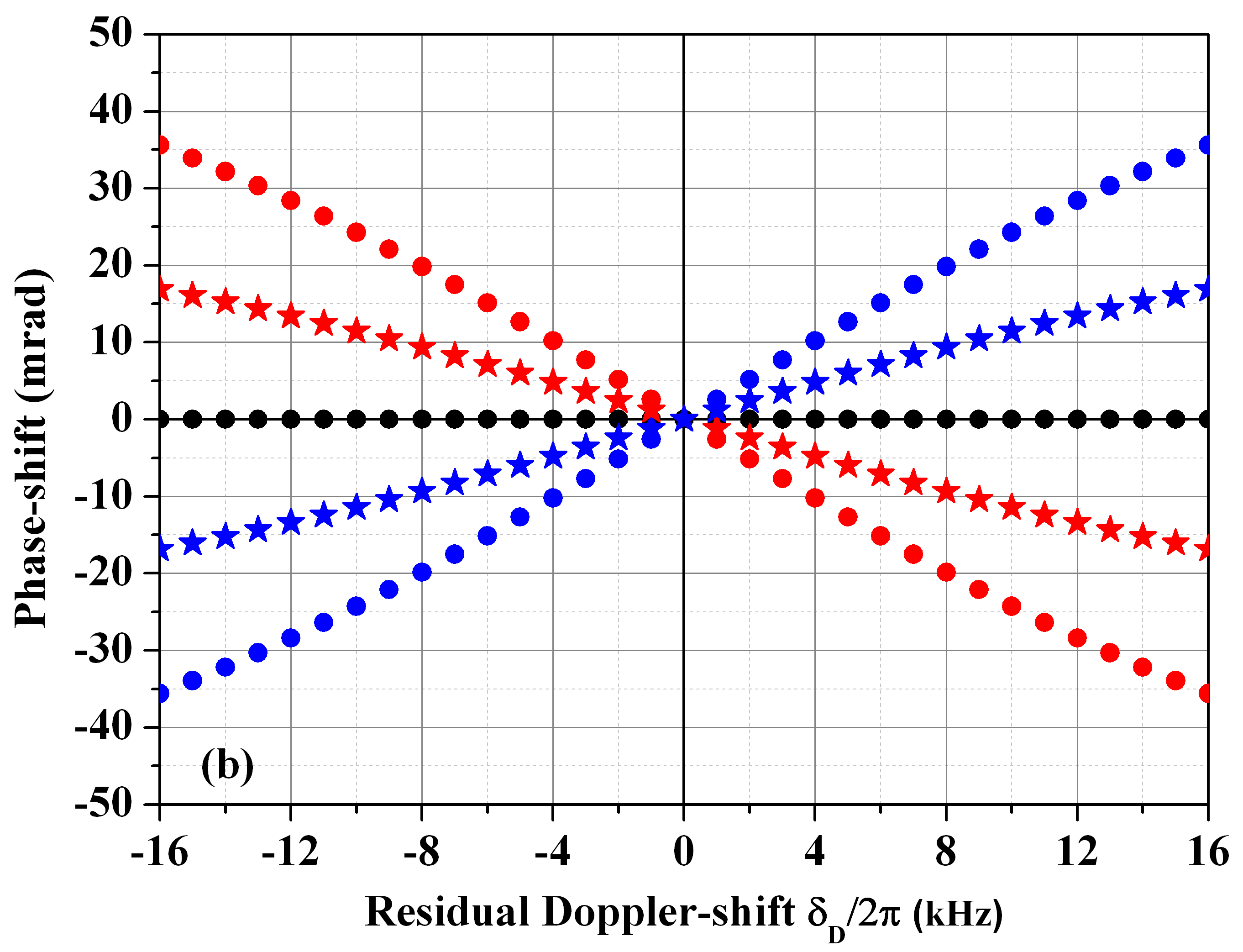}}
\caption{(color online). (a) and (b) Numerical tracking of frequency-shifts of $\Delta\textup{E}_{\textup{HMZ1}}(-)$ ($\Delta\textup{E}_{\textup{HMZ2}}(-)$) and $\Delta\textup{E}_{\textup{HBU1}}(-)$ ($\Delta\textup{E}_{\textup{HBU2}}(-)$) of error signals versus residual Doppler-shifts $\delta_{D}/2\pi$ for HMZ1 (HBU1) interferences (blue crosses, dots and stars) and HMZ2 (HBU2) interferences (red crosses, dots and stars). The black dots $\bullet$ are the zero mean value of the global sum of phase-shifted locking points. Pulse duration is $\tau=10\mu$s and the pulse area is $\Omega=\pi/2\tau$. A Rabi pulse area variation of $0\%\leq\Delta\theta/\theta\leq10\%$ is tolerated between the first set of composite pulses and the last set of composite pulses as in~\cite{Gillot:2016}.}
\label{fig-HMZ-HBU-shifts}
\end{figure}
\indent The Mach-Zehnder (MZ) configuration is reported from the first line of Tab.~\ref{protocol-table-3}. This geometry can be sensitive both to local acceleration and rotation (see subsection S4-3 from appendix).
We show MZ matter-wave interferences versus the laser frequency chirp $\delta g$ ignoring rotation for simplicity in Fig.~\ref{fig-MZ}(a).
The laser frequency is linearly scanned during the three-pulse interferometer for four different free evolutions times. When the laser frequency chirp becomes equal the local acceleration, the free fall inducing a Doppler-shift is canceled and interferences become independent to a modification of the free evolution time~\cite{Cheinet:2006,Farah:2014,Menoret:2018,Karcher:2020}.

Several numerical plots of the center of fringes versus a residual Doppler shift $\delta_{D}/2\pi$ ($\delta_{D}\equiv\textup{kv}_{z}$) is reported in Fig.~\ref{fig-MZ}(b) under a $\Delta\theta/\theta\leq15\%$ pulse area error between the first beam splitter pulse and the last one. The composite phase-shift generated by the MZ scheme has a dispersive lineshape as it can be observed from Fig.~\ref{fig-MZ}(b).
If the relative asymmetry of the pulse area between the first pulse and the last pulse is vanishing, thus the phase-shift is canceled consistent with the analytical expression derived from Eq.~\ref{eq:MZ-phase-shift}. The plots with blue dots corresponds to the MZ type interferometer exhibiting a dispersive phase-shift about 24 mrad for a Doppler-shift of 10 kHz for a $\Delta\theta/\theta=10\%$ in accordance with~\cite{Gillot:2016}. The same dispersive feature has been also derived about a symmetrical RB interferometer subjected to pulse area variation between sets of pulses in~\cite{Morel:2020}.
We note that such MZ asymmetry is also responsible for breaking the robustness of generalized hyper-Ramsey protocols used for clock interferometry. One possible way to get round of this detrimental effect is to apply auto-balanced hyper-interferometers with composite pulses as we will show later.

Butterfly BU1, BU2 and BU3 protocols are reported in Tab.~\ref{protocol-table-3}. This geometry has been proposed to measure rotation and gravity gradient while eliminating sensitivity to local acceleration when first and last free evolution times T are related to the intermediate free evolution time by taking T'$=$2T~\cite{Kleinert:2015,Gustavson:2000,Takase:2008} (see also the subsection S4-3 from appendix).

We have respectively reported, in Fig.~\ref{fig-BU}(a), (b) and (c), BU1, BU2 and BU3 matter-wave interferences versus the laser frequency chirp $\delta g$ for three different configurations of the last beam splitter either $\boldsymbol{90^{\circ\uparrow}}_{0}$ or $\boldsymbol{270^{\circ\uparrow}}_{0}$ and finally a composite pulse $\boldsymbol{180^{\circ\uparrow}}_{\pi}\boldsymbol{90^{\circ\uparrow}}_{0}$.
The BU1 protocol exhibits a linear sensitivity to residual Doppler-shifts as shown in Fig.~\ref{fig-BU}(d) that is consistent with the measurement from~\cite{Tan:2020}. According to Eq.~\ref{eq:BU-phase-shift},
the laser detuning contribution to the interferometric phase-shift is expected to be linear for a Butterfly or double loop configuration based on beam splitters with $\boldsymbol{90^{\circ'\uparrow}}_{0}$ and $\boldsymbol{90^{\circ\uparrow}}_{0}$ pulses.

By replacing the first $\boldsymbol{90^{\circ'\uparrow}}_{0}$ pulse by a $\boldsymbol{270^{\circ'\uparrow}}_{0}$ or the last $\boldsymbol{90^{\circ\uparrow}}_{0}$ pulse by a $\boldsymbol{270^{\circ\uparrow}}_{0}$ pulse, we observe in Fig.~\ref{fig-BU}(e) a nonlinear sensitivity to residual Doppler-shifts depending on the pulse area variation between pulses. Such a sensitivity is in fact remnant to the Hyper-Ramsey protocol for clocks previously studied. The modification from a linear response to a flat one was in fact already observed in~\cite{Tan:2020} supposing no pulse area variation between pulses. Our work has established a remarkable feature by analyzing in details the effect of a $\boldsymbol{90^{\circ\uparrow}}_{0}\mapsto\boldsymbol{270^{\circ\uparrow}}_{0}$ composite pulse leading to a strong reduction of the phase-shift related to laser detunings. We have derived here the exact dependance to laser detunings with a new composite phase-shift through Eq.~\ref{eq:HBU-phase-shift} replacing Eq.~\ref{eq:BU-phase-shift}. As expected from Fig.~\ref{fig-BU}(f), by applying a new composite pulse sequence for the last beam splitter pulse as $\boldsymbol{270^{\circ\uparrow}}_{0}\mapsto\boldsymbol{180^{\circ\uparrow}}_{\pi}\boldsymbol{90^{\circ\uparrow}}_{0}$, a low sensitivity to residual Doppler-shifts is still recovered while the detrimental action of the pulse area variation is eliminated making the interferometric configuration more robust to residual Doppler-shifts.

\subsubsection{HMZ AND HBU PHASE-SHIFTED INTERFERENCES}

\indent Phase-shifted matter-wave interferences can offer a robust alternative technique to eliminate some distortions coupled to residual Doppler-shift and light-shift with pulse area variation between pulses.
We present in Fig.~\ref{fig-HMZ-HBU} some interferences that are produced by manipulating the relative laser phase $\varphi$ between pulses while chirping the laser frequency.
New HMZ1($\varphi$) and HMZ2($\varphi$), HBU1($\varphi$) and HBU2($\varphi$) protocols are reported in Tab.~\ref{protocol-table-3}).

Note we have found that additional combinations of HMZ1($\varphi$), HMZ2($\varphi$)and HBU1($\varphi$), HBU2($\varphi$) protocols can be also used where the first pulse $\boldsymbol{90^{\circ'\uparrow}}_{0}$ or the last pulse $\boldsymbol{90^{\circ\uparrow}}_{0}$ is replaced by a $\boldsymbol{270^{\circ'\uparrow}}_{0}$ or $\boldsymbol{270^{\circ\uparrow}}_{0}$ pulse.

First, we have considered a cooperative combination of schemes as HMZ1($\varphi$) and HMZ2($\varphi$) protocols by introducing a $\boldsymbol{180^{\circ\uparrow}}_{\pm\varphi}$ pulse on the right or a $\mapsto\boldsymbol{180^{\circ'\uparrow}}_{\pm\varphi}$ on the left side of the interferometer while including $\varphi=\pi/4,3\pi/4$ phase-step modulation.
A set of error signals is produced by an hybrid combination of these HMZ1($\varphi$) and HMZ2($\varphi$) interferometric signals as following:
\begin{subequations}
\begin{align}
\Delta\textup{E}_{\textup{HMZ1}}(-)&=\frac{1}{2}\left(\Delta\textup{E}_{\textup{HMZ1}(\pi/4)}-\Delta\textup{E}_{\textup{HMZ1}(3\pi/4)}\right)\label{eq:coop-HMZ-a}\\
\Delta\textup{E}_{\textup{HMZ2}}(-)&=\frac{1}{2}\left(\Delta\textup{E}_{\textup{HMZ2}(\pi/4)}-\Delta\textup{E}_{\textup{HMZ2}(3\pi/4)}\right)\label{eq:coop-HMZ-b}
\end{align}
\end{subequations}
The curves associated to cooperative protocols $\Delta\textup{E}_{\textup{HMZ1}}(-)$ (equivalently $\Delta\textup{E}_{\textup{HMZ2}}(-)$) are shown in Fig.~\ref{fig-HMZ-HBU}(a).
An expanded view of the two phase-shifted error signals is presented in Fig.~\ref{fig-HMZ-HBU}(c). The zero locking points of dispersive curves from Fig.~\ref{fig-HMZ-HBU}(c) are symmetrically shifted around $\delta g=g-g_{chirp}=0$ and all curves are crossing at $\delta g=g-g_{chirp}=0$ (indicated by black arrows) for different free evolution times $0.3$~s$\leq\textup{T}\leq0.36$~s.
By plotting the mean value of the phase-shift accumulated by $\Delta\textup{E}_{\textup{HMZ1}}(-)$ and $\Delta\textup{E}_{\textup{HMZ2}}(-)$ error signals versus the residual Doppler-shift, we can effectively track the correct value of the laser frequency chirp that is canceling the local acceleration while rejecting efficiently detrimental phase-shift from a pulse area variation between pulses.

Similarly, we have considered a combination of error signals generating a robust HBU interferometer using HBU1($\varphi$) and HBU2($\varphi$) protocols reported in Tab.~\ref{protocol-table-3}.
A set of hybrid error signals is produced by a combination of HBU1($\varphi$) and HBU2($\varphi$) interferometric signals as following:
\begin{subequations}
\begin{align}
\Delta\textup{E}_{\textup{HBU1}}(-)&=\frac{1}{2}\left(\Delta\textup{E}_{\textup{HBU1}(\pi/4)}-\Delta\textup{E}_{\textup{HBU1}(3\pi/4)}\right)\label{eq:coop-HBU-a}\\
\Delta\textup{E}_{\textup{HBU2}}(-)&=\frac{1}{2}\left(\Delta\textup{E}_{\textup{HBU2}(\pi/4)}-\Delta\textup{E}_{\textup{HBU2}(3\pi/4)}\right)\label{eq:coop-HBU-b}
\end{align}
\end{subequations}
These dispersive phase-shifted interferences, called $\Delta\textup{E}_{\textup{HBU1}}(-)$ (equivalently $\Delta\textup{E}_{\textup{HBU2}}(-)$) error signals, are presented in Fig.~\ref{fig-HMZ-HBU}(b).
An expanded view of the two phase-shifted error signals around the correct frequency chirp $\delta g=0$ is presented in Fig.~\ref{fig-HMZ-HBU} (d).
We have also verified that an additional contribution from any uncompensated asymmetric light-shifts of a few $\%$ between sets of composite pulses are still strongly reduced to the same level of correction.
We finally report the numerical tracking of the phase-shift accumulated by these HMZ1(2) and HBU1(2) dispersive error signals in Fig.~\ref{fig-HMZ-HBU-shifts}(a) and (b). Red and blue dots and stars are corresponding respectively to error signal protocols based on Eq.~\ref{eq:coop-HMZ-a}, Eq.~\ref{eq:coop-HMZ-b} and Eq.~\ref{eq:coop-HBU-a}, Eq.~\ref{eq:coop-HBU-b}. A plot of these anti-symmetrical shifts versus the residual Doppler-shift (black dots), is also confirming that phase-shifted interferences HMZ1(2) and HBU1(2) are cooperatively compensating a residual Doppler-shift contribution leading to a net zero mean value of the global shift.
An important result is that all dispersive error signals are always crossing together at the correct chirping frequency used to determinate the local acceleration for any value of the free evolution time. There are many ways to combine phase-shifted error signals to retrieve the correct frequency-chirp rate free from pulses inducing unwanted distortions. Note that the cooperative action of phase-shifted error signals removing residual frequency-shifts shares important similarities with the concept of synthetic frequency protocols developed for robust Ramsey interrogation of optical clock transitions by ref~\cite{Yudin:2016}.

\begin{table*}[t!!]
\renewcommand{\arraystretch}{1.7}
\begin{tabular}{|c|c|c|c|c|}
\hline
\hline
\hline
\textbf{C-protocol (frequency)} & N & $_{p}^{q}\Phi^{l}$, $_{p}^{q}\Phi^{s}$  \\
\hline
\hline
AB-R1 ($\varphi=\pm\pi/2$) & 4 & $\varphi_{L}+\phi_{L}$  \\
\hline
\hline
AB-R2 ($\varphi=\pm\pi/2$) & 4 & $\varphi_{L}+\phi_{L}$  \\
\hline
\hline
AB-HR3$_{\pi}$ ($\varphi=\pm\pi/2$) & 4 & $\varphi_{L}+\phi_{L}-\textup{Arg}[_{1}^{2}\beta(gg)]$   \\
\hline
\hline
AB-HR5$_{\pi}$ ($\varphi=\pm\pi/2$) & 4 & $\varphi_{L}+\phi_{L}-\textup{Arg}[_{1}^{4}\beta(gg)]$   \\
\hline
\hline
AB-R1/HR3$_{\pi}$ ($\varphi=\pm\pi/2$) & 4 &
\begin{tabular}{c}
$_{1}^{2}\Phi^{l}_{gg}=\varphi_{L}+\phi_{L}-\textup{Arg}[_{1}^{2}\beta(gg)]$\\
$_{1}^{1}\Phi^{s}_{gg}=\varphi_{L}+\phi_{L}$\\
\end{tabular}    \\
\hline
\hline
AB-GHHR ($\varphi=\pm\pi/4,\pm3\pi/4$) & 4$/$8 & $-\varphi'_{I,1}+2\varphi'_{II,1}-\varphi_{II,1}+\phi'_{I,1}-\phi_{II,1}$   \\
\hline
\hline
\hline
\textbf{AI-protocol (atomic recoil)} & N & $_{p}^{q}\Phi^{l}_{\overrightarrow{p}\mp\hbar\overrightarrow{k}}$, $_{p}^{q}\Phi^{s}_{\overrightarrow{p}\mp\hbar\overrightarrow{k}}$  \\
\hline
\hline
AB-RB1 ($\varphi=\pm\pi/4$) & 4 & $\varphi_{L}+\phi_{L}$  \\
\hline
\hline
AB-RB2 ($\varphi=\pm\pi/4$) & 4 & $\varphi_{L}+\phi_{L}$  \\
\hline
\hline
AB-HRB3$_{\pi}$ ($\varphi=\pm\pi/4$) & 4 &  \begin{tabular}{c}
                           $_{1}^{2}\Phi^{l}_{\overrightarrow{p}-\hbar\overrightarrow{k}}=_{1}^{2}\Phi^{s}_{\overrightarrow{p}-\hbar\overrightarrow{k}}=\varphi_{L}+\phi_{L}-\textup{Arg}\left[_{1}^{2}\beta^{I}_{gg}\cdot_{1}^{2}\beta^{II}_{ge}\right]$ \\
                           $_{1}^{2}\Phi^{l}_{\overrightarrow{p}+\hbar\overrightarrow{k}}=_{1}^{2}\Phi^{s}_{\overrightarrow{p}+\hbar\overrightarrow{k}}=\varphi_{L}+\phi_{L}-\textup{Arg}\left[_{1}^{2}\beta_{ge}^{I}\cdot_{1}^{2}\beta_{gg}^{II}\right] $
                         \end{tabular}
   \\
\hline
\hline
AB-HRB5$_{\pi}$ ($\varphi=\pm\pi/4$) & 4 & \begin{tabular}{c}
                           $_{1}^{4}\Phi^{l}_{\overrightarrow{p}-\hbar\overrightarrow{k}}=_{1}^{4}\Phi^{s}_{\overrightarrow{p}-\hbar\overrightarrow{k}}=\varphi_{L}+\phi_{L}-\textup{Arg}\left[_{1}^{4}\beta^{I}_{gg}\cdot_{1}^{4}\beta^{II}_{ge}\right]$ \\
                           $_{1}^{4}\Phi^{l}_{\overrightarrow{p}+\hbar\overrightarrow{k}}=_{1}^{4}\Phi^{s}_{\overrightarrow{p}+\hbar\overrightarrow{k}}=\varphi_{L}+\phi_{L}-\textup{Arg}\left[_{1}^{4}\beta_{ge}^{I}\cdot_{1}^{4}\beta_{gg}^{II}\right] $
                         \end{tabular}  \\
\hline
\hline
\hline
\textbf{AI-protocol (acceleration / rotation)} & N & $_{p}^{q}\Phi^{l}_{\overrightarrow{p}+\hbar\overrightarrow{k}}$, $_{p}^{q}\Phi^{s}_{\overrightarrow{p}+\hbar\overrightarrow{k}}$   \\
\hline
\hline
AB-MZ ($\varphi=\pm\pi/2$) & 4 & $\Phi_{MZ}$   \\
\hline
\hline
AB-BU ($\varphi=\pm\pi/2$) & 4 & $\Phi_{BU}$   \\
\hline
\hline
AB-HMZ ($\varphi=\pm\pi/4,\pm3\pi/4$) & 4/8 & $\Phi_{MZ}-\textup{Arg}\left[\beta_{1}'^{3}(gg)\cdot\beta_{2}^{4}(gg)\right]$   \\
\hline
\hline
AB-HBU ($\varphi=\pm\pi/4,\pm3\pi/4$) & 4/8 & $\Phi_{BU}-\textup{Arg}\left[\beta_{1}'^{3}(gg)\cdot\beta_{2}^{4}(gg)\right]$   \\
\hline
\hline
\end{tabular}
\centering%
\caption{Composite phase-shifts for fault-tolerant auto-balanced hyper-interferometers. Some reduced definitions are introduced into columns as $\varphi_{L}\equiv\varphi_{1}-\varphi'_{1}$ with $\phi_{L}\equiv\phi_{1}+\phi'_{1}$ for C-protocols (clocks) and $\varphi_{L}\equiv\varphi_{I,1}-\varphi'_{I,1}+\varphi_{II,1}-\varphi'_{II,1}$ with $\phi_{L}\equiv\phi'_{I,1}+\phi_{I,1}+\phi'_{II,1}+\phi_{II,1}$ for AI-protocols (Ramsey-Bordé interferometers).
The number of atomic state population measurements N required to build error signals is indicated. Dispersive error signals can be produced by $\pm\pi/2$ or $\pm\pi/4,\pm3\pi/4$ phase-steps for hyper-clocks through N$=4$ or N$=8$ measurements. Hyper-Ramsey-Bordé error signals HRB3$_{\pi}$ and HRB5$_{\pi}$ are produced with $\pm\pi/4$ phase-steps after N$=4$ measurements.}
\label{protocol-table-4}
\end{table*}
\begin{figure*}[t!!]
\center
\resizebox{8.5cm}{!}{\includegraphics[angle=0]{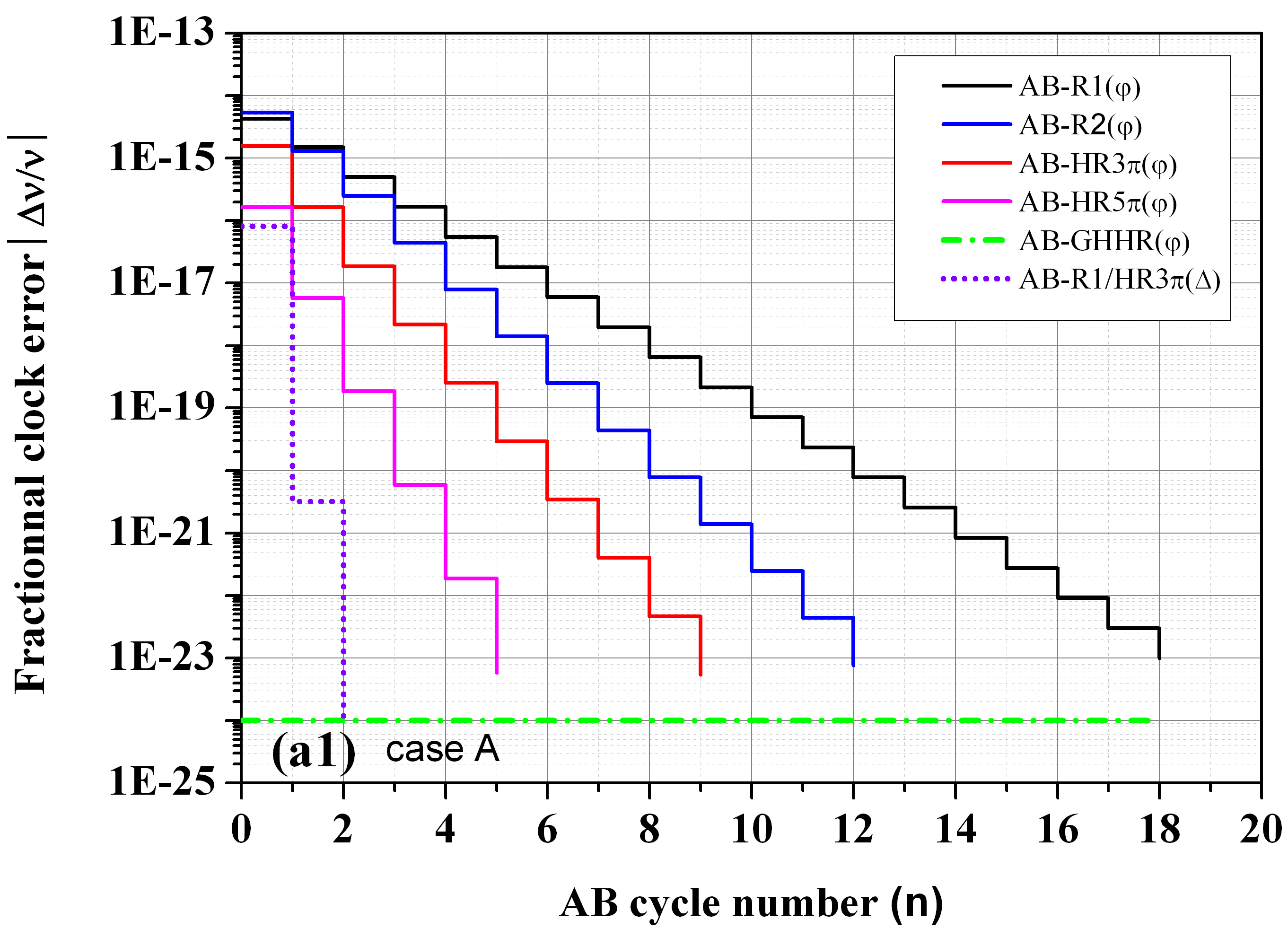}}\resizebox{8.5cm}{!}{\includegraphics[angle=0]{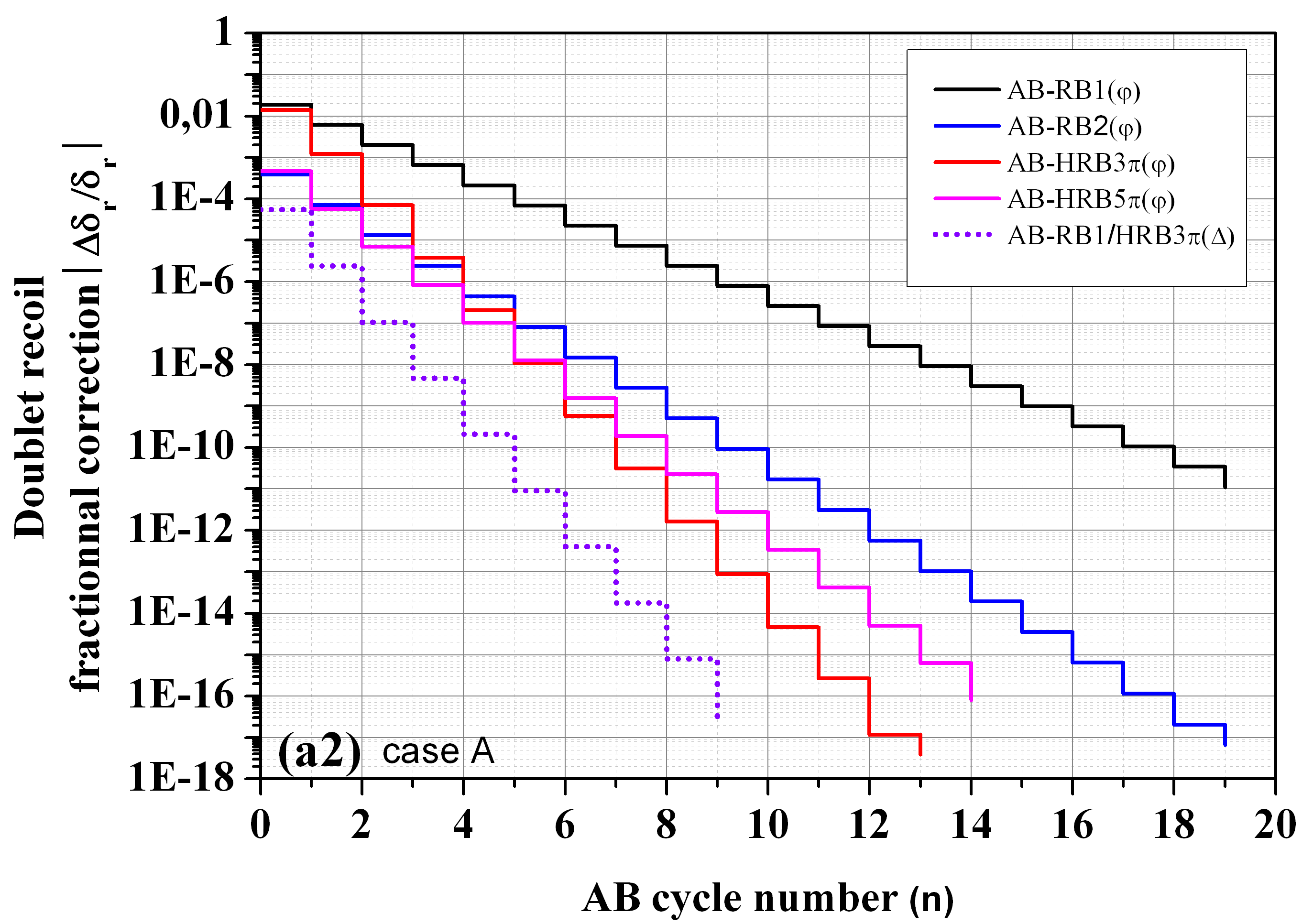}}
\resizebox{8.5cm}{!}{\includegraphics[angle=0]{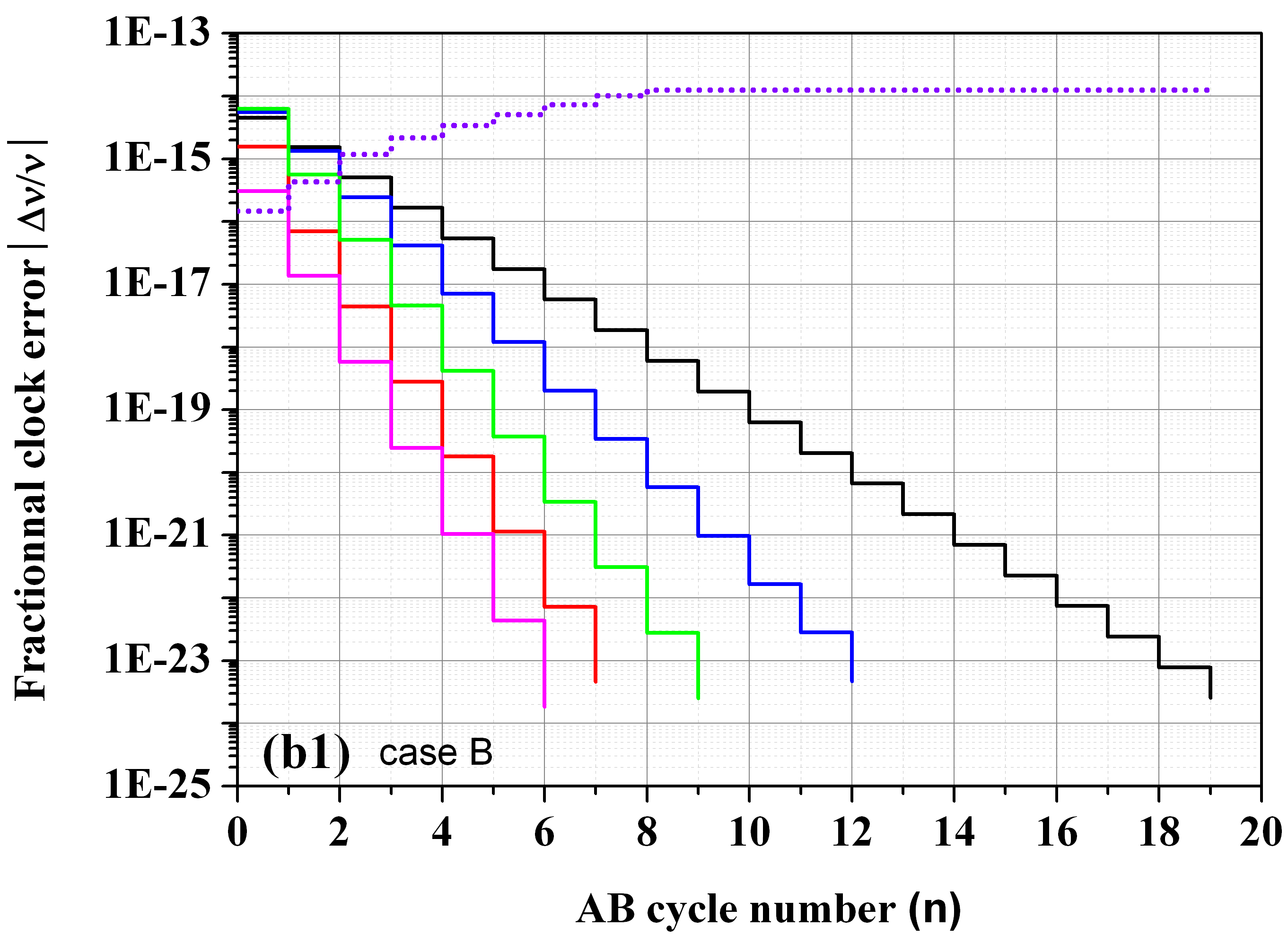}}\resizebox{8.5cm}{!}{\includegraphics[angle=0]{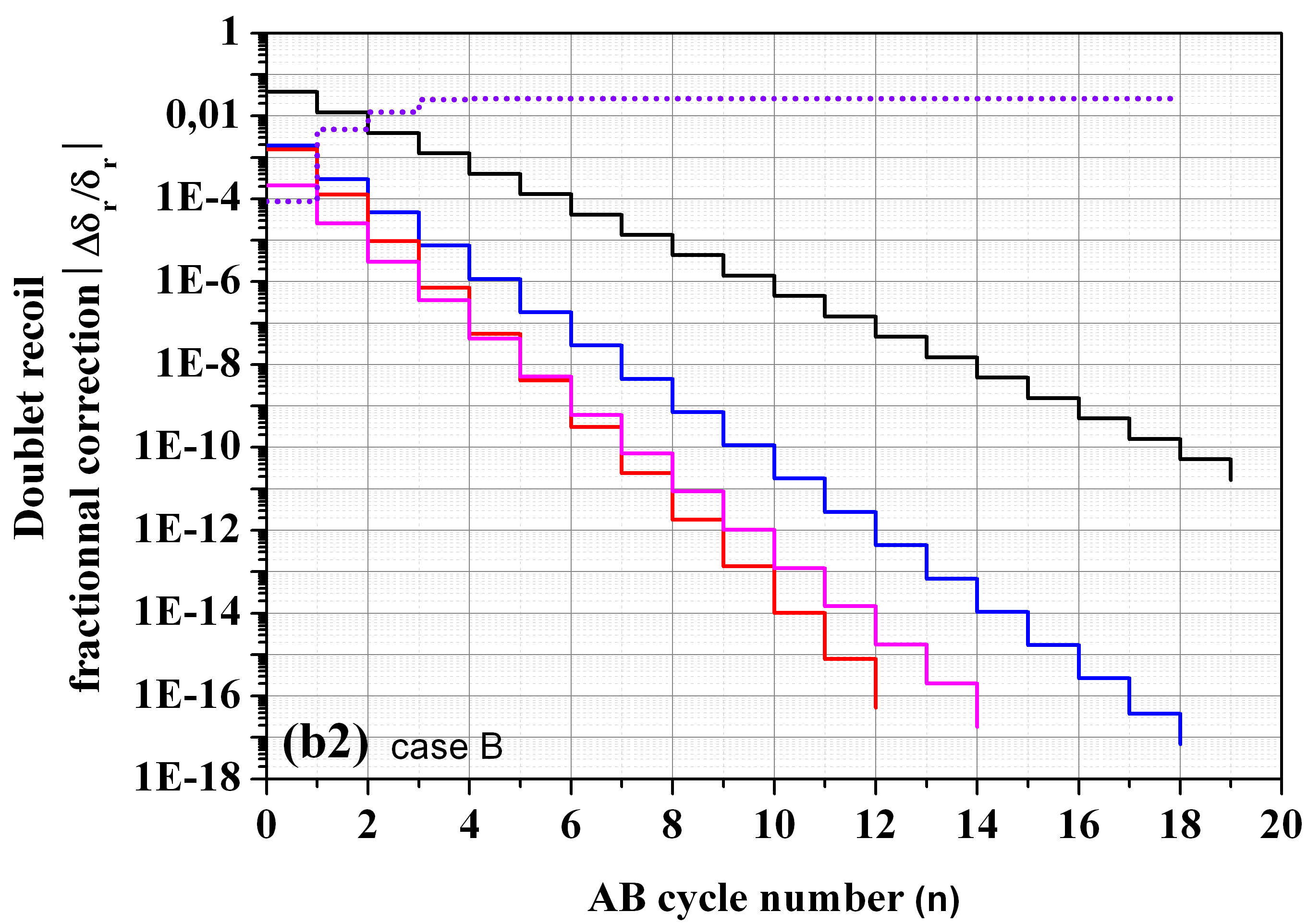}}
\caption{(color online). Relative fractional correction to the clock frequency (left panels) or to the recoil doublet measurement (right panels) based on fault-tolerant auto-balanced hyper-interferometers with composite pulses at T$=250$~pK. (a1) and (b1) auto-balanced hyper-clock configuration. (a2) and (b2) auto-balanced hyper-Ramsey-Bordé configuration. Cas A: the pulse area variation $\Delta\theta/\theta=\pm10\%$ is identical between $_{p}^{q}$M\textbf{I($\uparrow$)($\downarrow$)} and $_{p}^{q}$M\textbf{II($\downarrow$)($\uparrow$)} interaction zones while ignoring any phase-step distortion $\epsilon/2\pi=0$. Case B: the pulse area variation is different by $\Delta\theta/\theta=\pm10\%$ between $_{p}^{q}$M\textbf{I($\uparrow$)($\downarrow$)} and $_{p}^{q}$M\textbf{II($\downarrow$)($\uparrow$)} interaction zones including a phase-step distortion of $\epsilon/2\pi=1-10\%$. Other laser parameters as in Fig.~\ref{fig:recoil-doublet}.}
\label{fig:AB-interferometers}
\end{figure*}

\section{AUTO-BALANCED HYPER-INTERFEROMETERS WITH COMPOSITE PULSES}

\indent Today, optimal quantum control techniques are investigated for robust atomic interferometry~\cite{Malinovsky:2019,Goerz:2021} inspired by mathematical works based on optimization of computational algorithms~\cite{Krotov:1996,Khaneja:2005,Fouquieres:2011,Glaser:2015}.

We propose, here, a complementary approach for a very robust compensation technique enhancing immunity against various probing field-induced frequency-shifts of matter-wave interferences.
Auto-balanced Ramsey spectroscopy has been initially proposed in refs~\cite{Sanner_2018} and~\cite{Yudin:2018}, where the clock frequency stabilization is realized through the use of a feedback loop connected to two Ramsey protocols with a long T$_{l}$ and a short T$_{s}$ free evolution period. A specific adjustable laser parameter is used to extract some information about a technical pulse perturbation leading to a phase correction after one loop implementation that is then iterated again.
This approach can be seen as an active compensation of various deleterious effects than can strongly compromised the efficient robustness of clock-interferometers when technical pulse defects are themselves coupled to probe-induced phase-shifts and a possible error in phase-step modulation during the interrogation process.

One of the main limitation in atomic interferometry, at least for acceleration and rotation measurements based on a MZ or a double-loop interferometer is the wavefront distortion~\cite{Schkolnik:2015,Parker:2016,Trimeche:2017,Bade:2018}. Phases and the amplitudes of the laser beams present some inhomogeneities during the sequence of pulses because of the finite temperature of the atomic cloud~\cite{Karcher:2018}.

From the previous detailed analysis of this work, auto-balanced hyper-interferometers with composite pulses may be another way to actively reduce or compensate the wavefront deleterious action on quantum interferences. The correction is applied either to a clock scheme or to a laser frequency chirp method for interferometry which is then stabilized by auto-balanced servo-looping. The use of a specific composite pulse procotol (used as a noise filter) is accelerating the convergency rate to a robust level of fault-tolerant correction.
The set of auto-balanced coupled equations is introduced as following~\cite{Zanon-Willette:2018}:
\begin{eqnarray}
\xi \;
\rotatebox[origin=c]{90}{\LARGE{$ \curvearrowleft$}}\; \;
\begin{split}
\xi\cdot\textup{T}^{\kappa}_{l}-_{p}^{q}\Phi^{l}+\epsilon&=0\; \\
\xi\cdot\textup{T}^{\kappa}_{s}-_{p}^{q}\Phi^{s}+\epsilon&=0\;
\end{split}
\, \rotatebox[origin=c]{-90}{\LARGE{$ \curvearrowleft$}}
 \;_{p}^{q}\Phi^{s},
\label{eq:phase-frequency-equations}
\end{eqnarray}
Composite phase-shifts $_{p}^{q}\Phi^{l,s}$ may also include Doppler-shifts and atomic recoil for atomic interferometry and $\kappa$ is associated to the power T$^{\kappa}$ of the free evolution time. All composite phase-shifts are reported in Tab.~\ref{protocol-table-4}. The parameter $\epsilon$ is now introduced as a small error into the laser phase-step modulation due to a potential electronic servo-loop distortion.
The targeting atomic parameter $\xi$, which is depending on the protocol selected either for a clock configuration (c-protocol) or for an interferometer
configuration (AI-protocol), is listed below:
\begin{equation}
\xi\equiv\left\{
\begin{split}
&\delta, \kappa=1 \hspace{0.5cm}(frequency) \\
&2\left(\delta\pm\delta_{r}\right), \kappa=1  \hspace{0.5cm}(recoil) \\
&-\textbf{k}\cdot\left(\textbf{g}-\textbf{g}_{chirp}\right), \kappa=2  \hspace{0.5cm}(acceleration)\\
&-2\left(\textbf{k}\times\left(\textbf{g}-\textbf{g}_{chirp}\right)\right)\cdot\boldsymbol{\Omega}, \kappa=3  \hspace{0.5cm}(rotation).
\end{split}
\right.
\label{eq:parameter}
\end{equation}
The set of equations given by Eq.~(\ref{eq:phase-frequency-equations}) are converging after N loops to the following stable parameters:
\begin{equation}
\begin{split}
\xi\equiv0  \hspace{0.5cm} \text{and} \hspace{0.5cm} \Phi^{l,s}&\equiv0
\end{split}
\label{corrections}
\end{equation}
Any adjustable laser parameter can be one of the relative laser phase $\varphi'_{l},\varphi_{l}$, or pulse area modification through a pulse duration adjustement $\widetilde{\vartheta}'_{l},\widetilde{\vartheta}_{l}$ or any additional frequency-step compensation term $\Delta'_{l},\Delta_{l}$ applied during pulses~\cite{Yudin:2018}.

Our target is to optimize the convergency rate to the free parameter $\xi$ free from residual light-shift and technical pulse defects. We have reported in Fig.~\ref{fig:AB-interferometers}(a1) and (b1) the absolute fractional clock error $|\Delta\nu/\nu|$ and in Fig.~\ref{fig:AB-interferometers}(a2) and (b2) the relative correction to the recoil doublet of Ca versus the number of auto-balanced (AB) cycles. Two particular configurations of pulse area variation are taking into account. Case A is a pulse area variation of $\Delta\theta/\theta=\pm10\%$ between sequences of pulses ignoring a possible error in the phase-step modulation process $\epsilon=0$ to generate dispersive error signals. Case B is an assymetrical pulse area variation between sets of beam splitters including a possible error $\epsilon=1-10\%$ in the phase-step modulation process.

We observe in Fig.~\ref{fig:AB-interferometers}(a1) (a2) and (b1) (b2) that AB-R2 (AB-RB2), AB-HR3$_{\pi}$ (AB-HRB3$_{\pi}$) and AB-HR5$_{\pi}$ (AB-HRB5$_{\pi}$) protocols are converging faster in compensating a pulse area variation coupled to residual light-shifts compared to the initial auto-balanced Ramsey configuration AB-R1 (AB-RB1). We note that after a few cycles, some fractional clock errors can be largely different between protocols by several order of magnitude.
An hybrid protocol AB-R1/HR3$_{\pi}$ (AB-RB1/HRB3$_{\pi}$) is found to converge faster than all other protocols when using an additional frequency compensation step but is however fully unstable when a small distortion occurs on phase-steps generating dispersive error signals (divergence of the correction in Fig.~\ref{fig:AB-interferometers}(b1) and (b2)).
At least, the AB-GHHR clock protocol from Fig.~\ref{fig:AB-interferometers}(a1) is fully insensitive to any frequency correction when the pulse area variation is identical between the first and the last pulse (case A) while auto-balanced method efficiently corrects for any asymmetrical variation of the pulse area between the first pulse (or a set of composite pulses) opening the interferometer and the last pulse (or a set of composite pulses) closing the trajectories as shown in Fig.~\ref{fig:AB-interferometers}(b1).

Recently, the Hahn-Ramsey scheme was merged with auto-balanced Ramsey spectroscopy to probe quantum states of a single trapped ion
suppressing all inhomogeneous contributions and low frequency noise to phase accumulation other than the phase difference accumulated owing to the different free-precession times, unperturbed by laser beams~\cite{McCormick:2019}.

\section{CONCLUSIONS AND OUTLOOK}

\indent We have explored new applications of composite pulses from optical clocks to Ramsey-Bordé, Mach-Zehnder and Butterfly atom interferometers, and demonstrate that they might be designed to be more robust against pulse defects related to laser field amplitude variation coupled to residual Doppler-shifts and light-shifts between multiple interaction zones. We have introduced a new set of composite atomic beam-splitters with concatenated pulses tailored in frequency, duration and phase-steps to efficiently mitigate laser induced phase-shift distortions on optical Ramsey interferences.

The limits of actual atomic interferometry experiments raise the important question about inhomogeneities of phases and amplitudes of laser pulses used to diffract matter-waves related to the size of the atomic ensemble which is thermally expanding over time. Among these effects that have important consequences, the recoil determination can be modified in a distorted optical field~\cite{Bade:2018} and intensity-dependent phase-shifts are present through distortions in the wavefronts of the Bragg beams~\cite{Parker:2016}.
These error sources driven by fluctuation of coherent light excitation might be minimized by active control of laser wavefront~\cite{Trimeche:2017} and using colder atomic sources~\cite{Kovachy:2015,Karcher:2018}. For example, some of these systematics have been reduced through a spatial-filtering process by controlling the detection volume and limiting the spatial extent of the atom cloud before the interrogation protocol~\cite{Parker:2016}.

Generalized hyper-Ramsey-Bordé matter-wave interferometry is primarily dedicated to a new class of atomic interferometers using long-lived optical clock transitions in fermionic and bosonic alkaline-earth species inducing large momentum transfer~\cite{Hu:2017,Hu:2020,Abe:2021} but suffering from systematics coupled to undesirable ac Stark-shifts and magnetic-shift instabilities~\cite{Taichenachev:2006,Barber:2006,Baillard:2007,Kulosa:2015,Akatsuka:2008,Hobson:2016}. Auto-balanced techniques with composite pulses may be also of relevance to matter-wave control with Bragg diffraction and pulse shaping less susceptible to distortions related to ac-Stark and Zeeman shifts~\cite{Muller:2009,Altin:2013,Giese:2016}.

Advancing atomic and molecular coherent matter-wave manipulation with fault-tolerant auto-balanced hyper-interferometers~\cite{Sanner_2018,Yudin:2018} will bring atomic sensors to robust real-world application~\cite{Bongs:2019} from portable optical clocks to mobile gravimeters~\cite{Grotti:2018,Delahaye:2018,Wu:2019,Bidel:2018,Menoret:2018} as well as boosting performances of actual devices with a minimal experimental effort.
Combining such sensors with very recent quantum technologies through entanglement and spin squeezing~\cite{Hosten:2016,Salvi:2018,Brif:2020} will improve high-precision laser spectroscopy and metrology below a relative level of $10^{-18}$ in accuracy, opening new applications to track gravity induced phase-shifts in Ramsey interferometry~\cite{Abele:2010} or to detect gravitational waves in a frequency band between the LISA and LIGO detectors ~\cite{Abe:2021,Hogan:2011,Graham:2013,Kolkowitz:2016,Canuel:2018} while searching for new fundamental physics behind the standard model with a better accuracy~\cite{Safronova:2018}.

\section*{ACKNOWLEDGMENTS}

\indent  T.Z.W deeply thanks F. Pereira dos Santos, M. Cadoret, Oleg Prudnikov, E. Arimondo and Shau-Yu Lan for constructive comments about atomic interferometry,
M. Glass-Maujean for a careful reading of the manuscript and J. Ye for exciting discussion. I would here express a very deep acknowledgement to C.J. Bordé for interest in this work. V.I.Yudin was supported by the Russian Foundation for Basic Research (Grant Nos. 20-02-00505 and 19-32-90181) and Foundation for the Advancement of Theoretical Physics and Mathematics "BASIS".
A.V. Taichenachev acknowledges financial support from Russian Science Foundation through the grant 20-12-00081. T.Z.W. also acknowledges Sorbonne Université for a six months CRCT (Congés pour Recherches ou Convertions Thématiques) and MajuLab, CNRS-UCA-SU-NUS-NTU for supporting a six months collaborative project.

\section*{APPENDIX}

\indent We turn to look for an exact expression of the interaction matrix $_{p}^{q}$M using $p,q$ pulses applied around a single free evolution time.
We present our computational algorithm establishing matrix components $_{p}^{q}C_{gg},_{p}^{q}C_{ge}$ required to compute the transition probability generalizing the previous work by~\cite{Zanon-Willette:2019}.

\section*{S0: SYMMETRIC FUNCTIONS $S'_{p,k}$ AND $S_{q,k}$}

\indent We derive formulae and phase association rules to evaluate envelop terms $\alpha_{l}'^{p}(gg),\alpha_{l}^{q}(gg)$ and $\alpha_{l}'^{p}(ge)$ using some notations from~\cite{Hardy:2017} as following:
\begin{subequations}
\begin{align}
\alpha_{l}'^{p}(gg)&=\left(\prod_{l=\textup{1}}^{\textup{p}}\cos\widetilde{\vartheta}'_{l}e^{i\phi'_{\textup{l}}}\right)\cdot\left(\sum_{\textup{even k}\geq0}(-1)^{\frac{k}{2}}S'_{p,k}(gg)\right)\label{eq:alpha-ggp}\\
\alpha_{l}^{q}(gg)&=\left(\prod_{l=\textup{1}}^{\textup{q}}\cos\widetilde{\vartheta}_{l}e^{i\phi_{\textup{l}}}\right)\cdot\left(\sum_{\textup{even k}\geq0}(-1)^{\frac{k}{2}}S_{q,k}(gg)\right)\label{eq:alpha-ggq}\\
\alpha_{l}'^{p}(ge)&=\widetilde{\alpha}'\left(\prod_{l=\textup{1}}^{\textup{p}}\cos\widetilde{\vartheta}'_{l}e^{i\phi'_{\textup{l}}}\right)\cdot\left(\sum_{\textup{odd k}\geq1}(-1)^{\frac{k-1}{2}}S'_{p,k}(ge)\right)\label{eq:alpha-ge}
\end{align}
\end{subequations}
with $\widetilde{\alpha}'=-ie^{-i(\phi'_{\textup{p}}+\varphi'_{p}+\Xi'_{\textup{p}})}$ in Eq.~\ref{eq:alpha-ge}.
The convention is $S'_{p,0}=S_{q,0}=1$ in Eq.~\ref{eq:alpha-ggp} and Eq.~\ref{eq:alpha-ggq}.
Symmetric functions $S'_{p,k}$ and $S_{q,k}$ having respectively $\frac{p!}{k!(p-k)!}$ and $\frac{q!}{k!(q-k)!}$ elements are:
\begin{subequations}
    \begin{align}
S'_{p,k}(gg)&=\sum_{\tiny{\begin{array}{c}
       \textup{A}\subseteq\{1,2,3,...,p\}\\
     |\textup{A}|=k\\
  \end{array}}}e^{-i\Xi'_{\textup{A}}}\prod_{\tiny{\begin{array}{c}
      l\in\textup{A}\\
  \end{array}}}\tan\widetilde{\vartheta}'_{l}\label{eq:S-1}\\
S_{q,k}(gg)&=\sum_{\tiny{\begin{array}{c}
 \textup{A}\subseteq\{1,2,3,...,q\}\\
 |\textup{A}|=k\\
  \end{array}}
}e^{-i\Xi_{\textup{A}}}\prod_{\tiny{\begin{array}{c}
       l\in\textup{A}
  \end{array}}}\tan\widetilde{\vartheta}_{l}\label{eq:S-2}\\
S'_{p,k}(ge)&=\sum_{\tiny{\begin{array}{c}
       \textup{A}\subseteq\{1,2,3,...,p\}\\
     |\textup{A}|=k\\
  \end{array}}}e^{i\Xi'_{\textup{A}}}\prod_{\tiny{\begin{array}{c}
      l\in\textup{A}\\
  \end{array}}}\tan\widetilde{\vartheta}'_{l}\label{eq:S-3}
    \end{align}
\end{subequations}
Note that phase factors $\Xi'_{\textup{A}}$ as well as $\Xi_{\textup{A}}$ that are affected to the product in Eq.~\ref{eq:S-1} and Eq.~\ref{eq:S-2} are determined by all possible $\textup{k}$-combination of $\textup{l}$ elements in $\textup{A}$ ensemble with k even. The decomposition rules of $\Xi'_{\textup{A}}$ ($\Xi_{\textup{A}}$) are presented:
for $|\textup{A}|=2$, phase-factors are given by $\Xi'_{\textup{A}}\equiv\Xi'_{\textup{12}},\Xi'_{\textup{23}},...\Xi'_{\textup{13}},\Xi'_{\textup{24}},...$ where the decomposition is $\Xi'_{\textup{13}}=\Xi'_{\textup{12}}+\Xi'_{\textup{23}},...$.
For $|\textup{A}|=4$, we get $\Xi'_{\textup{A}}\equiv\Xi'_{\textup{1234}},\Xi_{\textup{2345}},...\Xi'_{\textup{1345}}...$ where $\Xi'_{\textup{1234}}=\Xi'_{\textup{12}}+\Xi'_{\textup{34}},...$ and so on.

Phase factors $\Xi'_{\textup{A}}$ affected to the product in Eq.~\ref{eq:S-3} are determined by all possible $\textup{k}$-combination of $\textup{l}$ elements in $\textup{A}$ ensemble with k odd. For $|\textup{A}|=1$, phase-factors are given by $\Xi'_{\textup{A}}\equiv\Xi'_{\textup{1}},\Xi'_{\textup{2}},\Xi'_{\textup{3}}...$. Here, by convention $\Xi'_{\textup{1}}=0$ and $\Xi'_{\textup{2}}\equiv\Xi'_{\textup{12}},\Xi'_{\textup{3}}\equiv\Xi'_{\textup{13}},...$ where the decomposition is $\Xi'_{\textup{12}}=\Xi'_{\textup{12}},\Xi'_{\textup{13}}=\Xi'_{\textup{12}}+\Xi'_{\textup{23}},\Xi'_{\textup{14}}=\Xi'_{\textup{12}}+\Xi'_{\textup{23}}+\Xi'_{\textup{34}},...$.
For $|\textup{A}|=3$, we get $\Xi'_{\textup{A}}\equiv\Xi'_{\textup{123}},\Xi_{\textup{124}},\Xi_{\textup{134}}...,\Xi_{\textup{234}},...$ where $\Xi'_\textup{123}=\Xi'_{\textup{1}}+\Xi'_{\textup{23}},\Xi'_{\textup{124}}=\Xi'_{\textup{1}}+\Xi'_{\textup{24}},...\Xi'_{\textup{234}}=\Xi'_{\textup{2}}+\Xi'_{\textup{34}},...$ and so on.
The basic structure of all phase-factors is $\Xi'_{l,l+1}=\varphi'_{l}-\varphi'_{l+1}+\phi'_{l}+\phi'_{l+1}$ for $\Xi'_{\textup{A}}$ and $\Xi_{\textup{l,l+1}}=\varphi_{l+1}-\varphi_{l}+\phi_{l}+\phi_{l+1}$ for $\Xi_{\textup{A}}$.

\section*{S1: SPINOR COMPONENTS WITH $\textit{p}=\textit{q}=4$}

\indent We present an interaction matrix $_{4}^{4}$M based on four pulses.
The matrix components $_{4}^{4}C_{gg},_{4}^{4}C_{ge}$ we need to evaluate are evaluated by the following elements:
\begin{subequations}
\begin{align}
\alpha_{1}'^{4}(gg)=&\left(\prod_{\textup{1}}^{p=\textup{4}}\cos\widetilde{\vartheta}'_{l}e^{i\phi'_{\textup{l}}}\right)\cdot\left(1-S'_{4,2}+S'_{4,4}\right)\\
\alpha_{1}^{4}(gg)=&\left(\prod_{\textup{1}}^{q=\textup{4}}\cos\widetilde{\vartheta}_{l}e^{i\phi_{\textup{l}}}\right)\cdot\left(1-S_{4,2}+S_{4,4}\right)\\
\alpha_{1}'^{4}(ge)=&-ie^{-i(\phi'_{4}+\varphi'_{4}+\Xi'_{4})}\\
&\times\left(\prod_{\textup{1}}^{p=\textup{4}}\cos\widetilde{\vartheta}'_{l}e^{i\phi'_{\textup{l}}}\right)\cdot\left(S'_{4,1}-S'_{4,3}\right)
\end{align}
\end{subequations}
We have for $S'_{4,k}(gg)$:
\begin{equation}
\begin{split}
S'_{4,0}=&1\\
S'_{4,2}=&e^{-i\Xi'_{12}}\tan\widetilde{\vartheta}'_{1}\tan\widetilde{\vartheta}'_{2}+e^{-i\Xi'_{23}}\tan\widetilde{\vartheta}'_{2}\tan\widetilde{\vartheta}'_{3}\\
&+e^{-i\Xi'_{34}}\tan\widetilde{\vartheta}'_{3}\tan\widetilde{\vartheta}'_{4}+e^{-i\Xi'_{13}}\tan\widetilde{\vartheta}'_{1}\tan\widetilde{\vartheta}'_{3}\\
&+e^{-i\Xi'_{14}}\tan\widetilde{\vartheta}'_{1}\tan\widetilde{\vartheta}'_{4}+e^{-i\Xi'_{24}}\tan\widetilde{\vartheta}'_{2}\tan\widetilde{\vartheta}'_{4}\\
S'_{4,4}=&e^{-i\Xi'_{1234}}\tan\widetilde{\vartheta}'_{1}\tan\widetilde{\vartheta}'_{2}\tan\widetilde{\vartheta}'_{3}\tan\widetilde{\vartheta}'_{4}
\end{split}
\label{eq:}
\end{equation}
and for $S'_{4,k}(ge)$:
\begin{equation}
\begin{split}
S'_{4,1}=&\tan\widetilde{\vartheta}'_{1}+e^{i\Xi'_{2}}\tan\widetilde{\vartheta}'_{2}\\
&+e^{i\Xi'_{3}}\tan\widetilde{\vartheta}'_{3}+e^{i\Xi'_{4}}\tan\widetilde{\vartheta}'_{4}\\
S'_{4,3}=&e^{i\Xi'_{123}}\tan\widetilde{\vartheta}'_{1}\tan\widetilde{\vartheta}'_{2}\tan\widetilde{\vartheta}'_{3}\\
&+e^{i\Xi'_{124}}\tan\widetilde{\vartheta}'_{1}\tan\widetilde{\vartheta}'_{2}\tan\widetilde{\vartheta}'_{4}\\
&+e^{i\Xi'_{134}}\tan\widetilde{\vartheta}'_{1}\tan\widetilde{\vartheta}'_{3}\tan\widetilde{\vartheta}'_{4}\\
&+e^{i\Xi'_{234}}\tan\widetilde{\vartheta}'_{2}\tan\widetilde{\vartheta}'_{3}\tan\widetilde{\vartheta}'_{4}
\end{split}
\label{eq:}
\end{equation}
We also have for $S_{4,k}(gg)$ elements:
\begin{equation}
\begin{split}
S_{4,0}=&1\\
S_{4,2}=&e^{-i\Xi_{12}}\tan\widetilde{\vartheta}_{1}\tan\widetilde{\vartheta}_{2}+e^{-i\Xi_{23}}\tan\widetilde{\vartheta}_{2}\tan\widetilde{\vartheta}_{3}\\
&+e^{-i\Xi_{34}}\tan\widetilde{\vartheta}_{3}\tan\widetilde{\vartheta}_{4}+e^{-i\Xi_{13}}\tan\widetilde{\vartheta}_{1}\tan\widetilde{\vartheta}_{3}\\
&+e^{-i\Xi_{14}}\tan\widetilde{\vartheta}_{1}\tan\widetilde{\vartheta}_{4}+e^{-i\Xi_{24}}\tan\widetilde{\vartheta}_{2}\tan\widetilde{\vartheta}_{4}\\
S_{4,4}=&e^{-i\Xi_{1234}}\tan\widetilde{\vartheta}_{1}\tan\widetilde{\vartheta}_{2}\tan\widetilde{\vartheta}_{3}\tan\widetilde{\vartheta}_{4}
\end{split}
\label{eq:}
\end{equation}
The corresponding complex phase factor $\beta_{1}'^{4}(gg),\beta_{1}^{4}(gg)$ leading to a phase-shift correction are now:
\begin{subequations}
\begin{align}
\beta_{1}'^{4}(gg)=&\frac{\tan\widetilde{\vartheta}'_{1}+e^{-i\Xi'_{12}}\frac{\tan\widetilde{\vartheta}'_{2}+e^{-i\Xi'_{23}}\frac{\tan\widetilde{\vartheta}'_{3}+e^{-i\Xi'_{34}}\tan\widetilde{\vartheta}'_{4}}
{1-e^{-i\Xi'_{34}}\tan\widetilde{\vartheta}'_{3}\tan\widetilde{\vartheta}'_{4}}}
{1-e^{-i\Xi'_{23}}\tan\widetilde{\vartheta}'_{2}\frac{\tan\widetilde{\vartheta}'_{3}+e^{-i\Xi'_{34}}\tan\widetilde{\vartheta}_{4}}
{1-e^{-i\Xi'_{34}}\tan\widetilde{\vartheta}'_{3}\tan\widetilde{\vartheta}'_{4}}}}
{1-e^{-i\Xi'_{12}}\tan\widetilde{\vartheta}'_{1}\frac{\tan\widetilde{\vartheta}'_{2}+e^{-i\Xi'_{23}}\frac{\tan\widetilde{\vartheta}'_{3}+e^{-i\Xi'_{34}}\tan\widetilde{\vartheta}'_{4}}
{1-e^{-i\Xi'_{34}}\tan\widetilde{\vartheta}'_{3}\tan\widetilde{\vartheta}'_{4}}}
{1-e^{-i\Xi'_{23}}\tan\widetilde{\vartheta}'_{2}\frac{\tan\widetilde{\vartheta}'_{3}+e^{-i\Xi'_{34}}\tan\widetilde{\vartheta}'_{4}}
{1-e^{-i\Xi'_{34}}\tan\widetilde{\vartheta}'_{3}\tan\widetilde{\vartheta}'_{4}}}}\label{eq:betaggp-14}
\\
\notag
\\
\beta_{1}^{4}(gg)=&\frac{\tan\widetilde{\vartheta}_{1}+e^{-i\Xi_{12}}\frac{\tan\widetilde{\vartheta}_{2}+e^{-i\Xi_{23}}\frac{\tan\widetilde{\vartheta}_{3}+e^{-i\Xi_{34}}\tan\widetilde{\vartheta}_{4}}
{1-e^{-i\Xi_{34}}\tan\widetilde{\vartheta}_{3}\tan\widetilde{\vartheta}_{4}}}
{1-e^{-i\Xi_{23}}\tan\widetilde{\vartheta}_{2}\frac{\tan\widetilde{\vartheta}_{3}+e^{-i\Xi_{34}}\tan\widetilde{\vartheta}_{4}}
{1-e^{-i\Xi_{34}}\tan\widetilde{\vartheta}_{3}\tan\widetilde{\vartheta}_{4}}}}
{1-e^{-i\Xi_{12}}\tan\widetilde{\vartheta}_{1}\frac{\tan\widetilde{\vartheta}_{2}+e^{-i\Xi_{23}}\frac{\tan\widetilde{\vartheta}_{3}+e^{-i\Xi_{34}}\tan\widetilde{\vartheta}_{4}}
{1-e^{-i\Xi_{34}}\tan\widetilde{\vartheta}_{3}\tan\widetilde{\vartheta}_{4}}}
{1-e^{-i\Xi_{23}}\tan\widetilde{\vartheta}_{2}\frac{\tan\widetilde{\vartheta}_{3}+e^{-i\Xi_{34}}\tan\widetilde{\vartheta}_{4}}
{1-e^{-i\Xi_{34}}\tan\widetilde{\vartheta}_{3}\tan\widetilde{\vartheta}_{4}}}}\label{eq:betaggq-14}\\
\beta_{1}'^{4}(ge)=&\frac{1}{\{\beta_{1}'^{4}(gg)\}^{*}}
\end{align}
\end{subequations}
We give the decomposition of phase factor expressions as following:
\begin{equation}
\begin{split}
\Xi'_{1}=&0\\
\Xi'_{2}=&\Xi'_{12}\\
\Xi'_{3}=&\Xi'_{13}\\
\Xi'_{4}=&\Xi'_{14}\\
\Xi'_{123}=&\Xi'_{1}+\Xi'_{23}\\
\Xi'_{124}=&\Xi'_{1}+\Xi'_{24}\\
\Xi'_{134}=&\Xi'_{1}+\Xi'_{34}\\
\Xi'_{234}=&\Xi'_{2}+\Xi'_{34}
\end{split}
\label{eq:}
\end{equation}
\begin{equation}
\begin{split}
\Xi'_{12}=&\varphi'_{1}-\varphi'_{2}+\phi'_{1}+\phi'_{2}\\
\Xi'_{23}=&\varphi'_{2}-\varphi'_{3}+\phi'_{2}+\phi'_{3}\\
\Xi'_{34}=&\varphi'_{3}-\varphi'_{4}+\phi'_{3}+\phi'_{4}\\
\Xi'_{13}=&\Xi'_{12}+\Xi'_{23}\\
\Xi'_{24}=&\Xi'_{23}+\Xi'_{34}\\
\Xi'_{14}=&\Xi'_{12}+\Xi'_{23}+\Xi'_{34}\\
\Xi'_{1234}=&\Xi'_{12}+\Xi'_{34}\\
\end{split}
\label{eq:}
\end{equation}
and:
\begin{equation}
\begin{split}
\Xi_{12}=&\varphi_{2}-\varphi_{1}+\phi_{1}+\phi_{2}\\
\Xi_{23}=&\varphi_{3}-\varphi_{2}+\phi_{2}+\phi_{3}\\
\Xi_{34}=&\varphi_{4}-\varphi_{3}+\phi_{3}+\phi_{4}\\
\Xi_{13}=&\Xi_{12}+\Xi_{23}\\
\Xi_{24}=&\Xi_{23}+\Xi_{34}\\
\Xi_{14}=&\Xi_{12}+\Xi_{23}+\Xi_{34}\\
\Xi_{1234}=&\Xi_{12}+\Xi_{34}
\end{split}
\label{eq:}
\end{equation}
See also \cite{Footnote} as another way to obtain $_{4}^{4}C_{ge}$ from $_{4}^{4}C_{gg}$.

\section*{S2: GHHR COMPONENTS WITH $p=1$, $q=1$ AND TWO FREE EVOLUTION TIME ZONES T$_{I}$, T$_{II}$}

\indent To obtain a full analytical solution of the complex amplitude related to the GHHR transition probability with only three pulses (one of the two intermediate pulses from Fig.~\ref{fig:GHHR-sequence} has a vanishing pulse area), we can simplify a few matrix elements from Eq.~\ref{eq:coefficient-function-GHHR} and Eq.~\ref{eq:amplitude-function-GHHR}. To get round of any divergence when taking one pulse area to zero, we decide to null the second laser pulse area denoted as $\widetilde{\vartheta}_{\textup{I,1}}\equiv0$ where we get $\alpha_{1}^{1}(gg)^{I}=\alpha_{1}^{1}(gg)^{I*}=1$ and $\beta_{1}^{1}(gg)^{I}=\beta_{1}^{1}(gg)^{I*}=0$ (we avoid the divergency of the other $\beta_{1}'^{1}(ge)^{II}$ pulse element from Eq.~\ref{eq:beta-function-GHHR}).
We therefore decompose previous matrix elements and obtain:
\begin{equation}
\begin{split}
_{1}^{1}\alpha^{I*}_{ge}\cdot_{1}^{1}\alpha^{II}_{eg}\cdot_{1}^{1}\beta_{ge}^{I*}&=0\\
_{1}^{1}\alpha^{I}_{gg}\cdot_{1}^{1}\alpha^{II}_{gg}\cdot_{1}^{1}\beta_{gg}^{I}&=0\\
_{1}^{1}\alpha^{I*}_{ge}\cdot_{1}^{1}\alpha^{II}_{eg}\cdot_{1}^{1}\beta_{ge}^{II}&=\alpha_{1}'^{1}(ge)^{I*}\alpha_{1}'^{1}(ge)^{II}\alpha_{1}^{1}(gg)^{II}\cdot_{1}^{1}\beta_{ge}^{II}\\
_{1}^{1}\alpha^{I}_{gg}\cdot_{1}^{1}\alpha^{II}_{gg}\cdot_{1}^{1}\beta_{gg}^{II}&=\alpha_{1}'^{1}(gg)^{I}\alpha_{1}'^{1}(gg)^{II}\alpha_{1}^{1}(gg)^{II}\cdot_{1}^{1}\beta_{gg}^{II}\\
_{1}^{1}\alpha^{I*}_{ge}\cdot_{1}^{1}\alpha^{II}_{eg}&=\alpha_{1}'^{1}(ge)^{I*}\alpha_{1}'^{1}(ge)^{II}\alpha_{1}^{1}(gg)^{II}\\
_{1}^{1}\alpha^{I}_{gg}\cdot_{1}^{1}\alpha^{II}_{gg}&=\alpha_{1}'^{1}(gg)^{I}\alpha_{1}'^{1}(gg)^{II}\alpha_{1}^{1}(gg)^{II}\\
_{1}^{1}\beta_{ge}^{II}&=\beta_{1}'^{1}(ge)^{II}\beta_{1}^{1}(gg)^{II}\\
_{1}^{1}\beta_{gg}^{II}&=\beta_{1}'^{1}(gg)^{II}\beta_{1}^{1}(gg)^{II}
\end{split}
\end{equation}
The $C_{g}(t)$ amplitude related to the transition probability is thus:
\begin{equation}
\begin{split}
C_{g}(t)=&Ae^{i\left(d_{+}\textup{T}_{+}/2\right)}-Be^{-i\left(d_{+}\textup{T}_{+}/2+_{1}^{1}\Phi^{II}_{gg}\right)}\\
&-Ce^{i\left(d_{-}\textup{T}_{-}/2-_{1}^{1}\Phi^{II}_{gg}\right)}-De^{-i\left(d_{-}\textup{T}_{-}/2\right)}
\end{split}
\label{eq:transition-probability-GHHR}
\end{equation}
where $d_{+}\textup{T}_{+}=\delta_{I}\textup{T}_{I}+\delta_{II}\textup{T}_{II}$, $d_{-}\textup{T}_{-}=\delta_{I}\textup{T}_{I}-\delta_{II}\textup{T}_{II}$
and:
\begin{equation}
\begin{split}
A&=_{1}^{1}\alpha^{I}_{gg}\cdot_{1}^{1}\alpha^{II}_{gg}\\
&=\cos\widetilde{\vartheta}'_{\textup{I,1}}\cos\widetilde{\vartheta}'_{\textup{II,1}}\cos\widetilde{\vartheta}_{\textup{II,1}}e^{i(\phi'_{\textup{I,1}}+\phi'_{\textup{II,1}}+\phi_{\textup{II,1}})}\\
B&=_{1}^{1}\alpha^{I*}_{ge}\cdot_{1}^{1}\alpha^{II}_{eg}\cdot_{1}^{1}\beta_{ge}^{II}\\
&=\sin\widetilde{\vartheta}'_{\textup{I,1}}\cos\widetilde{\vartheta}'_{\textup{II,1}}\sin\widetilde{\vartheta}_{\textup{II,1}}e^{i(\varphi'_{I,1}-\varphi'_{II,1}+\phi_{\textup{II,1}})}\\
C&=_{1}^{1}\alpha^{I}_{gg}\cdot_{1}^{1}\alpha^{II}_{gg}\cdot_{1}^{1}\beta_{gg}^{II}\\
&=\cos\widetilde{\vartheta}'_{\textup{I,1}}\sin\widetilde{\vartheta}'_{\textup{II,1}}\sin\widetilde{\vartheta}_{\textup{II,1}}e^{i(\phi'_{\textup{I,1}}+\phi'_{\textup{II,1}}+\phi_{\textup{II,1}})}\\
D&=_{1}^{1}\alpha^{I*}_{ge}\cdot_{1}^{1}\alpha^{II}_{eg}\\
&=\sin\widetilde{\vartheta}'_{\textup{I,1}}\sin\widetilde{\vartheta}'_{\textup{II,1}}\cos\widetilde{\vartheta}_{\textup{II,1}}e^{i(\varphi'_{I,1}-\varphi'_{II,1}+\phi_{\textup{II,1}})}
\end{split}
\label{eq:coefficients-GHHR}
\end{equation}
After simplification, we get the transition amplitude expression:
\begin{equation}
\begin{split}
C_{g}(t)=&\widetilde{A}\left(1-\tan\widetilde{\vartheta}'_{\textup{I,1}}\tan\widetilde{\vartheta}_{\textup{II,1}}e^{-i\left(d_{+}\textup{T}_{+}+\Phi^{+}_{gg}\right)}\right)\\
&\times\left\{1-\widetilde{B}\left(1+\frac{\tan\widetilde{\vartheta}'_{\textup{I,1}}}{\tan\widetilde{\vartheta}_{\textup{II,1}}}e^{-i\left(d_{-}\textup{T}_{-}+\Phi^{-}_{gg}\right)}\right)\right\}
\end{split}
\label{eq:GHHR}
\end{equation}
with:
\begin{equation}
\begin{split}
\widetilde{A}&=Ae^{i\left(d_{+}\textup{T}_{+}/2\right)}\\
\widetilde{B}&=\frac{Ce^{i\left(d_{-}\textup{T}_{-}/2-d_{+}\textup{T}_{+}/2+\Phi^{II}_{gg}\right)}}{1-\tan\widetilde{\vartheta}'_{\textup{I,1}}\tan\widetilde{\vartheta}_{\textup{II,1}}e^{-i\left(d_{+}\textup{T}_{+}+\Phi^{+}_{gg}\right)}}\\
\Phi^{II}_{gg}&=\varphi_{II,1}-\varphi'_{II,1}+\phi'_{\textup{II,1}}+\phi_{\textup{II,1}}\\
\Phi^{+}_{gg}&=-\varphi'_{I,1}+\varphi'_{II,1}+\phi'_{\textup{I,1}}+2\phi'_{\textup{II,1}}+\phi_{\textup{II,1}}\\
\Phi^{-}_{gg}&=-\varphi'_{I,1}+2\varphi'_{II,1}-\varphi_{II,1}+\phi'_{\textup{I,1}}-\phi_{\textup{II,1}}
\end{split}
\label{eq:envelops}
\end{equation}

\section*{S3: GHRB COMPONENTS WITH $\textit{p}=1,\textit{q}=2$}

\indent For a general purpose, we explicitly give interaction matrix components $_{1}^{2}$M\textbf{I($\uparrow$)($\downarrow$)} and $_{1}^{2}$M\textbf{II$(\downarrow$)($\uparrow$)} inside successive building-blocks that are including pairs of traveling waves with opposite orientation to close the interferometer.
We have for the first $_{1}^{2}$M\textbf{I($\uparrow$)($\downarrow$)} interaction zone, the following components:
\begin{equation}
\begin{split}
_{1}^{2}\alpha^{I}_{gg}&=\alpha_{1}'^{1I}(gg)\cdot\alpha_{1}^{2I}(gg)\\
_{1}^{2}\alpha^{I}_{ge}&=\alpha_{1}'^{1I}(ge)\cdot\alpha_{1}^{2I}(gg)\\
_{1}^{2}\beta^{I}_{gg}&=\beta_{1}'^{1I}(gg)\cdot\beta_{1}^{2I}(gg)\\
_{1}^{2}\beta^{I}_{ge}&=\beta_{1}'^{1I}(ge)\cdot\beta_{1}^{2I}(gg)
\end{split}
\label{eq:}
\end{equation}
where we have:
\begin{equation}
\begin{split}
\alpha_{1}'^{1I}(gg)&=\cos\widetilde{\vartheta}'_{1}e^{i\phi'_{\textup{1}}}\\
\alpha_{1}^{2I}(gg)&=\cos\widetilde{\vartheta}_{1}\cos\widetilde{\vartheta}_{2}e^{i(\phi_{\textup{1}}+\phi_{\textup{2}})}\cdot\left(1-S_{2,2}\right)\\
\beta_{1}'^{1I}(gg)&=\tan\widetilde{\vartheta}'_{1}\\
\beta_{1}^{2I}(gg)&=\frac{\tan\widetilde{\vartheta}_{1}+e^{-i\Xi_{12}}\tan\widetilde{\vartheta}_{2}}
{1-e^{-i\Xi_{12}}\tan\widetilde{\vartheta}_{1}\tan\widetilde{\vartheta}_{2}}
\end{split}
\label{eq:}
\end{equation}
and
\begin{equation}
\begin{split}
\alpha_{1}'^{1I}(ge)&=-i\sin\widetilde{\vartheta}'_{1}e^{-i(\varphi'_{\textup{1}}+\phi'_{\textup{1}})}e^{i\phi'_{\textup{1}}}\\
\beta_{1}'^{1I}(ge)&=\frac{1}{\tan\widetilde{\vartheta}'_{1}}
\end{split}
\label{eq:}
\end{equation}
with a laser detuning definition given by $\delta_{I}^{\uparrow\downarrow}=\delta\mp\textup{kv}_{z}-\delta_{r}+\Delta_{I}$ for wavevector orientation.\\

We have for the second $_{1}^{2}$M\textbf{II$(\downarrow$)($\uparrow$)} interaction zone with opposite wavevectors the following components:
\begin{equation}
\begin{split}
_{1}^{2}\alpha^{II}_{gg}&=\alpha_{1}'^{1II}(gg)\cdot\alpha_{1}^{2II}(gg)\\
_{1}^{2}\alpha^{II}_{ge}&=\alpha_{1}'^{1II}(ge)\cdot\alpha_{1}^{2II}(gg)\\
_{1}^{2}\beta^{II}_{gg}&=\beta_{1}'^{1II}(gg)\cdot\beta_{1}^{2II}(gg)\\
_{1}^{2}\beta^{II}_{ge}&=\beta_{1}'^{1II}(ge)\cdot\beta_{1}^{2II}(gg)
\end{split}
\label{eq:}
\end{equation}
where laser detunings are given by $\delta_{II}^{\downarrow\uparrow}=\delta\pm\textup{kv}_{z}+3\delta_{r}+\Delta_{II}$ for the down-shifted frequency component and $\delta_{II}^{\downarrow\uparrow}=\delta\pm\textup{kv}_{z}-\delta_{r}+\Delta_{II}$ for the up-shifted component by the atomic recoil.\\
We have used definitions from~\cite{Borde:1984,Salomon:1984} for arbitrary transverse Doppler-shift orientation and momentum quantization along each path of the interferometer.

\section*{S4: HMZ AND HBU COMPONENTS WITH $\textit{p}=3,\textit{q}=4$}

\indent Some analytic results from section S1 are required again to evaluate a single interaction matrix $_{3}^{4}$M\textbf{$(\uparrow$)} including a composite set of $p=3$ pulses for the left side of the HMZ (\textbf{MI$(\uparrow)$}) and a composite set of $q=4$ pulses for the right side of the HMZ (\textbf{MII$(\uparrow)$}). For the HMZ (MZ) case, the two free evolution times are equal $\textup{T}_{I}=\textup{T}_{II}=\textup{T}$ while for the HBU (BU) case, the intermediate free evolution time $\textup{T'}$ is twice the other ones as $\textup{T'}=2\textup{T}$ to eliminate the clock detuning variable $\delta\textup{T}$.
The matter-wave interferometric signal can be thus directly computed leading to the complex expression:
\begin{equation}
\begin{split}
\textup{P}_{\textup{HMZ/HBU}}=\left|_{3}^{4}\alpha_{gg}\left[1-\left|_{3}^{4}\beta_{gg}\right|e^{-i_{3}^{4}\Phi(gg)}\right]\right|^{2}
\end{split}
\label{eq:}
\end{equation}
where
\begin{subequations}
\begin{align}
_{3}^{4}\alpha_{gg}=&\alpha_{1}'^{3}(gg)\cdot\alpha_{1}^{4}(gg)\\
_{3}^{4}\beta_{gg}=&\beta_{1}'^{3}(gg)\cdot\beta_{1}^{4}(gg)
 \end{align}
\end{subequations}
 with
\begin{subequations}
\begin{align}
\alpha_{1}'^{3}(gg)=&\left(\prod_{\textup{1}}^{3}\cos\widetilde{\vartheta}'_{l}e^{i\phi'_{\textup{l}}}\right)\cdot\left(1-S'_{3,2}\right)\label{eq:HMZalpha-13}\\
\alpha_{1}^{4}(gg)=&\left(\prod_{\textup{1}}^{q=\textup{4}}\cos\widetilde{\vartheta}_{l}e^{i\phi_{\textup{l}}}\right)\cdot\left(1-S_{4,2}+S_{4,4}\right)\label{eq:HMZalpha-14}
\end{align}
\end{subequations}
and
\begin{subequations}
\begin{align}
\beta_{1}'^{3}(gg)=&\frac{\tan\widetilde{\vartheta}'_{1}+e^{-i\Xi'_{12}}\frac{\tan\widetilde{\vartheta}'_{2}+e^{-i\Xi'_{23}}\tan\widetilde{\vartheta}'_{3}}
{1-e^{-i\Xi'_{23}}\tan\widetilde{\vartheta}'_{2}\tan\widetilde{\vartheta}'_{3}}}
{1-e^{-i\Xi'_{12}}\tan\widetilde{\vartheta}'_{1}\frac{\tan\widetilde{\vartheta}'_{2}+e^{-i\Xi'_{23}}\tan\widetilde{\vartheta}'_{3}}
{1-e^{-i\Xi'_{23}}\tan\widetilde{\vartheta}'_{2}\tan\widetilde{\vartheta}'_{3}}}\label{eq:HMZbetagg-13}
\\
\notag
\\
\beta_{1}^{4}(gg)=&\frac{\tan\widetilde{\vartheta}_{1}+e^{-i\Xi_{12}}\frac{\tan\widetilde{\vartheta}_{2}+e^{-i\Xi_{23}}\frac{\tan\widetilde{\vartheta}_{3}+e^{-i\Xi_{34}}\tan\widetilde{\vartheta}_{4}}
{1-e^{-i\Xi_{34}}\tan\widetilde{\vartheta}_{3}\tan\widetilde{\vartheta}_{4}}}
{1-e^{-i\Xi_{23}}\tan\widetilde{\vartheta}_{2}\frac{\tan\widetilde{\vartheta}_{3}+e^{-i\Xi_{34}}\tan\widetilde{\vartheta}_{4}}
{1-e^{-i\Xi_{34}}\tan\widetilde{\vartheta}_{3}\tan\widetilde{\vartheta}_{4}}}}
{1-e^{-i\Xi_{12}}\tan\widetilde{\vartheta}_{1}\frac{\tan\widetilde{\vartheta}_{2}+e^{-i\Xi_{23}}\frac{\tan\widetilde{\vartheta}_{3}+e^{-i\Xi_{34}}\tan\widetilde{\vartheta}_{4}}
{1-e^{-i\Xi_{34}}\tan\widetilde{\vartheta}_{3}\tan\widetilde{\vartheta}_{4}}}
{1-e^{-i\Xi_{23}}\tan\widetilde{\vartheta}_{2}\frac{\tan\widetilde{\vartheta}_{3}+e^{-i\Xi_{34}}\tan\widetilde{\vartheta}_{4}}
{1-e^{-i\Xi_{34}}\tan\widetilde{\vartheta}_{3}\tan\widetilde{\vartheta}_{4}}}}\label{eq:HMZbetagg-14}
\end{align}
\end{subequations}

\subsubsection*{S4-1: HMZ COMPONENTS}

\indent MZ and HMZ geometries use a $\boldsymbol{180}_{\varphi_{1}}$ pulse to recombine atomic wave-packets. The $2\times2$ matrix associated to the mirror pulse action in matter-wave interferometry is:
\begin{equation}
\begin{split}
\left(
  \begin{array}{cc}
    0 & -i e^{-i(\varphi_{1})}\sin\widetilde{\vartheta}_{1} \\
    -i e^{i(\varphi_{1})}\sin\widetilde{\vartheta}_{1} & 0 \\
  \end{array}
\right)
\end{split}
\label{eq:matrix-MZ-mirror}
\end{equation}
where diagonals elements are zero because non-overlapping matter-waves give no interferometric signal.
From Eq.\ref{eq:HMZalpha-14} and Eq.\ref{eq:HMZbetagg-14}, we apply $\cos\widetilde{\vartheta}_{1}e^{i\phi_{\textup{1}}}\mapsto0$ with $\phi_{\textup{1}}=0$ while keeping $\sin\widetilde{\vartheta}_{1}$.
We finally get modified expressions for $\alpha_{1}^{4}(gg)\mapsto\alpha_{2}^{4}(gg)$ and $\beta_{1}^{4}(gg)\mapsto\beta_{2}^{4}(gg)$ within section IV.B of the main text.

\subsubsection*{S4-2: HBU COMPONENTS}

\indent Again, BU and HBU geometries use a $\boldsymbol{180^{\uparrow}}_{\varphi_{0}}\dashv\delta^{\uparrow}\textup{T'}\vdash\boldsymbol{180^{\uparrow}}_{\varphi_{1}}$  time delayed double pulse to exchange and recombine atomic wave-packets. The overall $2\times2$ matrix associated to this particular sequence of pulses mirror in matter-wave interferometry is:
\begin{equation}
\begin{split}
\left(
  \begin{array}{cc}
  \begin{array}{c}
-e^{-i\left(\varphi_{0}-\varphi_{1}\right)}\sin\widetilde{\vartheta}_{0}\sin\widetilde{\vartheta}_{1}
  \end{array}
& 0 \\
    0 &   \begin{array}{c}
-e^{i\left(\varphi_{0}-\varphi_{1}\right)}\sin\widetilde{\vartheta}_{0}\sin\widetilde{\vartheta}_{1}
  \end{array}
 \\
  \end{array}
\right)
\end{split}
\label{eq:matrix-BU-mirror}
\end{equation}
This time, within Eq.\ref{eq:HMZalpha-14} and Eq.\ref{eq:HMZbetagg-14}, we apply the transformation $\cos\widetilde{\vartheta}_{1}e^{i\Phi_{\textup{1}}}\mapsto-e^{-i\left(\varphi_{0}-\varphi_{1}\right)}\sin\widetilde{\vartheta}_{0}\sin\widetilde{\vartheta}_{1}$ where $\Phi_{\textup{1}}\mapsto-\left(\varphi_{0}-\varphi_{1}\right)$ while taking $\sin\widetilde{\vartheta}_{1}\mapsto0$.
We finally get modified expressions for $\alpha_{1}^{4}(gg)\mapsto\alpha_{2}^{4}(gg)$ and $\beta_{1}^{4}(gg)\mapsto\beta_{2}^{4}(gg)$ within section IV.B of the main text.

\subsubsection*{S4-3: SENSITIVITY TO ACCELERATION AND ROTATION}

\indent The local laser phase that atoms are experiencing is:
\begin{equation}
\begin{split}
\Delta\Phi=\Delta_{prop}+\Delta_{laser}
\end{split}
\label{eq:}
\end{equation}
where we have for a HMZ interferometer:
\begin{subequations}
\begin{align}
\Delta_{prop}=&-\textbf{k}\left(\textbf{z}_{1'}-2\textbf{z}_{1}+\textbf{z}_{2}\right)\\
\Delta_{laser}=&-\varphi'_{1}+2\varphi_{1}-\varphi_{2}+\phi'_{1}-\phi_{2}\\
&-\textup{Arg}\left[\beta_{1}'^{3}(gg)\beta_{2}^{4}(gg)\right]
\end{align}
\label{eq:phase-shift_HMZ}
\end{subequations}
and for a HBU interferometer:
\begin{subequations}
\begin{align}
\Delta_{prop}=&-\textbf{k}\left(\textbf{z}_{1'}-2\textbf{z}_{0}+2\textbf{z}_{1}-\textbf{z}_{2}\right)\\
\Delta_{laser}=&-\varphi'_{1}-2\varphi_{1}+2\varphi_{0}+\varphi_{2}+\phi'_{1}+\phi_{2}\\
&-\textup{Arg}\left[\beta_{1}'^{3}(gg)\beta_{2}^{4}(gg)\right]
\end{align}
\label{eq:phase-shift_HBU}
\end{subequations}
The wave-packet trajectory of the mass center under acceleration and rotation is $\textbf{z}_{i}=\textbf{z}_{0}+\textbf{v}_{0}t_{i}+\frac{1}{2}\textbf{g}t_{i}^{2}-\left(\textbf{k}\times\textbf{g}\right)\cdot\boldsymbol{\Omega}~t_{i}^{3}/6$.
After calculation, we get the full HMZ phase-shift $\Delta\Phi_{HMZ}$ and the full HBU phase-shift $\Delta\Phi_{HBU}$ expressions:
\begin{subequations}
\begin{align}
\Delta\Phi_{HMZ}=&-\textbf{k}_{l}\cdot\textbf{g}\textup{T}^{2}-\left(\textbf{k}_{l}\times\textbf{g}\right)\cdot\boldsymbol{\Omega}\textup{T}^{3}\\
&-\varphi'_{1}+2\varphi_{1}-\varphi_{2}+\phi'_{1}-\phi_{2}\\
&-\textup{Arg}\left[\beta_{1}'^{3}(gg)\beta_{2}^{4}(gg)\right]\\
\Delta\Phi_{HBU}=&-2\left(\textbf{k}_{l}\times\textbf{g}\right)\cdot\boldsymbol{\Omega}\textup{T}^{3}\\
&-\varphi'_{1}-2\varphi_{1}+2\varphi_{0}+\varphi_{2}+\phi'_{1}+\phi_{2}\\
&-\textup{Arg}\left[\beta_{1}'^{3}(gg)\beta_{2}^{4}(gg)\right]
\end{align}
\label{eq:phase-shift_HMZ_HBU}
\end{subequations}


\begin{thebibliography}{1}
\bibitem{Footnote} By applying a direct transformation
to the last pulse $\widetilde{\vartheta}'_{p}$ of $_{p}^{q}C_{gg}$ as following:
\begin{equation}
\begin{split}
\cos\widetilde{\vartheta}'_{p}&\mapsto-i\sin\widetilde{\vartheta}'_{p}e^{-i(\phi'_{\textup{p}}+\varphi'_{p})}\\
\sin\widetilde{\vartheta}'_{p}&\mapsto i\cos\widetilde{\vartheta}'_{p}e^{-i(\phi'_{\textup{p}}+\varphi'_{p})}
\end{split}
\end{equation}
we directly get the complex component $_{p}^{q}C_{ge}$ of the composite wavefunction.
\end{thebibliography}
\end{document}